\documentclass[longauth]{aa} 

\usepackage{lscape} 

\usepackage{subcaption}
\usepackage{graphicx}
\usepackage{txfonts}
\usepackage[colorlinks=true,citecolor=blue]{hyperref}
\DeclareUnicodeCharacter{2212}{-}

\def\fxuv{$F_{\rm XUV}$}
\def\mobs{$\dot{m}_{\rm obs}$}
\def\mtheory{$\dot{m}_{\rm theory}$}
\def\vesc{$v_{\rm esc}$}

\def\hd73{HD\,73583}

\begin{document}

   \title{The MOPYS project: A survey of 70 planets in search of extended \ion{He}{i} and H atmospheres}

    \subtitle{No evidence of enhanced evaporation in young planets}

   \titlerunning{The MOPYS project}

   \author{
J.~Orell-Miquel\inst{\ref{ins:iac},\ref{ins:ull}} \and
F.~Murgas\inst{\ref{ins:iac},\ref{ins:ull}} \and
E.~Pall\'e\inst{\ref{ins:iac},\ref{ins:ull}} \and
M.~Mallorqu\'in\inst{\ref{ins:iac},\ref{ins:ull}} \and 
M.~L\'opez-Puertas\inst{\ref{ins:iaa}} \and
M.~Lamp\'on\inst{\ref{ins:iaa}} \and
J.~Sanz-Forcada\inst{\ref{ins:madrid}} \and 
L.~Nortmann\inst{\ref{ins:gottingen}} \and
S.~Czesla\inst{\ref{ins:tautenburg}} \and 
E.~Nagel\inst{\ref{ins:gottingen}} \and 
I.~Ribas\inst{\ref{ins:ICE},\ref{ins:IEEC}} \and 
M.~Stangret\inst{\ref{ins:inaf}} \and 
J.~Livingston\inst{\ref{ins:astrobiology_tokio},\ref{ins:NAOJ},\ref{ins:SOKENDAI}} \and 
E.~Knudstrup\inst{\ref{ins:Sweden},\ref{ins:Denmark}} \and 
S.\,H.~Albrecht\inst{\ref{ins:Denmark}} \and 
I.~Carleo\inst{\ref{ins:iac},\ref{ins:ull}} \and 
J.\,A.~Caballero\inst{\ref{ins:madrid}} \and 
F.~Dai\inst{\ref{ins:NASA_Sagan},\ref{ins:California},\ref{ins:Caltech}} \and 
E.~Esparza-Borges \inst{\ref{ins:iac},\ref{ins:ull}} \and
A.~Fukui\inst{\ref{ins:uni_tokio},\ref{ins:iac}} \and 
K.~Heng\inst{\ref{ins:LMU}} \and 
Th.~Henning\inst{\ref{ins:MPIA}} \and 
T.~Kagetani\inst{\ref{ins:dep_tokio}} \and 
F.~Lesjak\inst{\ref{ins:gottingen}} \and 
J. P.~de Leon\inst{\ref{ins:dep_tokio}} \and 
D.~Montes\inst{\ref{ins:complutense}} \and 
G.~Morello\inst{\ref{ins:Sweden},\ref{ins:iac}} \and 
N.~Narita\inst{\ref{ins:uni_tokio},\ref{ins:astrobiology_tokio},\ref{ins:iac}} \and 
A.~Quirrenbach\inst{\ref{ins:LSW}} \and 
P.\,J.~Amado\inst{\ref{ins:iaa}} \and 
A.~Reiners\inst{\ref{ins:gottingen}} \and 
A.~Schweitzer\inst{\ref{ins:Hamburg}} \and 
J.\,I.~Vico Linares\inst{\ref{ins:caha}} 
          }

   \institute{
        \label{ins:iac}Instituto de Astrofísica de Canarias (IAC), 38205 La Laguna, Tenerife, Spain\\
        \email{jom@iac.es}
        \and
        \label{ins:ull}Departamento de Astrofísica, Universidad de La Laguna (ULL), 38206 La Laguna, Tenerife, Spain
        \and
        \label{ins:iaa}Instituto de Astrof{\'i}sica de Andaluc{\'i}a (IAA-CSIC), Glorieta de la Astronom{\'i}a s/n, E-18008 Granada, Spain
        \and
        \label{ins:madrid}Centro de Astrobiolog\'ia (CSIC-INTA), Camino Bajo del Castillo s/n, Villanueva de la Ca\~nada, E-28692 Madrid, Spain
        \and
        \label{ins:gottingen}Institut f\"ur Astrophysik und Geophysik, Georg-August-Universit\"at, Friedrich-Hund-Platz 1, 37077 G\"ottingen, Germany
        \and
        \label{ins:tautenburg}Th\"uringer Landessternwarte Tautenburg, Sternwarte 5, 07778 Tautenburg, Germany
        \and
        \label{ins:ICE}Institut de Ci\`encies de l'Espai (ICE, CSIC), Campus UAB, Can Magrans s/n, 08193 Bellaterra, Barcelona, Spain
        \and
        \label{ins:IEEC}Institut d’Estudis Espacials de Catalunya (IEEC), 08034 Barcelona, Spain
        \and
        \label{ins:inaf}Osservatorio Astronomico di Padova, Vicolo dell’Osservatorio 5, 35122, Padova, Italy
        \and
        \label{ins:astrobiology_tokio}Astrobiology Center, 2-21-1 Osawa, Mitaka, Tokyo 181-8588, Japan
        \and
        \label{ins:NAOJ}National Astronomical Observatory of Japan, 2-21-1 Osawa, Mitaka, Tokyo 181-8588, Japan
        \and
        \label{ins:SOKENDAI}Astronomical Science Program, Graduate University for Advanced Studies, SOKENDAI, 2-21-1, Osawa, Mitaka, Tokyo, 181-8588, Japan
        \and
        \label{ins:Sweden}Department of Space, Earth and Environment, Chalmers University of Technology, 412 93, Gothenburg, Sweden
        \and
        \label{ins:Denmark}Stellar Astrophysics Centre, Department of Physics and Astronomy, Aarhus University, Ny Munkegade 120, DK-8000 Aarhus C, Denmark
        \and
        \label{ins:NASA_Sagan}NASA Sagan Fellow
        \and
        \label{ins:California}Division of Geological and Planetary Sciences, 1200 E California Blvd, Pasadena, CA, 91125, USA
        \and
        \label{ins:Caltech}Department of Astronomy, California Institute of Technology, Pasadena, CA 91125, USA
        \and
        \label{ins:uni_tokio}Komaba Institute for Science, The University of Tokyo, 3-8-1 Komaba, Meguro, Tokyo 153-8902, Japan
        \and
        \label{ins:LMU}Faculty of Physics, Ludwig Maximilian University, Scheinerstrasse 1, D-81679, Munich, Bavaria, Germany
        \and
        \label{ins:MPIA}Max-Planck-Institute f\"ur Astronomie, K\"onigstuhl 17, D-69117 Heidelberg, Germany
        \and
        \label{ins:dep_tokio}Department of Multi-Disciplinary Sciences, Graduate School of Arts and Sciences, The University of Tokyo, 3-8-1 Komaba, Meguro, Tokyo
        \and
        \label{ins:complutense}Departamento de F\'{i}sica de la Tierra y Astrof\'{i}sica  and IPARCOS-UCM (Instituto de F\'{i}sica de Part\'{i}culas y del Cosmos de la UCM),  Facultad de Ciencias F\'{i}sicas, Universidad Complutense de Madrid, E-28040, Madrid, Spain
        \and
        \label{ins:LSW}Landessternwarte, Zentrum f\"ur Astronomie der Universit\"at Heidelberg, K\"onigstuhl 12, D-69117 Heidelberg, Germany
        \and
        \label{ins:Hamburg}Hamburger Sternwarte, Gojenbergsweg 112, 21029 Hamburg, Germany
        \and
        \label{ins:caha}Centro Astron\'omico Hispano en Andaluc\'ia, Observatorio Astron\'omico de Calar Alto, Sierra de los Filabres, 04550 G\'ergal, Almer\'ia, Spain 
        }

   \date{Received 31 January 2024 / Accepted 30 May 2024}

 
  \abstract
  {
  During the first billion years of their life, exoplanet atmospheres are modified by different atmospheric escape phenomena that can strongly affect the shape and morphology of the exoplanet itself. These processes can be studied with Ly$\alpha$, H$\alpha$, and/or \ion{He}{i} triplet observations.
  We present high-resolution spectroscopy observations from CARMENES and GIARPS checking for \ion{He}{i} and H$\alpha$ signals in 20 exoplanetary atmospheres: V1298\,Tau\,c, K2-100\,b, HD\,63433\,b, HD\,63433\,c, HD\,73583\,b, HD\,73583\,c, K2-77\,b, TOI-2076\,b, TOI-2048\,b, HD\,235088\,b, TOI-1807\,b, TOI-1136\,d, TOI-1268\,b, TOI-1683\,b, TOI-2018\,b, MASCARA-2\,b, WASP-189\,b, TOI-2046\,b, TOI-1431\,b, and HAT-P-57\,b. We report two new high-resolution spectroscopy \ion{He}{i} detections for TOI-1268\,b and TOI-2018\,b, and a  H$\alpha$ detection for TOI-1136\,d. Furthermore, we detect hints of \ion{He}{i} for HD\,63433\,b, and H$\alpha$ for HD\,73583\,b and c, which need to be confirmed.
  The aim of the Measuring Out-flows in Planets orbiting Young Stars (MOPYS) project is to  understand the evaporating phenomena and test their predictions from the current observations.
  We compiled a list of 70 exoplanets with \ion{He}{i} and/or H$\alpha$ observations, from this work and the literature, and we considered the \ion{He}{i} and H$\alpha$ results as proxy for atmospheric escape.
  Our principal results are that 0.1--1Gyr  planets do not exhibit more \ion{He}{i} or H$\alpha$ detections than older planets, and evaporation signals are more frequent for planets orbiting $\sim$1--3\,Gyr  stars.
  We provide new constraints  to the cosmic shoreline, the empirical division between rocky planets and planets with atmosphere, by using the evaporation detections and we explore the capabilities of a new dimensionless parameter, $R_{\rm He}/R_{\rm Hill}$, to explain the \ion{He}{i} triplet detections. Furthermore, we present a statistically significant upper boundary for the \ion{He}{i} triplet detections in the $T_{\rm eq}$ versus $\rho_{\rm p}$ parameter space. Planets located above that boundary are unlikely to show \ion{He}{i} absorption signals.
  } 


   \maketitle
%

\section{Introduction}
\label{Sect: Introduction}

\textcolor{red}{\bf NOTE: This accepted version of the manuscript has a different numbering of figures, tables and appendices than the published version. Please refer to the published version for proper referencing.}\\

Along with their stars, planets evolve and change over time. During their early stages of formation, in particular, they undergo severe changes in their physical and orbital properties due to internal and external forces (\citealp{Baruteau2016_yplanets}). Exoplanets form embedded in the protoplanetary disc, from where they can accrete H and He, which are the major constituents of their gaseous envelopes (\citealp{Guenther2023}). Following the disc dispersion in $\sim$1--10\,Myr (\citealp{Haisch2001_disk_time, Hernandez2007_disk_timescale, Baruteau2016_yplanets}), different phenomena occur favouring the mass-loss processes in the planetary atmospheres:
i) the gas disc dissipates leaving the exoplanet and its H/He-rich atmosphere exposed to the direct radiation of the host star (\citealp{Fedele_2010, Barenfeld_2016, Dawson_Johnson_2018});
ii) stars are more active and have their highest levels of X-ray and extreme ultraviolet (XUV, 1--912\,\AA) energy irradiation until $\sim$\,100\,Myr (\citealp{Jackson2012_100Myr});
iii) exoplanets are undergoing contraction and cooling processes until $\sim$1\,Gyr (\citealp{Ginzburg2016_corepower, Ginzburg2018_corepower, Gupta2020_corepower});
iv) initial gas envelopes tend to be inflated by internal and external heat, resulting in more extended atmospheres (\citealp{JorgeSanz2011, Owen2016_boiloff}, and references therein); and
v) the XUV radiation supports the population of metastable \ion{He}{i} (\citealp{JorgeSanz2008}).
Therefore, at this stage of the system’s evolution the conditions are optimal for detecting the young primordial extended atmospheres.

Photo-evaporation processes are stronger when they are driven by X-ray radiation, which is more significant during the saturated luminosity state (from $\lesssim$100\,Myr until $\sim$1\,Gyr for late spectral types; \citealp{Jackson2012_100Myr}).
After this timescale, photo-evaporation is mainly driven, progressively, by the extreme ultraviolet (EUV, 100-912\,\AA) flux, and its effects on the exoplanet atmosphere are less significant (\citealp{JorgeSanz2011, Owen_Jackson_2012, Owen_Wu_2013}).
Furthermore, atmospheric escape can also be driven by the heat released from the cooling core of the planet. This mechanism, called core-powered evaporation, plays an important role, and this process can act approximately  on  gigayear  timescales (\citealp{Ginzburg2016_corepower, Ginzburg2018_corepower, Gupta2020_corepower}).

Both escape mechanisms have some overlap in time and may act simultaneously in shaping the final atmosphere of mature exoplanets.
In this context, the study of planets at the early stages of evolution is crucial for a better comprehension of the different processes, and to prove theory predictions, such as planet formation and migration, giant planets' gas accretion, or escape of the primary atmospheres of rocky-core planets with gas envelopes  (e.g. \citealp{Baruteau2016_yplanets, Owen_Wu_2017, Dawson_Johnson_2018}).
In particular, core-powered and photo-evaporation processes have been suggested to explain the formation of the radius gap in the small (1--4\,R$_{\oplus}$) planets radius distribution (e.g. \citealp{Fulton2017_radiusgap, Fulton2018_radiusgap, VanEylen2018_radius_gap, VanEylen2021_radiusgap_M}), although formation and migration mechanisms have also been proposed to explain the valley \citep{Rafa_2022Sci}.

Exoplanets can also experience atmospheric mass loss during their later lifetime because photo-evaporation and core-powered mechanisms can act at longer timescales than $\sim$0.1--1\,Gyr, but at  reduced total mass-loss rates (\citealp{JorgeSanz2011,Owen_Jackson_2012}).
According to stellar and planetary evolution models and assuming that high mass-loss rates are easier to detect, the number of evaporation detections should be higher in the $<$100Myr period, where the X-ray photo-evaporation and core-powered mechanisms act together (\citealp{JorgeSanz2011, Owen_Wu_2013, Ginzburg2016_corepower}). After $\sim$1\,Gyr, the number of detections should decrease as the core-powered atmospheric escape is weakening during the first few gigayears  (\citealp{Ginzburg2016_corepower, Ginzburg2018_corepower}). As a rough approximation, we should expect a plateau of atmospheric mass-loss detections after the approximately gigayear timescale.
At the time photo-evaporation is mainly driven by the EUV flux (\citealp{JorgeSanz2011}), the planet properties are already fixed. For gas giants and Jupiter-mass planets, the impact of this late-evaporation on the planet mass is close to null (\citealp{Owen_Wu_2013}), while for low-mass planets only small changes in their masses are expected (\citealp{Lopez_2012, Owen_Jackson_2012, Owen_Wu_2013}). Therefore, the detection of hydrodynamic escape in old planets ($\gtrsim$1\,Gyr)  may not imply a significant change in their total mass and it may have no evolutionary consequences. Hydrodynamic escape by Roche lobe overflow can be considered  an exception (see e.g. \citealp{Koskinen2022_Roche}). The middle atmosphere of extremely close-in planets (with orbital periods of $\lesssim$1\,d) extends to the Roche lobe due to the stellar gravitational tide, producing a very strong atmospheric mass-loss rate. The Roche lobe mechanism can be assumed stellar-age independent, and may not introduce a detectable change in the number of detections across stellar age.

The study of atmospheric escape in exoplanet atmospheres is mainly performed with the observations of three evaporation tracers: the \ion{H}{} Lyman-$\alpha$ line (Ly$\alpha$) at 1216\,$\AA$ at   ultraviolet (UV) wavelengths, the \ion{H}{} Balmer-$\alpha$ line (H$\alpha$) at 6564\,$\AA$ at optical wavelengths, and the \ion{He}{i} triplet at 10833\,$\AA$ at near-infrared (NIR) wavelengths. All the wavelengths in this work are referenced in a vacuum.

The Ly$\alpha$ was the first line used for probing evaporating atmospheres, when \citet{VidalMadjar2003_Lya} detected its absorption in the upper atmosphere of HD\,209458\,b.
However, the Ly$\alpha$ line has some limitations. First,  UV observations cannot be carried out from ground-based facilities, relegating its study to space observations with the \textit{Hubble Space Telescope} (HST) instrument  Space Telescope Imaging Spectrograph (STIS).
Second, Ly$\alpha$ is strongly affected by the interstellar medium extinction, limiting its observation to only the closest stars, due to their large relative motions.
Finally,  the \ion{H}{} in the upper layers of Earth's atmosphere (geocorona) have Ly$\alpha$ emission.
In practice, the core of the Ly$\alpha$ line cannot be observed, and only the broad wings of the line are accessible. This has led to a very limited number of planetary atmospheres being accessible to Ly$\alpha$ observations.

The H$\alpha$ line was used by \citet{KELT-9_Ha} to prove the evaporation of KELT-9\,b's atmosphere. The core and wings of this optical line can be observed from space, but also from ground-based facilities, allowing  line profile characterisation when high-resolution spectrographs are used. However, the H$\alpha$ line has some caveats as well:
relatively low mass-loss rates can produce large Ly$\alpha$ absorption signals, while remaining undetected in H$\alpha$ (e.g. GJ\,436\,b, \citealp{GJ436b_Lya,Ehrenreich2015_GJ436b_Lya,GJ436b_Ha});
stellar lines in the visible are more sensitive to stellar activity than those in the NIR (e.g. \ion{H}{}Paschen lines and \ion{He}{i} triplet), making their analysis very challenging in the presence of such activity (\citealp{Fuhrmeister2020_He_variability, AU_Mic_Enric, Howard2023_He_flares}).

The \ion{He}{i} triplet was proposed by \citet{Seager2000_He_inici}, and later modelled by \citet{Oklopcic2018_He_inici}, as an alternative to search for evidence of atmospheric escape in the NIR.
The first detections came almost simultaneously from ground- and space-based observations (\citealp{Spake2018_WASP-107b_HST, Nortmann_WASP-69_He, HAT-P-11b_He_Allart}). Moreover, ground-based high-resolution spectroscopy observations allowed   several physical parameters to be retrieved from line profile fitting (\citealp{Lampon_2020_HD209, Lampon_2021_HD187_GJ3470, Lampon_2021_regimenes, Lampon_2023_varios_planetas}). Since stellar lines in the NIR are less sensitive to stellar activity (but not exempted from it; e.g. \citealp{Spake2018_WASP-107b_HST, HD189733b_He}) than the optical lines, the \ion{He}{i} triplet is a good tracer to explore young exoplanet atmospheres.

Unfortunately, the \ion{He}{i} triplet has some disadvantages too. First, there are telluric absorption and emission lines surrounding the NIR triplet wavelengths that can overlap with the planet signal depending on the observing epoch.
Second,  the \ion{He}{i} ionisation wavelength cutoff ($\lambda$\,$\leq$\,504\,$\AA$) is lower than that of \ion{H}{i} ($\lambda$\,$\leq$\,912\,$\AA$). Finally,  the lifetime of the helium metastable state is $\sim$2.2\,h \citep{Drake1971_He_lifetime}, and a strong and constant XUV irradiation is needed to maintain the helium metastable state population detectable (\citealp{JorgeSanz2008}).

This work presents the first results of the Measuring Out-flows in Planets orbiting Young Stars (MOPYS) survey. The MOPYS project's aim is to provide observational constraints to the timescales of atmospheric evolution and mass loss processes, and test the predictions of planetary evolution models by comparing them to the evaporation tracers observations. In particular, we are interested in testing the mass-loss processes in planet formation at different timescales.
Assuming that young planets undergo higher mass-loss rates than older ones, and higher mass-loss rates are, in general, easier to detect, we search for signs of evaporation using the \ion{He}{i} triplet and/or H$\alpha$ line observations as proxy, focusing on transmission spectroscopy observations of $\lesssim$1\,Gyr  exoplanets.
Moreover, we study the detection rate between young and old planet populations in our sample, and compare the observed \ion{He}{i} triplet signals with predicted photo-evaporation mass-loss rates. We scheduled the observations such that telluric contamination of the \ion{He}{i} triplet is minimised, taking advantage of a favourable barycentric velocity of the Earth. The majority of the high-resolution spectrographs that we used allowed us to observe the \ion{He}{i} triplet and H$\alpha$ simultaneously: CARMENES (\citealp{Quirrenbach_2014, Quirrenbach_2020}), HARPS-N+GIANO-B (GIARPS; \citealp{GIARPS}), and HARPS+NIRPS (\citealp{HARPS_Mayor2003, NIRPS}). Although H$\alpha$ might be more affected by the probable stellar activity, the simultaneous observation of the two lines increases the capability to detect ongoing evaporation. For consistency, we complemented the database of \ion{He}{i} triplet and H$\alpha$ observations with some Ly$\alpha$ observations from the literature when possible, although this work does not focus on this UV line.

This manuscript is organised as follows. We describe the observational datasets and the spectroscopic analysis methodology in Section\,\ref{Sect: Observations and analysis}. The atmospheric results for each planet are detailed in Section\,\ref{sect: Results}, while Section\,\ref{sec:literature} introduces further H and/or \ion{He}{i} atmospheric results from the literature. In Sect.\,\ref{Sect: det non up criteria} we detail our criteria to consider atmospheric detections. In Section\,\ref{Sect: Discussion} we discuss our main findings regarding atmospheric evaporation and the \ion{He}{i} triplet. The conclusions of this work can be found in Section\,\ref{sect: Conclusions}.

\section{Observations and data analysis}
\label{Sect: Observations and analysis}

\subsection{Spectroscopic observations}

\begin{table*}
\caption[width=\textwidth]{ 
\label{table - log observations} Observing log of the transits analysed in this work. Columns from left to right: name of the planet, spectrograph used, date of starting night of the observations, starting and ending time of observations, central time of the scheduled transit $T_{\rm c}$, 1$\sigma$ uncertainty of $T_{\rm c}$, median exposure time of the observations, number of spectra fully taken between the first ($T_1$) and fourth ($T_4$) contacts ($N_{\rm T_{14}}$) compared to the total number of spectra taken during the observations ($N_{\rm obs}$), median signal-to-noise ratio (S/N) value at the spectral orders of H$\alpha$ and/or \ion{He}{I} lines, respectively.
}
\centering
\resizebox{\textwidth}{!}{%

\begin{tabular}{l c c c c c c c c c c}

\hline \hline 
\noalign{\smallskip} 


Planet & Instrument & Date & Start & End & $T_{\rm c}$ & $\sigma_{\rm T_c}$ & $T_{\rm exp}$ & $N_{\rm T_{14}}$/$N_{\rm obs}$ & Median\\
  &   &  &  [UT] &   [UT] &  [UT] &   [min] & [min] &  & S/N\\

\noalign{\smallskip}
\hline
\noalign{\smallskip}

MASCARA-2\,b & CARMENES & 2017\,Aug\,23 & 21:09 & 02:36 & 23:53 & 0.5 & 3.3 & 50/70 & 101/86\\
\hline
\noalign{\smallskip}
K2-100\,b & CARMENES & 2020\,Jan\,07 & 21:21  & 01:18  & 23:18 & 1 & 3.3 & 23/60 & 28/28 \\
\hline
\noalign{\smallskip}
V1298 Tau\,c & CARMENES & 2020\,Jan\,05 & 18:27  & 01:08  & 20:21 & 16 & 20 & 12/20  & 97/104\\ 
\hline
\noalign{\smallskip}
TOI-1431\,b & GIANO-B & 2020\,May\,31 & 23:46  & 05:14  & 01:14 & 1 & 1.6 & 50/116  & 44 \\
\hline
\noalign{\smallskip}
TOI-2048\,b & CARMENES & 2021\,Jun\,07 & 21:11  & 03:04  & 23:27 & 8 & 10 & 18/32  & 32/39 \\
\hline
\noalign{\smallskip}
HD\,63433\,b & CARMENES & 2021\,Nov\,01 & 02:08  & 05:10  & 02:43 & 2 & 1 & 79/111  & 40/43 \\ 
& CARMENES & 2022\,Nov\,27 & 22:48  & 03:10  & 01:00 & 2.5 & 5 & 26/36  & 111/124 \\
\hline
\noalign{\smallskip}
HD\,63433\,c & CARMENES & 2023\,Feb\,18 & 21:40  & 03:00  & 00:23 & 4.4 & 5 & 41/54  & 142/139 \\
\hline
\noalign{\smallskip}
HD\,73583\,b & HARPS-N & 2022\,Jan\,12 & 23:44  & 03:33  & 01:31 & 1 & 10 & 12/23  & 86 \\ 
\hline
\noalign{\smallskip}
HD\,73583\,c & GIARPS & 2023\,Jan\,04 & 23:41  &  05:56  & 01:59 & 1 & 10 & 20/37  & 69/72 \\ 
\hline
\noalign{\smallskip}
HD\,235088\,b & CARMENES & 2022\,Aug\,06 & 21:00  & 01:18  & 23:07 & 0.8 & 5 & 28/44  & 72/86 \\
\hline
\noalign{\smallskip}
K2-77\,b & CARMENES & 2022\,Sep\,27 & 23:40  & 04:28  & 03:21 & 9 & 10 & 14/26  &  23/29  \\
\hline
\noalign{\smallskip}
TOI-2046\,b & GIARPS & 2022\,Sep\,29 & 19:51  & 01:05  & 21:39 & 9 & 10 & 13/30  & 22/21  \\
\hline
\noalign{\smallskip}
TOI-1807\,b & CARMENES &  2021\,Dec\,16 & 02:48  & 05:39  & 05:34 & 1 & 10 & 3/16  & 32/40  \\
 & CARMENES & 2022\,Dec\,23 & 02:05  & 05:20 & 03:45 & 2 & 6.6 & 8/25  & 61/77  \\
\hline
\noalign{\smallskip}
TOI-1136\,d & HARPS-N & 2021\,May\,14 & 21:29  &  03:55  & 00:36 & 7 & 15 & 15/26  & 43 \\ 
 & CARMENES & 2023\,Jan\,30 & 23:41  &  06:15  & 03:17 & 3 & 15 & 15/24  & 91/88 \\
\hline
\noalign{\smallskip}
TOI-1268\,b & GIANO-B & 2023\,Feb\,24 & 23:15  & 06:50  & 04:02 & 0.5 & 10 & 21/42 & 39 \\ 
\hline
\noalign{\smallskip}
TOI-2076\,b & CARMENES & 2022\,May\,11 & 21:55  & 03:29  & --\,$^{(a)}$ & -- & 5 & 0/54 & 72/85 \\
 & CARMENES & 2023\,Apr\,13 & 22:01  & 04:35  & 23:37 & 6 & 10 & 15/32 & 80/90 \\
\hline
\noalign{\smallskip}
TOI-1683\,b & CARMENES & 2022\,Sep\,11 & 00:33  & 01:15  & 01:40 & 3.7 & 6.6 & 3/6 & 26/33 \\
 & GIANO-B & 2023\,Sep\,19 & 01:05  & 05:55  & 02:09 & 3.8 & 10 & 8/24 & 39 \\
 & GIANO-B & 2023\,Sep\,22 & 00:57  & 05:43  & 03:32 & 3.8 & 10 & 8/26 & 41 \\
\hline
\noalign{\smallskip}
WASP-189\,b & GIANO-B & 2019\,May\,06 & 21:42 & 04:46 & 01:18 & 0.1 & 1.6 & 95/156 & 67 \\
\hline
\noalign{\smallskip}
HAT-P-57\,b & CARMENES & 2018\,Jul\,08 & 21:24  & 02:56  & 00:05 & 6 & 10 & 18/30 & /47 \\
\hline
\noalign{\smallskip}
TOI-2018\,b & GIARPS & 2022\,Apr\,09 & 23:34  & 06:00  & 00:12 & 3.3 & 5 & 19/67 & 29/53 \\
 & GIARPS & 2022\,Jun\,15 & 21:52  & 02:33  & 22:17 & 3 & 5 & 11/45 & 33/40 \\

\noalign{\smallskip} 
\hline \hline

\end{tabular}

}
\tablefoot{ $^{(a)}$ No planetary transit was observed. }
\end{table*}

In this work we analysed 16 transits observed with the Calar Alto high-Resolution search for M dwarfs with Exoearths with Near-infrared and optical \'Echelle Spectrographs (CARMENES, \citealp{Quirrenbach_2014, Quirrenbach_2020}) spectrograph located at the Calar Alto Observatory, Almer\'ia, Spain. CARMENES has two spectral channels: the optical channel (VIS), which covers the wavelength range from 0.52--0.96\,$\mu$m with a resolving power of $\mathcal{R}$\,=\,94\,600, and the near-infrared channel (NIR), which covers 0.96--1.71\,$\mu$m with a resolving power of $\mathcal{R}$\,=\,80\,400. The targets were observed with both channels simultaneously.

Fibre A was used to observe the targeting star, while fibre B was placed on blank sky in order to monitor the sky emission lines (fibres A and B are separated by 88\,arcsecs in the east-west direction). The observations were reduced using the CARMENES pipeline \texttt{caracal} \citep{Caballero2016}, and both fibres were extracted with the flat-optimised extraction algorithm (\citealp{FOX_extraction}).

We also analysed 11 transits observed with the High Accuracy Radial velocity Planet Searcher for the Northern hemisphere (HARPS-N, \citealp{HARPS-N}) and/or GIANO-B (\citealp{GIANO-B}) spectrographs mounted on the 3.6m \textit{Telescopio Nazionale Galileo} (TNG) at Roque de los Muchachos Observatory, La Palma, Spain. HARPS-N is an optical spectrograph which covers the wavelength range from 0.383--0.693\,$\mu$m with a resolving power of $\mathcal{R}$\,=\,115\,000. HARPS-N spectra were extracted using the standard Data Reduction Software (DRS) pipeline \citep{HARPS-N_DRS}. GIANO-B (\citealp{GIANO-B}) is a near-infrared spectrograph which covers the wavelength range from 0.95--2.45\,$\mu$m with a resolving power of $\mathcal{R}$\,$\simeq$\,50\,000. When possible, the observations were done in GIARPS mode (\citealp{GIARPS}), which allows the simultaneous use of HARPS-N and GIANO-B spectrographs.

GIANO-B observations were carried out with the nodding acquisition mode, where the object is observed at two different predefined positions on the slit (A and B) following an ABAB pattern (\citealp{GIARPS}). The nodding technique enable to monitor the sky with the slit position that is not on the object, and then efficiently subtract the thermal background and telluric emission lines. The GIANO-B spectra were wavelength calibrated, and extracted using the \texttt{GOFIO} pipeline (\citealp{GOFIO_GIANO}).

Table\,\ref{table - log observations} shows the observing log of the planetary transits analysed in this work, indicating the instrument used in each case. We computed the central time of transit ($T_{\rm c}$) and its uncertainty ($\sigma_{T_{\rm c}}$) with the \textit{Transit and Ephemeris Service} tool from NASA Exoplanet Archive\footnote{\url{https://exoplanetarchive.ipac.caltech.edu/cgi-bin/TransitView/nph-visibletbls?dataset=transits}} and the parameters from Table\,\ref{table - TRANSIT PARAMETERS}.
The core of the MOPYS observations were conducted as part of the 21B-3.5-004, 22A-3.5-007, 22B-3.5-009, 23A-3.5-009, and 23B-3.5-002 observing programmes (PI J.\,Orell-Miquel) with CARMENES, and CAT21B\_61, CAT22A\_9, CAT22B\_43, and CAT23A\_100 observing programmes (PI J.\,Orell-Miquel) at TNG. K2-100\,b, V1298\,Tau\,c, HD\,235088\,b, and TOI-2048\,b were observed using GTO time by the CARMENES consortium.

We processed the VIS and NIR CARMENES, and HARPS-N observations with \texttt{serval}\footnote{\url{https://github.com/mzechmeister/serval}} \citep{SERVAL}, which derives the relative radial velocities (RVs) and several activity indicators: the chromatic radial velocity index (CRX), the differential line width (dLW), and the H$\alpha$, \ion{Na}{I}\,D1 and D2 and \ion{Ca}{II}\,IRT line indices. Moreover, for the HARPS-N datasets we also used the YABI tool,\footnote{Available at \url{http://ia2-harps.oats.inaf.it:8000}.} an online version of the HARPS-N DRS pipeline, to derive absolute RVs and spectral activity indicators: cross-correlation function (CCF) full width at half maximum (FWHM), CCF constrast (CTR), bisector (BIS), and Mont-Wilson S-index.
For the targets only observed with GIANO-B (because the GIARPS mode was not possible), we inspected the \ion{H}{}\,Paschen lines in the NIR as stellar activity indicators. We constructed the stellar light curve of the \ion{H}{}\,Paschen\,$\beta$ (Pa-$\beta$, 12821.6\,\AA), \ion{H}{}\,Paschen\,$\gamma$ (Pa-$\gamma$, 10941.1\,\AA), and \ion{H}{}\,Paschen\,$\delta$ (Pa-$\delta$, 10052.1\,\AA) lines.
Abrupt and/or strong variations in the time evolution of the activity indices or lines may indicate stellar activity or flares during the transit, and could compromise or challenge the detection of planetary signals, in particular absorption lines in the visible part of the spectrum \citep{AU_Mic_Enric, Orell2023}.

RV measurements during a transit can be used to look for the Rossiter-McLaughlin (RM; \citealp{Rossiter_1924, McLaughlin_1924}) effect. During the crossing of a planet in front of its host star, the planet blocks different parts of the stellar disc. That produces an RV anomaly during the transit known as the RM effect. The detection of the RM effect in RV time series taken during a planetary transit enables the confirmation of the presence of the transiting planet and also helps to determine the orbital configuration and architecture of the planetary system. This technique has been successfully applied to young planets unveiling the architecture of its planetary systems, such as AU\,Mic\,b \citep{AU_Mic_Enric} or DS Tuc A\,b \citep{DS_Tuc_Ab_Benatti}.
Because we used the CARMENES instrumental configuration where fibre B points at the sky, there were no simultaneous Fabry-P\'erot calibrations during the observations. Without them, the RV time evolution is dominated by the instrument drift during the night.

\subsection{Photometric observations}
\label{Sect: Photommetric observations}

We refined the ephemerides and planetary properties of TOI-2048\,b, HD\,73583\,b \& c, K2-77\,b, TOI-1807\,b, TOI-1683\,b, TOI-2076\,b, TOI-2018\,b, and TOI-1268\,b using photometric data.

From the \textit{Transiting Exoplanet Survey Satellite} (TESS; \citealp{TESS_Ricker}), we analysed the 2-minute cadence TESS simple aperture photometry (SAP; \citealp{SAP}) using \texttt{juliet}\footnote{\url{https://juliet.readthedocs.io/en/latest/index.html}} (\citealp{juliet}). This \texttt{python} library is based on other public packages for transit light curve (\texttt{batman}, \citealp{batman}), and for Gaussian process (GP; \texttt{celerite}, \citealp{celerite}) modelling and uses nested sampling algorithms (\texttt{dynesty}, \citealp{dynesty}; \texttt{MultiNest}, \citealp{MultiNest, PyMultiNest}) to explore all the parameter space.
In the fitting procedure, we adopted a quadratic limb-darkening law with the ($q_1$,$q_2$) parameterisation introduced by \citet{Kipping2013}, and we considered the uninformative sample ($r_1$,$r_2$) parameterisation introduced in \citet{Espinoza2018} to explore the impact parameter of the orbit ($b$) and the planet-to-star radius ratio ($p$\,$=$\,${R}_{\mathrm{p}}/{R}_{\star}$) values. We modelled the photometric variability of the young host stars adding to fit the \texttt{celerite} GP exponential or \texttt{celerite} GP quasi-periodic kernels. We followed the same procedure applied to analyse the HD\,235088 system in \cite{Orell2023}.

One transit of TOI-1268\,b and TOI-2018\,b each were observed by the multi-colour imager MuSCAT2 (\citealp{Narita2019}) mounted at the 1.52-m Telescopio Carlos S\'anchez (TCS) in the Teide Observatory in Tenerife, Spain. MuSCAT2 obtained simultaneous photometric data of the transits in four bands (Sloan $g$, $r$, $i$, $z_s$). The exposure times for each band were optimised for each night, CCD and target to avoid the saturation of the target and comparison stars in the field. Standard data reduction, aperture photometry, and transit modelling including systematic effects was performed by the MuSCAT2 custom pipeline (\citealp{Parviainen2019}).

One partial transit of TOI-2076\,b was observed on the night of 14 May 2023 with the Sinistro imager mounted on one of the two 1-m telescopes (Dome B) operated by Las Cumbres Observatory (LCO; \citealp{2013PASP..125.1031B}) at the Teide Observatory in Tenerife, Spain. The observations were performed through the rp-band filter, in the full-frame mode, and with an exposure time of 10\,s. To avoid detector saturation, the telescope was defocused such that the FWHM of the stellar point spread function was 7"--8". The observation started at 22:00\,UT, about 1.6 hours before the expected ingress time, and halted at 01:03\,UT, in the middle of the expected transit, due to a technical problem. The obtained images were processed by the {\tt BANZAI} pipeline \citep{McCully:2018} for dark and flat corrections. The light curve of TOI-2076 was then extracted by aperture photometry using a custom pipeline \citep{2011PASJ...63..287F}, and the transit modelling was performed with \texttt{juliet}.

More details of each transit analysis can be found in their corresponding sections (TOI-1268\,b: Sect.\,\ref{Sect TOI-1268}, TOI-2018\,b: Sect.\,\ref{Sect TOI-2018}, and TOI-2076\,b: Sect.\,\ref{Sect TOI-2076}).

\subsection{Telluric correction}
\label{Sect: Telluric correction}

The main objective of these observations was the analysis of the \ion{He}{I} triplet at $\sim$10833\,\AA. However, the \ion{He}{I} lines are surrounded by emission and absorption telluric lines.
There is an H$_2$O absorption line at 10835\,$\AA$ and there are four emission lines of hydroxyl (OH) at 10832.1\,$\AA$, 10832.4\,$\AA$, 10834.2\,$\AA$, 10834.3\,$\AA$ (\citealp{Oliva_OH}), although the  last two are detected in the spectra as a single peak. Due to the Earth's orbital motion the relative positions of the planet and telluric spectral lines change with the epoch.
Thus, we only scheduled the observations during transiting epochs that minimise overlapping and mitigate the contamination over the \ion{He}{I} triplet (see \citealp{Orell2022, Spake_He_GJ1214} for further details).

When analysing the transits presented in this work, we encountered another source of contamination of the \ion{He}{I} triplet. We detected an emission line at $\sim$10833\,$\AA$ in few transits where we needed to extend the observations into the astronomical twilight. This emission line only appears in the last few spectra and its strength increases towards the end of the night. We presume it is scattered sunlight from the incoming dawn.

Because each spectrograph has its own particularities, we handled with the telluric emission and absorption contamination differently for each of the instruments, see next subsections.

\subsubsection{CARMENES telluric correction}
\label{Sect: CARMENES Telluric correction}

We corrected the CARMENES VIS and NIR spectra from telluric absorptions following the approach described in \citet{Evangelos_telurics} with the \texttt{molecfit} package in version 1.5.9 (\citealp{molecfit_1,molecfit_2}). Then, we corrected the \ion{He}{i} region of the CARMENES NIR data from telluric OH emission following the methodology described in previous \ion{He}{I} studies (e.g. \citealp{Nortmann_WASP-69_He, He_GJ3470b_Enric2020, HAT-P-32b_Ha_He_Czesla2022}), in particular \cite{Orell2022, Orell2023}.

We used fibre B to generate a synthetic emission model fitting simultaneously the three main OH peaks with three independent Gaussian profiles. The amplitude, central position, and standard deviation of the Gaussian profiles were set free, and we only introduced an initial guess for their central positions.
Prior to applying the emission model to fibre A, we accounted for the different efficiency of the two injection fibres. For each dataset, we computed the scaling factor between fibres comparing the strongest OH peak from the co-added spectra of each fibre. We assumed the scaling factor to be constant during the night. Finally, for each pair of target and sky spectra, we divided the science spectra by its particular OH emission model, multiplied by the nightly scaling factor.
When we detected the solar \ion{He}{I} emission line in fibre B, we added a fourth Gaussian profile in the fitting procedure.

\subsubsection{HARPS-N telluric correction}
\label{Sect: HARPSN Telluric correction}

We corrected the HARPS-N spectra from telluric absorptions with \texttt{molecfit} (v4.2) via its implementation in the \texttt{SLOPpy} (Spectral Lines Of Planets with python, \citealp{SLOPpy}) package. We only used \texttt{SLOPpy} for the purpose of running \texttt{molecfit} easily on HARPS-N data, and correct those spectra from telluric absorptions.

Although the wavelength solution from the original HARPS-N data is based on wavelengths in air, the \texttt{molecfit} correction provides a wavelength solution in the vacuum. Because CARMENES and GIANO-B wavelengths are in the vacuum, we also used the vacuum wavelengths for the HARPS-N data for consistency.


\subsubsection{GIANO-B telluric correction}
\label{Sect: GIANO Telluric correction}

Due to our planning of the observations, the H$_2$O absorption line is always far from the \ion{He}{i} triplet. Thus, we decided to not correct the GIANO-B spectra from telluric absorption. If a particular residual map or transmission spectrum shows strong telluric residuals, we simply masked that spectral region.

The sky emission lines are corrected by the \texttt{GOFIO} pipeline, taking advantage of the nodding technique. This procedure is very efficient in removing emission lines that have similar strength in consecutive exposures. However, the \ion{He}{I} emission increases too quickly between exposures for the ABAB nodding procedure to provide an accurate correction for the emission contamination. We inspected the individual spectra and masked the affected wavelength region of particular spectra.

\subsection{Transmission spectrum analysis}

\begin{table*}
\caption[width=\textwidth]{
\label{table - TRANSIT PARAMETERS}
Transit and system parameters used to compute the transmission spectra for each planet analysed in this work.
}
\centering
\resizebox{\textwidth}{!}{%

\begin{tabular}{l c c c c c c c c c}

\hline \hline 
\noalign{\smallskip} 

       & $P$ & $T_0$\,$^{(a)}$ & $T_{14}$\,$^{(b)}$ & $T_{12}$\,$^{(b)}$ & $a_{\rm p}$ & $i_{\rm p}$ & $\gamma$ & $K_{\star}$ & $K_{\rm p}$\,$^{(c)}$ \\
Planet & [d] & [d] & [h] & [min] & [au] & [deg] & [km\,s$^{-1}$] & [m\,s$^{-1}$] & [km\,s$^{-1}$] \\
\noalign{\smallskip}
\hline
\noalign{\smallskip}


K2-100\,b &  1.6739035\,(4)\,$^{(2)}$ &  140.71941\,(27)\,$^{(2)}$ & 1.60$\pm$0.01\,$^{(1)}$ & 2.8$\pm$0.8\,$^{(1)}$ & 0.0301\,(14)\,$^{(2)}$ & 81.27$\pm$0.37\,$^{(2)}$  & $+$34.393\,(3)\,$^{(2)}$ & 10.6$\pm$3.0\,$^{(2)}$ & 193$\pm$9 \\ \noalign{\smallskip}

MASCARA-2\,b &  3.4741070\,(19)\,$^{(1)}$ &  909.5906\,(3)\,$^{(2)}$ &  3.57$\pm$0.02\,$^{(1)}$ &  29$\pm$1\,$^{(1)}$ &  0.0542\,(21)\,$^{(1)}$ &  86.15$^{+0.28}_{-0.27}$\,$^{(1)}$ &  $-$21.07\,(3)\,$^{(2)}$ &  322.51\,$^{(3)}$ &  170$\pm$7\,$^{(3)}$\\ \noalign{\smallskip}

V1298 Tau\,c & 8.249071\,(58)\,$^{(1)}$ &  1854.3479\,(11)\,$^{(1)}$ & 4.66$\pm$0.12\,$^{(2)}$  & 12\,$^{(2)}$ & 0.0841\,(13)\,$^{(3)}$  &  88.5$^{+0.9}_{-0.7}$\,$^{(2)}$  & $+$14.64\,(14)\,$^{(4)}$  &  4$^{+5}_{-3}$\,$^{(3)}$ & 111$\pm$2 \\ \noalign{\smallskip}

TOI-1431\,b & 2.6502409\,(41)\,$^{(1)}$ &  1739.17728\,(11)\,$^{(1)}$ & 2.489$\pm$0.009\,$^{(1)}$ &  0.716$\pm$0.022\,$^{(1)}$ &  0.0465\,(17)\,$^{(1)}$ &   80.30$^{+0.18}_{-0.17}$\,$^{(1)}$ &  $-$25.154\,$^{(1)}$ & 294.1$\pm$1.1\,$^{(1)}$  & 188.15\,$^{(2)}$ \\ \noalign{\smallskip}

TOI-2048\,b & 13.790546\,(55)\,$^{(1)}$ &  1739.1123\,(27)\,$^{(1)}$ & 3.5$\pm$0.1\,$^{(1)}$ &  6.9$^{+1.2}_{-0.7}$\,$^{(1)}$ &  0.1078\,(80)\,$^{(1)}$ &   89.41$\pm$0.35\,$^{(1)}$ &  $-$7.6\,(2)\,$^{(2)}$ & $\sim$2.8\,$^{(1)}$  & 85$\pm$6 \\ \noalign{\smallskip}

HD\,63433\,b & 7.107789\,(10)\,$^{(2)}$ &  1916.45142\,(32)\,$^{(2)}$ &  3.22$\pm$0.03\,$^{(1)}$ &  9$\pm$1\,$^{(1)}$ &  0.0719\,(44)\,$^{(1)}$ & 89.4$^{+0.4}_{-0.6}$\,$^{(1)}$ & $-$15.81\,(10)\,$^{(1)}$  & 1.4$^{+1.4}_{-0.9}$\,$^{(3)}$  & 110$\pm$6 \\ \noalign{\smallskip}

HD\,63433\,c & 20.543888\,(46)\,$^{(2)}$ &  1844.05824\,(48)\,$^{(2)}$ &  4.07$\pm$0.03\,$^{(1)}$ &  8$\pm$1\,$^{(1)}$ &  0.1458\,(62)\,$^{(1)}$ & 89.15$^{+0.07}_{-0.20}$\,$^{(1)}$ & $-$15.81\,(10)\,$^{(1)}$  & 3.6$^{+1.1}_{-1.0}$\,$^{(3)}$  & 77$\pm$3 \\ \noalign{\smallskip}

HD\,73583\,b & 6.3980580\,(26)\,$^{(1)}$ &   2592.56287\,(25)\,$^{(1)}$ & 2.100$^{+0.015}_{-0.013}$\,$^{(1)}$  &  6.9$\pm$0.2\,$^{(1)}$ & 0.0618\,(20)\,$^{(1)}$  & 88.35$\pm$0.07\,$^{(1)}$  &  +21.52\,$^{(3)}$ &  4.37$^{+1.5}_{-1.3}$\,$^{(2)}$ & 105$\pm$3 \\ \noalign{\smallskip} 

HD\,73583\,c & 18.879300\,(48)\,$^{(1)}$ &   2949.58243\,(77)\,$^{(1)}$ & 3.60$^{+0.03}_{-0.05}$\,$^{(1)}$  &  6.9$\pm$0.20\,$^{(1)}$ & 0.1270\,(40)\,$^{(1)}$  & 89.96$\pm$0.03\,$^{(1)}$  &  +21.52\,$^{(3)}$ &  2.89$^{+0.53}_{-0.51}$\,$^{(2)}$ & 73.2$\pm$2.3 \\ \noalign{\smallskip} 

HD\,235088\,b &  7.4341394\,(60)\,$^{(1)}$ &  2798.4635\,(56)\,$^{(1)}$ & 2.700$\pm$0.025\,$^{(1)}$  & 4.4$\pm$0.5\,$^{(1)}$  & 0.0725\,(35)\,$^{(1)}$  & 88.85$\pm$0.30\,$^{(1)}$  & $-$27.370\,(2)\,$^{(2)}$  &  $\sim$2.5\,$^{(1)}$ & 106$\pm$5 \\ \noalign{\smallskip}

K2-77\,b & 8.200139\,(60)\,$^{(1)}$ &  2522.6338\,(35)\,$^{(1)}$ &  2.68$\pm$0.22\,$^{(1)}$ &  6.3$^{+1.5}_{-1.1}$\,$^{(1)}$  &  0.0751\,(45)\,$^{(1)}$ &  88.7$^{+0.7}_{-0.4}$\,$^{(1)}$ &  +7.35\,(20)\,$^{(2)}$ &  $\sim$3.4\,$^{(1)}$ &  100$\pm$6 \\ \noalign{\smallskip}

TOI-2046\,b & 1.4971842\,(6)\,$^{(1)}$ & 1792.2767\,(23)\,$^{(1)}$ &  2.410\,(32)\,$^{(1)}$ & 21$\pm$1\,$^{(2)}$ &  0.0267\,(18)\,$^{(2)}$ & 83.6$\pm$0.9\,$^{(1)}$ &  $−$9.480\,(43)\,$^{(1)}$ & 374.7$\pm$7.8\,$^{(1)}$  & 193$\pm$13 \\ \noalign{\smallskip}

TOI-1807\,b & 0.54937084\,(65)\,$^{(1)}$ & 2664.06930\,(75)\,$^{(1)}$ & 0.970$\pm$0.022\,$^{(1)}$ & 1.45$^{+0.25}_{-0.18}$\,$^{(1)}$ & 0.0121\,(9)\,$^{(1)}$ & 81.7$\pm$1.8\,$^{(1)}$ &  $-$7.33$\pm$0.59\,$^{(3)}$ & 2.39$^{+0.45}_{-0.46}$\,$^{(2)}$ &  235$\pm$6 \\ \noalign{\smallskip}

TOI-1136\,d & 12.51937\,(41)\,$^{(1)}$ & 2349.525\,(5)\,$^{(1)}$ & 4.12$\pm$0.11\,$^{(1)}$ &  11.0$\pm$0.5 \,$^{(1)}$ &  0.1057\,(46)\,$^{(1)}$ & 89.41$\pm$0.28\,$^{(1)}$ &  $+$6.91$\pm$0.33\,$^{(2)}$ & $\sim$2.2$\pm$0.6\,$^{(1)}$ &  92$\pm$4 \\ \noalign{\smallskip}

TOI-1268\,b & 8.1577094\,(45)\,$^{(1)}$ & 3000.66841\,(14)\,$^{(2)}$ & 4.001$\pm$0.025\,$^{(1)}$ &  21$\pm$2\,$^{(1)}$ &  0.0711\,(63)\,$^{(1)}$ & 88.63$^{+0.32}_{-0.30}$\,$^{(1)}$ &  $+$3.79$\pm$0.14\,$^{(3)}$ & 31.7$^{+2.5}_{-2.6}$\,$^{(1)}$ &  105$\pm$8 \\ \noalign{\smallskip}

TOI-2076\,b & 10.355183\,(65)\,$^{(1)}$ & 3079.5495\,(45)\,$^{(2)}$ & 3.251$\pm$0.03\,$^{(3)}$ &  8.5$\pm$0.2\,$^{(3)}$ &   0.0682\,(13)\,$^{(3)}$ & 88.9$\pm$0.11\,$^{(3)}$ &  $-$13.19$\pm$0.21\,$^{(4)}$ & $\sim$2\,$^{(3)}$ &  71.6$\pm$1.3 \\ \noalign{\smallskip}

TOI-1683\,b & 3.057541\,(14)\,$^{(1)}$ & 2522.7001\,(12)\,$^{(1)}$ & 1.43$^{+0.07}_{-0.05}$\,$^{(1)}$ &  5.0$^{+0.8}_{-0.6}$\,$^{(1)}$ &  0.0368\,(23)\,$^{(1)}$ & 86.80$\pm$0.38\,$^{(1)}$ &  $+$38.4$\pm$0.3\,$^{(2)}$ & $\sim$5\,$^{(3)}$ &  130$\pm$8 \\ \noalign{\smallskip}

WASP-189\,b &  2.7240330\,(42)\,$^{(1)}$ & 1926.5416960\,(65)\,$^{(1)}$ & 4.351$\pm$0.026\,$^{(2)}$ &  18.6$\pm$6.2\,$^{(2)}$ &   0.05053\,(98)\,$^{(1)}$ & 84.03$\pm$0.14\,$^{(1)}$ &  $-$24.465$\pm$0.012\,$^{(1)}$ & 182$\pm$13\,$^{(1)}$ &  200$\pm$4 \\ \noalign{\smallskip}

HAT-P-57\,b & 2.4652950\,(32)\,$^{(1)}$ & 5113.48127\,(48)\,$^{(a , 1)}$ & 3.499$\pm$0.019\,$^{(1)}$ &  19.1$\pm$0.8\,$^{(1)}$ &  0.0406\,(11)\,$^{(1)}$ & 88.26$\pm$0.85\,$^{(1)}$ &  $-$5.99$\pm$0.35\,$^{(1)}$ & $<$215\,$^{(1)}$ &  180$\pm$5 \\ \noalign{\smallskip}

TOI-2018\,b & 7.435583\,(22)\,$^{(1)}$ & 2746.4287\,(21)\,$^{(2)}$ & 2.36$\pm$0.09\,$^{(1)}$ &  6.5$\pm$0.5\,$^{(1)}$ &  0.0609\,(22)\,$^{(1)}$ & 88.52$\pm$0.22\,$^{(1)}$ &  $-$25.617$\pm$0.002\,$^{(3)}$ & 4.4$\pm$1.0\,$^{(1)}$ &  89.1$\pm$3.3 \\ \noalign{\smallskip}

\noalign{\smallskip} 
\hline \hline

\end{tabular}

}
\tablebib{
K2-100b: $^{(1)}$ \citet{K2-100_Stefansson_2018}, $^{(2)}$ \citet{K2-100_Barragan_2019}. 
MASCARA-2\,b: $^{(1)}$ \citet{MASCARA-2b_Lund} , $^{(2)}$ \citet{MASCARA-2b_Talens}, $^{(3)}$ \citet{MASCARA2_Nuria}. 
V1298 Tau\,c: $^{(1)}$ J.\,Livingston, priv. comm., $^{(2)}$ \citet{V1298Tau_paper}, $^{(3)}$ \citet{V1298Tau_c_Alejandro}, $^{(4)}$ \citet{V1298Tau_Gaidos}.
TOI-1431\,b: $^{(1)}$ \citet{TOI-1431_2021}, $^{(2)}$ \citet{TOI-1431_Monika}.
TOI-2048\,b: $^{(1)}$ This work (App.\,\ref{App: TOI-2048 juliet}), $^{(2)}$ \cite{TOI2048b_Newton}
HD\,63433\,b: $^{(1)}$ \citet{HD63433_Mann_discovery}, $^{(2)}$ \citet{HD63433_Zhang}, $^{(3)}$ \citet{Manu_HD63433} 
HD\,63433\,c: $^{(1)}$ \citet{HD63433_Mann_discovery}, $^{(2)}$ \citet{HD63433_Zhang}, $^{(3)}$ \citet{Manu_HD63433} 
HD\,73583\,b: $^{(1)}$ \cite{Barragan_TOI560b}, $^{(2)}$ \citet{ElMufti_TOI-560}.
HD\,73583\,c: $^{(1)}$ \cite{Barragan_TOI560b}, $^{(2)}$ \citet{ElMufti_TOI-560}.
HD\,235088\,b: $^{(1)}$ \cite{Orell2023}, $^{(2)}$ \textit{Gaia} DR2 (\citealp{GAIA_DR2}).
K2-77\,b: $^{(1)}$ This work (App.\,\ref{App: K2-77 juliet}), $^{(2)}$ \cite{K2-77_Gaidos}.
TOI-2046\,b: $^{(1)}$ \cite{TOI-2046b_Kabath}, $^{(2)}$ Computed using this table parameters.
TOI-1807\,b: $^{(1)}$ This work (App.\,\ref{App: TOI-1807 juliet}), $^{(2)}$ \cite{TOI-1807_Nardiello}, $^{(3)}$ \textit{Gaia} DR2 (\citealp{GAIA_DR2}).
TOI-1136\,d: $^{(1)}$ \cite{TOI-1136_system}, $^{(2)}$ \textit{Gaia} DR2 (\citealp{GAIA_DR2}).
TOI-1268\,b: $^{(1)}$ \cite{TOI-1268_Subjak2022}, $^{(2)}$ This work (Sect.\,\ref{Sect TOI-1268}), $^{(3)}$ \textit{Gaia} DR2 (\citealp{GAIA_DR2}).
TOI-2076\,b: $^{(1)}$ \cite{Zhang_young_planets}, $^{(2)}$ This work (Sect.\,\ref{Sect TOI-2076}), $^{(3)}$ \cite{TOI-2076_Osborn22}, $^{(4)}$ \textit{Gaia} DR2 (\citealp{GAIA_DR2}).
TOI-1683\,b: $^{(1)}$ This work (Sect.\,\ref{App: TOI-1683 juliet}) , $^{(2)}$ \textit{Gaia} DR2 (\citealp{GAIA_DR2}), $^{(3)}$ \cite{Zhang_young_planets}.
WASP-189\,b: $^{(1)}$ \citet{WASP-189b_Lendl} , $^{(2)}$ \citet{WASP-189b_Anderson}.
HAT-P-57\,b: $^{(1)}$ \citet{HAT-P-57_Hartmann}.
TOI-2018\,b: $^{(1)}$ \cite{TOI-2018_Dai23}, $^{(2)}$ This work (Sect.\,\ref{Sect TOI-2018}), $^{(3)}$ \textit{Gaia} DR2 (\citealp{GAIA_DR2}).
}
\tablefoot{
$^{(a)}$ $T_0$ given in BJD\,$-$\,2\,457\,000, except for HAT-P-57\,b that is BJD\,$-$\,2\,450\,000.
$^{(b)}$ $T_{14}$ is the total transit duration between the first ($T_1$) and fourth ($T_4$) contacts, and $T_{12}$ is the duration of the ingress--egress.
$^{(c)}$ Calculated from $K_{\rm p} = 2 \pi ~ a_{\rm p} ~ P^{-1} \sin{i_{\rm p}}$ using the parameters in this table, when there is no reference.

}
\end{table*}

We analysed the spectroscopic observations via the well-established transmission spectroscopy technique (e.g. \citealp{Wyttenbach_2015, Nuria_2017}) successfully applied in \cite{Orell2022, Orell2023}. The planetary and stellar parameters needed to compute the transmission spectra for each planet are shown in Table\,\ref{table - TRANSIT PARAMETERS}.

The telluric corrected spectra are normalised by a first degree polynomial. We used a blue and red region near the line of interest and free of tellurics to fit the polynomial. The spectra are shifted into the stellar rest frame, accounting for the Earth's barycentric movement, the stellar systemic velocity ($\gamma$), and the stellar Doppler shift induced by the planet.
Then, we created a high signal-to-noise ratio (S/N) stellar spectrum (Master-Out; MO) by calculating the mean of all the spectra taken completely during out-of-transit, and we divided all the spectra by the MO. With this step, the stellar spectrum is removed from the spectroscopic time series. Next, we moved the spectra to the planetary rest frame using the formula
\begin{equation}
v_{\rm p}(t) = K_{\rm p} \sin{(2 \pi \phi(t)),}
\label{eq: Planet rest frame}
\end{equation}
where $K_{\rm p}$ is the radial velocity semi-amplitude that the star induces to the planet, and $\phi$ is the orbital phase.
We neglected orbital eccentricity because the values provided for the planets in the literature are usually poorly constrained and the impact for all but one case is of the same order as the uncertainties derived for the signal shifts. The exception to this is TOI-1268 b for which we used the \texttt{radvel} (\citealp{radvel}) package to account for the eccentricity shift during the transit.

To compute the transmission spectrum (TS), we only consider those spectra taken entirely between the first ($T_1$) and fourth ($T_4$) contacts. We averaged them using the inverse of the squared propagated errors (1/$\sigma^2$) as weights. If the residual map contains strong telluric residuals and/or stellar line variability near the inspected line, we mask those affected regions to compute the TS. However, each night is different, and we explain further details of their particularies in the data analysis section (Sect.\,\ref{sect: Results}), where needed.

When the TS shows a clear absorption at the expected position, we fitted the feature with a Gaussian profile using the MultiNest algorithm (\citealp{MultiNest}) via its {\tt python} implementation \texttt{PyMultinest} \citep{PyMultiNest}.
Otherwise, when the residual map or the transmission spectrum do not show evidences of absorption from the planetary atmosphere, we derived conservative limits for the planetary absorption. We estimated a 3$\sigma$ upper limit absorption peak as three times the root-mean-squared (RMS) value of flat spectral region in the continuum of the spectrum close to the line(s) of interest. The exact spectral range might be different in each case to avoid including strong residuals or very scattered regions. The equivalent width (EW) of a line is a more appropriate measure because it is independent from the instrumental resolution, but it is difficult to compute when there is no line to measure. We give the upper limits in terms of absorption peak and we translate them into EWs (or from EW to absorption peak) as it is explained in App.\,\ref{App: He data base}.

To analyse the data from GIANO-B, we followed some of the recommendations from \cite{GIANO_Helium}, \cite{WASP-80_He_Fossati}, and \cite{GIANO_He_survey}. We considered the science spectra taken at positions A and B as different instruments, and we combined them after computing both TS. GIANO-B is known to be affected by a fringing pattern that affects high S/N spectra (\citealp{GIANO_Helium}).
However, the lower S/N of our spectra, due to its maximum exposure time of 600\,s and the faintness of our targets, prevented us to detect hints of the fringing pattern. The MO spectra obtained from GIANO-B did not show any clear sinusoidal pattern, as seen in \citet[Fig.\,2]{GIANO_Helium}. Thus, we decided to not correct the fringing pattern, as it was done in \citet{WASP-80_He_Fossati}. Lastly, the wavelength solution of GIANO-B may not be stable during the night. \cite{GIANO_Helium} found a typical instrument drift of half of the GIANO-B pixel element during four hours of observations. That could be an issue for the cross-correlation technique, where excellent precision in the wavelength solution is needed (\citealp{Brogi2018_GIANO_CC}). However, the transmission spectra technique does not require such extreme precision, and hence we did not apply any correction to our GIANO-B observations, which have lower S/N than in \cite{GIANO_Helium}.

Another aspects to take into account with transit observations are the Rossiter-McLaughlin (RM) and center-to-limb-variation (CLV) effects, which can interfere or mimic single line absorptions (\citealp{Nuria2021_HD209ESPRESSO}). Because these effects scale with $R_{\rm p}/R_{\star}$, they have larger impact on gas giant transmission observations. Previous \ion{He}{i} triplet line works proved that they have negligible contributions on the \ion{He}{i} results (e.g. \citealp{Nortmann_WASP-69_He, GIANO_He_survey}). Here, we computed the RM and CLV models for HD\,63433\,b ($R_{\rm p}$\,=\,2.1\,$R_{\oplus}$) and TOI-1136\,d ($R_{\rm p}$\,=\,4.6\,$R_{\oplus}$) whose contribution to the H$\alpha$ and \ion{He}{i} triplet signals is negligible. These analyses and results are presented in Sect.\,\ref{sect: Results}.
The models of RM and CLV effects were created with the \texttt{Turbospectum2019} \citep{Turbospectrum2019_Plez}, using the Kurucz ATLAS9 \citep{ATLAS_2003} and VALD3 \citep{VALD3_Ryabchikova} line lists. The stellar models were created for 21 limb-darkening angles assuming local thermodynamical equilibrium (LTE). In the next step, taking into account the system parameters, we calculated the stellar models for orbital phases from $-$0.05 to 0.05, remembering that the planet covers different stellar regions with different limb-darkening during its transit. Next, all models were divided by the out-of-transit spectrum, creating the RM and CLV effect model for each orbital phase and wavelength. 
However, it is necessary to first detect the RM in the RV time series to compute the models, and the effect is not detected for all our targets.

\section{Transmission spectroscopy results}
\label{sect: Results}

\begin{figure*}[h!]
    \centering
    \includegraphics[width=\hsize]{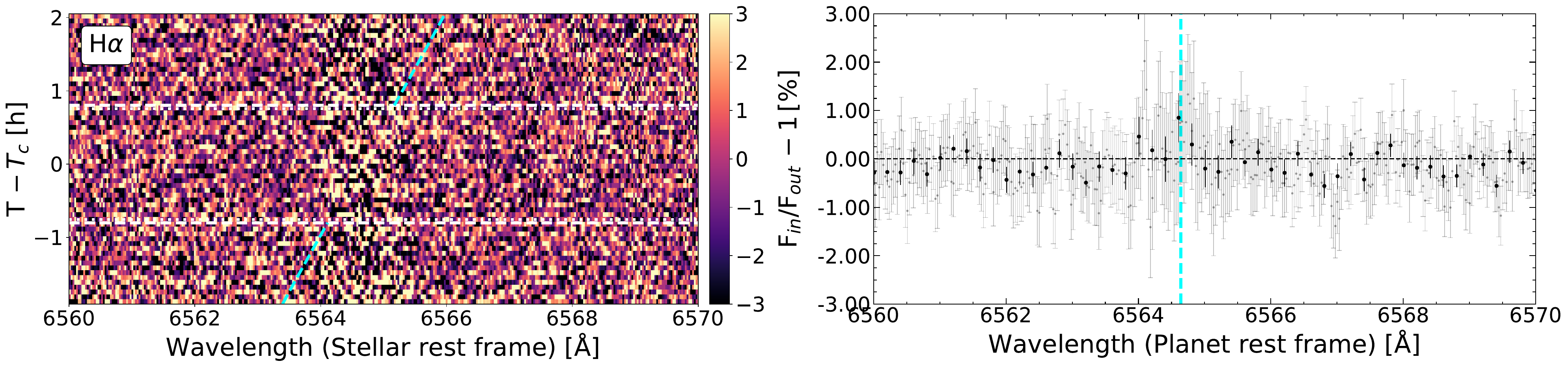}
    \includegraphics[width=\hsize]{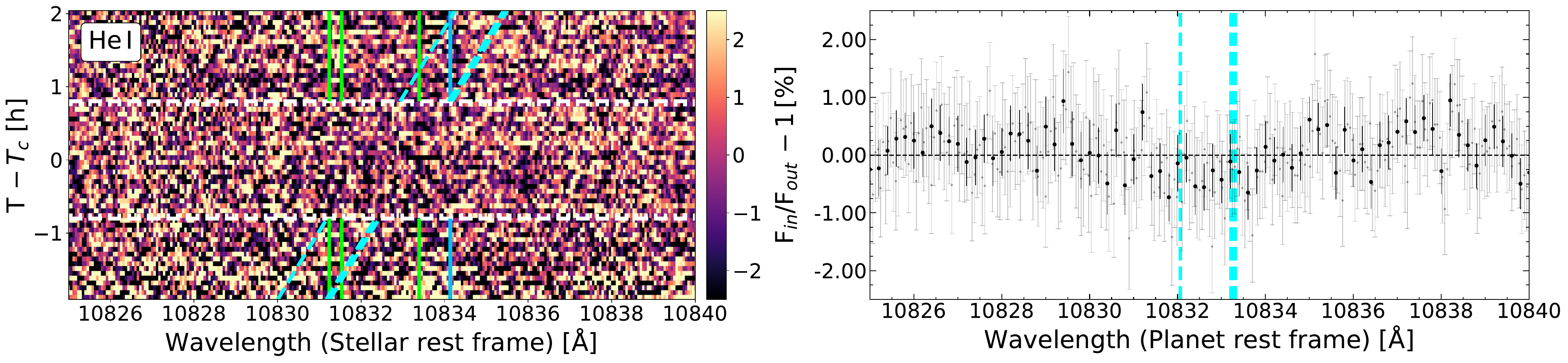}
    \caption{\label{Fig: TS K2-100}
    Residuals maps and transmission spectra around the H$\alpha$ line (\textit{top panels}) and \ion{He}{I} NIR triplet (\textit{bottom panels}) lines for K2-100\,b observations with CARMENES.
    \textit{Left panels}: Residual maps in the stellar rest frame. Time since mid-transit time ($T_{\rm c}$) is shown on the vertical axis, wavelength is on the horizontal axis, and relative absorption is colour-coded. The dashed and dotted white horizontal lines indicate the different contacts during the transit. The dashed cyan tilted lines indicate the predicted trace of the planetary signals. The solid green vertical lines indicate the position of the OH emission telluric lines. The solid blue line indicates the position of the H$_2$O absorption telluric line.
    \textit{Right panels}: Planet transmission spectra (TS) in the planet rest frame. We show the original data in light grey and the data binned by 0.2\,\AA\ in black. When an absorption signal is fitted, a red line and shaded region show the best Gaussian fit model with its $1\sigma$ uncertainties. The dotted cyan vertical lines indicate the H$\alpha$ (\textit{top}) and the \ion{He}{I} triplet (\textit{bottom}) lines positions. All the wavelengths in this figure are referenced in a vacuum.
    }
\end{figure*}

In this section we present the results and the details of the residuals maps and TS calculations planet by planet. We present an example of a residuals map and TS figure in Figure\,\ref{Fig: TS K2-100}, and the figures for the rest of transits without detections are shown in Appendix\,\ref{App: Additional figures}. Furthermore, the time evolution of the activity indicators derived from the analysed transits can be found in Appendix\,\ref{App: Additional figures} as well.

\subsection{K2-100\,b}

The resulting residual maps and TS centred in the spectral regions of the H$\alpha$ and \ion{He}{i} triplet are shown in Figure\,\ref{Fig: TS K2-100}. The time evolution of all the activity indices is mainly flat, and there is no evidence of strong stellar variability during the transit.

The \ion{He}{I} triplet TS shows a flat spectrum, while the H$\alpha$ TS is flat with some emission features from stellar variations in the line core during the transit. We found no significant absorption in either of the lines, and we placed a 3$\sigma$ upper limit to the excess absorptions of 1.4\,\% and 1.3\,\% for H$\alpha$ and \ion{He}{I}, respectively.

K2-100\,b was already observed with the InfraRed Doppler (IRD) spectrograph on the 8.2-m Subaru telescope ($\mathcal{R}$\,=\,70\,000; \citealp{IRD_spectrograph}). \citet{K2-100b_Gaidos_He} derived an EW 99\% confidence detection upper limit of 5.7\,m$\AA$. From Fig.\,8 therein, we estimated an absorption depth of $\sim$1.2\%, which is consistent with the upper limit derived from the CARMENES observations.

\subsection{MASCARA-2\,b}

\cite{MASCARA2_Nuria} already detected H$\alpha$ in MASCARA-2\,b using CARMENES and HARPS-N observations. The H$\alpha$ absorption measured with CARMENES VIS data was $-$0.85$\pm$0.03\,\%, and while combining 3 nights with HARPS-N was $-$0.68$\pm$0.06\,\%.

Here, we inspected the CARMENES NIR spectra, which were not analysed in \cite{MASCARA2_Nuria}, looking for \ion{He}{I} excess absorption. Figure\,\ref{Fig: TS MASCARA-2} presents the residual map and TS around the \ion{He}{I} triplet.
Because some telluric OH variability was still remaining in the residual map, we masked the affected regions. 
Despite the structure in the final TS, due to the masked regions, we found no significant planetary absorption. We computed a $3\sigma$ upper limit to the \ion{He}{I} excess absorption of 0.5\,$\%$.

\subsection{V1298\,Tau\,c}

V1298\,Tau planetary system is known to have transit time variations (TTVs) that can complicate the analysis of high-resolution spectroscopy observations (\citealp{V1298Tau_paper}). According to non-linear ephemerids for the V1298\,Tau system (J.\,Livingston, priv. comm.), a nearly complete transit of planet c was observed the night of 5 of January 2020. Unfortunately, we missed pre-transit observations which are critical to compute a good MO and to check for stellar variability during the observations.

For this transit, the \texttt{serval} H$\alpha$ index shows a clear decrease only from about mid-transit to egress, and an increase at the end of the observations. We did not include those last spectra in the computation of the MO spectrum, although they are plotted in the residual maps. The other \texttt{serval} products do not show signs of strong stellar activity (see Fig.\,\ref{Fig: serval 1}).

The H$\alpha$ residual map (Fig.\,\ref{Fig: TS V1298Tau c} left) shows a clear absorption region coincident with the \texttt{serval} H$\alpha$ index decrease, suggesting the stellar origin of the signal. We computed the TS only with the non-affected spectra from the first half of the transit. We fitted a $-$1.10\,$\pm$\,0.13\,\% feature at the H$\alpha$ position, which is consistent with the 3$\sigma$ upper limit derived from the nearby spectral region (1.1\,\%). The priors and posterior values from the fit are listed in Table\,\ref{table - V1298Tau_c Ha priors and posteriors}, and the posterior distributions are shown in Fig.\,\ref{Fig: V1298Tau_c nested Halpha Helium}.

The first half of the transit in the \ion{He}{I} NIR triplet residual map (Fig.\,\ref{Fig: TS V1298Tau c} left) shows an extended absorption region, which does not coincide in time with the H$\alpha$ variability detected. In the second half of the transit, there is a vertical absorption feature in the stellar rest frame that does not follow the planetary track when we shift the spectra into the planet rest frame. This feature seems to be related to telluric residuals. We computed the TS with those unaffected spectra from the first half of the transit. The \ion{He}{I} TS has a deep ($-$3.75\,$\pm$\,0.12\,\%) and broad (1.02\,$\pm$\,0.04\,\AA) absorption feature, which is larger than the 3$\sigma$ upper limit (1.1\,\%). The priors and posterior values from the fit are listed in Table\,\ref{table - V1298Tau_c He priors and posteriors}, and the posterior distributions are shown in Fig.\,\ref{Fig: V1298Tau_c nested Halpha Helium}.

To explore the origin of the H$\alpha$ and \ion{He}{i} features, we computed the TS from the unaffected spectra (start phase), the centre of the transit (centre phase), the end of the transit (end phase), and post-transit. We compare those TS with the MO spectra in Fig.\,\ref{Fig: V1298Tau_line_profile}. The start TS from H$\alpha$ and \ion{He}{i} lines show a different behaviour from centre and end. Furthermore, the comparison between TS and MO evidences the stellar line profile variation during the observations. Therefore, we cannot confidently attribute the detected features to V1298\,Tau\,c atmosphere. Moreover, the H$\alpha$ TS in Fig.\,\ref{Fig: V1298Tau_line_profile} show a tiny bump at $\sim$6563\,$\AA$ which can be attributed to the star and its youth.
The surprisingly large $\Delta v$ ($-$28.7\,$\pm$\,1.1\,km\,s$^{-1}$) is another reason for not claiming the significant \ion{He}{i} signal as a detection.

The H$\alpha$ line of V1298\,Tau\,c was previously explored by \citet{V1298Tau_c_Feinstein} and \citet{V1298Tau_c_Halpha}. \citet{V1298Tau_c_Feinstein} studied with the Gemini/GRACES spectrograph the behaviour of the H$\alpha$ and \ion{Ca}{II} infrared triplet during a transit, finding significant variations of the H$\alpha$ during the observations. \citet{V1298Tau_c_Halpha} analysed one transit looking for H$\alpha$ absorption with the PEPSI spectrograph, but the observations were affected by stellar variability. Furthermore, \citet{V1298Tau_He_Vissapragada2021} explored the \ion{He}{I} triplet of planet c with the Habitable-zone Planet Finder (HPF) spectrograph, but the in-transit spectra were affected by a stellar flare. Lastly, a transit of planet c was observed with narrow-band helium filter photometry, but \citet{V1298Tau_He_Vissapragada2021} did not detect the transit during the observations.

Unfortunately, our observations join the previous unsuccessful attempts to analyse the atmosphere of V1298\,Tau\,c. Due to the many complications that affected the observations, we adopted the absorption peaks from the H$\alpha$ and \ion{He}{I} features as upper limits of any possible planetary absorption. Our upper limits to H$\alpha$ and \ion{He}{I} excess absorption are 1.1\,\% and 3.7\,\%, respectively.

\subsection{TOI-1431\,b}

\citet{TOI-1431_Monika} analysed two HARPS-N and one EXPRES transits, finding no signs of atomic or molecular absorption. A tentative detection of H$\alpha$ appeared only during the first HARPS-N dataset (0.33\,$\pm$\,0.07\,\%), but it was considered as stellar variability because it was not reproduced in the other two visits. We consider that absorption feature as an upper limit for the purpose of this work.

Here, we looked for \ion{He}{I} excess absorption in the GIANO-B spectra from the first night of \citet{TOI-1431_Monika}. The residual map and TS are shown in Fig.\,\ref{Fig: TS TOI-1431}. The TS is mainly flat and we placed a 3$\sigma$ upper limit to the \ion{He}{I} excess absorption of 0.4\,\%.

\subsection{TOI-2048\,b}

\cite{TOI2048b_Newton} noted that ground-based follow-up observations found a tentative transit detection $\sim$20\,min later than expected. We analysed the new available TESS data to derive updated ephemerides. The details of the photometric fit are explained in Appendix\,\ref{App: TOI-2048 juliet}. We refined TOI-2048\,b ephemerides and planet properties (Table\,\ref{table - TOI-2048 juliet priors and posteriors}), and confirmed the detected delay from ground-based observations. Furthermore, because TOI-2048\,b does not have mass measurements, we estimated its probable mass from its radius (\citealp{Wolfgang_2016}) to predict its semi-amplitude $K_{\star}$. This parameter is required for a proper measurement of the TS. We forecasted a mass of 9.4\,$\pm$\,1.0\,M$_\oplus$ and a $K_{\star}$ of 2.8\,$\pm$\,0.3\,m\,s$^{-1}$. The details of all these calculations are explained in App.\,\ref{App: TOI-2048 juliet} as well.

The time evolution of all the activity indicators is mainly flat except for the last four H$\alpha$ index points. We excluded those spectra from the MO combination, although they are plotted in the residual maps. Furthermore, we masked a region near the \ion{He}{I} triplet because some residuals still remained after correcting the OH emission. We obtained similar results with or without applying the OH emission correction.

H$\alpha$ shows a narrow absorption feature (see Fig.\,\ref{Fig: TS TOI-2048}), but we consider it too narrow to have a planetary origin (it is only one binned point). Moreover, there are emission signals around the absorption signal in the residual map, and the absorption peak is of the order of the 3$\sigma$ upper limit of 1.5\,\%, and is difficult to distinguish from the continuum in the final TS. We found no significant absorption features around \ion{He}{I} triplet (Fig.\,\ref{Fig: TS TOI-2048}). We placed a 3$\sigma$ upper limit to the \ion{He}{I} excess absorption of $\sim$1\,\%.

\subsection{HD\,63433\,b}
\label{Sect HD63433b}

The night on 1 November 2021 was cloudy until $\sim$01:40\,UT, and we could only cover the transit partially, without pre-transit information. The observations are a bit noisy because the exposures were set to 1\,minute and we missed pre-transit observations due to bad weather conditions.
Then, we observed a complete transit one year later, on 27 November 2022. The time evolution of all the activity indicators is mainly flat, without evidence of stellar activity during the observations in both nights.
Although the RM effect was detected by \cite{HD63433_Mann_discovery}, the RM and CLV contribution to the H$\alpha$ and \ion{He}{i} triplet transmission spectra is well within the error bars of the TS (see Fig.\,\ref{Fig: HD63433b RM}), so we decided to not correct the data from these marginal effects.

The H$\alpha$ transmission spectrum from the first partial transit displays an absorption feature of $-$1.21$^{+0.22}_{-0.24}$\,\% (Fig.\,\ref{Fig: TS HD63433 B}). Table\,\ref{table - HD63433b priors and posteriors} shows the priors and posterior values from the fit and Fig.\,\ref{Fig: HD63433b nested Halpha} shows the posterior distribution. However, we did not recover that H$\alpha$ absorption in our second visit, which covered the entire transit. We consider the H$\alpha$ feature does not have a planetary origin. Due to the different quality around the H$\alpha$ from both nights, we did not combine the two datasets. The residual map and TS for the H$\alpha$ from the second transit is shown in Fig.\,\ref{Fig: TS HD63433 B}. We placed a 3$\sigma$ upper limit to the H$\alpha$ excess absorption of 0.4\,\%.

The residual maps and TS for the \ion{He}{I} triplet from both nights separately are shown in Fig.\,\ref{Fig: TS HD63433 B}. The \ion{He}{I} triplet was completely uncontaminated from tellurics, and both transmission spectra show some residuals from the stellar \ion{Si}{I} line at 10830\,$\AA$. Due to the different quality of the nights, we did not combine them and focused on the second event which is less noisy. The TS presents a $\sim$0.25\,\% excess absorption feature just at the expected position, but there is noise structure of similar amplitude. Finally, we adopted a conservative 3$\sigma$ upper limit of 0.4\,\%.

Our results on the \ion{He}{I} triplet are consistent with the previous upper limit of $\sim$0.5\,\% reported by \cite{HD63433_Zhang}. Their partial transit also showed a $\sim$0.2\% excess absorption, but they found a correlation between the stellar \ion{He}{I} and \ion{H}{}\,Paschen\,$\gamma$ lines. 
We also compared the \ion{H}{}\,Paschen lines with the \ion{He}{I} line, but we could not explain the \ion{He}{I} triplet feature as stellar line variability.
Further high-S/N observations of HD\,63433\,b are required to confirm the interesting feature of $\sim$0.2\% around the \ion{He}{I} triplet found in this work and \cite{HD63433_Zhang}.

Although the study of Ly$\alpha$ observations is not the focus of this work, it is worth mentioning \cite{HD63433_Zhang} presented HST/STIS observations of HD 63433\,b around the Ly$\alpha$ line, but no absorption was found for this planet.\\

\subsection{HD\,63433\,c}

The time evolution of the \texttt{serval} activity indicators is flat, except the H$\alpha$ line index that displays a narrow peak close to mid-transit, which is also visible in the H$\alpha$ residual map (Fig.\,\ref{Fig: TS HD63433 C}). Therefore, we excluded those spectra to compute the H$\alpha$ TS, which is mainly flat. The H$\alpha$ TS displays a too narrow absorption feature that originates only from the first half of the transit. Thus, we attributed this signal to a stellar origin rather than to absorption from the planet. We put a 3$\sigma$ upper limit of 0.4\,\% to any possible H$\alpha$ excess absorption from the planet's atmosphere.

The \ion{He}{I} triplet residual map and TS do not show features from the planetary atmosphere, or from stellar variability (Fig.\,\ref{Fig: TS HD63433 C}).
We placed a 3$\sigma$ upper limit to the \ion{He}{I} excess absorption of 0.4\,\%, consistent with the previous upper limit of $\sim$0.5\% reported by \cite{HD63433_Zhang} with Keck/NIRSPEC.

\cite{HD63433_Zhang} detected an excess absorption of Ly$\alpha$ in one visit of HD\,63433\,c atmosphere with HST/STIS. However, we do not detect the evaporating atmosphere of HD\,63433\,c via H$\alpha$ in the VIS, or the \ion{He}{I} triplet in the NIR. Similar situation applies to, e.g. GJ\,436\,b where Ly$\alpha$ was detected (\citealp{GJ436b_Lya}), but not H$\alpha$ or \ion{He}{I} triplet (\citealp{GJ436b_Ha, Nortmann_WASP-69_He, GJ436b_Symulation}).

\subsection{HD\,73583\,b}

We analysed the new TESS data to update the ephemerides on \hd73\ planetary system (see App.\ref{App: HD73583 juliet}) and we used them to analyse the transit of HD\,73583\,b observed with HARPS-N. The MO spectra was computed without the last three spectra, which show different values in the \texttt{serval} H$\alpha$ indicator and YABI S-index. The time evolution for the other activity indicators is mainly flat.

The H$\alpha$ residual map and TS are shown in Fig.\,\ref{Fig: TS HD 73583 B}, where the last spectra clearly exhibit higher flux in the H$\alpha$ line core.
The TS shows a small absorption feature at the H$\alpha$ position, which seems to come from the second half of the transit according to the residual map. The signal is $-$0.46$\pm$0.16\,\% deep (significance $<$3$\sigma$), and has a net shift of $-$6.1$^{+3.2}_{-2.5}$\,km\,s$^{-1}$. The priors and posteriors from the fit are shown in Table\,\ref{table - HD73583c Ha priors and posteriors}, and the posterior distributions are plotted in Fig.\,\ref{Fig: HD73583 b and c nested Ha}.
However, it is hard to link that absorption to the planet atmosphere for two reasons: $i$) the absorption comes mostly from the second half of transit (see the residual map on Fig.\,\ref{Fig: TS HD 73583 B}), and $ii$) the shift of the feature is in the opposite direction of the \ion{He}{I} detection from \cite{Zhang_TOI560b}.
We considered the $\sim$0.5\% absorption feature as the upper limit for H$\alpha$ absorption.

Because GIANO-B was not on the telescope the night of the observations, we could not explore the \ion{He}{i} triplet. \cite{Zhang_TOI560b} detected an excess absorption of 0.68$\pm$0.08\,\% from one full and one partial transits of \hd73\,b with Keck/NIRSPEC. These observations were re-analysed in \cite{Zhang_young_planets} reporting a new excess absorption measurement of 0.72$\pm$0.08\,\%. The \ion{He}{I} triplet signal has a blue shift, rather than the usual red-shift reported in the other \ion{He}{i} detections.

\subsection{HD\,73583\,c}

As we mention in the previous section, the TESS photometric analysis for the HD\,73583 planetary system is shown in App.\,\ref{App: HD73583 juliet}. Using the values from \cite{Barragan_TOI560b}, the uncertainty on HD\,73583\,c $T_{\rm c}$ would be 50\,minutes, while we now obtained a precision of $\sim$1\,min.

The time evolution of the \texttt{serval} H$\alpha$ and dLW indicators show a peak at the beginning of the transit that extends until mid-transit. Thus, to compute the TS, we only considered the spectra from the second half of the transit. The other activity indicators show a scattered but flat time evolution. The H$\alpha$ residual map and TS are shown in Fig.\,\ref{Fig: TS HD 73583 C}, where the H$\alpha$ variability is visible in the residual map.
The TS has a small feature at the expected position. We fitted an absorption of $-$0.54$^{+0.13}_{-0.14}$\,\%, and we present the prior and posterior values in Table\,\ref{table - HD73583c Ha priors and posteriors}, and the posterior distributions are shown in Fig.\,\ref{Fig: HD73583 b and c nested Ha}. Although the fitted absorption is significant ($\sim$3.8$\sigma$), we consider that value as an upper limit or a tentative detection. A second transit with better S/N and no stellar activity will confirm these results.

The \ion{He}{i} triplet residual map and TS is shown in Fig.\,\ref{Fig: TS HD 73583 C}. 
There is no clear evidence of a planetary trace on the residual map or planetary absorption detected on the TS. We placed a 3$\sigma$ upper limit to the \ion{He}{I} excess absorption of 0.5\,\%.

\subsection{HD\,235088\,b}

We observed one transit of this sub-Neptune-sized planet during this project and its results were analysed in depth in \cite{Orell2023}. We detected a \ion{He}{I} blueshifted absorption signal of $-$0.91$\pm$0.11\,\%, confirming the previous detection from \cite{Zhang_young_planets} of $-$0.64$\pm$0.06\,\% absorption. The residual maps and transmission spectra for the H$\alpha$ and \ion{He}{I} triplet are shown in Fig.\,\ref{Fig: TS TOI-1430} (adapted from \citealp{Orell2023}).

In this work we adopted the results presented in \cite{Orell2023}, where an age of 600--800 Myr was estimated for HD\,235088.

\subsection{K2-77\,b}

Because K2-77\,b ephemerides are based on \textit{K2} photometry, we decided to analyse the available TESS data and improve the uncertainties on $T_{\rm c}$ for this target (see App.\,\ref{App: K2-77 juliet}).

According to our results, the last spectrum from the CARMENES observations was taken during the egress, missing the post-transit coverage to compute the MO spectrum. Moreover, the pre-transit observations have low S/N due to higher airmass.

The difficulties to calculate the MO spectrum are visible in the residual maps and TS for both H$\alpha$ and \ion{He}{I} triplet lines (Fig.\,\ref{Fig: TS K2-77}). We could only derive broad 3$\sigma$ upper limits for H$\alpha$ and \ion{He}{I} excess absorptions of 2.5\,\% and 2.7\,\%, respectively.

\subsection{TOI-2046\,b}
\label{Sect: TOI-2046b}

We set GIARPS mode observations with similar exposure times than for HARPS-N and GIANO-B. Due to the faintness of the host star (J\,=\,10.4\,mag), we used GIANO-B's maximum time exposure of 600\,s for both spectrographs. The spectra have relatively low S/N around the H$\alpha$ and the \ion{He}{I} triplet.
Also, the \texttt{GOFIO} pipeline could not properly remove the biggest telluric OH emission peak, and we decided to mask those telluric affected regions. The masked regions are far enough from the \ion{He}{i} line to ensure that they do not affect the final results.

H$\alpha$ and \ion{He}{I} residual maps and TS are shown in Figures\,\ref{Fig: TS TOI-2046}. We only could derive very broad 3$\sigma$ upper limits for H$\alpha$ and \ion{He}{I} excess absorptions of 5\,\% and 2.9\,\%, respectively.

\subsection{TOI-1807\,b}

We analysed the newly available TESS data to derive updated ephemerids, and refine the $T_{\rm c}$. The details of the photometric fit are explained in App.\,\ref{App: TOI-1807 juliet}, and our results for the planet properties are shown in Table\,\ref{table - TOI-1807 juliet priors and posteriors}. 
However, we planned the observations with older ephemerides (\citealp{TOI-1807_Nardiello}), missing part of the transit on 16 December 2022.
We observed a second full transit on 23 December 2022, also with better S/N. The H$\alpha$ and \ion{He}{I} residual maps and TS for both events are shown in Figures\,\ref{Fig: TS TOI-1807}. The second visit does not show evidences of absorption features.
We only used the data from the second transit to compute upper limits due to the different quality between both observations. We derived a 3$\sigma$ upper limit of 0.95\,\% for H$\alpha$ excess absorption, and an upper limit of 0.80\,\% for the \ion{He}{I} triplet.

A transit of this ultra-short period (USP) Earth-like density planet was already observed with IRD, reporting an upper limit to the \ion{He}{I} triplet EW of 4\,m$\AA$ \citep[$\sim$0.4\,\% absorption]{TOI-1807_TOI-2076_Gaidos2022}. Because our upper limit is less constraining than the previous observations, we used the results from \citet{TOI-1807_TOI-2076_Gaidos2022} in the discussion in Section\,4.

\subsection{TOI-1136\,d}
\label{Sect TOI-1136d}

\begin{figure*}[h!]
    \centering
    \includegraphics[width=\hsize]{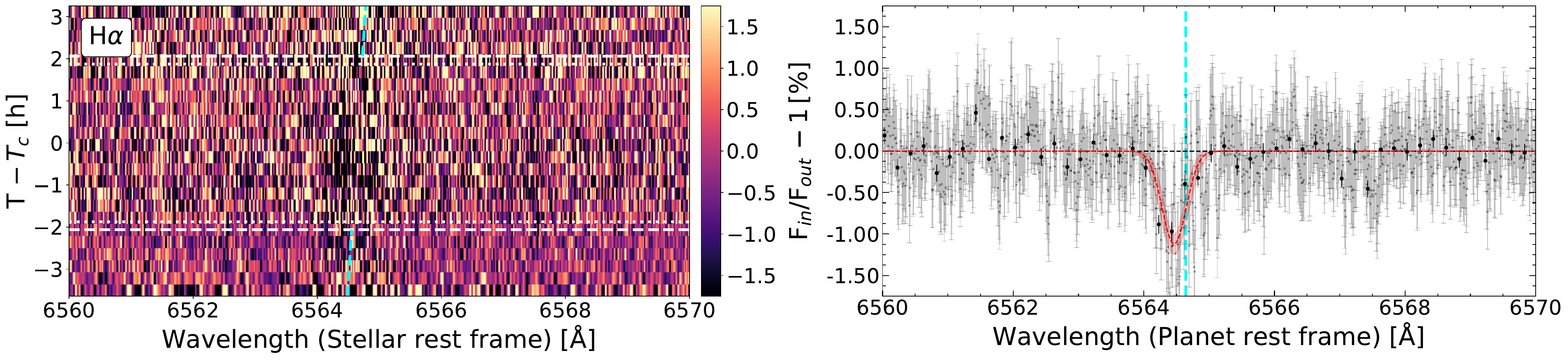}
    \includegraphics[width=\hsize]{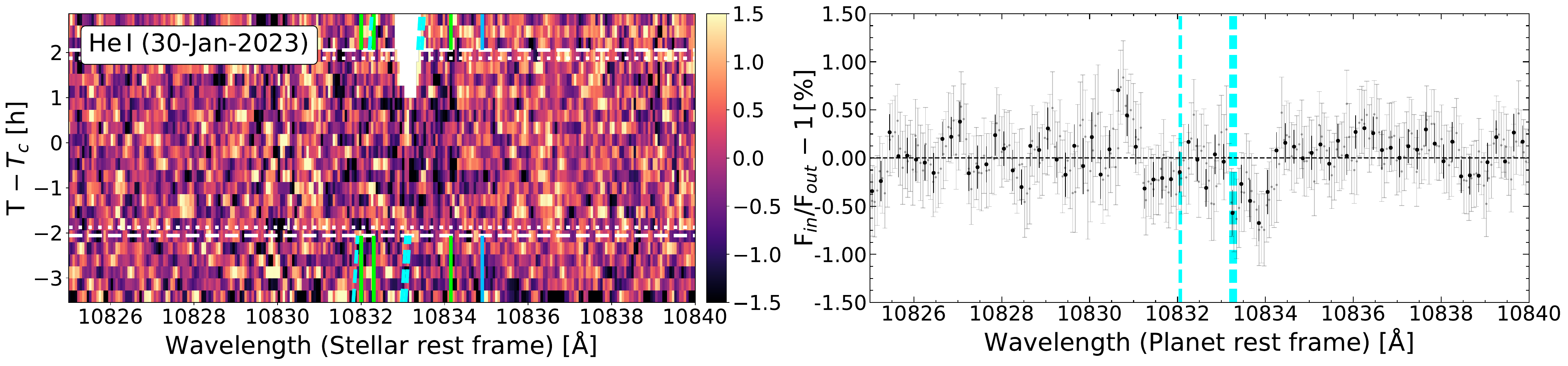}
    \caption{\label{Fig: TS TOI-1136 detections}
    Same as Fig.\,\ref{Fig: TS K2-100}, but  for TOI-1136\,d. H$\alpha$ results (top panels) are the combination of HARPS-N and CARMENES VIS observations. \ion{He}{i} triplet results (bottom panels) are from CARMENES NIR observations.
    }
\end{figure*}

\begin{figure}[]
    \centering
    \includegraphics[width=\hsize]{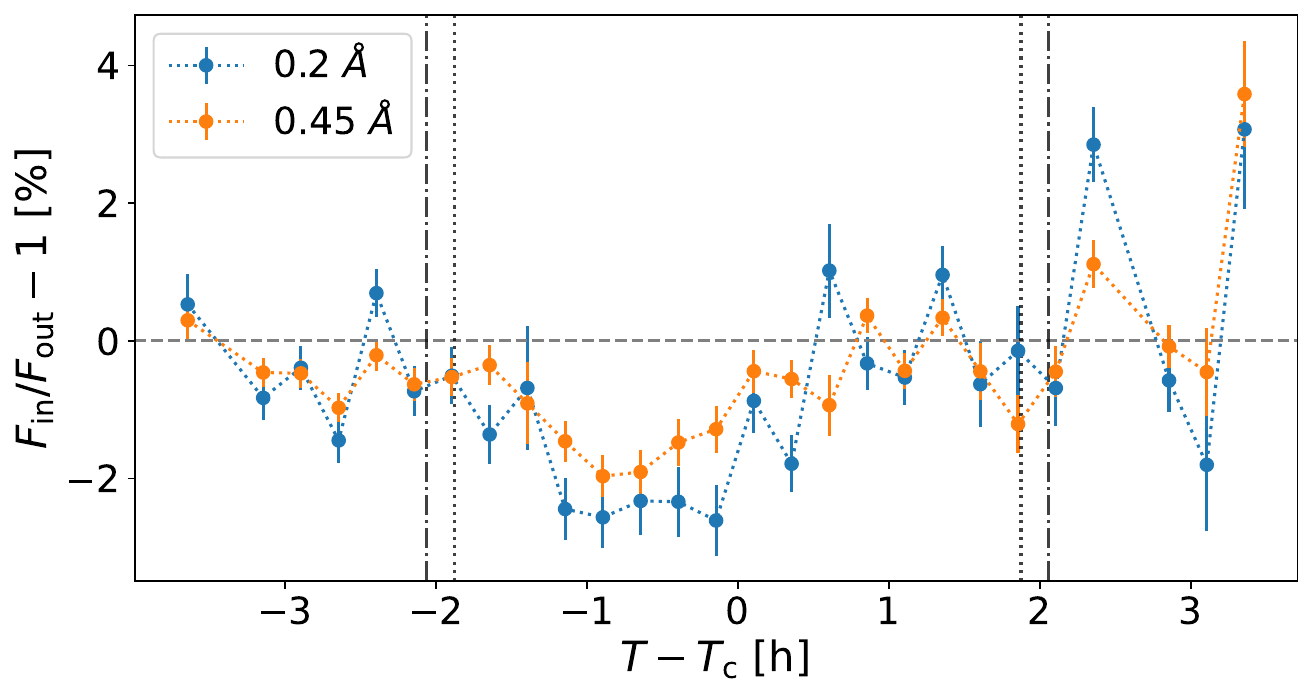}
    \caption{Transit light curve of H$\alpha$ line from the combined observations of TOI-1136\,d. The H$\alpha$ light curve was constructed integrating the counts of the residual map in the planet rest frame around $\lambda_0$ using $\sigma$ (blue) and FWHM (orange) wavelength band passes from Table\,\ref{table - TOI-1136d Ha priors and posteriors}. The vertical lines represent the different contacts during the transit.}
    \label{Fig: TOI-1136d TLC Halpha}
\end{figure}

One HARPS-N transit of TOI-1136\,d was observed to check for the planetary RM signature in the RV time series \citep{TOI-1136_system}. In this work, we inspected the dataset looking for H$\alpha$ planetary absorption signal. Furthermore, we scheduled a second transit with CARMENES to inspect the whole spectral range. We used nonlinear ephemerides to determine the $T_{\rm c}$ for the second transit (F.\,Dai, priv. comm.). The RM and CLV effects over the final TS of H$\alpha$ and \ion{He}{i} triplet lines are negligible (Fig.\,\ref{Fig: TOI-1136d RM}). Thus, we did not correct the data from these marginal effects.

The residual maps and TS of the H$\alpha$ for the individual nights are shown in Figure\,\ref{Fig: TS TOI-1136}, while the final results are shown in Figure\,\ref{Fig: TS TOI-1136 detections}. HARPS-N results show a noisy TS with a small absorption feature. However, a similar feature is also detected in the CARMENES H$\alpha$ TS, which has higher S/N.
The results from both nights confirm the H$\alpha$ detection on the atmosphere of TOI-1136\,d. The combined H$\alpha$ residual map and TS are shown in Fig.\,\ref{Fig: TS TOI-1136}. We fitted the signal obtaining an absorption of $-$1.12$^{+0.12}_{-0.13}$\,\%. Table\,\ref{table - TOI-1136d Ha priors and posteriors} presents the priors and posteriors, and Fig.\,\ref{Fig: TOI-1136d nested Ha} shows the posterior distributions.
Although the signal is a bit deeper in the first half of the transit, the H$\alpha$ transit light curve (Fig.\,\ref{Fig: TOI-1136d TLC Halpha}) shows an absorption consistent with the transit duration without evidences of tail-like structure.

For the \ion{He}{I} triplet, because the CARMENES observations were performed close to the end of the night, the last spectra are affected by solar \ion{He}{I} emission. Although we corrected the emission line, some residuals still persisted. Then, we decided to mask the affected spectral range only on particular spectra. The \ion{He}{I} residual map and TS are shown in Fig.\,\ref{Fig: TS TOI-1136}, and do not show clear evidences of \ion{He}{I} absorption. We placed a 3$\sigma$ upper limit to the \ion{He}{I} triplet excess absorption of $\sim$0.5\,\%.

\subsection{TOI-1268\,b}
\label{Sect TOI-1268}

\begin{figure*}[]
    \centering
    \includegraphics[width=\hsize]{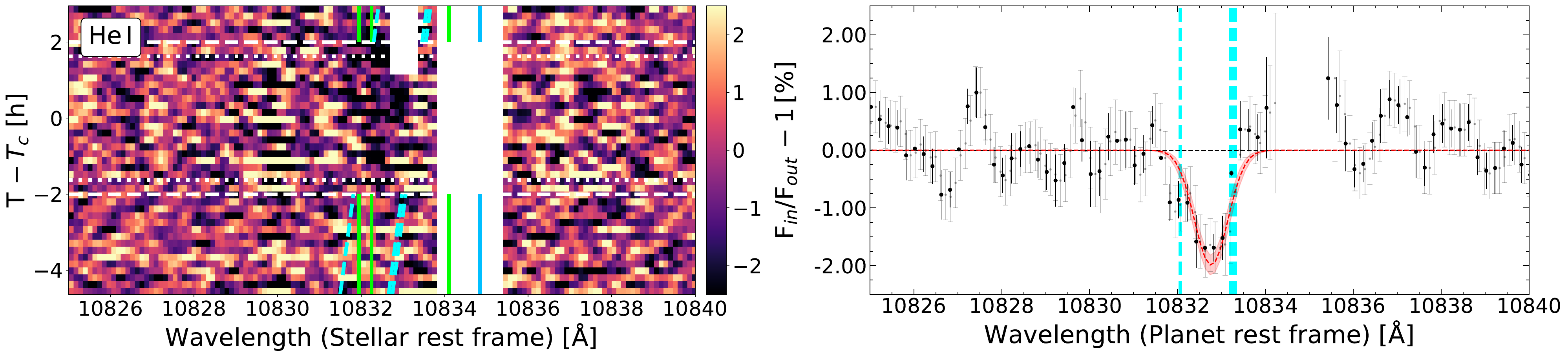}
    \caption{\label{Fig: TS TOI-1268}
    Same as Fig.\,\ref{Fig: TS K2-100}, but  for TOI-1268\,b \ion{He}{i} triplet observations with GIANO-B.
    }
\end{figure*}

\begin{figure}[]
    \centering
    \includegraphics[width=\hsize]{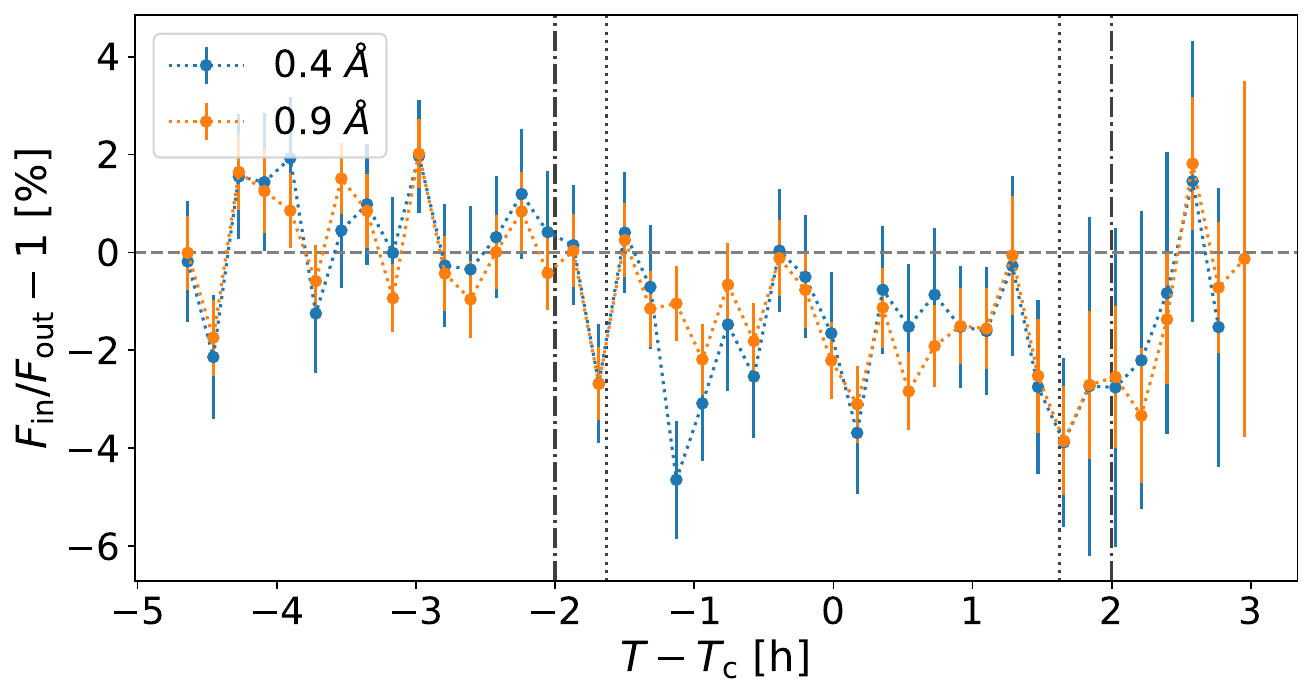}
    \caption{Transit light curve of \ion{He}{I} triplet from TOI-1268\,b detection. The \ion{He}{I} light curve was constructed integrating the counts of the residual map in the planet rest frame around $\lambda_0$ using $\sigma$ (blue) and FWHM (orange) wavelength band passes from Table\,\ref{table - TOI-1268 He priors and posteriors}. The vertical lines represent the different contacts during the transit.}
    \label{Fig: TOI-1268b TLC He}
\end{figure}

The transit of TOI-1268\,b on the 24 of February 2023 was only followed spectroscopically by GIANO-B, and we missed the information about H$\alpha$, and the stellar variability or activity from the visible wavelength range covered by HARPS-N.
MuSCAT2 covered the same transit photometrically with the exposure times initially set to $g=5$\,s, $r$=10\,s, $i$=10\,s, and $z_s$=15\,s and later modified to  $g$=5\,s, $r$=10\,s, $i$=8\,s, and $z_s$=12\,s to avoid the saturation of the target star. Our MuSCAT2 transit fit ha a central time of $T_{\rm c}$\,=\,2460000.66841\,$\pm$\,0.00014\,BJD.

The four bands of the MuSCAT2 photometry do not show the presence of strong stellar activity or the planet crossing in front of any big starspot(s). Moreover, the Pa-$\beta$, Pa-$\gamma$, and Pa-$\delta$ lines are mainly flat with no correlation with the stellar \ion{He}{i} line.
Because our observations ended close to the twilight, our last spectra are affected by \ion{He}{i} sunlight contamination. Because this kind of emission increases quickly between exposures as the observations are approaching the twilight, it is not well corrected by the ABAB procedure. We decided to mask those regions affected by the \ion{He}{i} emission only in the selected spectra.

The residual map and TS are shown in Fig.\,\ref{Fig: TS TOI-1268}, where the residual map shows an absorption region at the expected position of the planetary trace. The signal is well detected in the residual map and the TS, where we fitted a blue-shifted absorption of $-$2.00$^{+0.15}_{-0.16}$\,\% (EW\,=\,19.1$\pm$1.9\,m\AA). Table\,\ref{table - TOI-1268 He priors and posteriors} presents the priors and posteriors, and Fig.\,\ref{Fig: TOI-1268 nested He} shows the posterior distributions. These high-resolution spectroscopy results confirm the \ion{He}{i} detection using narrow-band photometry from \cite{TOI-1268_He_phot}.
We also constructed the \ion{He}{i} triplet transit light curve, shown in Fig.\,\ref{Fig: TOI-1268b TLC He}. The absorption seems to extend further than the end of the white light transit hinting an He tail, that needs to be confirmed in further observations.

\subsection{TOI-2076\,b}
\label{Sect TOI-2076}

In this section we present the analyses and results from one CARMENES transit of TOI-2076\,b observed in 2023. Previously in May 2022, we missed a transit of TOI-2076\,c due to imprecise ephemerides from \cite{toi-2076_Hedges}. Nowadays, the periods of TOI-2076 planets are well-constrained and refined (\citealp{TOI-2076_Osborn22, Zhang_young_planets}), and we could confirm that none of the three planets were transiting the night of 11 May 2022. However, we take advantage of those observations to study the behaviour of the star between epochs. The values derived from the \texttt{serval} activity indicators for the two datasets are consistent, and there are no differences between epochs. The levels of stellar activity of both visits are very similar and TOI-2076\,b transit seems unaffected by stellar activity o variability.
Due to bad weather conditions at the beginning of the night of 13 April 2023, we missed the pre-transit baseline and a very small part of the transit. Moreover, the median S/N of the in-transit spectra is lower ($\sim$60) than after transit ($\sim$90).

The TTVs of the TOI-2076 system (\citealp{TOI-2076_Osborn22}) make the scheduling of its transits complicated and increase the probability of missing part of them. Thus, we complemented our observations on 2023 with a photometric follow-up. We could not perform simultaneous observations on 13 April 2023, but we detected TOI-2076\,b ingress one month after (14 May 2023). The details of the photometric analysis and their results are explained in App.\,\ref{App: TOI-2076}.

The H$\alpha$ residual map and TS are shown in Fig.\,\ref{Fig: TS TOI-2076}. The TS close to the H$\alpha$ has consistently negative values that could be considered as a planetary absorption. Although the lack of pre-transit baseline could be the origin of the signal, as for HD\,63433\,b transit on 1 November 2021 (Sect.\,\ref{Sect HD63433b}). Then, a full transit coverage is needed to confirm this tentative signal. We set a 3$\sigma$ upper limit of 0.7\% to H$\alpha$ exoplanet absorption.

The \ion{He}{I} triplet residual map and TS are shown in Fig.\,\ref{Fig: TS TOI-2076}. We do not detect clear evidences of \ion{He}{I} planetary absorption. We looked for correlations of the \ion{He}{I} NIR line with the \ion{He}{I}\,D3 (5877.2\,$\AA$), and \ion{H}{}\,Paschen lines, with no conclusive results. We could only put a 3$\sigma$ upper limit of $\sim$1\% for any \ion{He}{I} triplet planetary absorption.

TOI-2076\,b atmosphere was already targeted looking for \ion{He}{i}. The same partial transit was observed from the same mountain with two different 8-m telescopes with their high-resolution spectrographs. In both cases, a consistent feature was detected but there are differences in the interpretation of the signal. \citet{Zhang_young_planets} with Keck/NIRSPEC reported a \ion{He}{i} detection with an excess absorption of 1.01$\pm$0.05\,\% (EW\,=\,10.0$\pm$0.7). On the other hand, \cite{TOI-1807_TOI-2076_Gaidos2022} fitted a significant excess EW of 8.5$\pm$1.4\,m$\AA$, but their analyses did not rule out the stellar origin of the signal.
Although CARMENES has the capability to detect such \ion{He}{i} excess absorptions (e.g. \citealp{HD189733b_He, HD209458_He_Alonso, Orell2023}), the S/N of our transit only allows us to put an upper limit at $\sim$1\%, which is at the same level of the previous detected feature. Further observations in good weather conditions are needed to firmly confirm the TOI-2076\,b \ion{He}{I} signal. For the purpose of this work, we  use the positive results from \cite{Zhang_young_planets} in the discussion.

We also used the dataset from May 2022 to check for H/He structures around TOI-2076, recently reported in other exoplanets (\citealp{HAT-P-32b_He_Tail, HAT-P-67b_He_toroide}). However, we obtained very similar results when we included those spectra in the analyses of TOI-2076\,b H$\alpha$ and \ion{He}{I} lines.

\subsection{TOI-1683\,b}
\label{Sect TOI-1683}

We scheduled three transits of TOI-1683\,b for the MOPYS project, one with CARMENES and two with GIARPS. CARMENES transit was stopped due to bad weather and the few taken spectra are useless for atmospheric analyses. The two other transits were observed only with GIANO-B due to problems with the GIARPS mode, and we could not observe until ingress for the first GIANO-B visit (19 September 2023).

\cite{Zhang_young_planets} reported an age of 500$\pm$150\,Myr, based on gyrochronology relations. Here, we derived a new age estimation of 2$^{+1.3}_{-0.9}$\,Gyr based on gyrochronology, stellar kinematics, and stellar \ion{Li}{i} analyses, indicating that TOI-1683 is not a young star. Hence, we do not consider TOI-1683\,b as a young planet. The details of our stellar age analyses are explained in Appendix\,\ref{App: TOI-1683 juliet}. Furthermore, we derived new ephemerides for TOI-1683\,b using the available TESS data from Sectors\,19, 43, and 44 in the MAST archive. The photometric analyses and results are shown in Appendix\,\ref{App: TOI-1683 juliet} as well.

\citet{Zhang_young_planets} observed one transit in bad weather conditions with Keck/NIRSPEC, and reported a \ion{He}{I} detection of 0.84$\pm$0.17\,\%. Although we also observe in poor weather and seeing conditions, we do not detect a significant \ion{He}{i} excess absorption in the individual or combined nights from GIANO-B. The residual maps and TS from the individual and combined nights are shown in Fig.\,\ref{Fig: TS TOI-1683}. We placed a 3$\sigma$ upper limit from our two visits with GIANO-B to the \ion{He}{i} excess absorption of 0.7\,\%, which is $\sim$1$\sigma$ consistent with \citet{Zhang_young_planets} detection. We use the positive detection from \cite{Zhang_young_planets} in the discussion.

\subsection{WASP-189\,b}
\label{Sect WASP-189}

\cite{UHJ_Monika} already analysed one HARPS-N and two HARPS observations finding a H$\alpha$ absorption feature of $-$0.13$\pm$0.02\,\%. However, because the line was affected by the RM effect, they were cautious claiming a detection. Here, we inspected the GIANO-B data taken simultaneously to those HARPS-N observations.

The \ion{Si}{I} and \ion{He}{I} stellar lines are not detectable in the spectra and the only visible spectral features are the H$_2$O telluric absorption lines, which we masked in the residual map and TS, shown in Fig.\,\ref{Fig: TS WASP-189}. Although the TS has some structure, there is no evidence of \ion{He}{i} triplet absorption. We set a 3$\sigma$ upper limit of 0.3\,\% to the \ion{He}{I} absorption on WASP-189\,b.

\subsection{HAT-P-57\,b}
\label{Sect HAT-P-57}

HAT-P-57\,b is an `adolescent' planet (1.00$^{+0.67}_{-0.51}$\,Gyr; \citealp{HAT-P-57_Hartmann}), where \cite{UHJ_Monika} found an H$\alpha$ absorption feature ($-$0.7$\pm$0.2\,\%). However, the signal is narrow compared to other H$\alpha$ detections from the literature and it is surrounded by pulsations from the host star.

We analysed one archival CARMENES NIR transit to inspect the \ion{He}{i} triplet. Although we masked some regions of strong tellurics, the residual map and TS in Fig.\,\ref{Fig: TS HATP57} do not show any significant planetary absorption signal. We set a 3$\sigma$ upper limit of 1\,\% to the planetary \ion{He}{I} absorption.

\subsection{TOI-2018\,b}
\label{Sect TOI-2018}

\begin{figure*}[]
    \centering
    \includegraphics[width=\hsize]{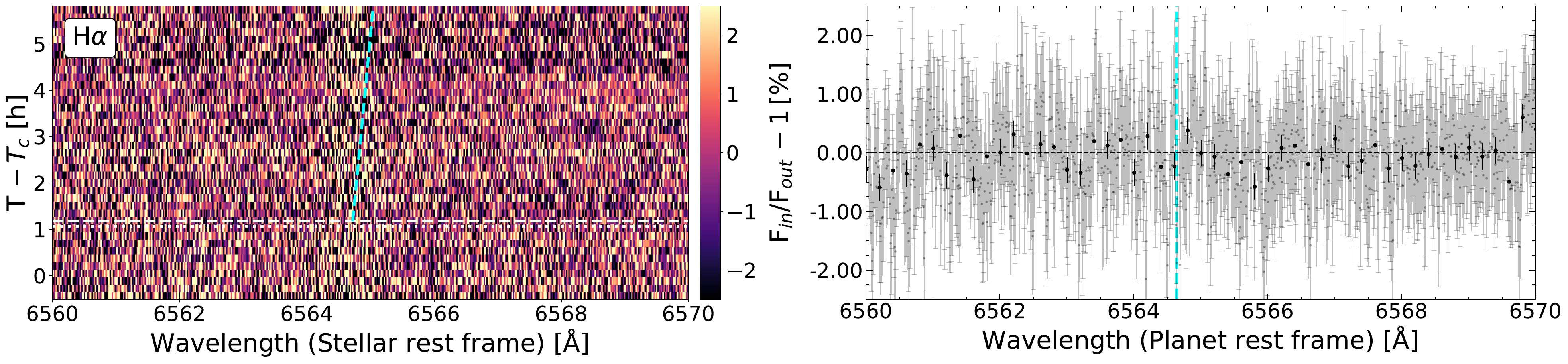}
    \includegraphics[width=\hsize]{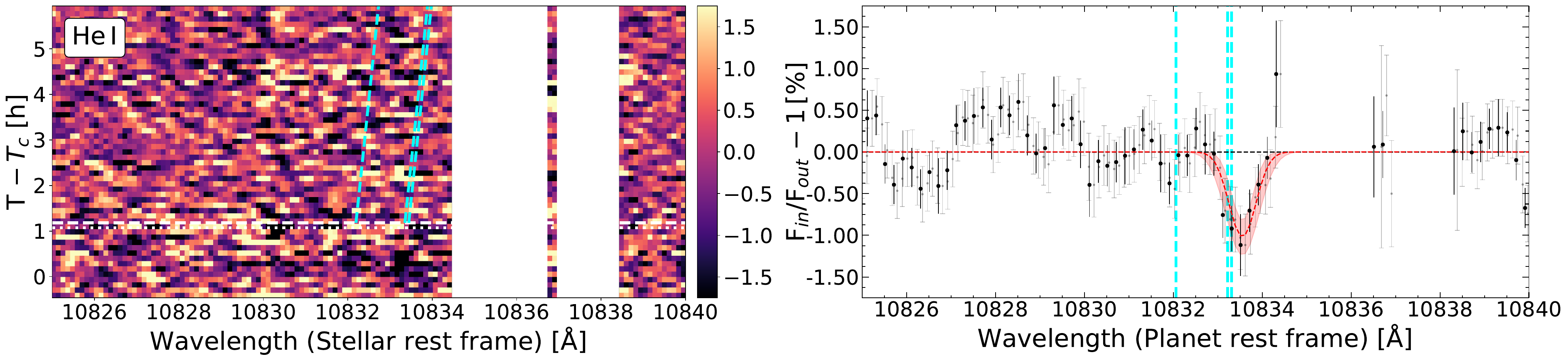}
    \caption{\label{Fig: TS TOI-2018 RESULTS}
    Same as Fig.\,\ref{Fig: TS K2-100}, but for TOI-2018\,b combined observations of H$\alpha$ with HARPS-N and \ion{He}{i} triplet with GIANO-B.
    }
\end{figure*}

\begin{figure}[]
    \centering
    \includegraphics[width=\hsize]{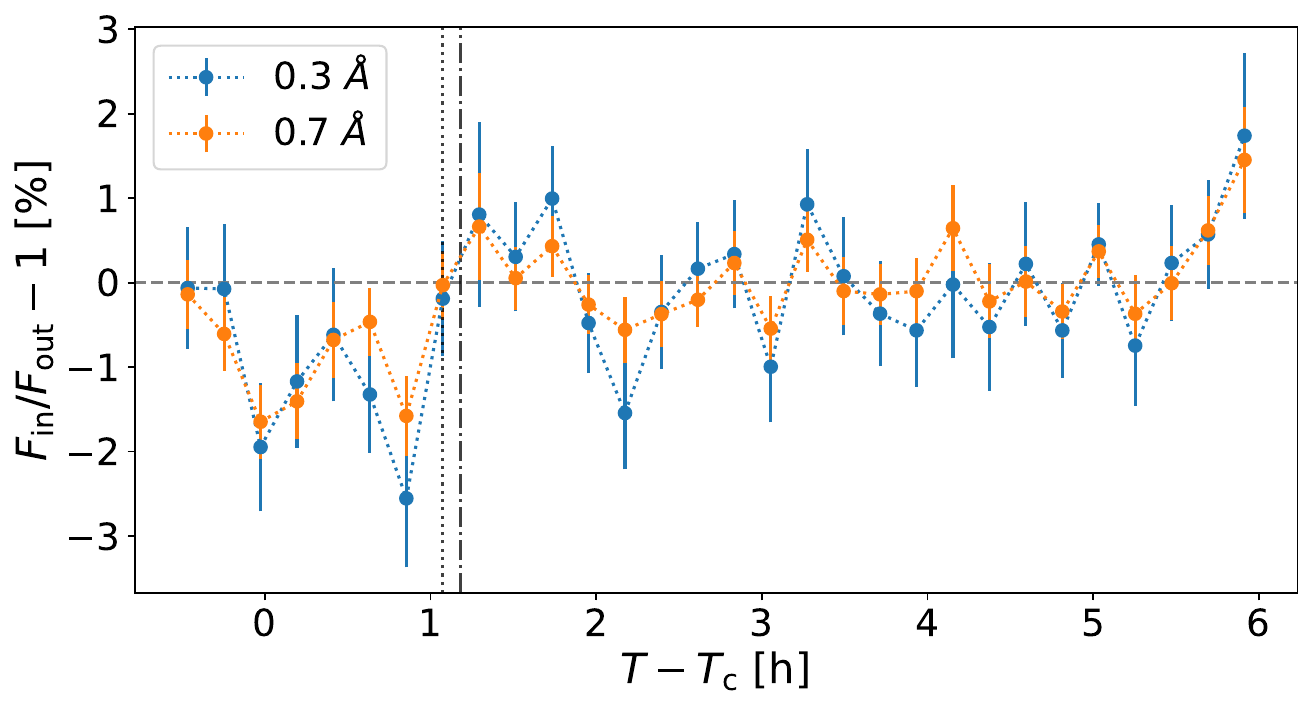}
    \caption{Transit light curve of \ion{He}{I} triplet from the combined nights of TOI-2018\,b. The \ion{He}{I} light curve was constructed integrating the counts of the residual map in the planet rest frame around $\lambda_0$ using $\sigma$ (blue) and FWHM (orange) wavelength band passes from Table\,\ref{table - TOI-2018 He priors and posteriors}. The vertical lines represent the different contacts during the observation.}
    \label{Fig: TOI-2018b TLC He}
\end{figure}

The planet candidate TOI-2018\,b was initially alerted to the community as a `young planet', but final analyses of its age were ambiguous and it was not possible to confirm an age below 1\,Gyr (2.4$^{+0.8}_{-0.2}$; \citealp{TOI-2018_Dai23}). Although we can not include TOI-2018\,b in our sample of young planets, the results obtained from two partial transits observed with GIARPS deserve to be included in this work.

The GIARPS transit on the night of 15 June 2023 was photometrically followed with MuSCAT2, observing the transit in four bands with the exposure times set to $g$=15\,s, $r$=15\,s, $i$=10\,s, and $z_s$=10\,s. Our MuSCAT2 transit fit found a central time of $T_{\rm c}$\,=\,2459746.4287$\pm$0.0021\,BJD, confirming that we missed partially the transits in both visits.

The H$\alpha$ residual map and TS from the individual nights and their combination are shown in Figures\,\ref{Fig: TS TOI-2018 HALPHA} and \ref{Fig: TS TOI-2018 RESULTS}, respectively. The TS is flat, and the 3$\sigma$ upper limit for H$\alpha$ excess absorption is set to 1.5\,\%.

The \ion{He}{i} triplet residual map and TS from the individual nights and their combination are shown in Figures\,\ref{Fig: TS TOI-2018 HE} and \ref{Fig: TS TOI-2018 RESULTS}, respectively. Although the quality of the two nights is different, the results of both nights are consistent within 2$\sigma$, and we detect a consistent excess absorption of $\sim$1\,\% at the expected position of the planetary \ion{He}{i} signal. When we combine both nights, we confirm the detection of a red-shifted \ion{He}{i} absorption of $-$1.02$^{+0.19}_{-0.22}$\,\% (EW\,=\,7.8$\pm$1.5\,m\AA). All the nested sampling material for the individual and combined datasets is shown in App.\,\ref{App: TOI-2018 material}.
The \ion{He}{i} triplet transit light curve (Figure\,\ref{Fig: TOI-2018b TLC He}) shows an absorption signal consistent with the transit duration, with no evidence of an extended He signal. Although the \ion{He}{i} signal is consistent and well detected in the two parcial transits, a full transit observation will help to confirm our detection and the study of TOI-2018\,b atmospheric evaporation.

\section{Additional literature evaporation tracers observations}
\label{sec:literature}

\begin{table*}
\caption[width=\textwidth]{
\label{table - planet results}
Compilation of H$\alpha$ and \ion{He}{I} observations from high-resolution spectroscopy facilities of young ($\lesssim$1\,Gyr) transiting exoplanets, from the literature and this work (Sect. \ref{sect: Results}).
A cross ($\times$) indicates when the stellar activity prevented to measure any planetary absorption or derive upper limits. 
}
\centering
%

\begin{tabular}{ l c c c l c c  }

\hline \hline 
\noalign{\smallskip} 

Planet  & H$\alpha$\,[$\%$] & \ion{He}{I}\,[$\%$] & \,\,\,\,\,\,\,\,\,\,\,\,\,\,\,\,\,\,\,\,\,\,\,\,\,\,\,\,\,\,\,\,\,\,\,\,\, & Planet  & H$\alpha$\,[$\%$] & \ion{He}{I}\,[$\%$] \\

\noalign{\smallskip}
\hline
\noalign{\smallskip}

\hline \noalign{\smallskip}
MASCARA-2\,b  & 0.85$\pm$0.03\,$^{(1)}$ & $<$0.5 & & TOI-2046\,b  & $<$5 & $<$2.4 \\ 
\hline \noalign{\smallskip}
K2-100\,b  & $<$1.4 & $<$1.3 & & TOI-1807\,b  & $<$0.95 & $<$0.80 \\
 &  & $\lesssim$1.2\,$^{(2)}$ & &              &       & $\lesssim$0.4\,$^{(8)}$ \\
\hline \noalign{\smallskip}
TOI-1431\,b  & $<$0.33\,$^{(3)}$ & $<$0.4 & & TOI-2048\,b & $<$1.5 & $<$1.0 \\ 
\hline \noalign{\smallskip}
HD\,63433\,b  & $<$0.4 & $<$0.4 & & HD\,63433\,c  & $<$0.4 & $<$0.4 \\
 &  & $\lesssim$0.5\,$^{(4)}$  & &  & -- & $\lesssim$0.5\,$^{(4)}$  \\
\hline \noalign{\smallskip}
HD\,73583\,b  & $\lesssim$0.5 & 0.72$\pm$0.08\,$^{(5,6)}$ & & HD\,73583\,c  & $\lesssim$0.5 & $<$0.5 \\ 
\hline \noalign{\smallskip}
HD\,235088\,b  & $\times$\,$^{(7)}$ &  0.91$\pm$0.11\,$^{(7)}$ & & TOI-2076\,b  & $<$0.7 &  $\lesssim$1 \\
             &  &  0.64$\pm$0.06\,$^{(6)}$ & &   &  &  1.01$\pm$0.05\,$^{(6)}$ \\
\hline \noalign{\smallskip}
V1298 Tau\,b  & -- & $\lesssim$1.7\,$^{(20)}$ & & V1298 Tau\,c  & <1.1 & <3.7 \\   
\hline \noalign{\smallskip}
TOI-1268\,b  & -- & 2.00$^{+0.15}_{-0.16}$  & & TOI-1136\,d  &  1.12$^{+0.12}_{-0.13}$ & $<$0.5 \\ \noalign{\smallskip}
\hline \noalign{\smallskip}
WASP-189\,b  & 0.13$\pm$0.02\,$^{(9)}$ & <0.3 & & HAT-P-57\,b  & <0.7$\pm$0.2$^{(9)}$ & <1 \\ 
\hline \noalign{\smallskip}
K2-77\,b & $<$2.5 & $<$2.7  & & K2-25\,b  & -- & $<$1.7\,$^{(12)}$ \\
\hline \noalign{\smallskip}
AU Mic\,b  & $\times$\,$^{(13)}$ & $<$0.34\,$^{(14)}$ & & K2-136\,c  & -- & $<$2.3\,$^{(15)}$ \\ 
\hline \noalign{\smallskip}
DS Tuc\,b  & $\times$\,$^{(16)}$ & -- &  & WASP-52\,b  & 0.86$\pm$0.13\,$^{(17)}$ & 3.44$\pm$0.31\,$^{(18)}$ \\
\hline \noalign{\smallskip}
HAT-P-70\,b  & 1.56$\pm$0.15\,$^{(19)}$ & -- & & WASP-80\,b  & -- & <0.85\,$^{(21)}$ \\
\hline \noalign{\smallskip}
KELT-9\,b  & 1.15$\pm$0.05\,$^{(10)}$ & $<$0.33\,$^{(11)}$ &  & \\
\hline \noalign{\smallskip} \noalign{\smallskip} \noalign{\smallskip} \noalign{\smallskip}
TOI-2018\,b\,$^{(\dagger)}$ & $<$1.5 & 1.02$^{+0.19}_{-0.22}$ & & TOI-1683\,b\,$^{(\dagger)}$  & -- & <0.7 \\ 
 &  &   & &   &  & 0.84$\pm$0.17\,$^{(6)}$  \\

\hline
\hline

\end{tabular}

\tablebib{$^{(1)}$\,\citet{MASCARA2_Nuria}.
$^{(2)}$\,EW at 99\% confidence of 5.7\,m$\AA$ \citep{K2-100b_Gaidos_He}.
$^{(3)}$\,\citet{TOI-1431_Monika}.
$^{(4)}$\,\citet{HD63433_Zhang}.
$^{(5)}$\,\citet{Zhang_TOI560b}.
$^{(6)}$\,\citet{Zhang_young_planets}.
$^{(7)}$\,\citet{Orell2023}.
$^{(8)}$\,EW at 99\% confidence of 4\,m$\AA$ \citep{TOI-1807_TOI-2076_Gaidos2022}.
$^{(9)}$\,\cite{UHJ_Monika}.
$^{(10)}$\,\citet{KELT-9_Ha}.
$^{(11)}$\,\citet{Nortmann_WASP-69_He}.
$^{(12)}$\,EW at 99\% confidence of 17\,m$\AA$ \citep{K2-25_Gaidos}.
$^{(13)}$\,\citet{AU_Mic_Enric}
$^{(14)}$\,EW at 99\% confidence of 3.7\,m$\AA$ \citep{AU_Mic_Hirano_He}.
$^{(15)}$\,EW at 99\% confidence of 25\,m$\AA$ \citep{K2-136c_Gaidos_He}.
$^{(16)}$\,\citet{DS_Tuc_Ab_Benatti}.
$^{(17)}$\,\citet{WASP-52_Halpha_Chen}.
$^{(18)}$\,\citet{WASP-52_He_NIRSPEC}.
$^{(19)}$\,\citet{HAT-P-70_Halpha}.
$^{(20)}$\,\citet{V1298Tau_Gaidos}.
$^{(21)}$\,\citet{WASP-80_He_Fossati}
}

\tablefoot{$^{(\dagger)}$ TOI-2018\,b and TOI-1683\,b are included in the table, but they are likely not young planets. }

\end{table*}

\subsection{Young planets from the literature}
\label{Subsec: OTHER YOUNG PLANETS}

In the framework of the MOPYS project, we adopted the 1\,Gyr stellar age as the threshold to classify exoplanets as young ($\lesssim$1\,Gyr) or old ($\gtrsim$1\,Gyr). We used this nomenclature in the discussion in Section\,\ref{Sect: Discussion}. The 1\,Gyr threshold is mainly based on the core-powered evaporation timescale, which is longer than the strong photo-evaporation initial stage (X-ray driven).

Young exoplanets are an interesting population that have called the attention of different research groups. To put the young planet results obtained in this work into context, we inspected the ExoAtmospheres\footnote{\url{http://research.iac.es/proyecto/exoatmospheres/index.php}} database looking for other young exoatmospheric analyses. Table\,\ref{table - planet results} presents a compilation of literature high-resolution spectroscopy studies of young transiting exoplanet atmospheres targeting the H$\alpha$ or the \ion{He}{I} triplet.\footnote{We also acknowledge the \ion{He}{i} observations of K2-33\,b (\citealp{K2-33b_He_IRD}), which appeared during the peer review process of the manuscript.} Here, we give a short context to those observations from the literature included in Table\,\ref{table - planet results}.

KELT-9\,b: As the hottest exoplanet known to date, this planet attracted attention to probe and investigate its extreme atmosphere. \cite{KELT-9_Ha} detected an evaporating exosphere of H$\alpha$. However, \citet[see supplementary material section therein]{Nortmann_WASP-69_He} could only place an upper limit to the \ion{He}{I} triplet absorption.

K2-25\,b: The \ion{He}{I} triplet was observed using the IRD spectrograph, although the transit was contaminated by telluric OH emission. \citet{K2-25_Gaidos} reported a 99\% confidence upper limit to the transit-associated EW of 17\,m$\AA$.

AU Mic\,b: Due to the host star's youth, the H$\alpha$ observations with ESPRESSO were strongly affected by stellar activity, making it impossible to set an upper limit \citep{AU_Mic_Enric}. Recently, \citet{AU_Mic_Lyalpha} reported a Ly$\alpha$ detection but only in one of the two visits with HST/STIS.
IRD and Keck/NIRSPEC spectrographs obtained a 99\% confidence upper limit to the \ion{He}{I} EW of 4.4 and 3.7\,m$\AA$, respectively \citep{AU_Mic_Hirano_He}. \cite{Allart_He_survey} observed one transit with SPIRou spectrograph finding a significant \ion{He}{i} absorption feature (0.37\,$\pm$\,0.09\,\%). However, the authors finally reported a conservative 3$\sigma$ upper limit of 0.26\,\%, consistent with previous observations.

K2-136\,c: \citet{K2-136c_Gaidos_He} put a 99\% confidence upper limit to the \ion{He}{I} triplet EW of 25\,m$\AA$ with one transit observed with the IRD spectrograph. The \ion{He}{I} lines position in between the telluric OH emission and H$_2$O absorption lines complicated the calculation of the transmission spectra.

DS Tuc\,b: The H$\alpha$ was analysed with ESPRESSO and HARPS spectrographs observations. However, the transits were affected by stellar activity and an H$\alpha$ upper limit could not be set \citep{DS_Tuc_Ab_Benatti}.

V1298 Tau system: This 20-Myr old multi-planet system has seen different attempts to study the presence of \ion{He}{I} in their planetary atmospheres. \cite{V1298Tau_Gaidos} used the IRD spectrograph to observe, during different nights, the star alone and a transit of planet b. An increasing \ion{He}{I} absorption was detected during the transit, but the authors proposed other explanations besides the planetary absorption. Using the narrowband helium filter technique, \citet{V1298Tau_He_Vissapragada2021} observed the transits of planets b and d. For planet b they do not require extra absorption to explain the flux decrease,  while they found a tentative excess absorption combining two partial transits of planet d, but it requires a significant transit time variation.

WASP-52\,b: The planet was observed with ESPRESSO detecting H$\alpha$, and other atomic species as well \citep{WASP-52_Halpha_Chen}. The presence of \ion{He}{I} in its atmosphere has been studied, first with a tentative detection using narrow-band photometry \citep{WASP-52_He_2020, WASP-52_He_narrowband} and later confirmed with the Keck/NIRSPEC spectrograph \citep{WASP-52_He_NIRSPEC}. However, a recent paper by \cite{Allart_He_survey} did not find \ion{He}{i} absorption in two visits with the SPIRou spectrograph, and derived a 3$\sigma$ upper limit of 1.69\,\%.

HAT-P-70\,b: The atmosphere of this ultra-hot Jupiter was inspected by \cite{HAT-P-70_Halpha} using HARPS-N. They detected absorption coming from the H$\alpha$, H$\beta$, and H$\gamma$ lines, and many other atomic and molecular species as well.

WASP-80\,b: Its atmosphere has been targeted in search of \ion{He}{i} absorption but without success (\citealp{WASP-80_He_Fossati,Allart_He_survey}). \citet{Salz2015_WASP80_age} is the only reference which provides a formerly calculated age of $<$200\,Myr. However, the observed X-ray luminosity ($\log L_{\rm X}$\,$\sim$\,27.8\,erg\,s$^{-1}$ in \citealp{Salz2015_WASP80_age}, down to 27.5\,erg\,s$^{-1}$ in Sanz-Forcada et al. in prep.) yields a $\log L_{\rm X}/L_{\rm bol}$\,$\sim$\,$-5.0$, implying an age of $\sim$\,3\,Gyr using \citet{JorgeSanz2011} age--$L_{\rm X}$ relations. Moreover, the TESS light curve do not suggest a rotational period of $\lesssim$\,15\,d. Therefore, WASP-80 is probably older than 1\,Gry according to the gyrochronology method shown in Figure\,\ref{fig:Prot} ($G - J$\,=\,2.05).

\subsection{Older planets from the literature}
\label{Subsec: OTHER OLD PLANETS}

To put in context the young exoplanet \ion{He}{i}  (planets with ages $\lesssim$\,1\,Gyr) findings, we complemented them with old exoplanet high-resolution \ion{He}{i} observations (planets with age $\gtrsim$\,1\,Gyr) from the literature.
We have compiled a \ion{He}{i} triplet database, detailed in Table\,\ref{table - DATABASE}, which we constructed using the ExoAtmospheres database and literature results. For consistency, we only considered \ion{He}{i} triplet results from high-resolution spectrographs. That is, the narrow-band photometry detections of HAT-P-26\,b (\citealp{Vissapragada2022_Neptune_Desert}) and TOI-1420\,b (\citealp{TOI-1420b_Vissapragada2024}) were not included.\footnote{ We also acknowledge the \ion{He}{i} observations of TOI-1259A\,b (\citealp{TOI1259Ab_He_phot_spect}), which appeared during the peer review process of the manuscript.}
Although we analysed the H$\alpha$ and we cited some results on the Ly$\alpha$ as well, Table\,\ref{table - DATABASE} only reports the H (H$\alpha$ or Ly$\alpha$) observations for the planets with \ion{He}{i} triplet observations, except for the young hot Jupiter HAT-P-70\,b with H$\alpha$ detection (\citealp{HAT-P-70_Halpha}). The young Neptune DS\,Tuc\,b is not included in Table\,\ref{table - DATABASE} because no upper limit could be set from H$\alpha$ observations (\citealp{DS_Tuc_Ab_Benatti}) and no \ion{He}{i} triplet observations could be found in the literature.

Given the number of planets and details of each observation, we do not discuss the planets individually here. We refer to Table\,\ref{table - DATABASE} and the appropriate references.

\subsection{A note on He I variability}

The planetary \ion{He}{i} triplet is known to show variability in its strength but also in its detectability (e.g. \citealp{He_GJ3470b_Enric2020, HD189733b_He_variability_Zhang2022}). In this work, we presented two upper limits for TOI-2076\,b and TO1683\,b, for which \citet{Zhang_young_planets} reported \ion{He}{i} detections. Our upper limits are consistent at 1$\sigma$ with \citet{Zhang_young_planets} detections. Thus, \ion{He}{i} triplet variability may be one reason for our absence of planetary signals. Variability was also invoked by \citet{Orell2023} to explain the $\sim$2$\sigma$ deeper absorption found in HD\,235088\,b. We note that \citet{TOI-1807_TOI-2076_Gaidos2022} derived non-conclusive results for TOI-2076\,b \ion{He}{i} signal, and TOI-1683\,b transit from \citet{Zhang_young_planets} was performed in poor observing conditions. We want to stress the importance of re-observing targets as a sanity check to confirm previous results (detections or upper limits), but also to study the planetary \ion{He}{i} triplet variability.

On the other hand, \citet{Krolikowski2023_He_in_young_stars} analysed the variability of the \ion{He}{i} triplet of young stars. In particular, V1298\,Tau, K2-100, K2-136, K2-77, HD\,63433, and TOI-2048 were in the list of observed stars. They found that young stars show higher variability, with V1298\,Tau being extremely variable, which could explain the non-conclusive results from the \ion{He}{i} observations. The stellar \ion{He}{i} variability decreases rapidly and keeps constant for stars older than $\sim$300\,Myr (\citealp{Krolikowski2023_He_in_young_stars}). Although the stellar \ion{He}{i} triplet line is variable, the short-term variability might not have a significant impact on the transit observations performed during the same night (\citealp{Fuhrmeister2020_He_variability, Krolikowski2023_He_in_young_stars}). In fact, H$\alpha$ is more sensitive to stellar variability than the \ion{He}{i} triplet (\citealp{Fuhrmeister2020_He_variability}) and the analyses of those lines on AU\,Mic\,b and DS\,Tuc\,b are good examples (\citealp{AU_Mic_Enric, AU_Mic_Hirano_He, DS_Tuc_Ab_Benatti}).

\section{Criteria for detections, non-detections, and non-conclusive measurements}
\label{Sect: det non up criteria}

\begin{figure*}[h!]
    \sidecaption
    \includegraphics[width=12cm]{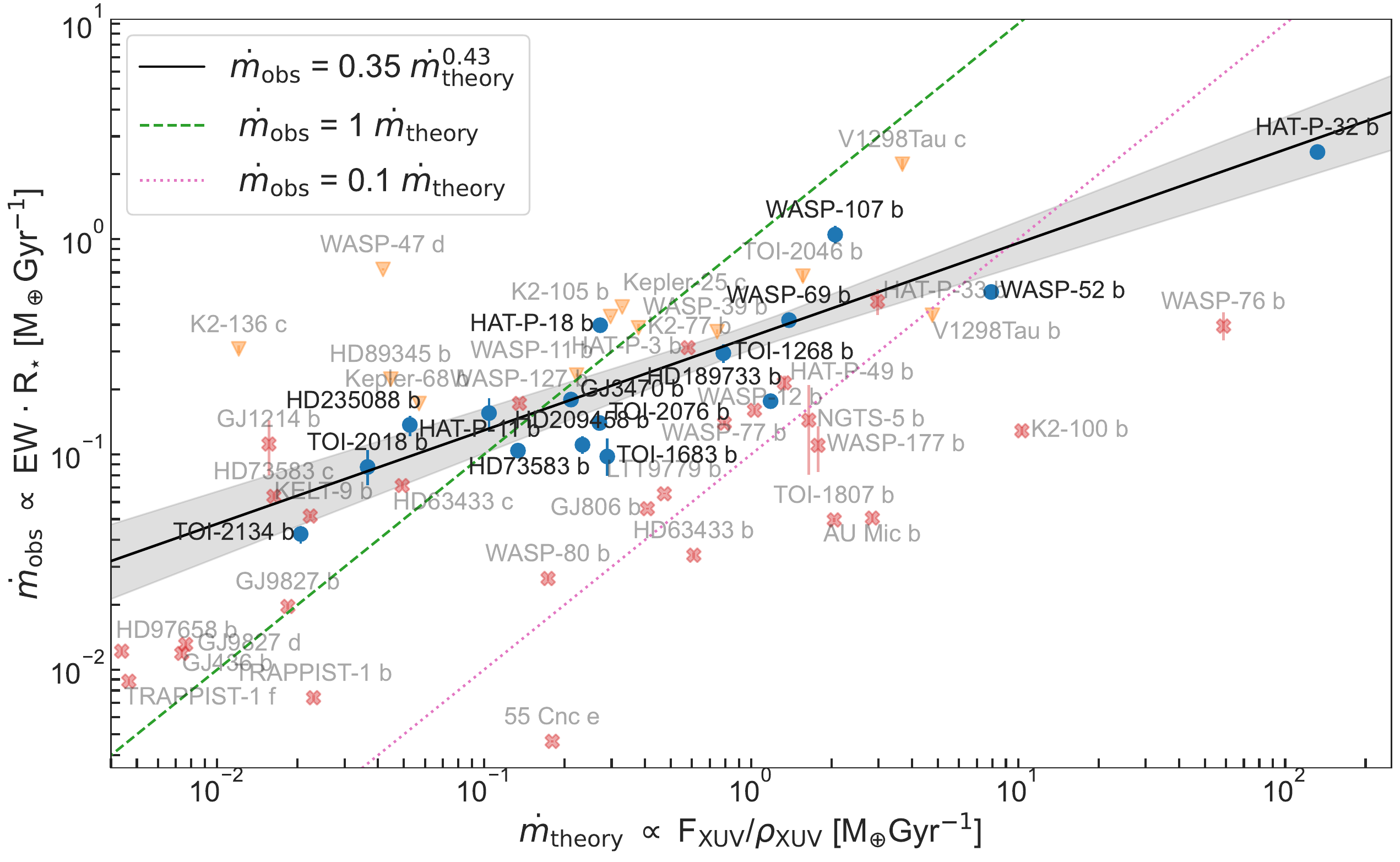}
    \caption{\label{Fig: Mdot obs vs theory} Relationship between the observed ($\dot{m}_{\rm obs}$) and the theoretical ($\dot{m}_{\rm theory}$) energy-limited mass-loss rates. We define XUV until the \ion{He}{i} ionisation range ($\lambda$\,=\,5\,--\,504\,\AA). The black line indicates the fitted relationship (shown in the legend) and the shaded area the 1$\sigma$ uncertainty.
    \ion{He}{i} observations are coded as blue circles for detections, red crosses for non-detections, orange down-pointing triangles for non-conclusive. 
    We did not plot the error bars of $\dot{m}_{\rm theory}$ due to the very large uncertainties associated with the \fxuv\ values and its calculation.
    Every planet has its name labelled: in black for detections and the others  in grey. The two new detections presented in this work, TOI-2018\,b and TOI-1268\,b, are in good agreement with the predicted trend.
    }
\end{figure*}

A problem we encountered when studying the \ion{He}{i} signal (and H$\alpha$ as well) was how to deal with the observations where only an upper limit value is given, which are the majority  of the cases (52 of 69 in Table\,\ref{table - DATABASE}). In this work, we classified the upper limit measurements in two groups: $a$) non-conclusive measurements (i.e. the upper limit value is large, and it does not actually constrain the presence of the atom in the exoplanet atmosphere) and $b$) non-detection (i.e. the upper limit value is low enough to confidently assume there is no significant presence of that atom in the exoplanet atmosphere).

To assign observations to either category, for the He observations, we used the relationship that the observed and theoretical mass-loss rates (\mobs\ and \mtheory, respectively) seem to follow, represented in Figure \,\ref{Fig: Mdot obs vs theory}. We present \mobs\ and \mtheory\ equations in Sect.\,\ref{Subsec: mass-loss rate}, and we further explore and discuss the trend in that section.
For the time being, the fitted line between \mobs--\mtheory\ seems to account for the differences between planetary characteristics, and populations (\citealp{Vissapragada2022_Neptune_Desert, TOI-2134b_Zhang}, and Sect.\,\ref{Subsec: mass-loss rate}). Therefore, we consider that relation to classify the \ion{He}{i} upper limits. Although other criteria might be chosen, it is used consistently for the whole sample.

We computed the \mobs\ upper limits from the \ion{He}{i} EW upper limit values. If the \mobs\ upper limit falls above the fitted line, the \mobs\ upper limit value is higher than the a priori expected signal, and so it is not constrained. Thus, we refer to those measurements as non-conclusive (orange down-pointing triangles in Fig.\,\ref{Fig: Mdot obs vs theory}).
On the other side, the \mobs\ upper limits falling below the line can be considered as non-detections (red crosses in Fig.\,\ref{Fig: Mdot obs vs theory}). We also considered as non-detections the \mobs\ upper limits that fall within 1$\sigma$ of the fit (grey shaded are in Fig.\,\ref{Fig: Mdot obs vs theory}).

We note that \fxuv\ is unknown for some of the exoplanets in our sample. For those planets with no measured \fxuv, we performed a similar procedure but in a 1$/\rho_{\rm XUV}$ versus \mobs\ diagram (not shown), which is a proxy for \mtheory\ versus \mobs\ . $\rho_{\rm XUV}$ is the planet density when considering as the radius its XUV radius, which is different from the planet density $\rho_{\rm p}$ (see Sect.\,\ref{Subsec: mass-loss rate}).

We did some exceptions when classifying some \ion{He}{i} observations. GJ\,1214\,b: we re-classified as a non-detection the tentative \ion{He}{i} signal reported by \citet{Orell2022} due to other upper limits reported in the literature (\citealp{delaRoche2020, Kasper_2020, Spake_He_GJ1214, Allart_He_survey}). V1298\,Tau\,b: because of the unclear origin of the signal detected by \citet{V1298Tau_Gaidos}, we set this observation as non-conclusive. V1298\,Tau\,c is in the same situation as planet b, and its \ion{He}{i} upper limit is non-conclusive. Moreover, the measurements for the V1298\,Tau planets are consistent with the stellar \ion{He}{i} line variability range derived for their host star (\citealp{Krolikowski2023_He_in_young_stars}). WASP-76\,b: its \ion{He}{i} feature was presented as upper limit in \citet{Nuria_2021_WASP-76b} but here we consider it as a non-detection due to the planet position well below the \mobs--\mtheory\ line.

For H$\alpha$, we simply considered the upper limits as non-detections. However, if the H$\alpha$ upper limit absorption is comparable to that of \ion{He}{i}, we used the \ion{He}{i} classification for both measurements.
We only considered few Ly$\alpha$ detections for some particular exoplanets. The classification and nomenclature described in this section (detection, non-detection, and non-conclusive) is used all across the manuscript.

We note that absoprtion measurements of the three lines (Ly$\alpha$, H$\alpha$, and \ion{He}{i} triplet) would be the ideal case to determine whether an exoplanet is undergoing strong atmospheric escape. However, these observations are not available for all targets, and may be hard (even impossible) to obtain for some individual stars (e.g. Ly$\alpha$ interstellar medium extinction, H$\alpha$ variability in active stars, or low \ion{He}{i} triplet population). When possible, we consider the results from the three lines to determine atmospheric escape detections.

\section{Discussion}
\label{Sect: Discussion}

We have organised the discussion of our results in the following manner: Section\,\ref{Subsec: young planets evaporation} gives a general view of the evaporation tracers across stellar age, planet radius, period, and mass. In Sect.\,\ref{Subsec: cosmic shoreline} we constrain the cosmic shoreline from \ion{He}{i} detections. In Section\,\ref{Subsec: mass-loss rate} we explore the relation between \ion{He}{i} detections and the energy-limited mass-loss rates. Finally, we explore further relations between \ion{He}{i} detections and different planet and stellar properties in Sections\,\ref{Subsec: Hill}, \ref{Subsec: stellarhost}, and \ref{Subsec: Density-Teq}.

\subsection{Evaporation tracers of planetary atmospheres across stellar age}
\label{Subsec: young planets evaporation}

\begin{figure}
    \resizebox{\hsize}{!}{\includegraphics{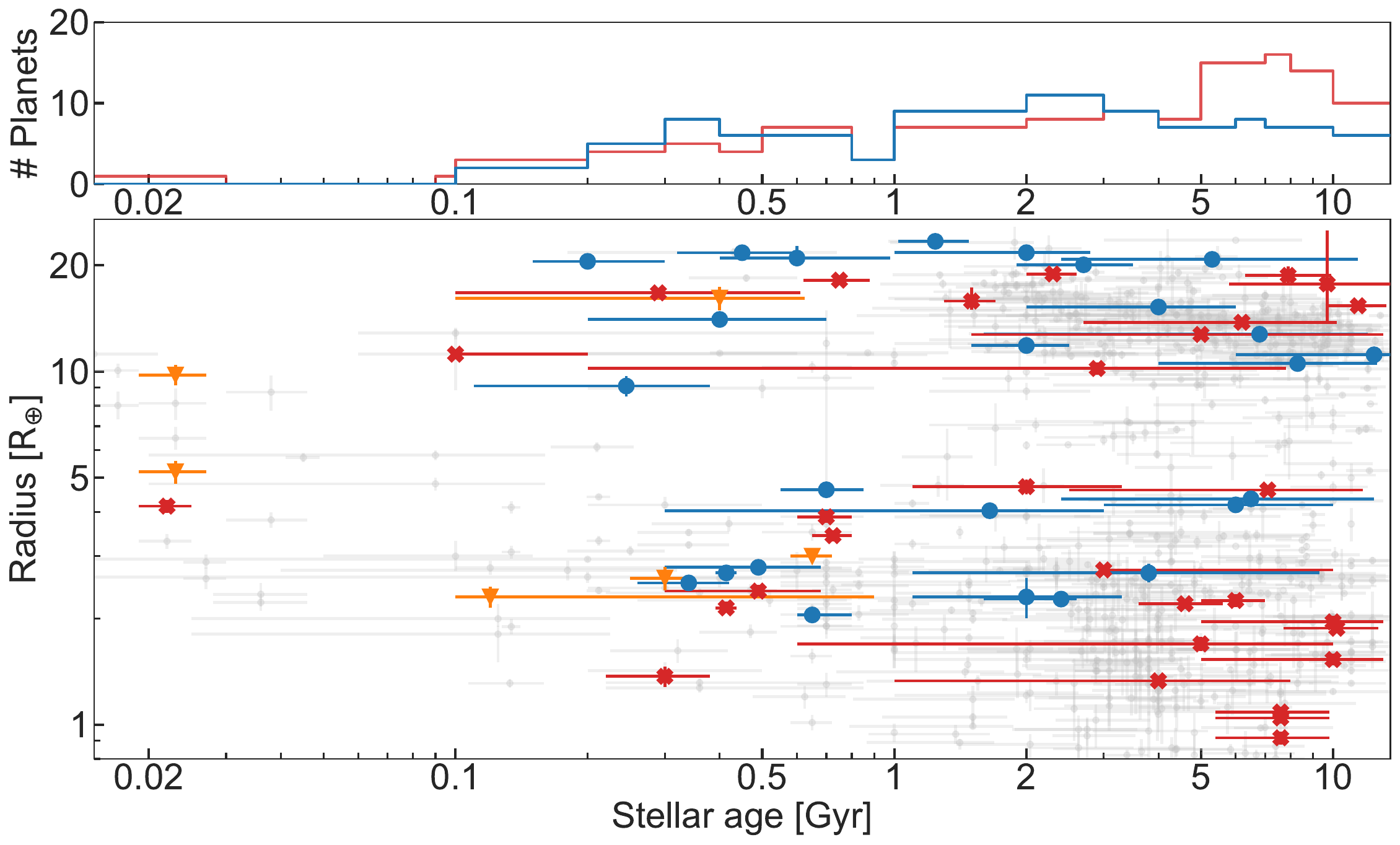}}
    \caption{\label{Fig: EVAPORATION AGE} Planetary radius vs stellar age diagram of evaporation (H and \ion{He}{i}) detections (blue circles), non-detections (red crosses), and non-conclusive observations (orange down-pointing  triangles). Planetary radii and stellar ages are from Table\,\ref{table - DATABASE}. The top panel shows  the summed histogram of evaporation detections (blue line) and non-detections (red line) across stellar age. The grey points represent all known planets with radius and age determined with precision better than 30\% and 50\%, respectively (data from NASA Exoplanet Archive).}
\end{figure}

\begin{figure}
    \resizebox{\hsize}{!}{\includegraphics{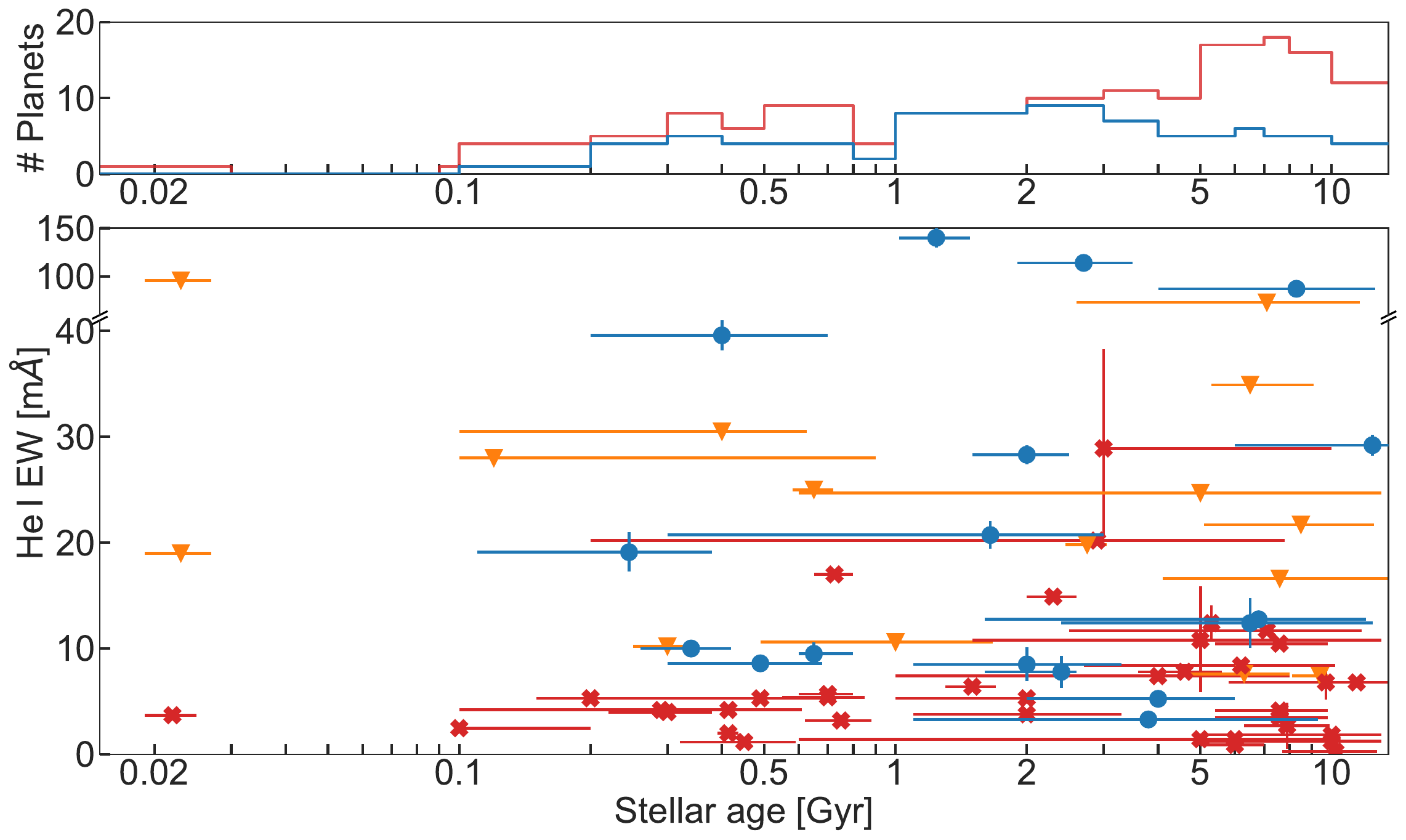}}
    \caption{\label{Fig: Age detections} Equivalent width (EW) of \ion{He}{i} detections (blue circles), non-detections (red crosses), and non-conclusive observations (orange down-pointing triangles)  as functions of stellar age. The top panel shows the summed histogram of \ion{He}{i} detections (blue line) and non-detections (red line) across stellar age bins.
    }
\end{figure}

\begin{figure*}[h!]
    \centering
    \includegraphics[width=\textwidth]{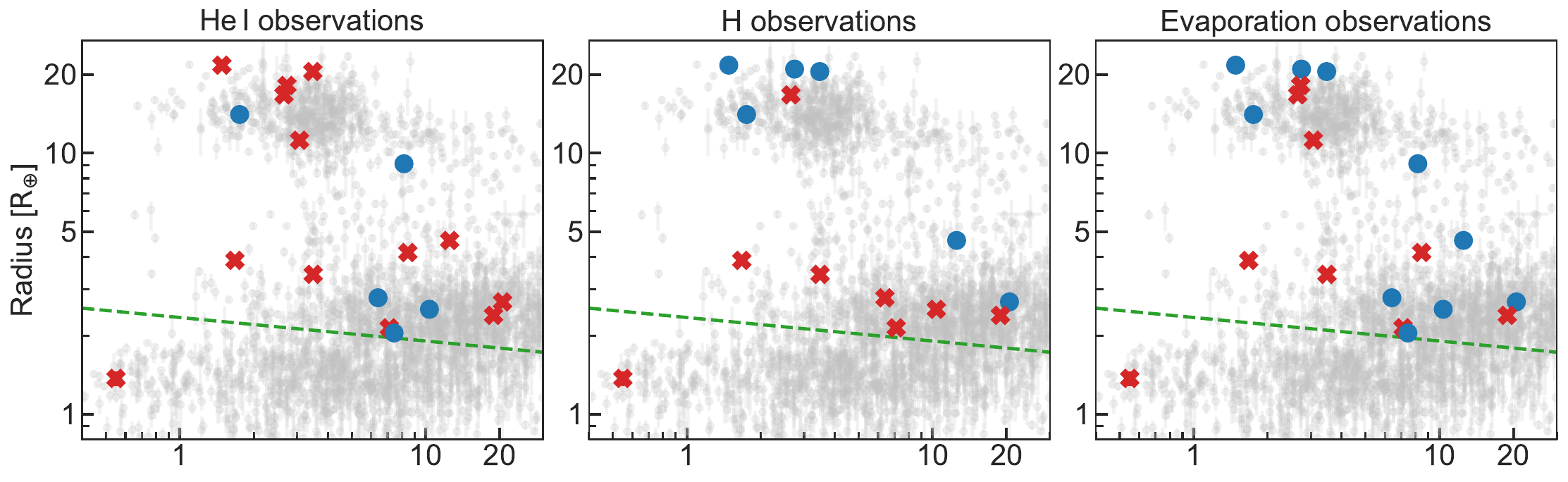}
    \includegraphics[width=\textwidth]{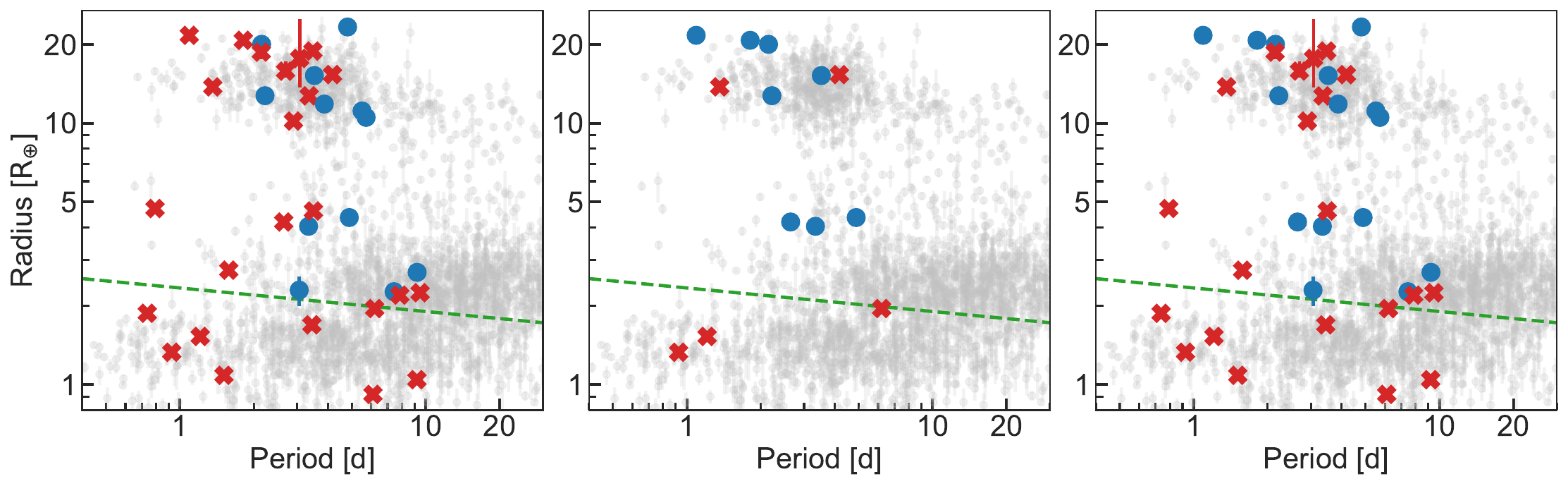}
    \caption{ Radius-period diagrams for planets with ages $\leq$\,1\,Gyr (top panels) and $>$\,1\,Gyr (bottom panels) with \ion{He}{i} (left panels) and H (H$\alpha$ or Ly$\alpha$, middle panels) observations, and evaporation (combination of \ion{He}{i} and/or \ion{H}{} observations, right panels). Detections and non-detections are marked as blue circles and red crosses, respectively. The radius gap \citep{VanEylen2018_radius_gap} is marked as a dashed green line. The grey points represent all known planets with period and radius determined with a precision better than 25\% (data from NASA Exoplanet Archive).}
    \label{Fig: EVAPORATION Period-Radius}
\end{figure*}

\begin{figure*}[h!]
    \centering
    \includegraphics[width=\textwidth]{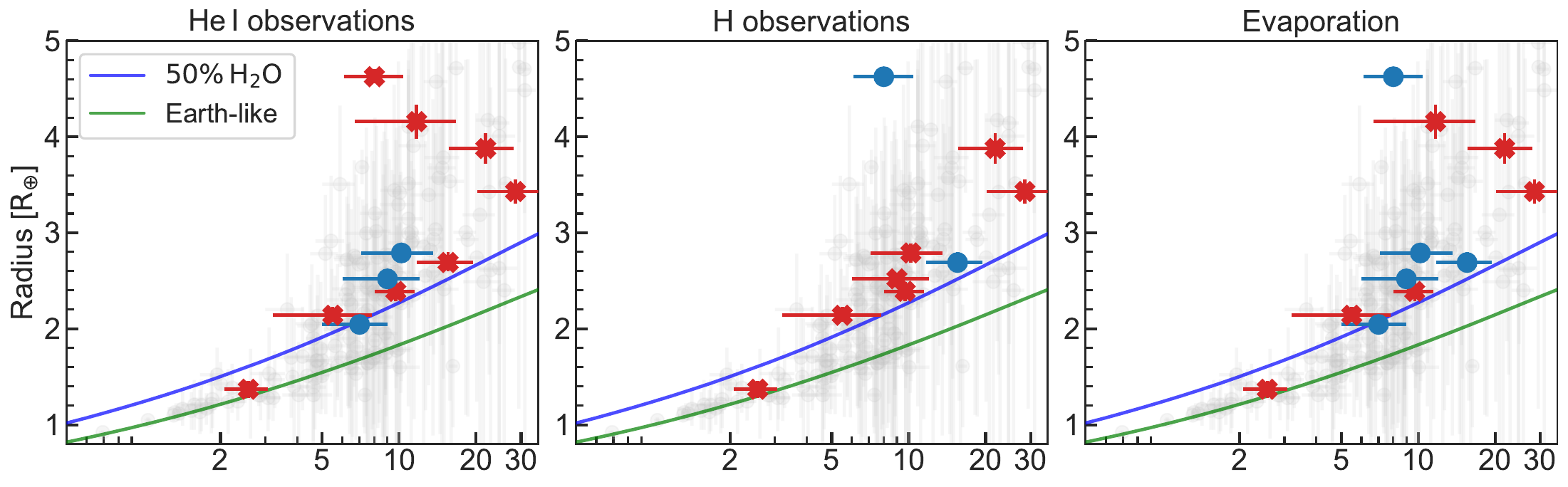}
    \includegraphics[width=\textwidth]{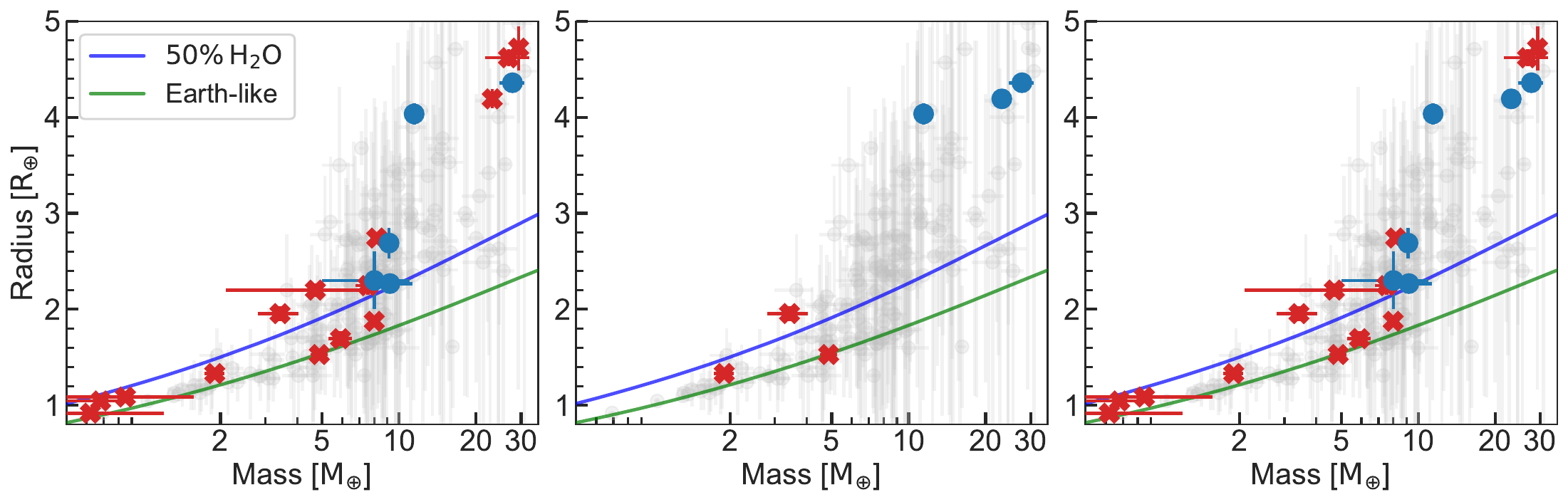}
    \caption{ Radius-mass diagrams for planets with ages $\leq$\,1\,Gyr (top panels) and $>$\,1\,Gyr (bottom panels) with \ion{He}{i} (left panels) and  H (H$\alpha$ or Ly$\alpha$, middle panels) observations, and evaporation (combination of \ion{He}{i} and/or \ion{H}{} observations, right panels). Detections and non-detections are marked as blue circles and red crosses, respectively. The solid lines are the theoretical models from \citet{Zeng_models} for Earth-like (green) and 50\%\,H$_2$O+50\%\,rocky (blue) compositions. The grey points represent all known planets with mass and radius determined with a precision better than 20\% (data from NASA Exoplanet Archive).}
    \label{Fig: EVAPORATION Mass-Radius}
\end{figure*}

Figure\,\ref{Fig: EVAPORATION AGE} displays the planetary radius versus stellar age, with the planets colour-coded according to their evaporation measurements. We considered the \ion{He}{i} triplet and/or H$\alpha$ absorption detections (and Ly$\alpha$ in some particular cases) as proxies of evaporation signs. Thus, an evaporation detection means that either \ion{He}{i} or H$\alpha$ or both have been positively detected.

Figure\,\ref{Fig: Age detections} is similar to Figure\,\ref{Fig: EVAPORATION AGE} but focuses only on the \ion{He}{i} triplet, and presents a comparison between the \ion{He}{i} EW signals and the stellar age. The planets are marked and colour-coded according to their \ion{He}{i} results. \cite{Allart_He_survey} reported that a trend between \ion{He}{i} and stellar age was noticeable in their sample of eleven planets, however the authors refrained from further conclusions due to their small sample size and the lack of precise stellar ages.
Although in this work we used a larger sample (18 young and 35 old planets, including detections and non-detections, but less homogeneous than the \citealp{Allart_He_survey} one), we do not notice a clear correlation between \ion{He}{i} EW and stellar age.

The first 100\,Myr are critical for the planetary atmospheric evolution, according to photo-evaporation models (see Introduction in Sect.\,\ref{Sect: Introduction}).
Unfortunately, there are only a few planets in Figures\,\ref{Fig: EVAPORATION AGE} and \ref{Fig: Age detections} with ages $\lesssim$\,100\,Myr. The youngest objects have ages of $\sim$20\,Myr (AU\,Mic\,b and V1298\,Tau\,c \& b) and then there is a lack of \ion{He}{I} and H$\alpha$ observations until $\sim$100\,Myr (WASP-80\,b, and K2-77\,b). WASP-80\,b, with a formerly calculated age of $<$200\,Myr \citep{Salz2015_WASP80_age}, has no detection of \ion{He}{I} triplet. However, a possible older age of WASP-80 (see Section\,\ref{Subsec: OTHER YOUNG PLANETS}) and the calculated $\log L_{\rm XUV}$ at the planet separation (Table\,\ref{table - DATABASE}) could explain the lack of detection of the \ion{He}{i} triplet. AU\,Mic\,b non-detection is harder to reconcile, but different scenarios can be invoked (e.g. stellar activity masking possible detections or H/He ratio). \ion{He}{i} and H$\alpha$ observations of V1298\,Tau\,b \& c resulted in non-conclusive measurements, mainly due to the strong stellar activity levels of the young host star. However, the masses derived for planets b and e indicate that their densities are similar to older planets, suggesting they are not inflated and that they contracted faster than expected (\citealp{V1298Tau_c_Alejandro}).

While the non-detections of WASP-80\,b, AU\,Mic\,b, V1298\,Tau\,b \& c do not seem to fit the predictions from photo-evaporation models (\citealp{Lopez_2012,Owen_Jackson_2012, Owen_Wu_2013, Owen_Wu_2017, Owen_Lai_2018, Dawson_Johnson_2018}), the sample of $\lesssim$\,100\,Myr-old planets is too small to draw conclusions. A larger sample of $\lesssim$150-Myr-old planets is needed to explore the photo-evaporation timescales.

From Fig.\,\ref{Fig: EVAPORATION AGE} and \ref{Fig: Age detections}, the youngest planet with \ion{He}{i} detection is TOI-1268\,b (245$\pm$135\,Myr), and the youngest planet with H$\alpha$ detection is MASCARA-2\,b (200$^{+100}_{-50}$\,Myr). The number of planets with evaporation detections increases until peaking at $\sim$2\,Gyr, and then it is roughly constant, but this is related to the age distribution in our sample. The proportion between detections and non-detections remains constant (within error bars) with age, except at old ages (>5\,Gyr) when non-detections dominate (see Figs.\,\ref{Fig: EVAPORATION AGE}, and \ref{Fig: Age detections}). This non-detections domination is more pronounced when considering only the \ion{He}{i} observations (Fig.\,\ref{Fig: Age detections}), and extended over all stellar ages, except over the 1--3\,Gyr range.

Our results on evaporation timescales agree with the conclusions from \citet{ParkeLoyd2020_photoevaporation} on the radius gap, and \citet{Christiansen2023_HotNeptunes_young} on the hot Neptune population (see Fig.\,6 in their work), and photo-evaporation mechanism is not more supported than the core-powered one. The fact that we do not see a decrease in the evaporation tracers, until very old ages ($\sim$5\,Gyr, see Figs.\,\ref{Fig: EVAPORATION AGE}, and \ref{Fig: Age detections}), might be marginally more consistent with the core-powered timescale of Gyr (\citealp{Ginzburg2016_corepower, Ginzburg2018_corepower, Gupta2020_corepower}).

\subsubsection{Radius--period diagram across stellar age}

Figure\,\ref{Fig: EVAPORATION Period-Radius} compares the population of young ($\leq$1\,Gyr, top panels) and old ($>$1\,Gyr, bottom panels) planets in a radius-period diagrams for only \ion{He}{i} triplet observations, only H (mainly H$\alpha$ with few Ly$\alpha$) observations, and both evaporation proxies. Figures\,\ref{Fig: EVAPORATION AGE}, \ref{Fig: Age detections}, and \ref{Fig: EVAPORATION Period-Radius} show no clear differences for the evaporation of the gas giant planets before and after 1\,Gyr. The detections of evaporation are evenly spread over the stellar ages, with more preference of H$\alpha$ detection than from \ion{He}{i} for the young gas giants.  We note that there are no evaporation detections of young and old planets below the radius gap  or for Earth-like planets, supporting rocky planets are not under extreme atmospheric mass-loss processes, at least after $\sim$300\,Myr which is the age of the youngest rocky planet in our dataset (TOI-1807\,b).

\citet{Allan2023_theory_He_young_planets} simulated how the \ion{He}{i} triplet planetary signal from a highly irradiated ($a_{\rm p}$\,=\,0.045\,AU) gas giant planet ($M_{\rm p}$\,=\,0.3\,M$_{\rm J}$) around a K-dwarf star changes with the stellar age. They assumed a typical H/He ratio of 98/2 (\citealp{Lampon_2020_HD209, Lampon_2021_HD187_GJ3470, Orell2023}). From their simulations, they derive excess absorption peaks of 4--7\,\% for young (16--550\,Myr) Hot Jupiters, while at 5\,Gyr the excess absorption would be of $\sim$1.5\%.
They state that a close-in ($a_{\rm p}$\,$<$\,0.1\,AU) planet with a radius of 1-2\,R$_{\rm J}$ transiting a $<$150-Myr-old K dwarf star would be the best target to test their evolution models.
Although the \ion{He}{i} triplet detection is favoured by the extreme radiation in XUV range ($\lambda$\,<\,504\,nm; \citealp{JorgeSanz2008}) and the TS simulations from \citet{Allan2023_theory_He_young_planets}, none of the three 20-Myr-old planets analysed here present a clear detection of \ion{He}{i}. 
WASP-80\,b ($a_{\rm p}$\,=\,0.035\,AU, $R_{\rm p}$\,=\,1\,R$_{\rm J}$, $M_{\rm p}$\,=\,0.5\,M$_{\rm J}$, $<$200\,Myr, spectral type $\sim$K7V) is the closest planet from Table\,\ref{table - DATABASE} to the simulated one in \citet{Allan2023_theory_He_young_planets} but it has very low \ion{He}{i} upper limits (\citealp{WASP-80_He_Fossati, Allart_He_survey}). WASP-52\,b might be in agreement with the simulations, although the largest \ion{He}{i} excess absorptions to date come from older planets (e.g. HAT-P-67\,b, \citealp{HAT-P-67b_He_toroide}; HAT-P-32\,b, \citealp{HAT-P-32b_Ha_He_Czesla2022}; and WASP-107\,b, \citealp{WASP-107b_He_Kirk2020}).
In Fig.\,\ref{Fig: EVAPORATION Period-Radius} WASP-52\,b is the only young gas giant planet with a \ion{He}{i} detection. The ratio of \ion{He}{i} detections to non-detections seems larger for old rather than young planets, but there are fewer differences when comparing the H detections to non-detections.

\subsubsection{Small planet evaporation across stellar age}

Figure\,\ref{Fig: EVAPORATION Mass-Radius} shows the radius-mass diagram for small planets ($R_{\rm p}$\,$<$5\,R$_{\oplus}$ and $M_{\rm p}$\,$<$\,30\,M$_{\oplus}$), also comparing evaporation proxies for young and old planets. Planets with $R_{\rm p}$\,$\sim$\,1.5--3\,R$_{\oplus}$ fall in a degenerated region of the mass-radius diagram, and their bulk compositions can be consistent with a large range of models, from water worlds (planets with a large water mass fraction) to  planets with rocky cores with H/He envelopes (\citealp{Zeng_models}). For these planets, in Figure\,\ref{Fig: EVAPORATION Mass-Radius}, we found a mixture of detections and non-detections, with no difference between young and old planets.

Planets with $R_{\rm p} > $3\,R$_{\oplus}$ are well above the water-rich composition line and are supposed to be gaseous with very light envelopes and low densities (\citealp{Rafa_2022Sci}). In our sample there are four young and five old puffy planets with evaporation observations (see right panels of Fig.\,\ref{Fig: EVAPORATION Mass-Radius}). While only 1 in 4 young planets has an evaporation detection, 3 out of 5 old planets have a detection, hinting that atmospheric escape of puffy sub-Neptunes is stronger at ages older than $\sim$\,1\,Gyr. Still, the numbers are small, and a larger sample is needed to confirm these findings.

For the planets above the radius valley, \citet{HeEnhancedNeptune} predicted a He enhancement due to the diffusive separation of the atmospheric constituents where the He and metals are preferentially retained while H is evaporated. The timescale of this mechanism is comparable to the planet lifetime ($\sim$10\,Gyr) and planets with $\lesssim$1\,Gyr have not had time to show the effects of favoured H evaporation (\citealp{HeEnhancedNeptune}). So, our young planet population is too young to suffer this differential escape process, and even some of our old planets could be considered young as well.
Then, the long timescale might explain why Figure\,\ref{Fig: EVAPORATION Mass-Radius} does not show more \ion{He}{i} detections for the old waterworlds than for the young ones.
\citet{HeEnhancedNeptune} predict that TOI-1235\,b will have an He enhanced atmosphere at 10\,Gyr, but \cite{Krishnamurthy2023_TOI1235b_GJ9827bd} put a very restricted \ion{He}{i} upper limit, although its age is poorly constrained (5$^{+5}_{-4.4}$\,Gyr). The other planets listed in \citet[Table\,1]{HeEnhancedNeptune} have no H/He observations to test their predictions.

Although \citet{HeEnhancedNeptune} focus only on the small planet population ($\lesssim$3\,$R_{\oplus}$), the diffusive separation and the subsequent preferential H evaporation could be the mechanism to explain some observational results from  Fig.\,\ref{Fig: EVAPORATION Period-Radius} and \ref{Fig: EVAPORATION Mass-Radius} regarding large and intermediate-sized planets.
i) There are no \ion{He}{i} detections in young puffy planets but there are for old ones. TOI-1136\,d is the only young puffy planet with H$\alpha$ detection, and the only three old puffy planets with H observations both show H evaporation.
ii) WASP-52\,b is the only young gas giant with \ion{He}{i} detection (and also has H$\alpha$ detection) while there are several detections on old gas giants, and H$\alpha$ is extensively detected in young and old gas giant planets.

\subsection{Cosmic shoreline from \ion{He}{i} observations}
\label{Subsec: cosmic shoreline}

\begin{figure*}[h]
    \centering
    \includegraphics[width=\hsize]{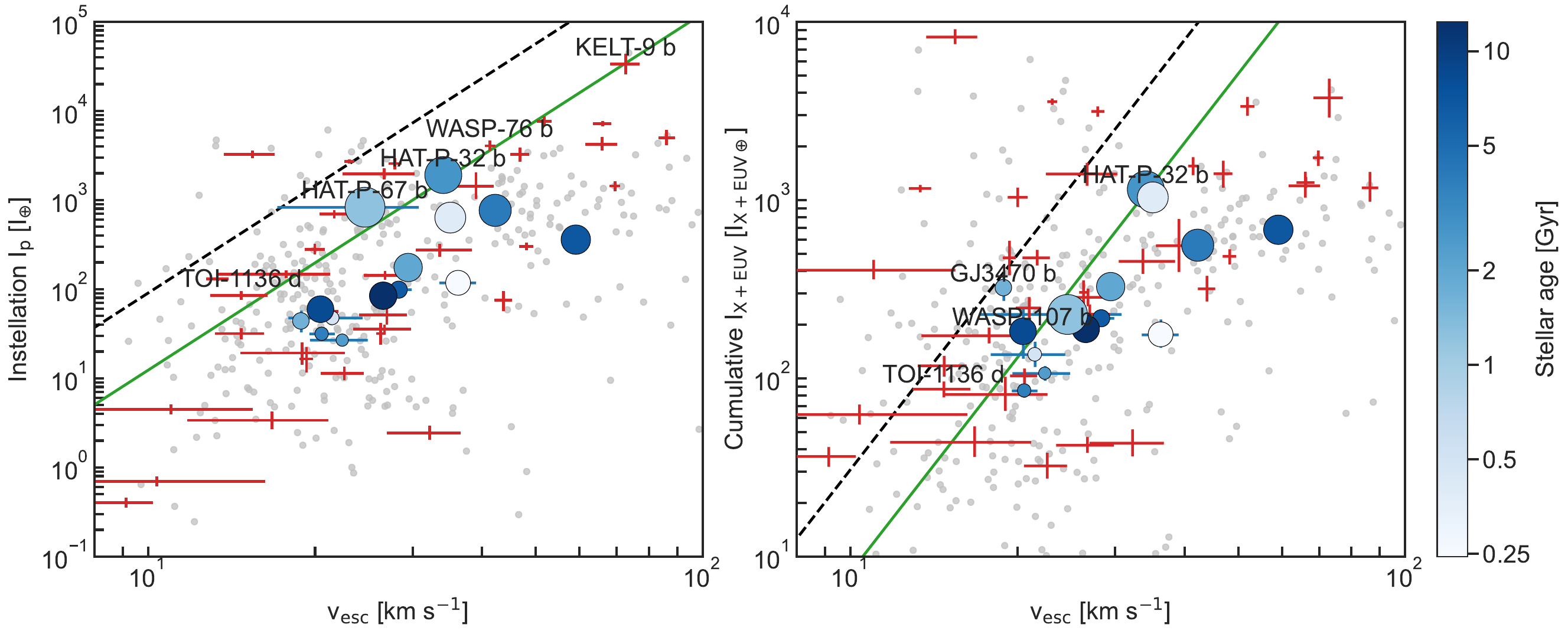}
    \caption{ Cosmic shoreline compared to \ion{He}{i} observations. \ion{He}{i} non-detections are denoted by red error bars with no marker. \ion{He}{i} detections are circles scaled by their R$_{\rm p}$ and colour-coded by their stellar age (lateral colour bar).
    Instellation relative to Earth vs escape velocity (\textit{left panel}) and cumulative instellation in the XUV range relative to Earth vs escape velocity (\textit{right panel}) graphics.
    We marked the cosmic shoreline, as I\,$\propto$\,\vesc$^4$, from \citet[solid green line]{Shoreline_2017} and constrained from the observations (dashed black line).
    Planets sitting above or on the cosmic shore and with H/He detections are labelled.
    \label{Fig: cosmic shoreline}}
\end{figure*}

The cosmic shoreline \citep{Shoreline_1998} is an empirical division found in the Solar System bodies that splits them between those that retain a certain amount of atmosphere and those that are purely bare rocks without atmosphere. The division was extended to the extrasolar planet population with success by \citet{Shoreline_2017}. Determining precisely the cosmic shoreline is important as it has often been used to predict the existence of an atmosphere for newly discovered planets, and to argue for atmospheric characterisation follow-up (e.g. with the James Webb Space Telescope). Observationally, extended atmospheres are relatively easy to detect via H and \ion{He}{I} absorptions. Here, we use these observations to better constrain the cosmic shoreline.

\citet{Shoreline_2017} found a power law between planet bolometric instellation ($I_{\rm p}$) and planet velocity escape \vesc\ that follows $I_{\rm p}$\,$\propto$\,\vesc$^4$. However, the radiation that shapes the exoplanet atmosphere is the X-rays and EUV stellar flux received by the planet during its early stages. \citet{Shoreline_2017} assumed the approximation of X-ray luminosity saturation (\citealp{Jackson2012_100Myr}) to estimate the total extreme radiation (X+EUV radiation) received by the planet. Following \citet[Eq.\,27]{Shoreline_2017}, we compute the cumulative X-rays and EUV instellation ($I_{\rm X+EUV}$; considering $\lambda$\,<\,100\,nm) as a function of the planet instellation ($I_{\rm p}$) and the stellar bolometric luminosity ($L_{\star}$):
\begin{equation}
I_{\rm X+EUV} = I_{\rm p} \, (L_{\star}/L_{\sun})^{-0.6}
\label{eq: total instellation XUV}
.\end{equation}
We computed \vesc\ as
\begin{equation}
v_{\rm esc} = \sqrt{ 2  \, G M_{\rm p}/ R_{\rm p} }
\label{eq: vesc}
,\end{equation}
where G is the universal gravitational constant.
The empirical relation found between $I_{\rm X+EUV}$ with \vesc\ also follows the same power law ($I_{\rm X+EUV}$\,$\propto$\,\vesc$^4$), as for $I_{\rm p}$.

\citet{Shoreline_2017} presented the relationships between $I_{\rm p}$ and \vesc , and between $I_{\rm X+EUV}$ and \vesc\ for the planets and small bodies of the Solar System along with the exoplanet population.
To test the predictions of the cosmic shoreline, we compared the proposed empirical relationships to the actual \ion{He}{i} detections. Figure\,\ref{Fig: cosmic shoreline} reproduces Figures\,1 and 2 from \citet{Shoreline_2017}, and includes also the \ion{He}{i} observations.
Because the cosmic shorelines are empirical relations, only the slope in logarithmic scale is determined. Thus, we get from \citet[Figs.\,1 and 2]{Shoreline_2017} the independent terms to plot the empirical equations.

By definition, all \ion{He}{i} (and H$\alpha$) detections must be below the cosmic shorelines plotted in Fig.\,\ref{Fig: cosmic shoreline}. However, we find seven gas planets with \ion{He}{i} (or H$\alpha$) detections in the region of supposedly bare rocky planets. Those are the ones labelled in Fig.\,\ref{Fig: cosmic shoreline} panels.
In Fig.\,\ref{Fig: cosmic shoreline} left panel, HAT-P-32\,b and HAT-P-67\,b are clearly located above the line, although their 1$\sigma$ uncertainties fall in the limit of the cosmic shoreline. TOI-1136\,d, WASP-76\,b, and KELT-9\,b which have \ion{He}{i} non-detections but H$\alpha$ detections, are also over or above the cosmic shoreline.
Moreover, in Fig.\,\ref{Fig: cosmic shoreline} right panel, HAT-P-32\,b and WASP-107\,b fall over the line, and GJ\,3470\,b and TOI-1136\,d are clearly above the cosmic shoreline for energy-limited regime. With the exception of TOI-1136\,d, the planets that contradict the shoreline are not young planets, with stellar ages $>$1\,Gyr.

\cite{Shoreline_2017} stated that the $I$\,$\propto$\,\vesc$^4$ and $I_{\rm X+EUV}$\,$\propto$\,\vesc$^4$ lines are `drawn in by hand to guide the eye'. Thus, the evaporation detections above the shoreline can help to constrain the independent terms of those by-hand equations. For that purpose, we considered the planet whose uncertainties have the largest separation from the by-hand line to compute the limits of the cosmic shoreline. We took the coordinates of the extreme uncertainty as a point to calculate the line equation. We took HAT-P-67\,b for the $I$\,$\propto$\,\vesc$^4$ line, and GJ\,3470\,b for $I_{\rm X+EUV}$\,$\propto$\,\vesc$^4$ although very similar results were obtained with TOI-1136\,d. Then, it is trivial to get the line equation with the slope and one point. The cosmic shoreline equations constrained from the evaporation detections are
\begin{equation}
\log(I/I_{\oplus}) = 4 \log(v_{\rm esc} \,[\rm {km\,s^{-1}}]) \, -2.04
\label{eq: I ves equation}
\end{equation}
and
\begin{equation}
\log(I_{\rm X+EUV}/I_{\rm X+EUV \oplus}) = 4 \log(v_{\rm esc} \,[{\rm km\,s^{-1}}]) \, -2.51
\label{eq: Ixuv ves equation}
.\end{equation}
They are shown in Figure\,\ref{Fig: cosmic shoreline} left and right panels as dashed black lines, respectively. 
In both cases, the cosmic shoreline moved to higher radiation levels reducing the amount of exoplanets with no atmosphere.
Further atmospheric observations and more precise measurements of the planets close to the shoreline will allow a better constraint of the observed cosmic shoreline.

\subsection{Observed versus theoretical mass-loss rates}
\label{Subsec: mass-loss rate}

\begin{figure}
    \centering
    \includegraphics[width=\hsize]{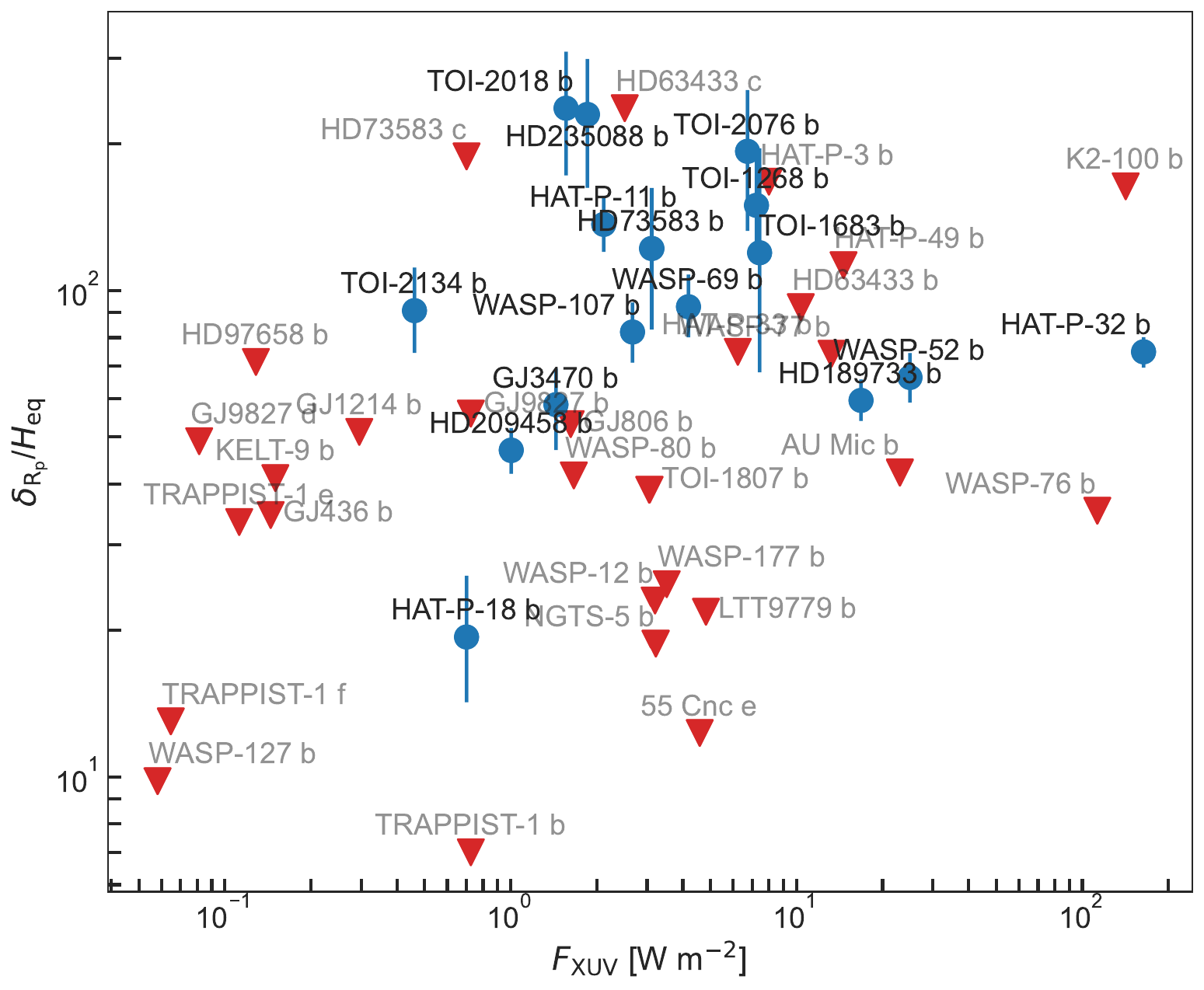}
    \caption{ \ion{He}{i} transmission signal strength for the planets with detections (blue circles with error bars  in black) and non-detections (red down-pointing triangles  in grey) as a function of the stellar \fxuv\ ($\lambda$\,=\,5\,--\,504\,\AA) at the planet distance.
    We show the equivalent height of the \ion{He}{I} atmosphere, $\delta_{Rp}$, normalised by the atmospheric scale height, $H_{\mathrm{eq}}$. Data from Table\,\ref{table - DATABASE} and references therein. 
    }
    \label{Fig: Lisa plot}
\end{figure}

Since the first \ion{He}{I} detection in planetary atmospheres, there have been many attempts to predict the presence of this line using other observable planetary parameters. \citet{Nortmann_WASP-69_He} performed one of the first attempts by computing $\delta_{Rp}/H_{\rm eq}$. That is: the equivalent height of the \ion{He}{i} absorbing atmosphere $\delta_{Rp}$\,=\,$( R_{\rm p}^{2} + D_{\rm He} \, R_{\star}^{2} )^{1/2} - R_{\rm p}$, where $D_{\rm He}$ is the \ion{He}{i} absorption peak in \%, divided by the atmospheric scale height $H_{\rm eq}$\,=\,$(k_{\rm B} T_{\rm eq})/(\mu g_{\rm p})$, where $k_{\rm B}$, $\mu$, and $g_{\rm p}$ are the Boltzmann constant, the mean molecular mass, and planet gravity, respectively. Figure\,4 in \citet{Nortmann_WASP-69_He} presented a correlation between the \ion{He}{i} absorption signal and \fxuv\ ($\lambda$\,=\,5\,--\,504\,\AA), but with a sample of only five planets. Figure\,\ref{Fig: Lisa plot} reproduces the same diagram with the current set of 43 \ion{He}{i} observations (detections and non-detections with \fxuv\ measurements). The trend is not that clear and there is no evident pattern between detections and non-detections, maybe due to the non-uniformity of the sample (e.g. different instruments and data analysis techniques).

Rather than using the strength of the  \ion{He}{I} absorption, \citet{Vissapragada2022_Neptune_Desert}, and later \citet{TOI-2134b_Zhang}, explored the relation between the mass-loss rates derived from the observations (\mobs) as a function of \fxuv/$\rho_{\rm p}$ ($\rho_{\rm p}$ is the planet density). While \citet{Vissapragada2022_Neptune_Desert} derived the observed mass-loss rates comparing the measured excess transit absorptions with a grid of Parker wind models, \citet{TOI-2134b_Zhang} estimated the observed mass-loss rates from an order-of-magnitude method. Furthermore, \citet{TOI-2134b_Zhang} found that the theoretical maximum energy-limited mass-loss rate (\mtheory) is proportional to \fxuv/$\rho_{\rm XUV}$ ($\rho_{\rm XUV}$ is defined below). However, they reported very similar results when using \fxuv/$\rho_{\rm p}$. Despite those differences, both works found a positive correlation between \mobs\ and \fxuv/$\rho_{\rm p}$, and, their energy-limited outflow efficiencies agree within uncertainties (\citet{Vissapragada2022_Neptune_Desert}: $\eta_0$\,=\,0.41$^{+0.16}_{-0.13}$; \citet{TOI-2134b_Zhang}: $\eta_0$\,=\,0.31\,$\pm$\,0.06).

Here, we focus on the relationship between \mobs\ and \mtheory\  while adding the two new \ion{He}{i} detections presented in this work (TOI-1268\,b and TOI-2018\,b), which is the main difference with the study performed by \citet{TOI-2134b_Zhang}. For consistency, we followed the indications from \citet{Zhang_young_planets, TOI-2134b_Zhang} to compute the \mobs--\mtheory\ diagram with our set of \ion{He}{i} observations.

We made the same assumptions as in \citet{Zhang_young_planets, TOI-2134b_Zhang} to calculate $\dot{m}_{\rm obs}$  with the order-of-magnitude method as
\begin{equation}
\dot{m}_{\rm obs} = \frac{m_{\rm e} \, m_{\rm He} \, c_{\rm s} \, c^2 \, \, EW \, R_{\star} }{0.25 \, f \, e^2 \, \lambda_0^2 \, \Sigma g_{l}f_{l}} \, \propto \, EW \cdot R_{\star}
\label{eq: mass-loss observed}
.\end{equation}Equation\,\ref{eq: mass-loss observed} is computed in cgs units, where $m_{\rm e}$ is the electron mass, $m_{\rm He}$ is the He atomic mass, $c_{\rm s}$ is the sound speed (assumed to be $c_{\rm s}$\,=\,10\,km\,s$^{-1}$), $c$ is the speed of light, EW is the equivalent width of the \ion{He}{i} signal, $R_\star$ is the stellar radius, 0.25 comes from assuming that 25\,\% of mass out-flow is He (atoms or ions), $f$ is the fraction of He atoms in the metastable ground state (assumed to be $f$\,=\,10$^{-6}$), $e$ is the electron charge, $\lambda_0$ is the \ion{He}{i} wavelength (fixed to 10833.3\,\AA), $\Sigma g_{l}f_{l}$ is the sum of the product of the degeneracy and oscillator strength over the three lines of the \ion{He}{i} triplet which is 1.62 \citep{TOI-2134b_Zhang}. 

The theoretical maximum energy-limited mass-loss rate ($\dot{m}_{\rm theory}$; \citealp{Caldiroli2022_xuv_evaporation, TOI-2134b_Zhang}), assuming that all the XUV flux received by the planet is spent on evaporation, is computed as
\begin{equation}
\dot{m}_{\rm theory} = \frac{ \pi \, R^3_{\rm XUV} \, F_{\rm XUV} }{G \, M_{\rm p}} = \frac{3 \, F_{\rm XUV}}{4\, G \, \rho_{\rm XUV}} \, \propto \, F_{\rm XUV}/\rho_{\rm XUV}
\label{eq: mass-loss theory}
,\end{equation}
where $F_{\rm XUV}$ is the XUV flux ($\lambda$\,=\,5\,--\,504\,\AA), $R_{\rm XUV}$ is the planetary XUV photosphere radius, G is the gravitational constant, $M_{\rm p}$ is the planet mass, and $\rho_{\rm XUV}$ is the planet density using the $R_{\rm XUV}$. We followed the equations and indications from \citet{Rxuv_calculation, Zhang_young_planets, TOI-2134b_Zhang} to estimate $R_{\rm XUV}$ as 
\begin{equation}
R_{\rm XUV} = \frac{ R_{\rm p} }{ 1 \, + \, \beta^{-1} \ln{(\rho_{\rm atm. XUV} / \rho_{\rm atm. phot} )} }
\label{eq: R XUV}
,\end{equation}
where
\begin{equation}
\rho_{\rm atm. phot} = \frac{P \, \mu}{ k_{\rm B} \, T_{\rm eq} } \,\,\,\, ; \,\,\,\, \beta \equiv \frac{G \, M_{\rm c} \, \mu}{ R_{\rm p} \, k_{\rm B} \, T_{\rm eq} }
\label{eq: rho phot beta}
.\end{equation}
In Eqs.\,\ref{eq: R XUV} and \ref{eq: rho phot beta}, $P$ is the pressure at the white-light planet radius (assumed $P$\,=\,100\,mbar), $\mu$ is the (dimensional) mean molecular mass (we assumed $\mu$\,=\,1.3\,$\times$\,$m_{\rm H}$; $m_{\rm H}$ is the H atomic mass), $k_{\rm B}$ is Boltzmann constant, $T_{\rm eq}$ is the equilibrium temperature, $M_{\rm c}$ is the planet mass core, and $\rho_{\rm atm. XUV}$ is the planet atmosphere's density at the $R_{\rm XUV}$ layer.
We approximated $M_{\rm c}$\,$\simeq$\,$M_{\rm p}$ and used $\rho_{\rm atm. XUV}$\,=\,10$^{-15}$\,g\,cm$^{-3}$. Our computed values for $\rho_{\rm XUV}$ are consistent within errors with those presented in \citet[Table\,3]{TOI-2134b_Zhang}.


Figure\,\ref{Fig: Mdot obs vs theory} presents our computed observed and theoretical mass-loss rates, recovering the predicted positive correlation for the \ion{He}{i} detections. For a quantitative analysis, we fitted a power law as \mobs\,=\,$\eta_0$\,(\mtheory)$^{\alpha}$. We used \texttt{scipy}'s orthogonal distance regression to fit the equation $\log_{10}($\mobs)\,=\,$\alpha \log_{10}($\mtheory)\,+\,$\log_{10}(\eta_0)$.
We obtained an energy-limited efficiency of $\eta_0$\,=\,0.35\,$\pm$\,0.04, which is consistent with the previous derived values (\citealp{Vissapragada2022_Neptune_Desert}: $\eta_0$\,=\,0.41$^{+0.16}_{-0.13}$; \citealp{TOI-2134b_Zhang}: $\eta_0$\,=\,0.31\,$\pm$\,0.06). TOI-1268\,b and TOI-2018\,b positions in the diagram agree with the previous detections, and they are over the fitted line.
For the linearity between log mass-loss rates, we got $\alpha$\,=\,0.43\,$\pm$\,0.05, also consistent with \citet[$\alpha$\,=\,0.50\,$\pm$\,0.08]{TOI-2134b_Zhang}. When using $\rho_{\rm p}$ instead of $\rho_{\rm XUV}$, we get slightly larger uncertainties for $\eta_0$ and $\alpha$ but consistent within uncertainties ($\eta_0$\,=\,0.39\,$\pm$\,0.06, $\alpha$\,=\,0.42\,$\pm$\,0.06).

If the evaporation on those planets was in the energy-limited regime, they should exhibit a linear relation between \mobs\ and \mtheory. Although we computed the mass-loss rates as an order-of-magnitude and with some approximations, $\alpha$ is $\sim$11$\sigma$ apart from 1.
We find a sub-linearity relation ($\alpha \sim 0.43 < 1$) and the efficiency of the \mobs\ decreases with \mtheory, as already pointed out by both \citet{V1298Tau_He_Vissapragada2021} and \citet{TOI-2134b_Zhang}. However, this is in agreement with the theory as models predict that as the \fxuv\ increases, the photo-evaporation escape begins to lose efficiency via radiative cooling (e.g. \citealp{MurrayClay2009_xuv_evaporation, Caldiroli2022_xuv_evaporation}). The H/He ratio or different hydrodynamical regimes (e.g. \citealp{Lampon_2020_HD209,Lampon_2021_HD187_GJ3470,Lampon_2021_regimenes,Lampon_2023_varios_planetas}) may also contribute to derive a sub-linear relation.

Along the \mobs--\mtheory\ sub-linearity, five planets (TOI-2134\,b, TOI-2018\,b, HD\,235088\,b, HAT-P-11\,b, and HAT-P-18\,b) show an unphysical $>$100\,\% photo-evaporation efficiency. \mtheory , as it is computed in Eq.\,\ref{eq: mass-loss theory}, assumes the photo-evaporation scenario.
However, there are other mechanisms that could enhance the atmospheric mass-loss, namely the core-powered and the Roche lobe overflow. The core-powered mechanism may contribute in the atmospheric escape of those planets as they are relatively young (except HAT-P-11\,b and HAT-P-18\,b). However, a core-powered contribution can not be confidently assumed in those planets, as other parameters as the H/He ratio, the stellar flux or the atmospheric heating efficiency may contribute to change the derived \mobs--\mtheory\ relation and yield unrealistic efficiencies.
On the other hand, Roche lobe overflow is not expected, as these planets are not extremely close-in. Non-thermal processes (e.g. ion pick-up or sputtering) may also help to achieving a $>$100\,\% efficiency in those planets (\citealp{non-thermic_paper}).

\subsection{\ion{He}{i} dependence with the Hill radius}
\label{Subsec: Hill}

\begin{figure*}[h]
    \sidecaption
    \includegraphics[width=12cm]{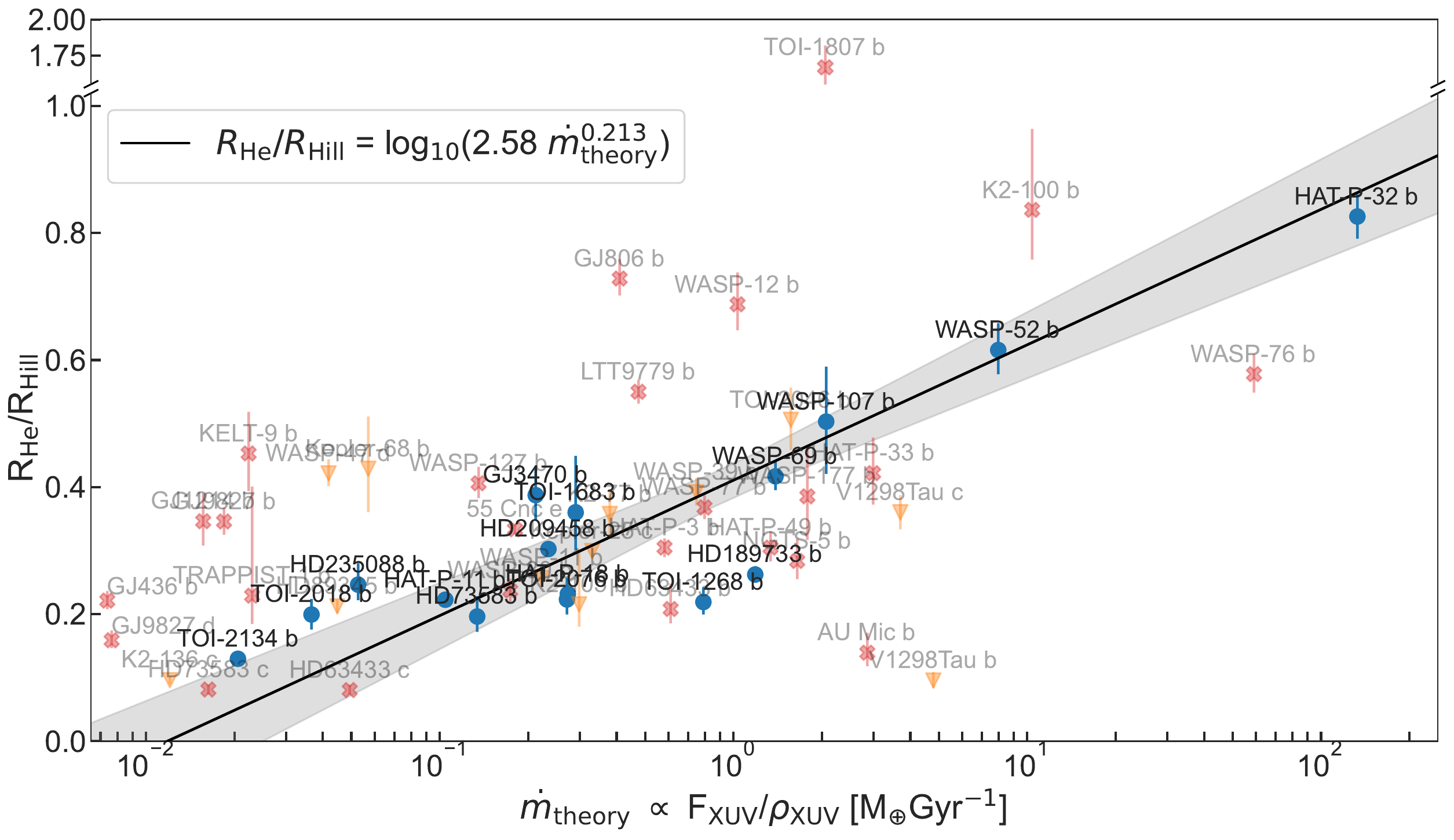}
    \caption{\label{Fig: Rhill vs m theory} Relationship between $R_{\rm He}/R_{\rm Hill}$  and the theoretical ($\dot{m}_{\rm theory}$) energy-limited mass-loss rate. We define XUV until the \ion{He}{i} ionisation range, $\lambda$\,$\in$\,5\,--\,504\,\AA. The black line indicates the fitted relationship (shown in the legend) and the dashed area the 1$\sigma$ uncertainty.
    \ion{He}{i} observations are coded as blue circles for detections, red crosses for non-detections, orange down-pointing triangles for non-conclusive. The colour-coding is according to the results from Sect.\,\ref{Subsec: mass-loss rate} (see Fig.\,\ref{Fig: Mdot obs vs theory}).
     We did not plot the error bars of $\dot{m}_{\rm theory}$, due to the very large uncertainties associated with the \fxuv\ values and its calculation.
     Every planet has its name labelled, in black for detections and the rest are in grey.
    }
\end{figure*}

Because the \fxuv\ is a critical factor for the detectability of extended \ion{He}{} atmospheres via the \ion{He}{i} triplet in the NIR (\citealp{JorgeSanz2008, Oklopcic2018_He_inici}), different relationships have been proposed, aimed at explaining the \ion{He}{i} detections, that take into account the \fxuv\ value, either directly (e.g. $\delta_{Rp}/H_{\rm eq}$--\fxuv ; see Fig.\,\ref{Fig: Lisa plot}) or indirectly (e.g. \mobs--\mtheory ; see Fig\,\ref{Fig: Mdot obs vs theory}).

Here, we explore other relationships that might explain the \ion{He}{i} signals but accounting also for the differences between system characteristics, as $\delta_{Rp}/H_{\rm eq}$.
We considered the dimensionless parameter $R_{\rm He}/R_{\rm Hill}$ which is the division between the apparent radius of the planet at the \ion{He}{i} triplet wavelength by the Hill radius of the planet. We computed the $R_{\rm He}$ as
\begin{equation}
R_{\rm He} = ( R_{\rm p}^{2} + D_{\rm He} \, R_{\star}^{2} )^{1/2}
\label{eq: R He}
,\end{equation}
where $D_{\rm He}$ is the \ion{He}{i} absorption. The equation for the $R_{\rm Hill}$ is
\begin{equation}
R_{\rm Hill} \simeq a_{\rm p} \, (1-ecc) \, ( M_{\rm p} / (3 (M_{\rm p} + M_{\star}))  )^{1/3}
\label{eq: R Hill}
.\end{equation}
As in Eq.\,\ref{eq: Planet rest frame}, we did not take into account planet's eccentricity.
Compared to \mobs , $R_{\rm He}/R_{\rm Hill}$ takes into account the \ion{He}{i} absorption peak instead of the EW of the signal, and the relation between masses of the system and the $a_{\rm p}$ instead of the star radius.
Figure\,\ref{Fig: Rhill vs m theory} presents $R_{\rm He}/R_{\rm Hill}$ as function of the \mtheory . As it is shown, \ion{He}{i} detections are aligned in the diagram, hinting that the dimensionless parameter $R_{\rm He}/R_{\rm Hill}$ is a good indicator to investigate \ion{He}{i} correlations. $R_{\rm He}/R_{\rm Hill}$--\mtheory\ have a Pearson correlation r\,=\,0.9034, and we obtained a very similar correlation with \mobs--\mtheory\ (r\,=\,0.8985).
For a more quantitative analysis, we fitted the trend that seems to be between $R_{\rm He}/R_{\rm Hill}$ and \mtheory . Using the same code as in Sect.\,\ref{Subsec: mass-loss rate}, we fitted the equation $R_{\rm He}/R_{\rm Hill}\,=\, A \, \log_{10}($\mtheory)\,+\,$\log_{10}(B)$ and we get A\,=\,0.213$\pm$0.027 and B\,=\,2.57$\pm$0.15.

We find that gas giant planets are able to fill a larger fraction of their Hill radius (or Hill sphere) than mini-Neptunes. However, \ion{He}{i} detections on mini-Neptunes have comparable $R_{\rm He}/R_{\rm Hill}$ values to those from larger planets such as HD\,189733\,b or TOI-1268\,b. This result could point to an observational bias for small planet observations where we are only detecting those exoplanets that their He signal is $R_{\rm He}/R_{\rm Hill}$\,$\gtrsim$0.2.
Moreover, compared to Fig.\,\ref{Fig: Mdot obs vs theory}, some planets move from one side to the other of the line defined by the \ion{He}{i} detections, finding planets coded as non-detections above the line or upper limits under the line. TOI-1807\,b is a clear example of a non-detection above the line. In particular, the planets with low \mtheory\ value seems to be consistently above the line, which could indicate another bias as well. But, this bias could be more related with the current capabilities of getting more precise measurements. Since they are likely low irradiated by \fxuv\ from the host star, the amount of atoms of He detectable via the NIR triplet might be also very low and very challenging thus for the current telescopes and instruments. The recent \ion{He}{i} detection on TOI-2134\,b (\citealp{TOI-2134b_Zhang}) is consistently in the lower corner on Figures\,\ref{Fig: Mdot obs vs theory} and \ref{Fig: Rhill vs m theory}, and might indicate the current limit in terms of signal detectability.

\subsection{\ion{He}{i} detections across stellar types}
\label{Subsec: stellarhost}

\begin{figure}
    \centering
    \includegraphics[width=\hsize]{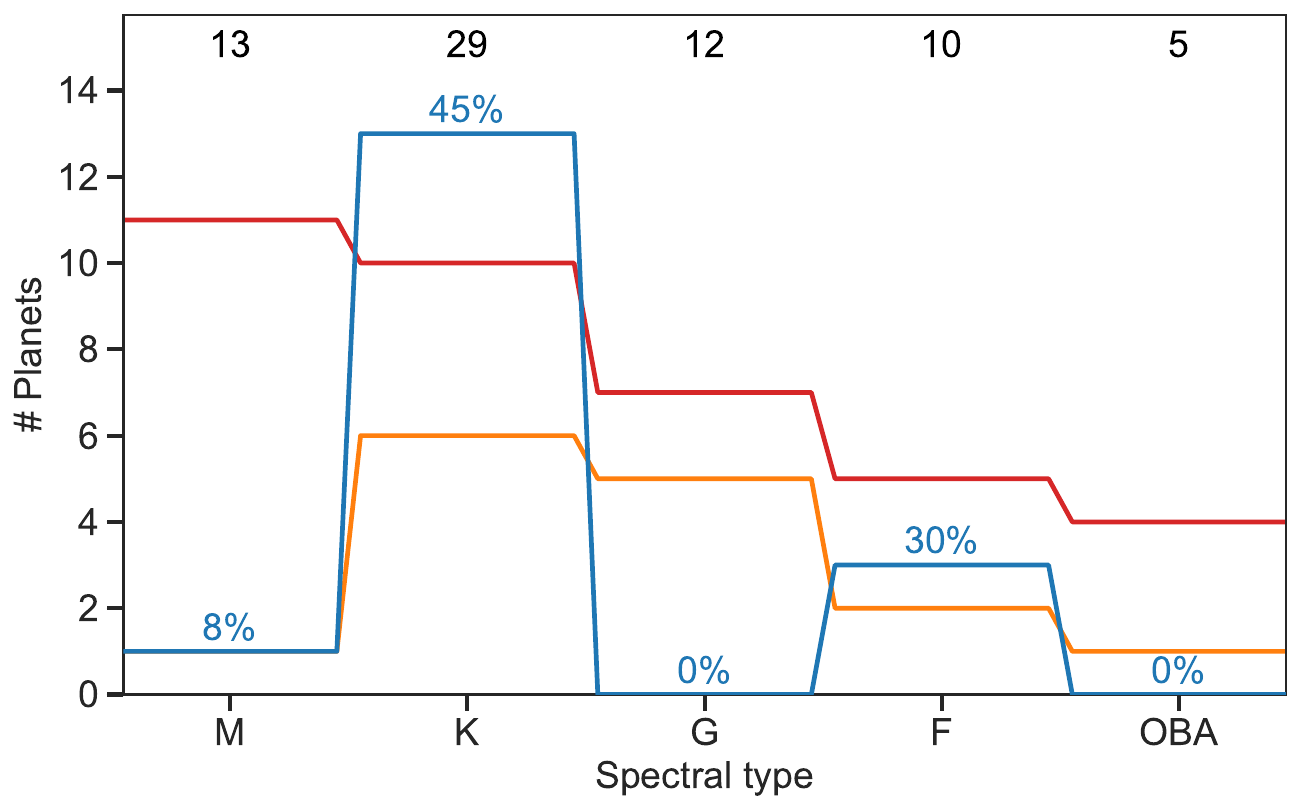}
    \caption{\label{Fig: histogram SpT} Histogram of \ion{He}{i} triplet observations as function of spectral types (M, K, G, F, and O, B, and A). The number of planets with detections, non-detections, and non-conclusive observations are shown as blue, red, and orange lines, respectively. We indicate the total number of inspected planets in black at the  top of each column. We indicate the percentage of detections for each spectral type in blue.
    }
\end{figure}

\citet{Oklopcic2019_He_Kstar} indicated that late-type stars are the most favourable host stars to detect the \ion{He}{i} triplet on their close-in transiting exoplanets, specially the K-type stars. We explored this theoretical prediction computing the histograms of detections, non-detections and non-conclusive \ion{He}{i} triplet observations as function of the different spectral types. Figure\,\ref{Fig: histogram SpT} shows that K-type exoplanet hosting stars are the most targeted ones (29 planets), but also the spectral type with a larger detection percentage (45\%). In fact, the two detections presented in this work, TOI-1268\,b and TOI-2018\,b, are both around K stars. For M-type stars, GJ\,3470\,b is the only clear \ion{He}{i} triplet detection (\citealp{He_GJ3470b_Enric2020, GJ3470b_He_Ninan2020}), along with the tentative detection on GJ\,1214\,b reported by \citealp{Orell2022}.
The three detections around F-type stars (HD\,209458\,b, \citealp{HD209458_He_Alonso}; HAT-P-67\,b, \citealp{HAT-P-67b_He_toroide}; and HAT-P-32\,b, \citealp{HAT-P-32b_Ha_He_Czesla2022}) are interesting because these planets are hot Jupiters, and HAT-P-67\,b and HAT-P-32\,b possess very extended \ion{He}{} structures (\citealp{HAT-P-67b_He_toroide, HAT-P-32b_He_Tail}). To date, there are no \ion{He}{i} detections of exoplanets orbiting G, A, B, or O stars.
To further understand the dependence on spectral type, we also computed the histogram of \ion{He}{i} detections and non-detections as function of the stellar mass (Fig.\,\ref{Fig: histogram Stellar mass}). The number of detections increases from 0.47\,$M_{\odot}$ (GJ\,3470) until peaking at 0.9\,$M_{\odot}$. Above 0.9\,$M_{\odot}$, non-detections are dominant over detections. Therefore, we can estimate a limit in host stellar mass at $\sim$0.9\,$M_{\odot}$ for the \ion{He}{i} detections, which deserves further investigation.

We note that all the \ion{He}{i} detections from Table\,\ref{table - DATABASE} are found in planets orbiting stars with $T_{\rm eff}$\,$\lesssim$\,6250\,K, which is the temperature at which the Kraft Break happens (\citealp{Kraft1967}). Exoplanet spin-orbit alignment seems to be related to the Kraft Break (e.g. \citealp{Winn2010_Kraft, Brown2017_Kraft, Attia2023_Kraft}). To date, the hottest host star with \ion{He}{i} detection is HAT-P-67 with $T_{\rm eff}$\,$\sim$\,6400\,K. The observations of other ten planets orbiting $\gtrsim$6250\,K stars resulted in upper limits only. The Kraft Break marks also the transition between stars with outer convection zones and those without. These convection zones are predicted to generate magnetic fields able to heat the chromosphere and the corona, increasing the XUV radiation (\citealp{Kraft_break_magnetic}). Therefore, stars without this heating mechanism are less likely to significantly populate the He metastable level of their exoplanets' atmospheres. The link between the stellar $T_{\rm eff}$, the flux in the XUV range and the \ion{He}{i} triplet detections deserves further investigation.

\begin{figure}
    \centering
    \includegraphics[width=\hsize]{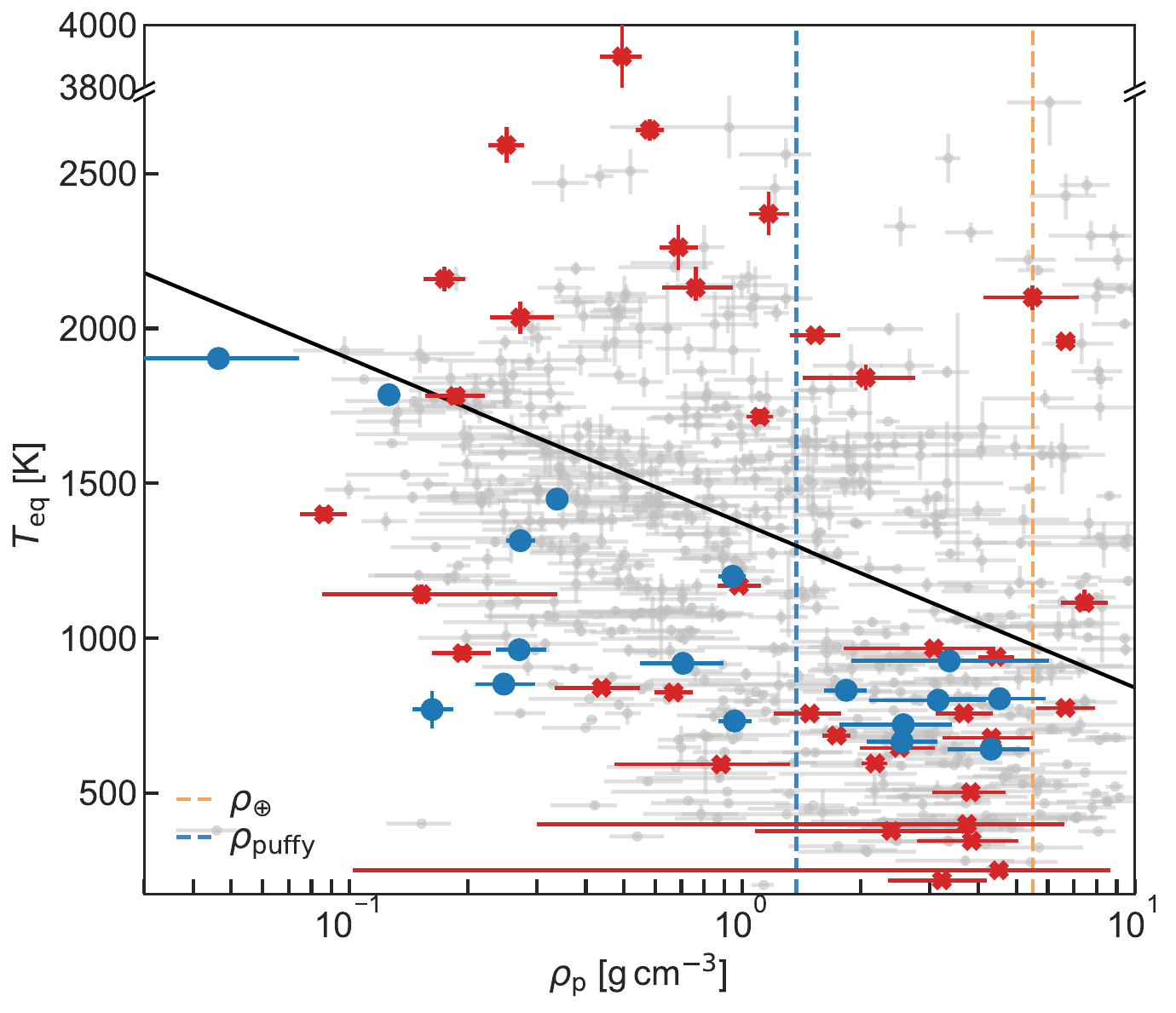}
    \caption{\label{Fig: Density Teq} Equilibrium temperature ($T_{\rm eq}$) vs planet density ($\rho_{\rm p}$) diagram for the \ion{He}{i} detections (blue circles) and non-detections (red crosses). The black line (drawn by-eye) indicates the hinted upper boundary for the \ion{He}{i} detections. The coloured dashed vertical lines indicate the Earth's density ($\rho_{\oplus}$; orange line at $\sim$5.5\,g\,cm$^{-3}$), and a representative density where the puffy planet population starts ($\rho_{\rm puffy}$; blue line at $\sim$1.3\,g\,cm$^{-3}$; \citealp{Rafa_2022Sci}). All known planets with $T_{\rm eq}$ and $\rho_{\rm p}$ determined with a precision better than 50\% are marked with grey dots (data from NASA Exoplanet Archive).
    }
\end{figure}

\subsection{\ion{He}{i} dependence on planetary parameters}
\label{Subsec: Density-Teq}

One of the aims of the MOPYS project was also to find a relation between planetary and stellar properties that could explain, and predict, the \ion{He}{i} detections and non-detections, but without the need to involve the strength of the signal itself. From the many explored relations and diagrams, only the equilibrium temperature ($T_{\rm eq}$) and the planet density ($\rho_{\rm p}$) showed some hints of such relationship. Figure\,\ref{Fig: Density Teq} presents the $T_{\rm eq}$ versus $\rho_{\rm p}$ diagram where the information about the \ion{He}{i} observations is encoded in the colour markers (blue circles are detections, red crosses are non-detections, and non-conclusive observations are not shown). Although both are planet properties, $T_{\rm eq}$ encapsulates information about the star via the $T_{\rm eff}$ and the $a_{\rm p}/R_{\star}$ parameters.

From Fig.\,\ref{Fig: Density Teq}, one clear result is that there are no \ion{He}{i} detections on exoplanets with $T_{\rm eq}$\,$\gtrsim$\,2000\,K. The hottest detection at $T_{\rm eq}$\,$\sim$\,2000\,K comes from the puffy planet HAT-P-67\,b. In fact, the data in Figure\,\ref{Fig: Density Teq} hint to a density/$T_{\rm eq}$ upper boundary for the detections. We draw `by-hand' this observational limit for a better visualisation, which follows $T_{\rm eq}$\,$\simeq$\,$-$500\,log$_{10}$($\rho_{\rm p}$) trend, approximately. If we assume that detections and non-detections are distributed randomly across the $T_{\rm eq}$--$\rho_{\rm p}$ parameter space, the probability of extracting 14 non-detections from a pool of 17 detections and 36 non-detections is only 0.16\,\%. Although it is an approximate calculation, the very low probability supports the robustness of the boundary. For planets falling above the line it is unlikely that one can detect \ion{He}{i}, as all the inspected planets resulted in non-detections. Below that line, \ion{He}{i} detections and non-detections are equally found.

At face value our results indicate that for high-density planets (from Earth-size to sub-Neptune-size) \ion{He}{i} absorptions are present only in cool atmospheres. A possible interpretation is that these type of planets quickly lose any H/He atmosphere when subjected to strong stellar instellation (assuming that $T_{\rm eq}$ is directly related at the population level to stellar flux). For low-density planets (gas giants) \ion{He}{i} detection are found up to $\sim$\,2000\,K.

Further \ion{He}{i} detections will confirm or disprove the observational boundary proposed here, and define its exact dependence on different planet parameters.

\section{Conclusions}
\label{sect: Conclusions}

This work shows the first results of the MOPYS survey, a project that aims to test the predictions of atmospheric evolution theories by confronting them with observations of evaporation proxies. We present CARMENES and GIARPS high-resolution transmission spectroscopy observations of 20 exoplanets. We analysed their atmospheres searching for \ion{He}{i} triplet absorption in the NIR, and H$\alpha$ in the VIS as well. We report two new detections of the \ion{He}{i} triplet for TOI-1268\,b and TOI-2018\,b, and a new H$\alpha$ detection for TOI-1136\,d. We also found hints of \ion{He}{i} on HD\,63433\,b, and of H$\alpha$ on \hd73\,b and c, which need further confirmation.

We complemented our target list with other planets from the literature with \ion{He}{i} triplet and H$\alpha$ (and Ly$\alpha$ for some particular planets as well) observations, for a total of 70 planets. We considered those lines as proxy of evaporation to test mass-loss theories. The main findings of our study are:

\begin{itemize}

    \item Our age distribution of evaporation detections does not favour either photo-evaporation or core-powered mass-loss as planet formation mechanisms.
    
    \item We find no trend in atmospheric evaporation with stellar age. Young ($<$1-Gyr-old) planets do not exhibit more \ion{He}{i} or H$\alpha$ detections than older planets.

    \item Evaporation (\ion{He}{i} or H$\alpha$) signals are more frequent or easier to detect for planets around stellar hosts within $\sim$1--3\,Gyr ages. 

    \item The fraction of planets that show \ion{He}{i} detections is much larger for K-type stars (45\%) than for any other spectral type.

    \item We find no evaporation detections of young and old planets below the radius gap, confirming rocky planets are not under extreme evaporation, at least after the first $\sim$300\,Myr.

    \item  We find hints that evaporation of puffy sub-Neptunes ($R_{\rm p}$\,$>$\,3\,$R_{\oplus}$) happens at  ages older than 1\,Gyr, although the number of planets is small, and a larger sample is needed to confirm this finding.
 
    \item We provide new constraints to the cosmic shoreline, by using the evaporation detections as evidences for the existence of planetary atmospheres. Our cosmic shoreline move to higher irradiation flux, reducing the parameter space of bare rocky planets.

    \item We present the He-related dimensionless parameter $R_{\rm He}/R_{\rm Hill}$, as a new valid parameter to study the \ion{He}{i} detections and upper limits.

    \item All the \ion{He}{i} detections are found in planets orbiting stars with $T_{\rm eff}$\,$\lesssim$\,6250\,K, which is the temperature at which the Kraft Break happens.

    \item We determine a statistically significant observational upper boundary for \ion{He}{i} detections in the $T_{\rm eq}$ versus $\rho_{\rm p}$ parameter space. The line decays at a rate of $T_{\rm eq}$\,$\simeq$\,$-$500\,log$_{10}$($\rho_{\rm p}$), approximately. Planets falling above that boundary are unlikely to show \ion{He}{i} absorption signals.

\end{itemize}

We encourage further evaporation observations (Ly$\alpha$, H$\alpha$, and \ion{He}{i} triplet) to increase and complete the sample presented in this work.
Although this work represents the biggest evaporation survey to date (specially focused on $<$1Gyr-old planets), some key questions of planet formation and atmospheric evaporation remained unanswered. In particular, the detection and atmospheric characterisation of very young planets ($<$100\,Myr) is key for discriminating between different mass-loss processes in planetary atmospheres. It is precisely because of their age that they are the most difficult planets to analyse their atmospheres.


\begin{acknowledgements}

CARMENES is an instrument at the Centro Astron\'omico Hispano en Andaluc\'ia (CAHA) at Calar Alto (Almer\'{\i}a, Spain), operated jointly by the Junta de Andaluc\'ia and the Instituto de Astrof\'isica de Andaluc\'ia (CSIC).

The authors wish to express their sincere thanks to all members of the Calar Alto staff for their expert support of the instrument and telescope operation.

CARMENES was funded by the Max-Planck-Gesellschaft (MPG), 
the Consejo Superior de Investigaciones Cient\'{\i}ficas (CSIC),
the Ministerio de Econom\'ia y Competitividad (MINECO) and the European Regional Development Fund (ERDF) through projects FICTS-2011-02, ICTS-2017-07-CAHA-4, and CAHA16-CE-3978, 
and the members of the CARMENES Consortium 
(Max-Planck-Institut f\"ur Astronomie,
Instituto de Astrof\'{\i}sica de Andaluc\'{\i}a,
Landessternwarte K\"onigstuhl,
Institut de Ci\`encies de l'Espai,
Institut f\"ur Astrophysik G\"ottingen,
Universidad Complutense de Madrid,
Th\"uringer Landessternwarte Tautenburg,
Instituto de Astrof\'{\i}sica de Canarias,
Hamburger Sternwarte,
Centro de Astrobiolog\'{\i}a and
Centro Astron\'omico Hispano-Alem\'an), 
with additional contributions by the MINECO, 
the Deutsche Forschungsgemeinschaft (DFG) through the Major Research Instrumentation Programme and Research Unit FOR2544 ``Blue Planets around Red Stars'', 
the Klaus Tschira Stiftung, 
the states of Baden-W\"urttemberg and Niedersachsen, 
and by the Junta de Andaluc\'{\i}a.

Based on observations collected at the Centro Astron\'omico Hispano-Alem\'an (CAHA) at Calar Alto, proposals 21B-3.5-004, 22A-3.5-007, 22B-3.5-009, 23A-3.5-009, and 23B-3.5-002, operated jointly by Junta de Andaluc\'ia and Consejo Superior de Investigaciones Cient\'ificas (IAA-CSIC). We used data from the CARMENES data archive at CAB (CSIC-INTA).

We acknowledge the use of the ExoAtmospheres database during the preparation of this work

This work is partly supported by JSPS KAKENHI Grant Number JPJP24H00017 and JSPS Bilateral Program Number JPJSBP120249910.
This paper is based on observations made with the MuSCAT2 instrument, developed by ABC, at Telescopio Carlos S\'anchez operated on the island of Tenerife by the IAC in the Spanish Observatorio del Teide.

We acknowledge financial support from the Agencia Estatal de Investigaci\'on (AEI/10.13039/501100011033) of the Ministerio de Ciencia e Innovaci\'on and the ERDF ``A way of making Europe'' through projects
PID2022-137241NBC4[1:4],
PID2019-109522GB-C5[1:4],
PID2021-125627OB-C31,
PID2022-141216NB-I00,
Ariel Postdoctoral Fellowship program of the Swedish National Space Agency (SNSA),
and the Centre of Excellence ``Severo Ochoa'' and ``Mar\'ia de Maeztu'' awards to the Instituto de Astrof\'isica de Canarias (CEX2019-000920-S), Instituto de Astrof\'isica de Andaluc\'ia (CEX2021-001131-S) and Institut de Ci\`encies de l'Espai (CEX2020-001058-M).

J.O-M agraeix el recolzament, suport i \`anims que sempre ha rebut per part de Padrina Conxa, Padrina Merc\`e, Jeroni, Merc\`e i m\'es familiars i amics.
J.O.M. gratefully acknowledge the inspiring discussions with Maite Mateu, Alejandro Almod\'ovar, and Joan Perell\'o, and the support from Guillem, Benet, and Montse.
This research has made use of resources from VallAlbaida-Mallorca collaboration. J.O.M. acknowledges the contributions and patience of Jorge Terol Calvo, i molt especialment a tu, Yess.


L.N. and F.L. acknowledge the support by the Deutsche Forschungsgemeinschaft (DFG, German Research Foundation) – Project number 314665159.
E.K. and S.H.A. acknowledge the support from the Danish Council for Independent Research through grant No.2032-00230B.
E.N. acknowledges the support by the DFG Research Unit FOR2544 ``Blue Planets around Red Stars''.
S.C. acknowledges the support of the DFG priority program SPP 1992 ``Exploring the Diversity of Extrasolar Planets'' (CZ 222/5-1).

\end{acknowledgements}


\bibliographystyle{aa}
\bibliography{references}


\begin{appendix}
\label{Sec:Appendix}

\section{Additional figures}
\label{App: Additional figures}

\begin{figure*}[h!]
    \centering
    \includegraphics[width=\hsize]{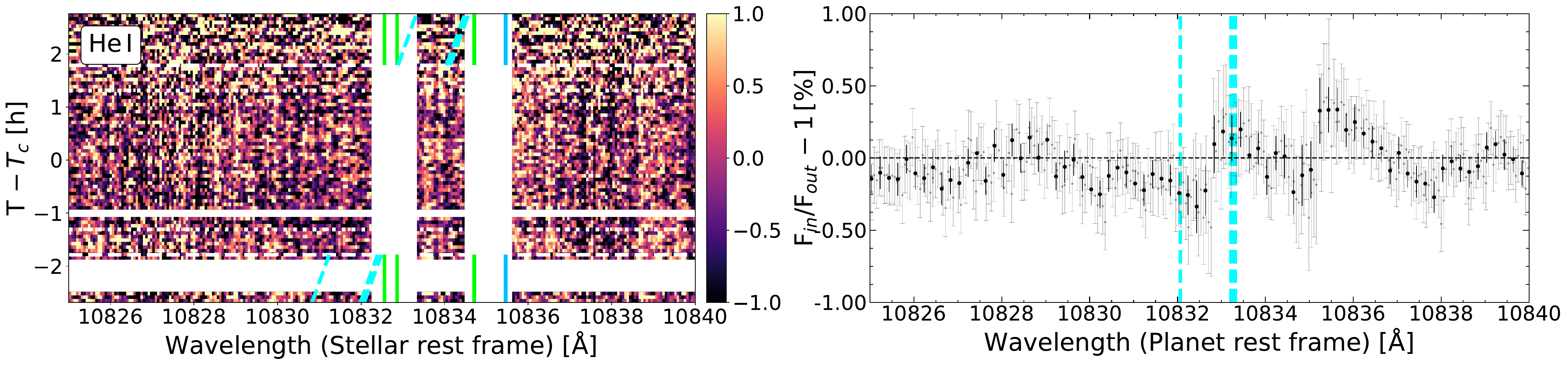}
    \caption{\label{Fig: TS MASCARA-2}
    Same as Fig.\,\ref{Fig: TS K2-100}, but  for MASCARA-2\,b observations with CARMENES.
    }
\end{figure*}

\begin{figure*}[h!]
    \centering
    \includegraphics[width=\hsize]{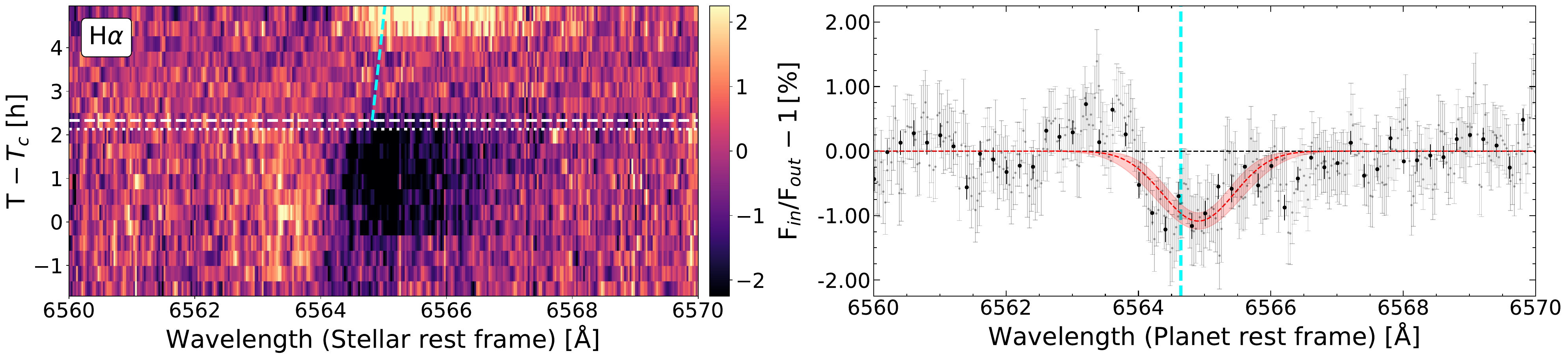}
    \includegraphics[width=\hsize]{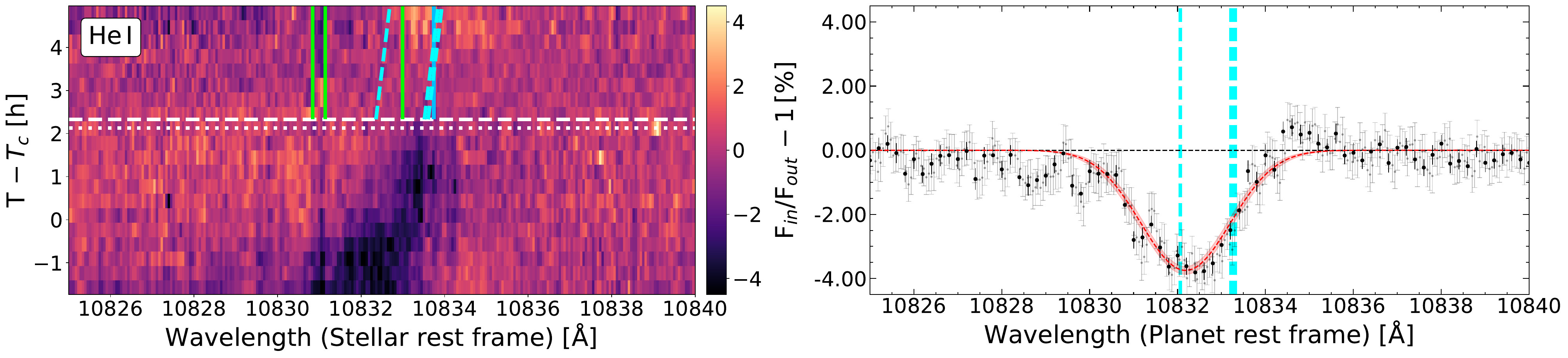}
    \caption{\label{Fig: TS V1298Tau c}
    Same as Fig.\,\ref{Fig: TS K2-100}, but  for V1298\,Tau\,c observations with CARMENES.
    }
\end{figure*}

\begin{figure*}[h!]
    \centering
    \includegraphics[width=\hsize]{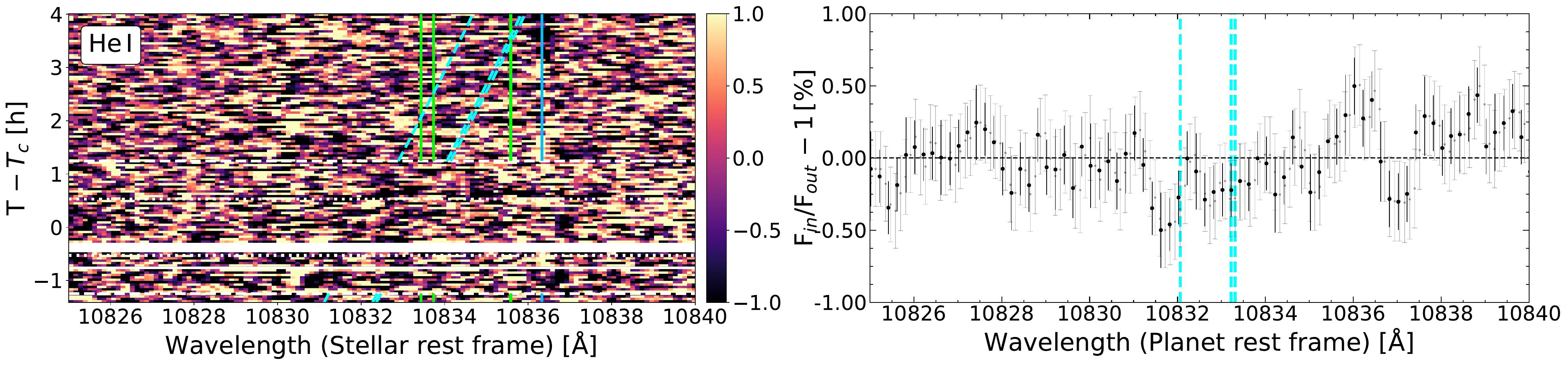}
    \caption{\label{Fig: TS TOI-1431}
    Same as Fig.\,\ref{Fig: TS K2-100}, but  for TOI-1431\,b observations with CARMENES.
    }
\end{figure*}

\begin{figure*}[h!]
    \centering
    \includegraphics[width=\hsize]{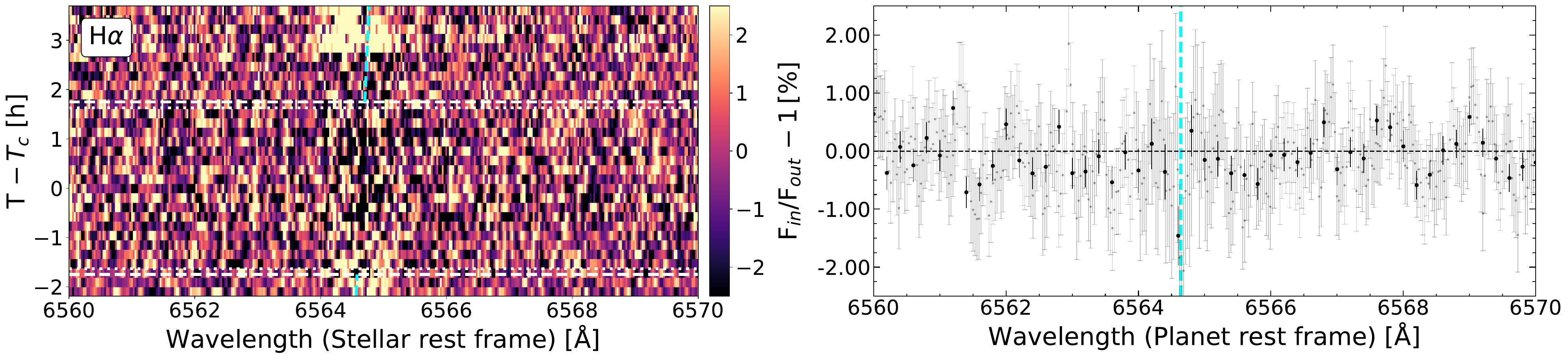}
    \includegraphics[width=\hsize]{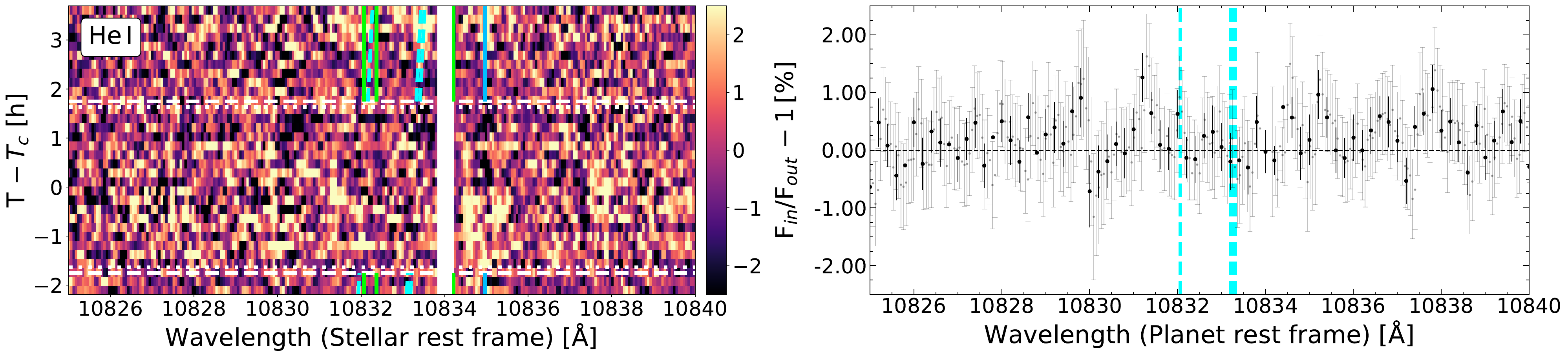}
    \caption{\label{Fig: TS TOI-2048}
    Same as Fig.\,\ref{Fig: TS K2-100}, but  for TOI-2048\,b observations with CARMENES.
    }
\end{figure*}

\begin{figure*}[h!]
    \centering
    \includegraphics[width=\hsize]{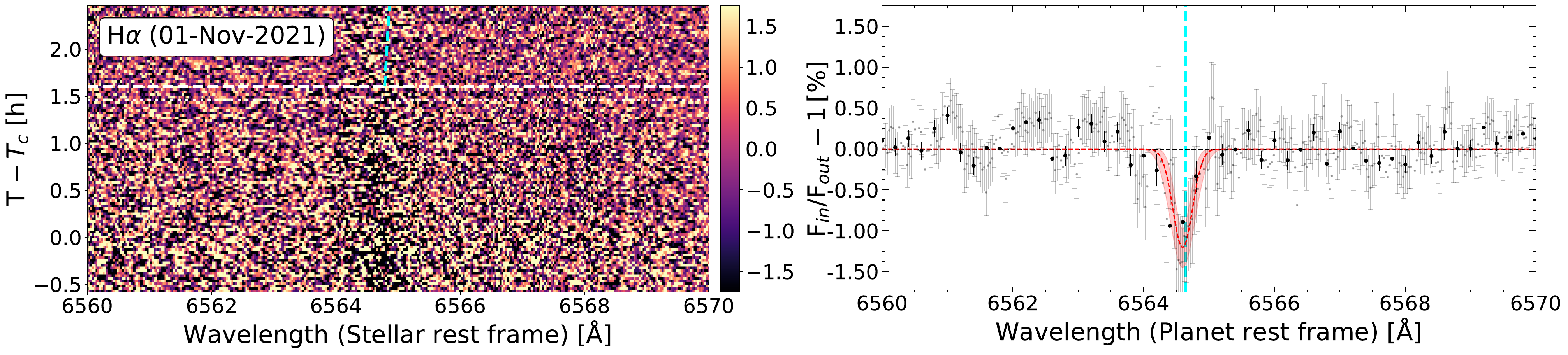}
    \includegraphics[width=\hsize]{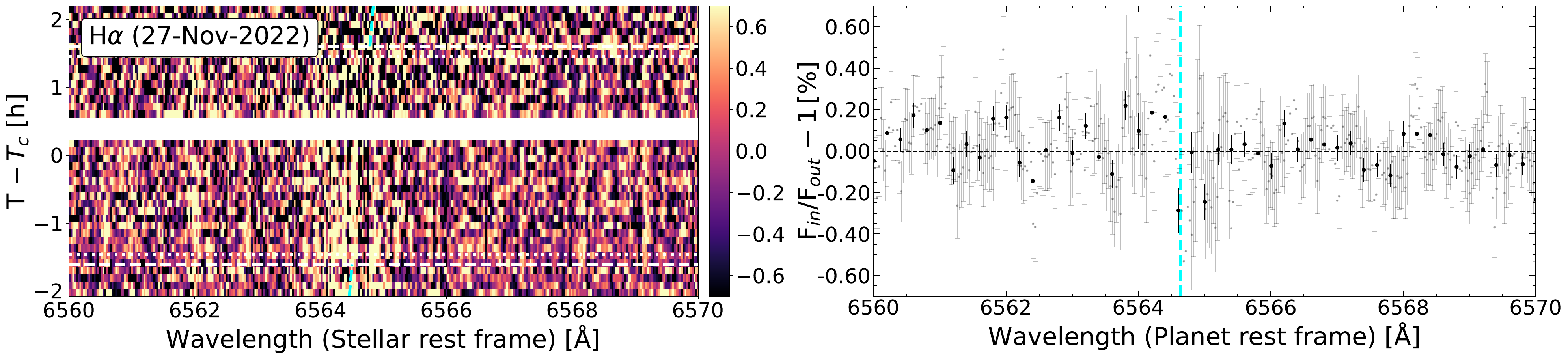}
    \includegraphics[width=\hsize]{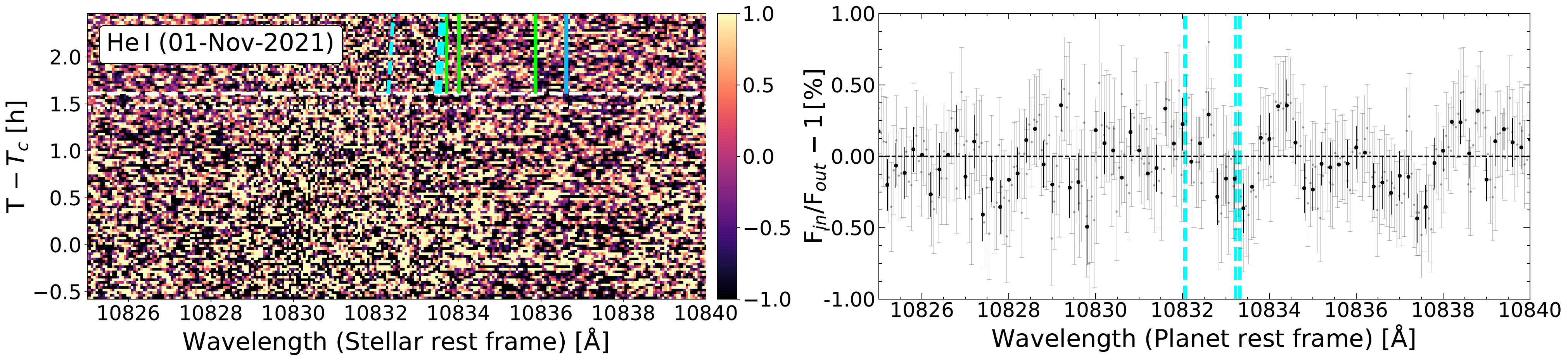}
    \includegraphics[width=\hsize]{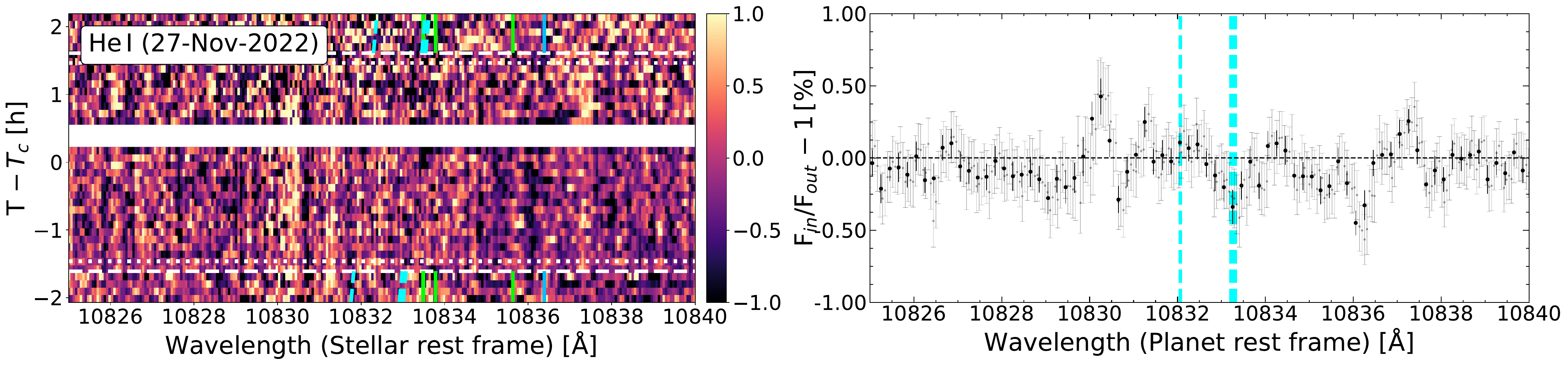}
    \caption{\label{Fig: TS HD63433 B}
    Same as Fig.\,\ref{Fig: TS K2-100}, but  for HD\,63433\,b observations with CARMENES.
    }
\end{figure*}

\begin{figure*}[h!]
    \centering
    \includegraphics[width=\hsize]{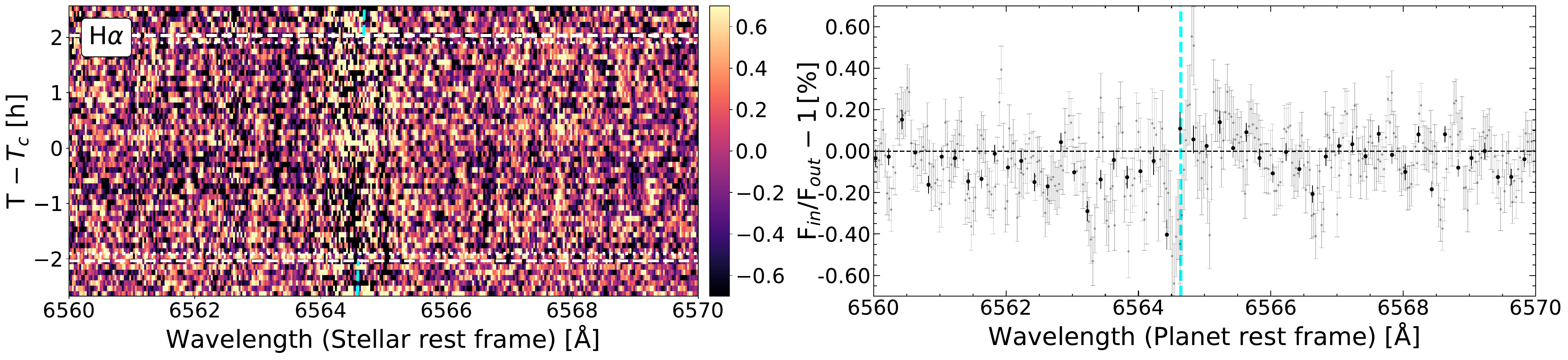}
    \includegraphics[width=\hsize]{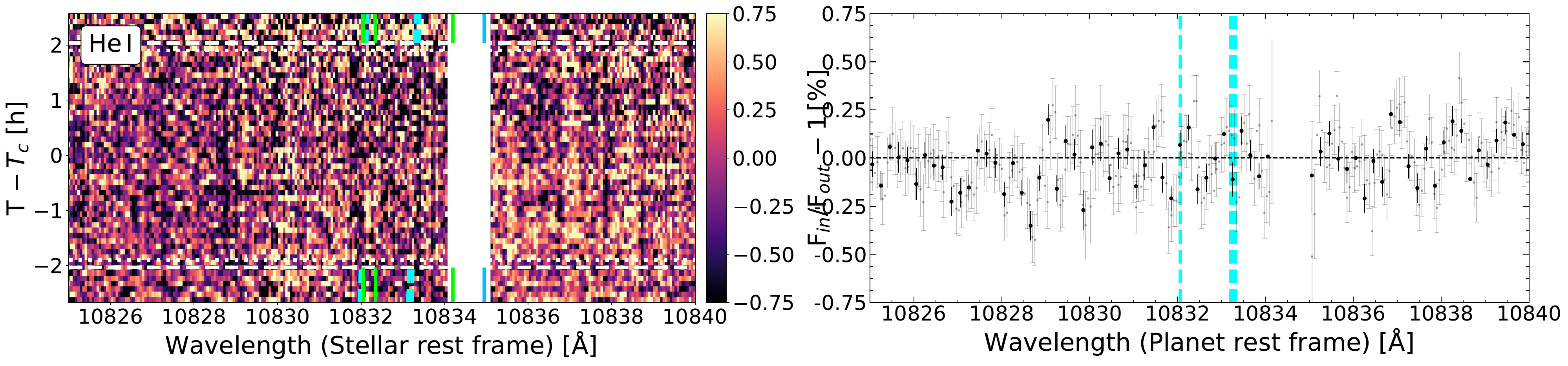}
    \caption{\label{Fig: TS HD63433 C}
    Same as Fig.\,\ref{Fig: TS K2-100}, but  for HD\,63433\,c observations with CARMENES.
    }
\end{figure*}

\begin{figure*}[h!]
    \centering
    \includegraphics[width=\hsize]{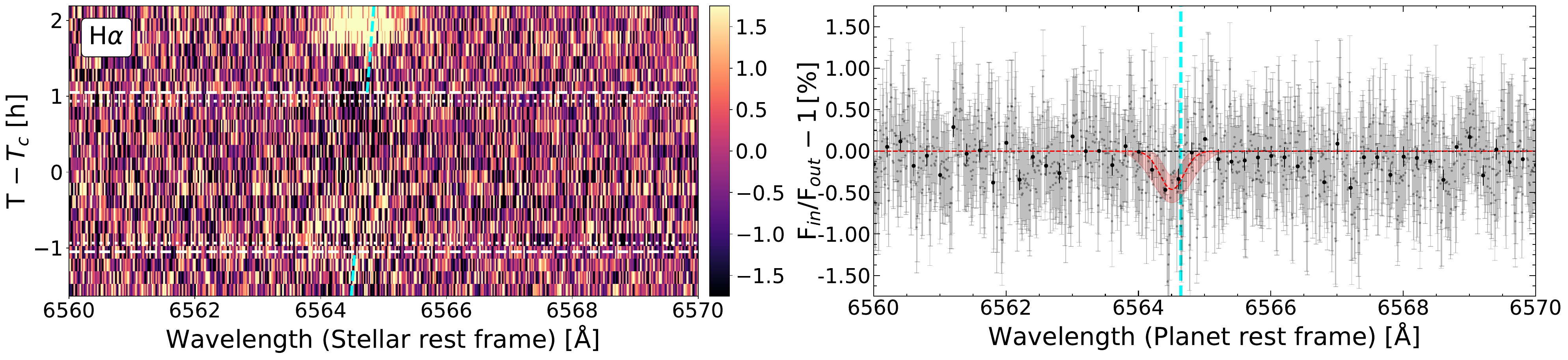}
    \caption{\label{Fig: TS HD 73583 B}
    Same as Fig.\,\ref{Fig: TS K2-100}, but  for HD\,73583\,b observations with HARPS-N.
    }
\end{figure*}

\begin{figure*}[h!]
    \centering
    \includegraphics[width=\hsize]{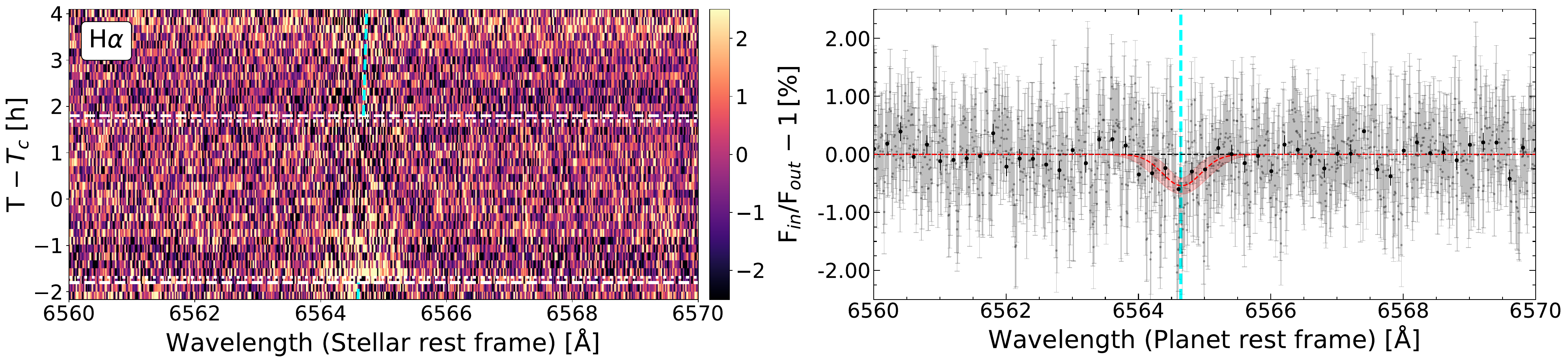}
    \includegraphics[width=\hsize]{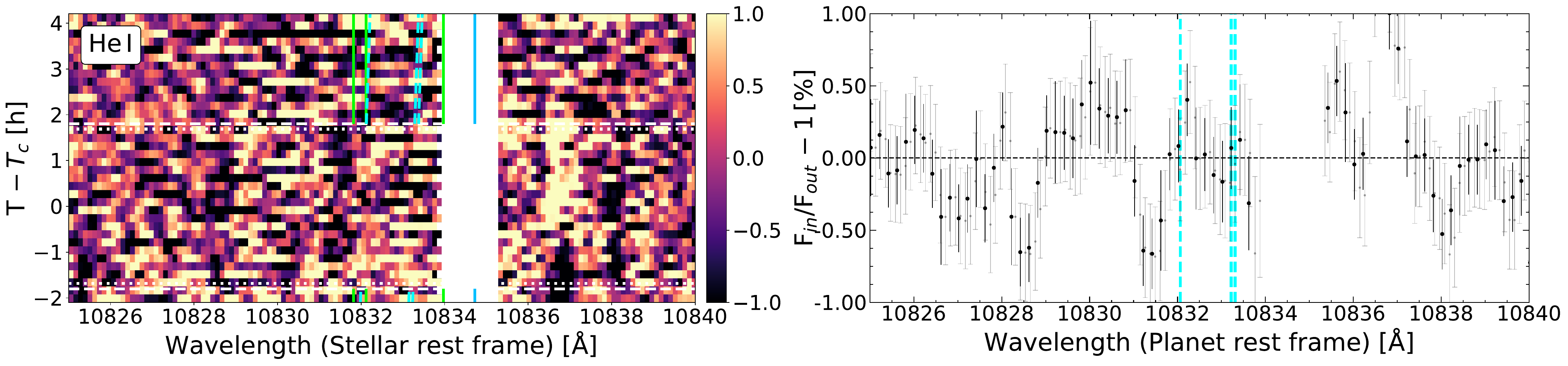}
    \caption{\label{Fig: TS HD 73583 C}
    Same as Fig.\,\ref{Fig: TS K2-100}, but  for HD\,73583\,c observations with GIARPS.
    }
\end{figure*}

\begin{figure*}[h!]
    \centering
    \includegraphics[width=\hsize]{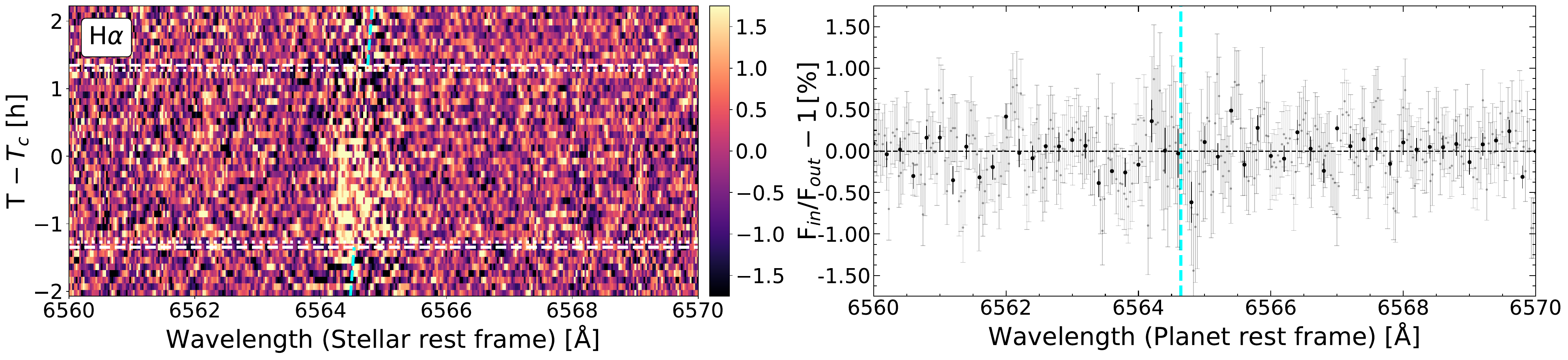}
    \includegraphics[width=\hsize]{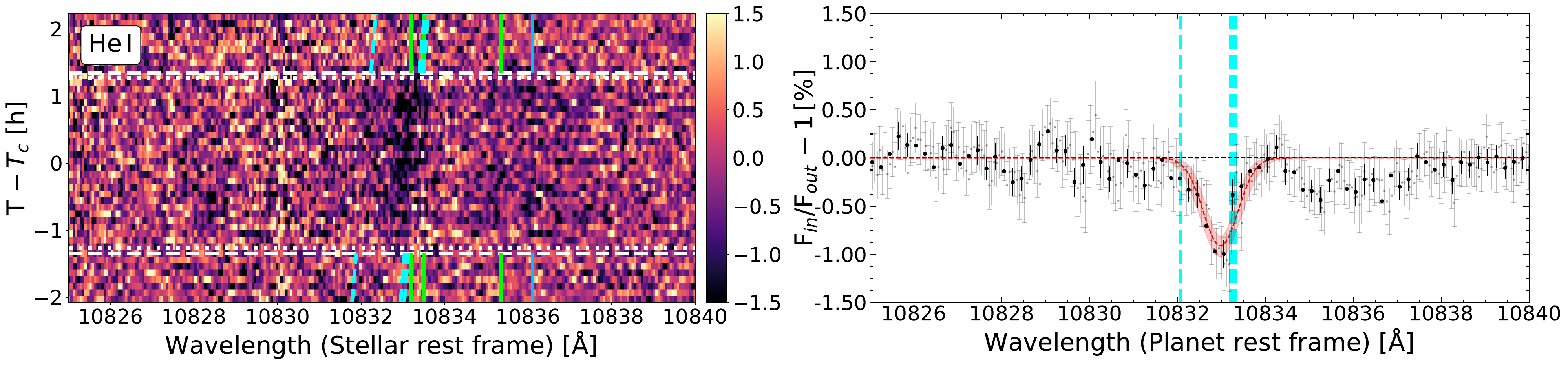}
    \caption{\label{Fig: TS TOI-1430}
    Same as Fig.\,\ref{Fig: TS K2-100}, but  for HD\,235088\,b observations with CARMENES.
    }
\end{figure*}

\begin{figure*}[h!]
    \centering
    \includegraphics[width=\hsize]{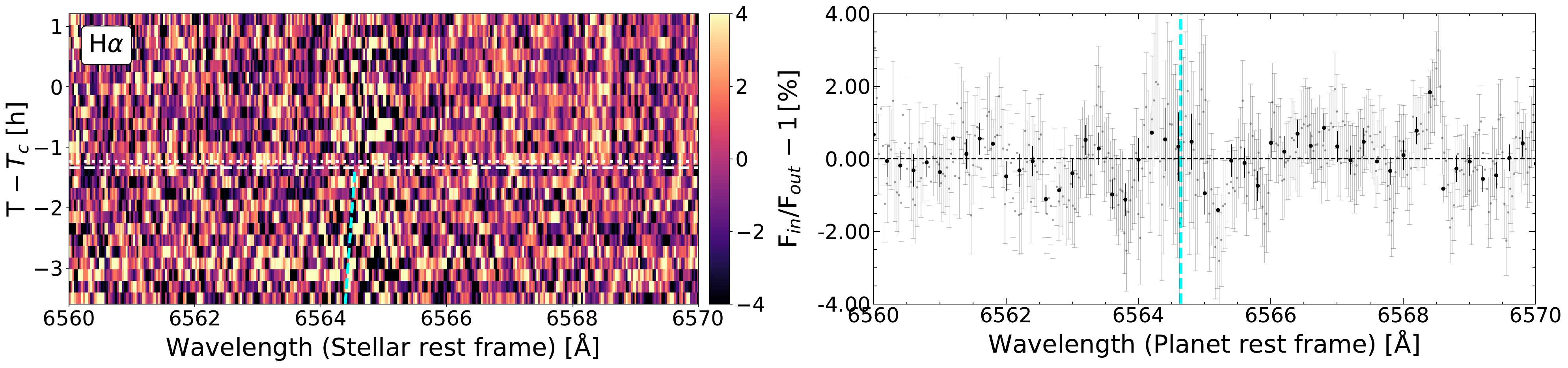}
    \includegraphics[width=\hsize]{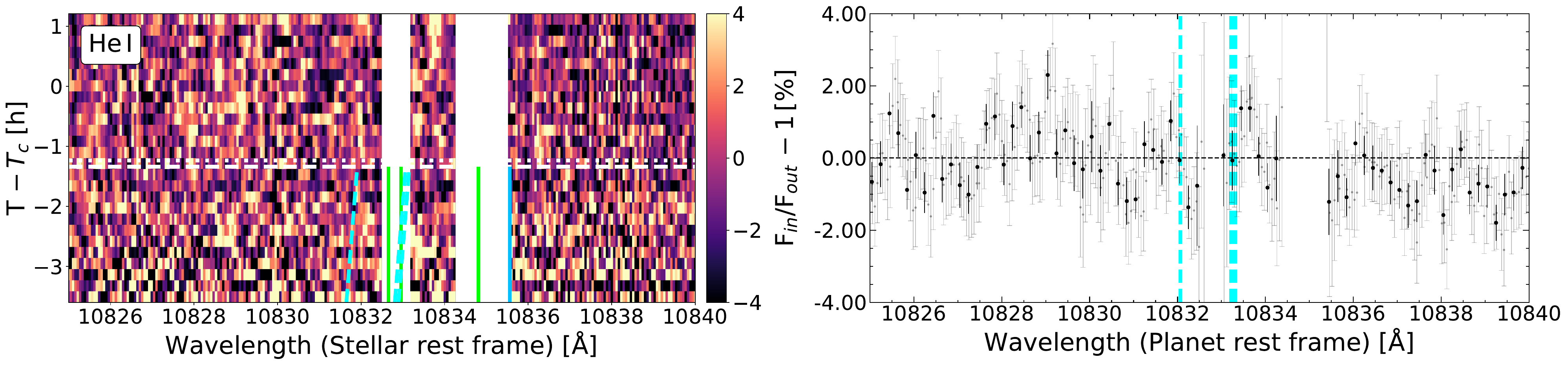}
    \caption{\label{Fig: TS K2-77}
    Same as Fig.\,\ref{Fig: TS K2-100}, but  for K2-77\,b observations with CARMENES.
    }
\end{figure*}

\begin{figure*}[h!]
    \centering
    \includegraphics[width=\hsize]{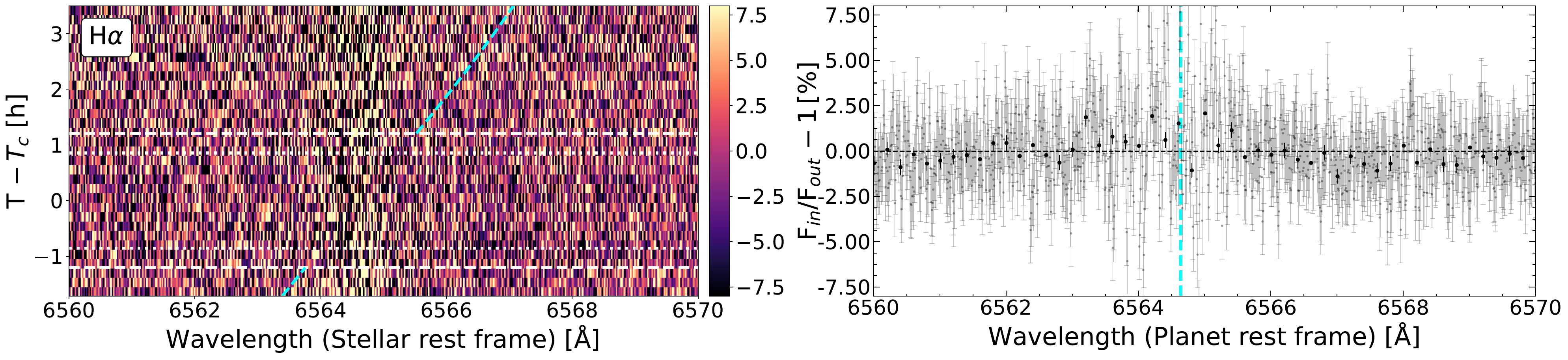}
    \includegraphics[width=\hsize]{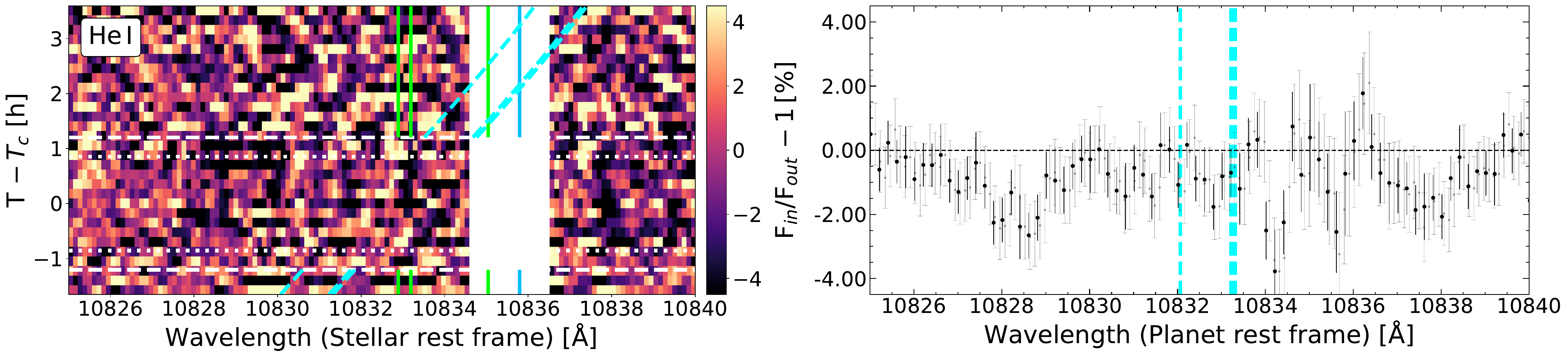}
    \caption{\label{Fig: TS TOI-2046}
    Same as Fig.\,\ref{Fig: TS K2-100}, but  for TOI-2046\,b observations with GIARPS.
    }
\end{figure*}

\begin{figure*}[h!]
    \centering
    \includegraphics[width=\hsize]{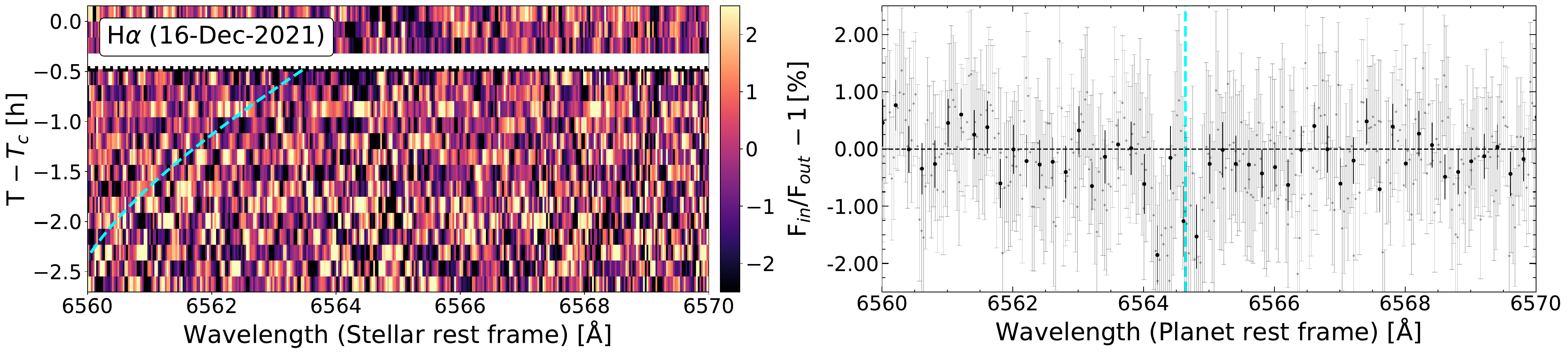}
    \includegraphics[width=\hsize]{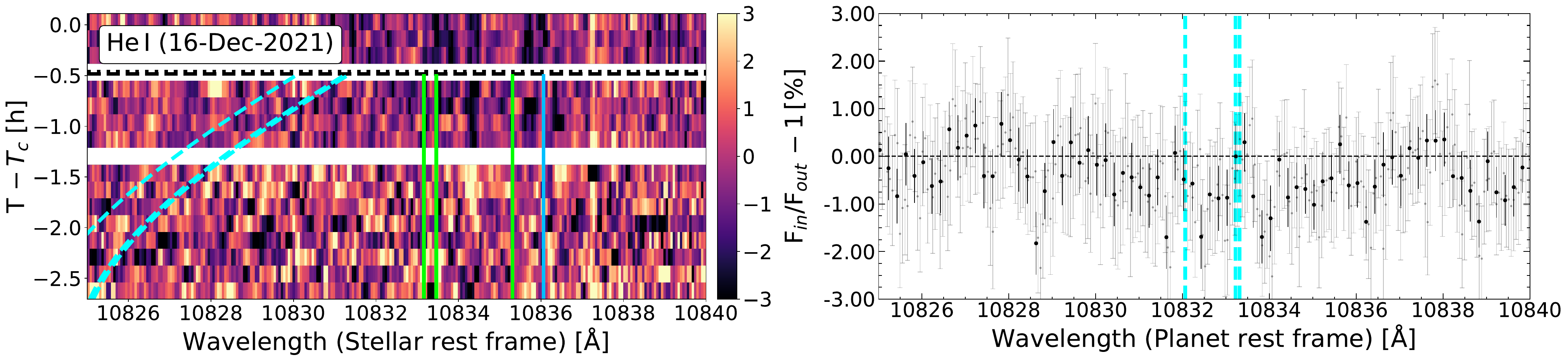}
    \includegraphics[width=\hsize]{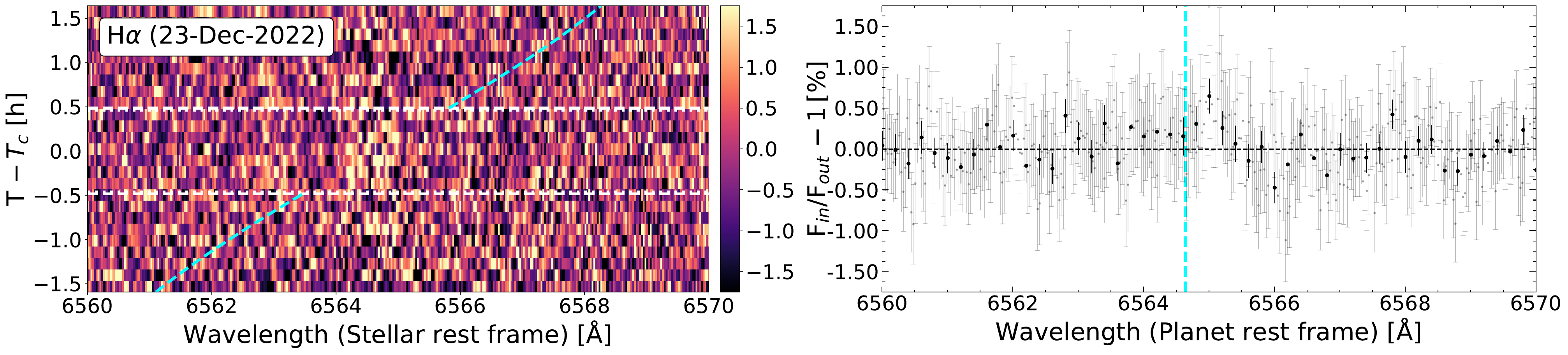}
    \includegraphics[width=\hsize]{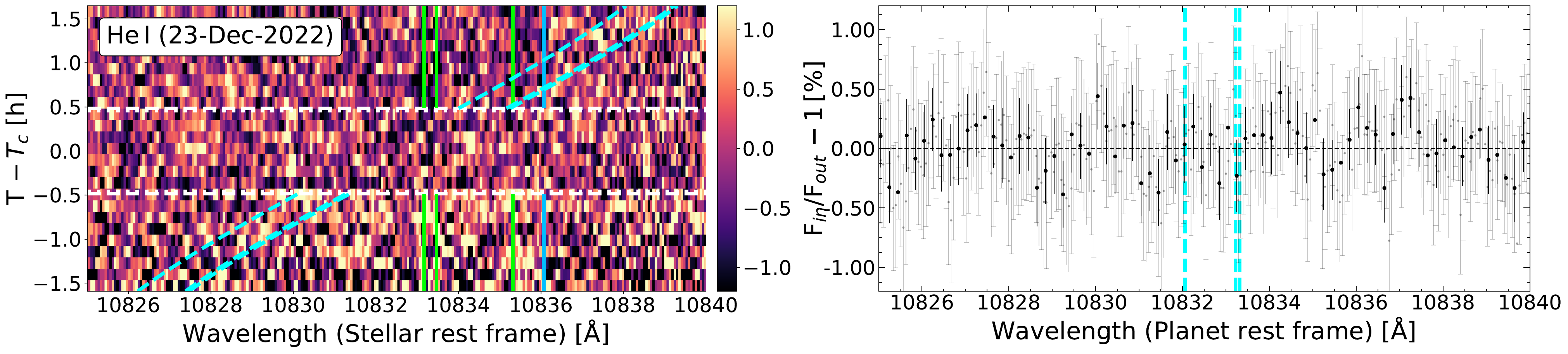}
    \caption{\label{Fig: TS TOI-1807}
    Same as Fig.\,\ref{Fig: TS K2-100}, but for TOI-1807\,b observations with CARMENES.
    }
\end{figure*}

\begin{figure*}[h!]
    \centering
    \includegraphics[width=\hsize]{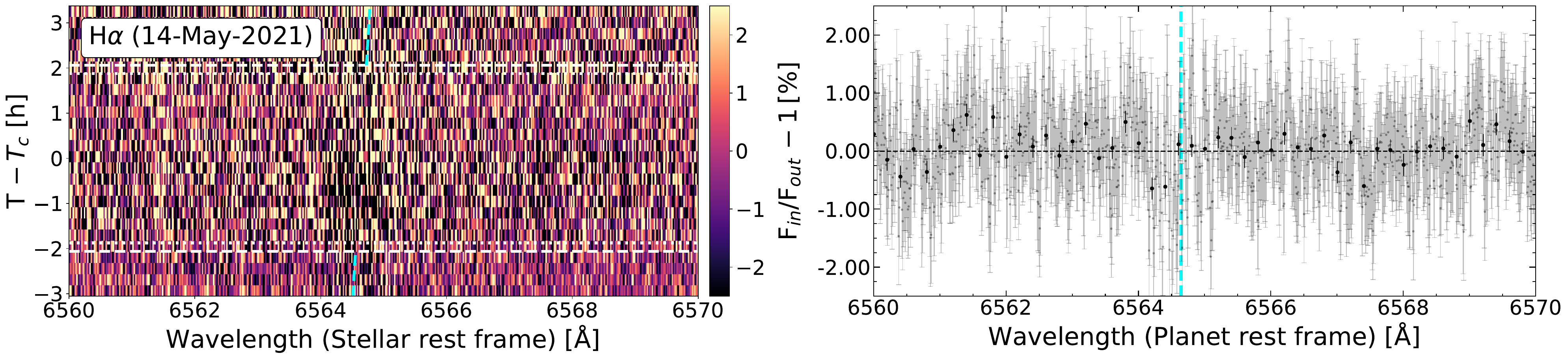}
    \includegraphics[width=\hsize]{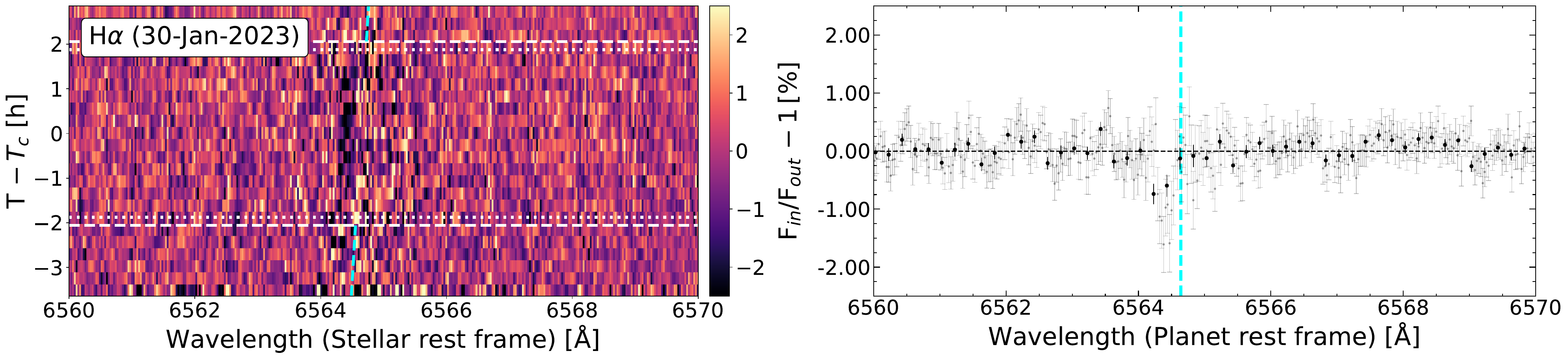}
    \caption{\label{Fig: TS TOI-1136}
    Same as Fig.\,\ref{Fig: TS K2-100} for TOI-1136\,d H$\alpha$ observations with HARPS-N (14-May-2021) and CARMENES VIS (30-Jan-2023).
    }
\end{figure*}

\begin{figure*}[h!]
    \centering
    \includegraphics[width=\hsize]{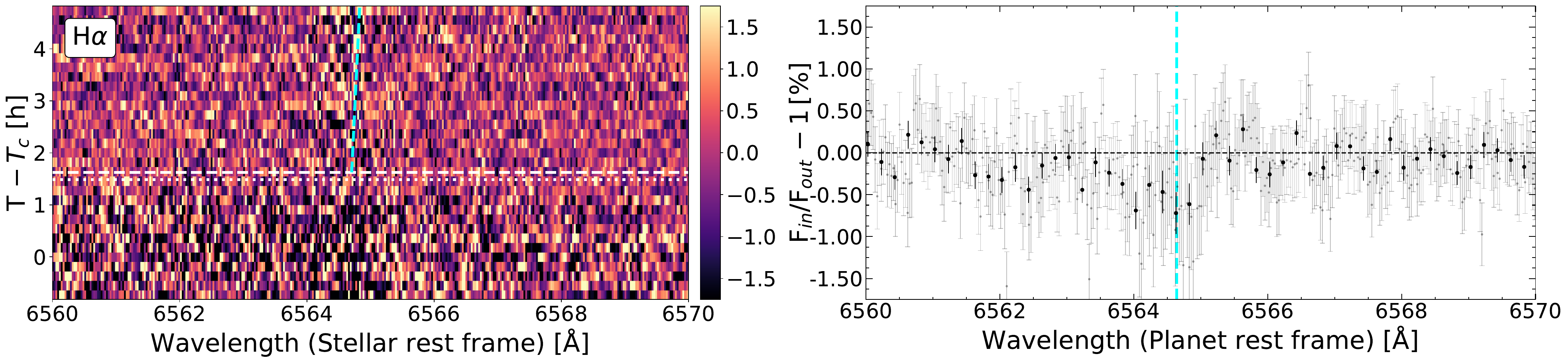}
    \includegraphics[width=\hsize]{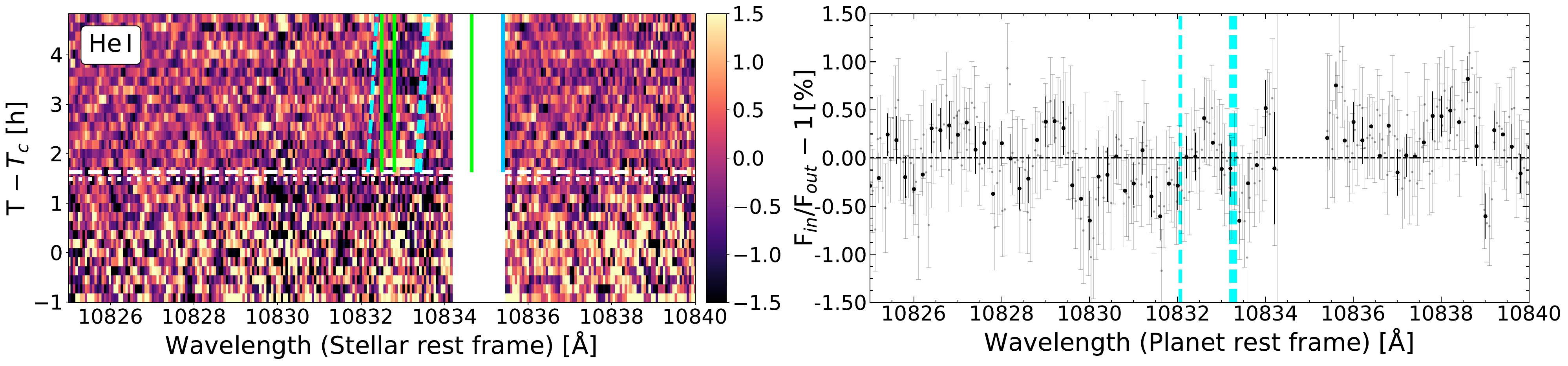}
    \caption{\label{Fig: TS TOI-2076}
    Same as Fig.\,\ref{Fig: TS K2-100}, but for TOI-2076\,b observations with CARMENES.
    }
\end{figure*}

\begin{figure*}[h!]
    \centering
    \includegraphics[width=\hsize]{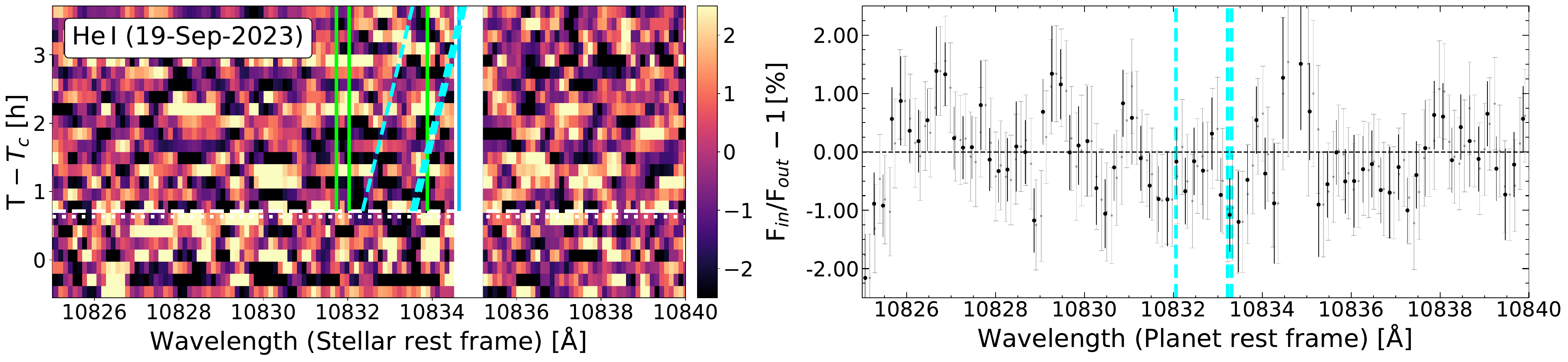}
    \includegraphics[width=\hsize]{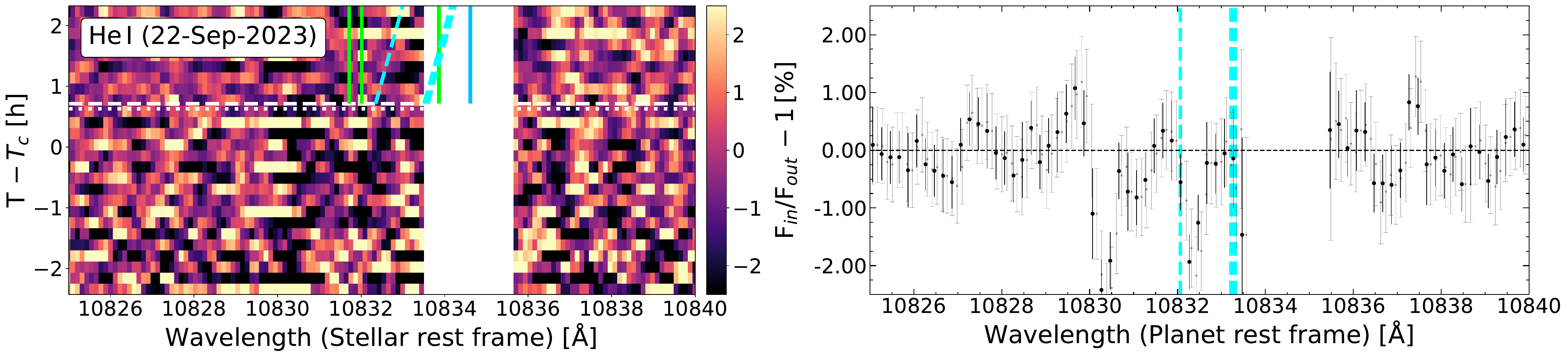}
    \includegraphics[width=\hsize]{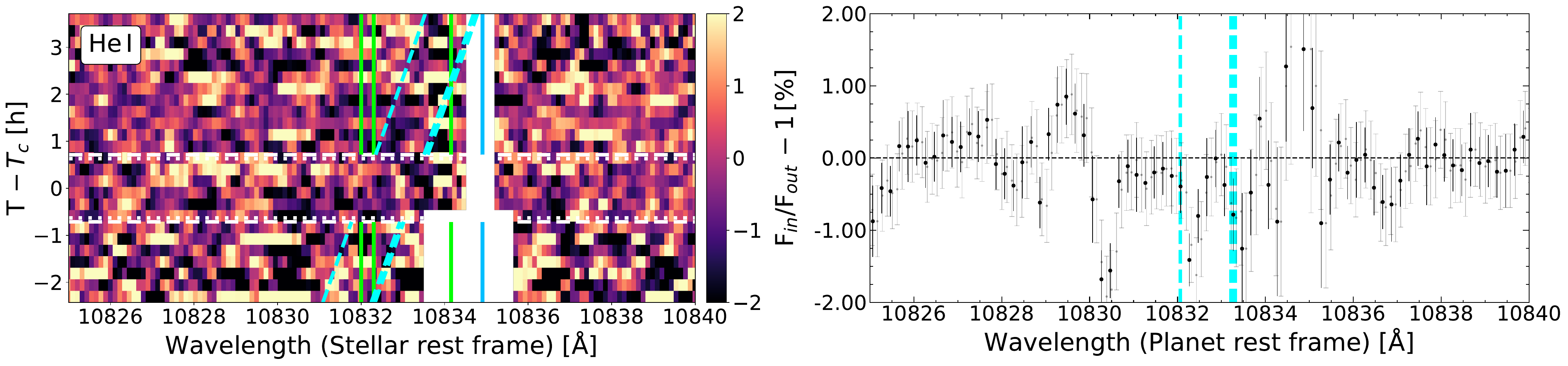}
    \caption{\label{Fig: TS TOI-1683}
    Same as Fig.\,\ref{Fig: TS K2-100}, but  for TOI-1683\,b observations with GIANO-B.
    }
\end{figure*}

\begin{figure*}[h!]
    \centering
    \includegraphics[width=\hsize]{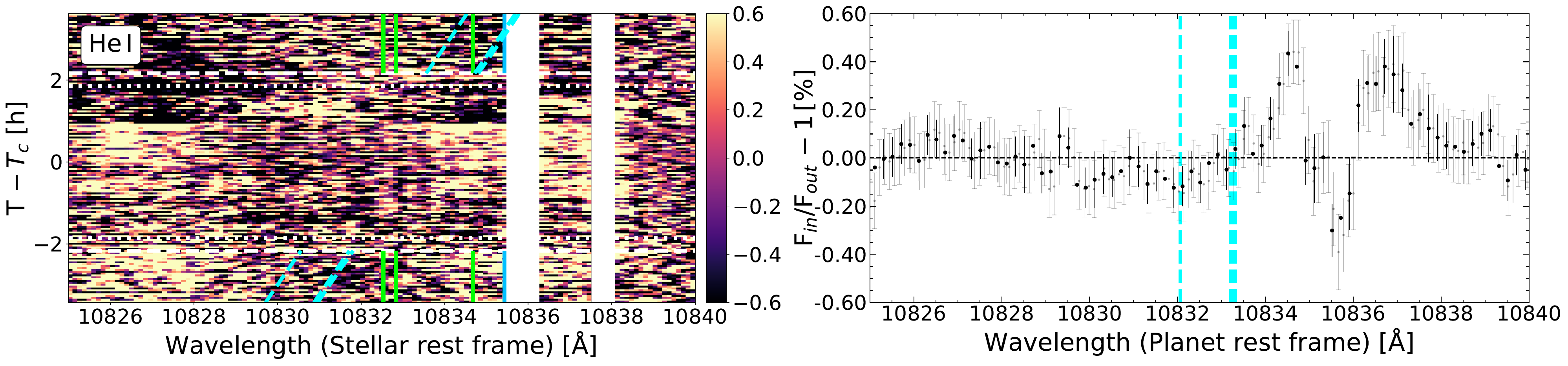}
    \caption{\label{Fig: TS WASP-189}
    Same as Fig.\,\ref{Fig: TS K2-100}, but for WASP-189\,b observations with GIANO-B.
    }
\end{figure*}

\begin{figure*}[h!]
    \centering
    \includegraphics[width=\hsize]{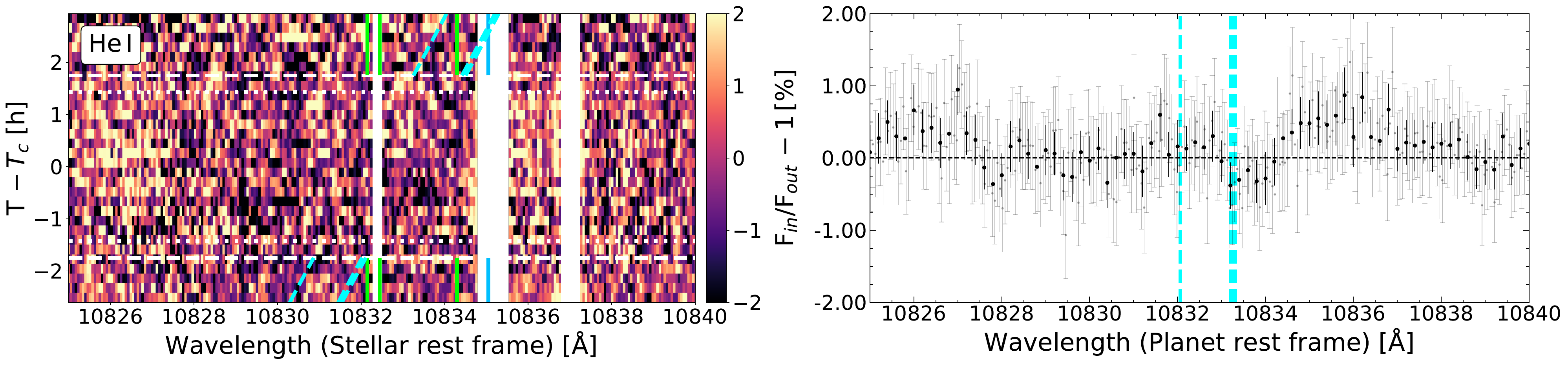}
    \caption{\label{Fig: TS HATP57}
    Same as Fig.\,\ref{Fig: TS K2-100}, but for HAT-P-57\,b observations with CARMENES.
    }
\end{figure*}

\begin{figure*}[h!]
    \centering
    \includegraphics[width=\hsize]{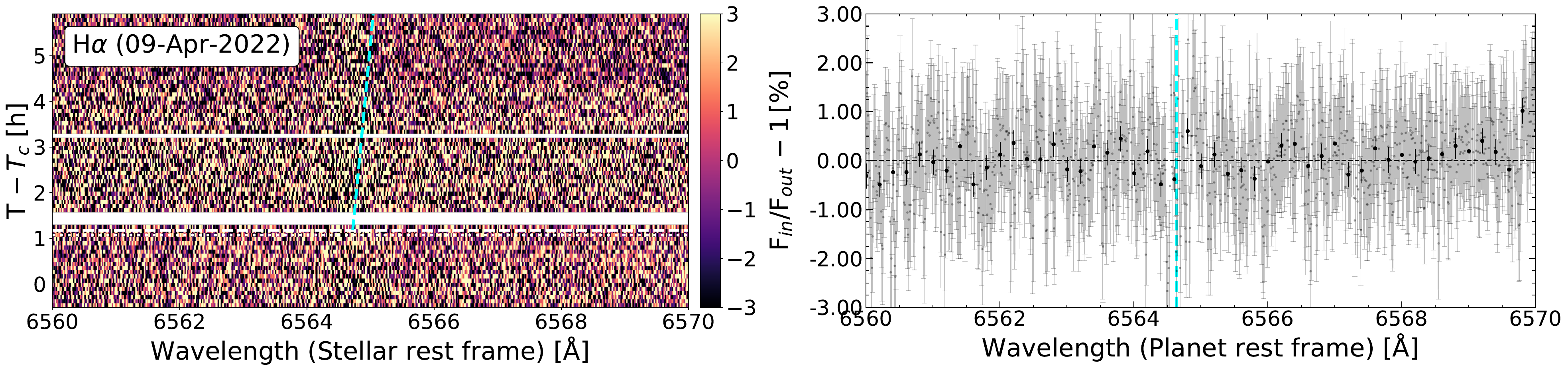}
    \includegraphics[width=\hsize]{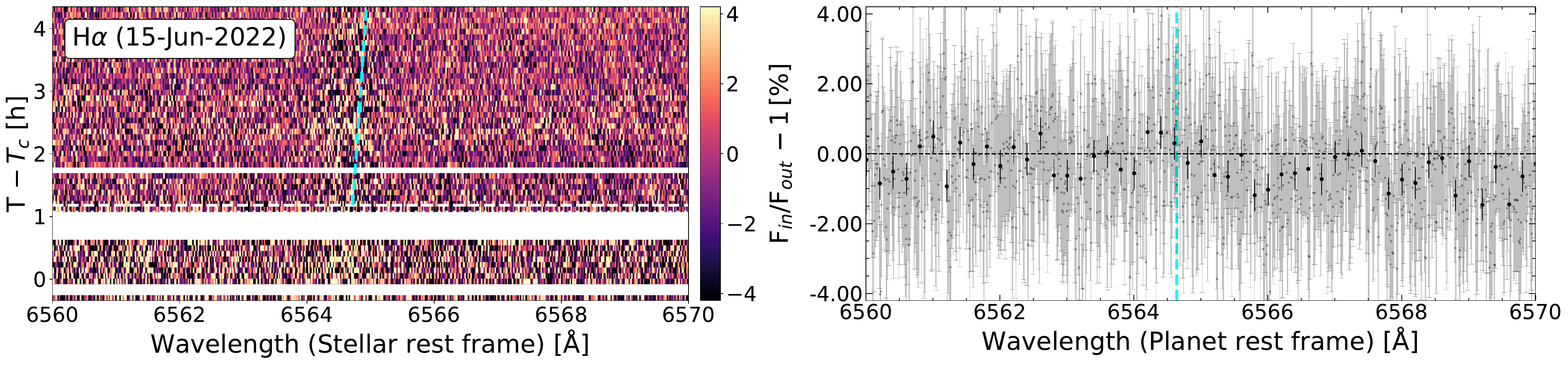}
    \caption{\label{Fig: TS TOI-2018 HALPHA}
    Same as Fig.\,\ref{Fig: TS K2-100}, but  for TOI-2018\,b individual observations of H$\alpha$ with HARPS-N.
    }
\end{figure*}

\begin{figure*}[h!]
    \centering
    \includegraphics[width=\hsize]{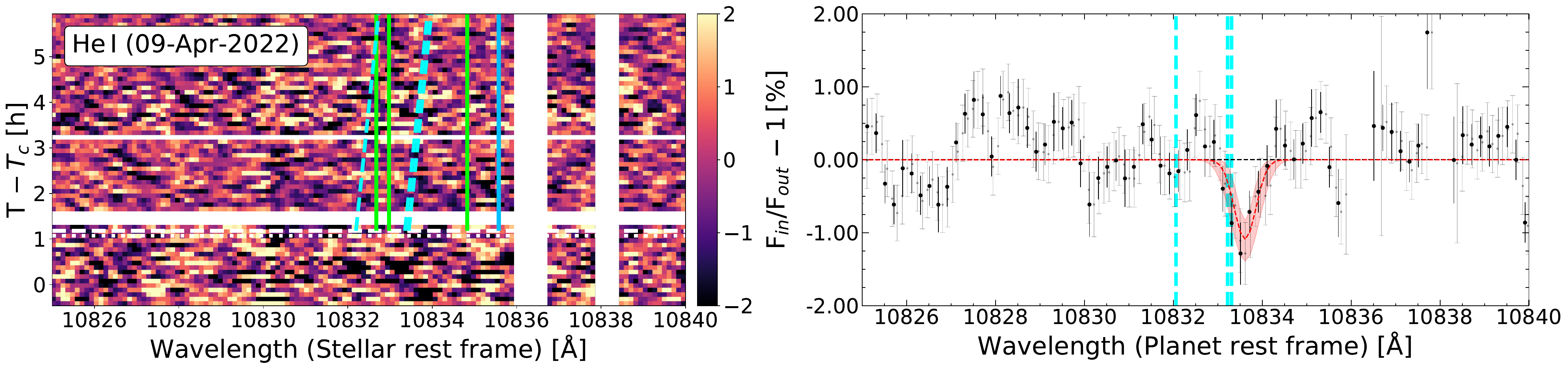}
    \includegraphics[width=\hsize]{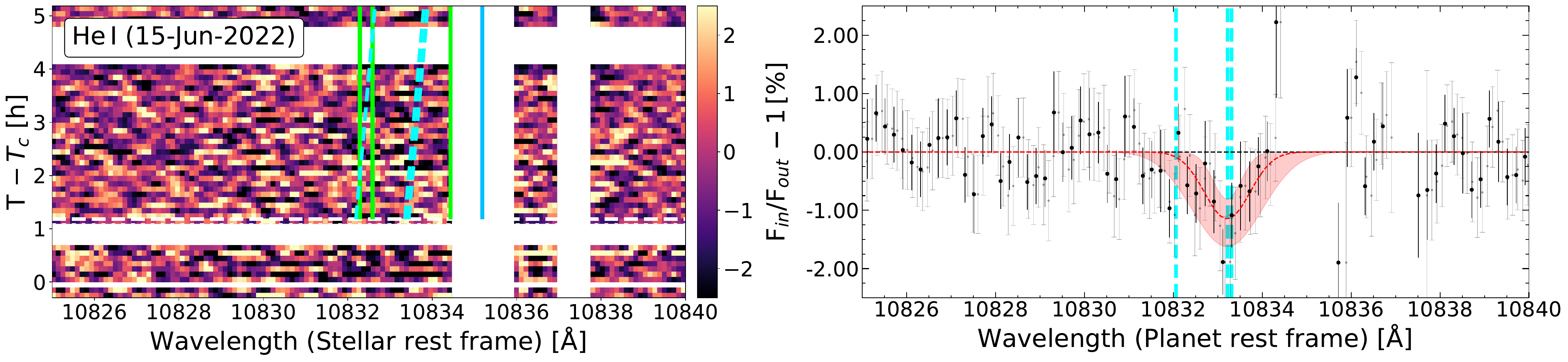}
    \caption{\label{Fig: TS TOI-2018 HE}
    Same as Fig.\,\ref{Fig: TS K2-100}, but for TOI-2018\,b idividual observations of \ion{He}{I} triplet with GIANO-B.
    }
\end{figure*}

\begin{figure*}[h!]
    \centering
    \begin{subfigure}{0.48\textwidth}
         \centering
         \includegraphics[width=\textwidth]{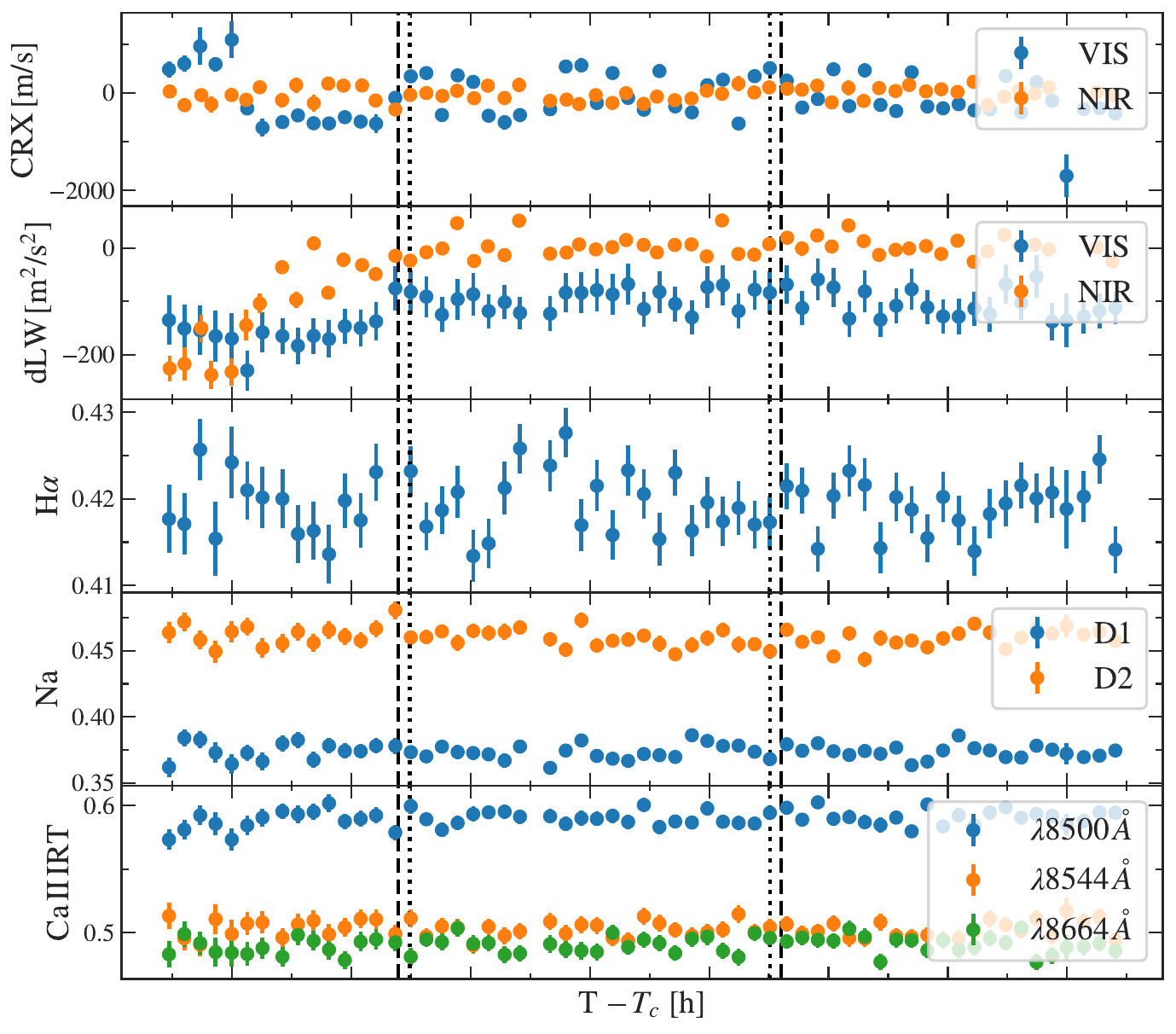}
    \end{subfigure}
    \hfill
    \begin{subfigure}{0.48\textwidth}
         \centering
         \includegraphics[width=\textwidth]{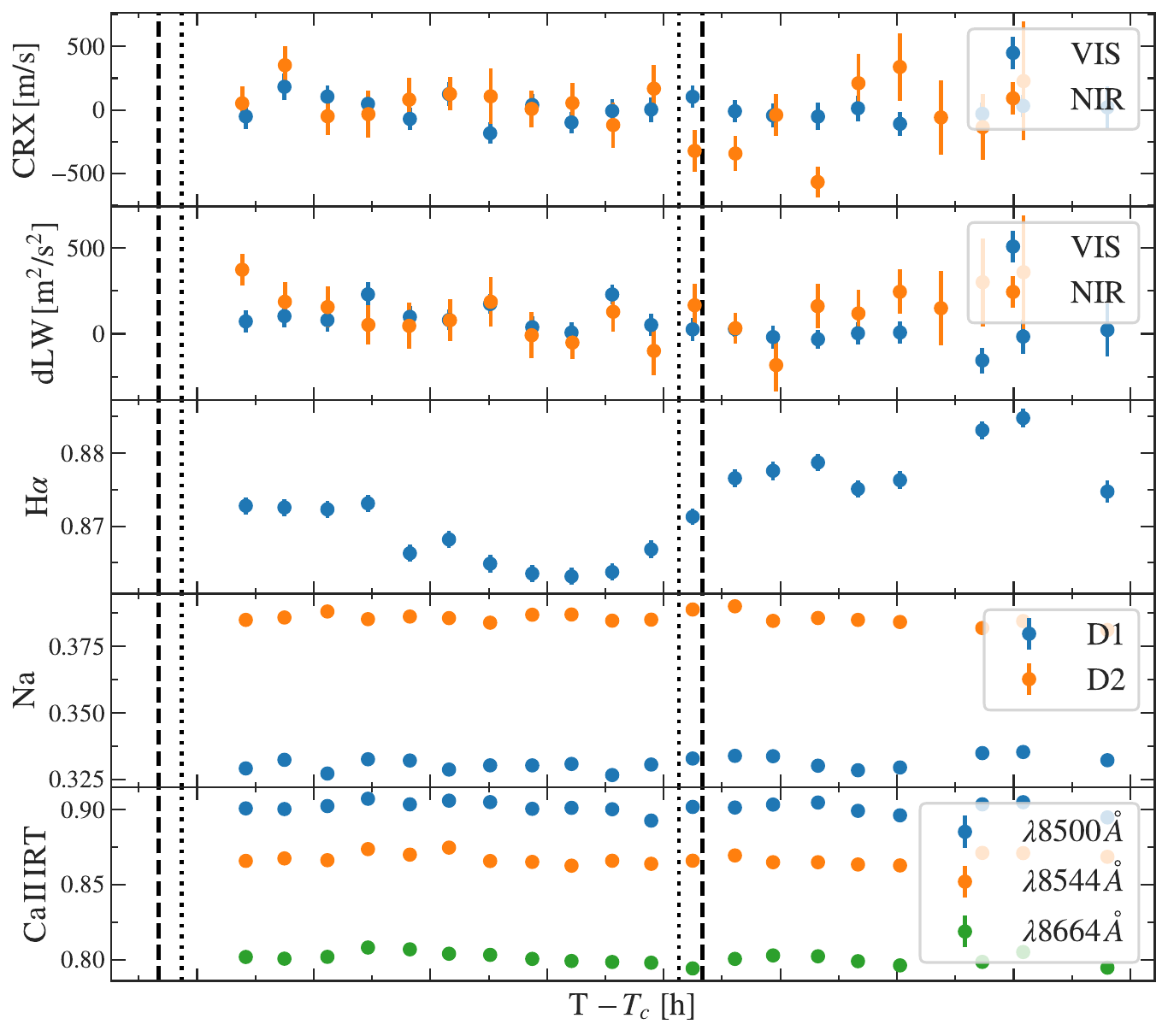}
    \end{subfigure}
    \hfill
    \begin{subfigure}{0.48\textwidth}
         \centering
         \includegraphics[width=\textwidth]{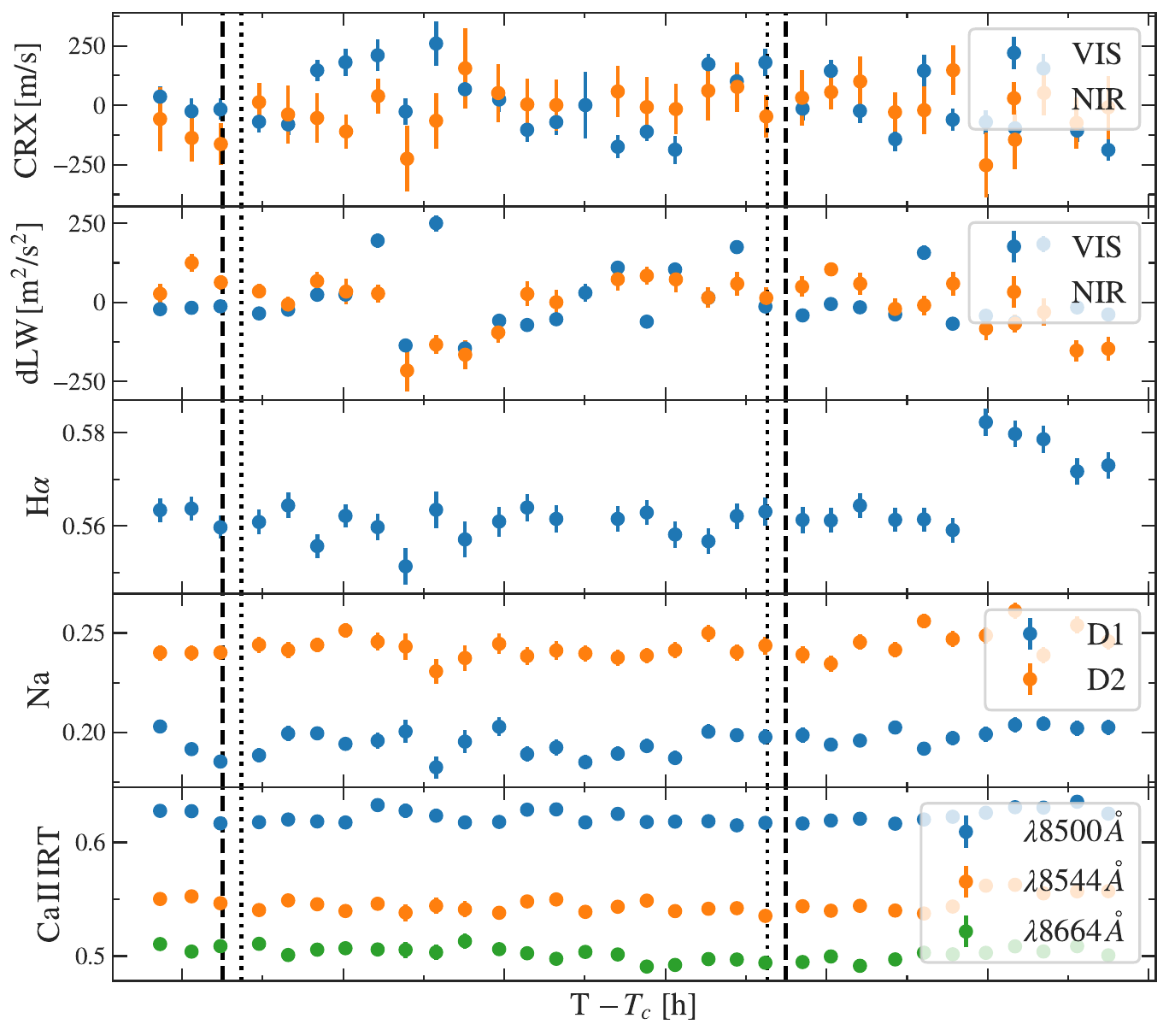}
    \end{subfigure}
    \hfill
    \begin{subfigure}{0.48\textwidth}
         \centering
         \includegraphics[width=\textwidth]{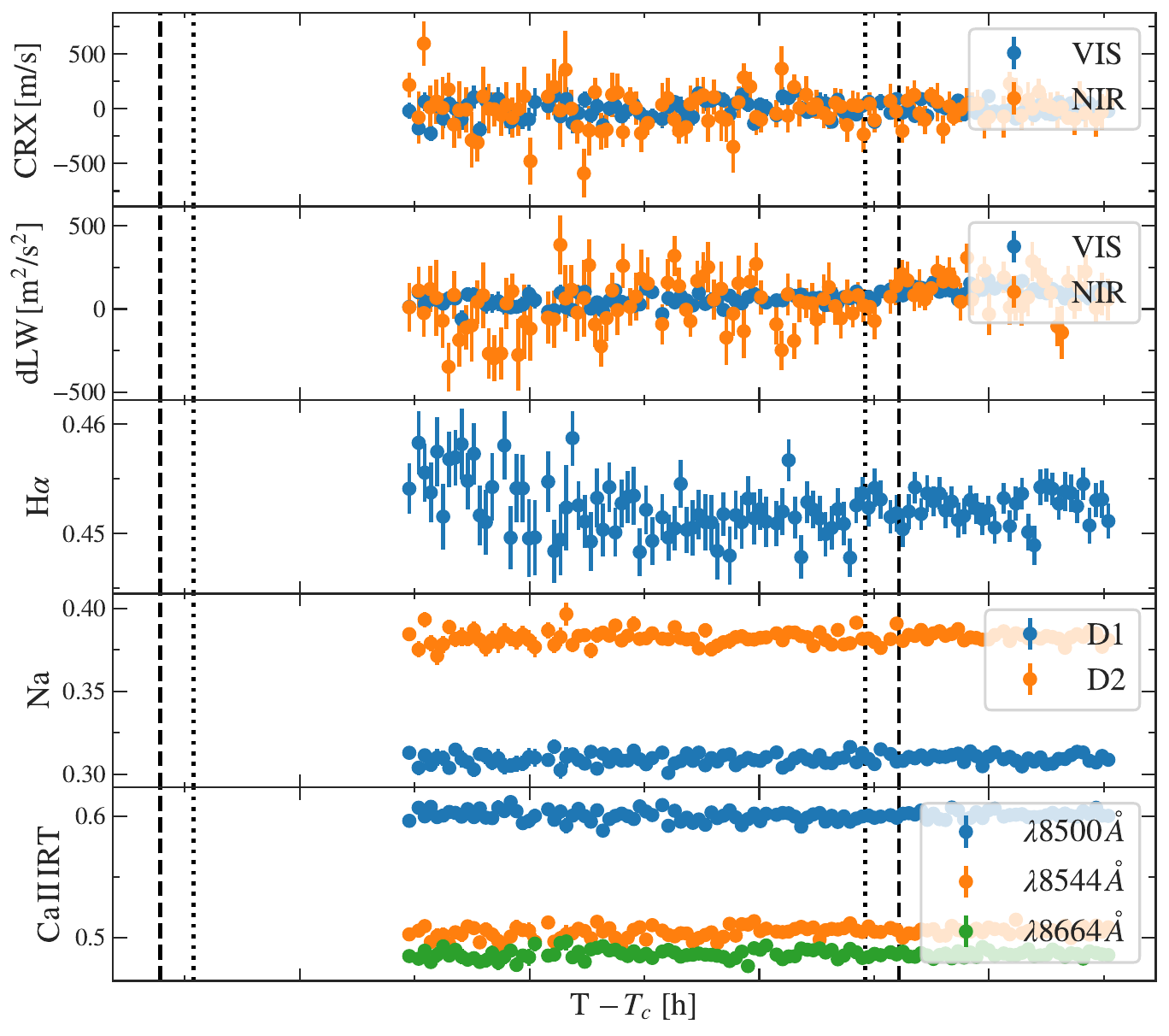}
    \end{subfigure}
    \hfill
    \begin{subfigure}{0.48\textwidth}
         \centering
         \includegraphics[width=\textwidth]{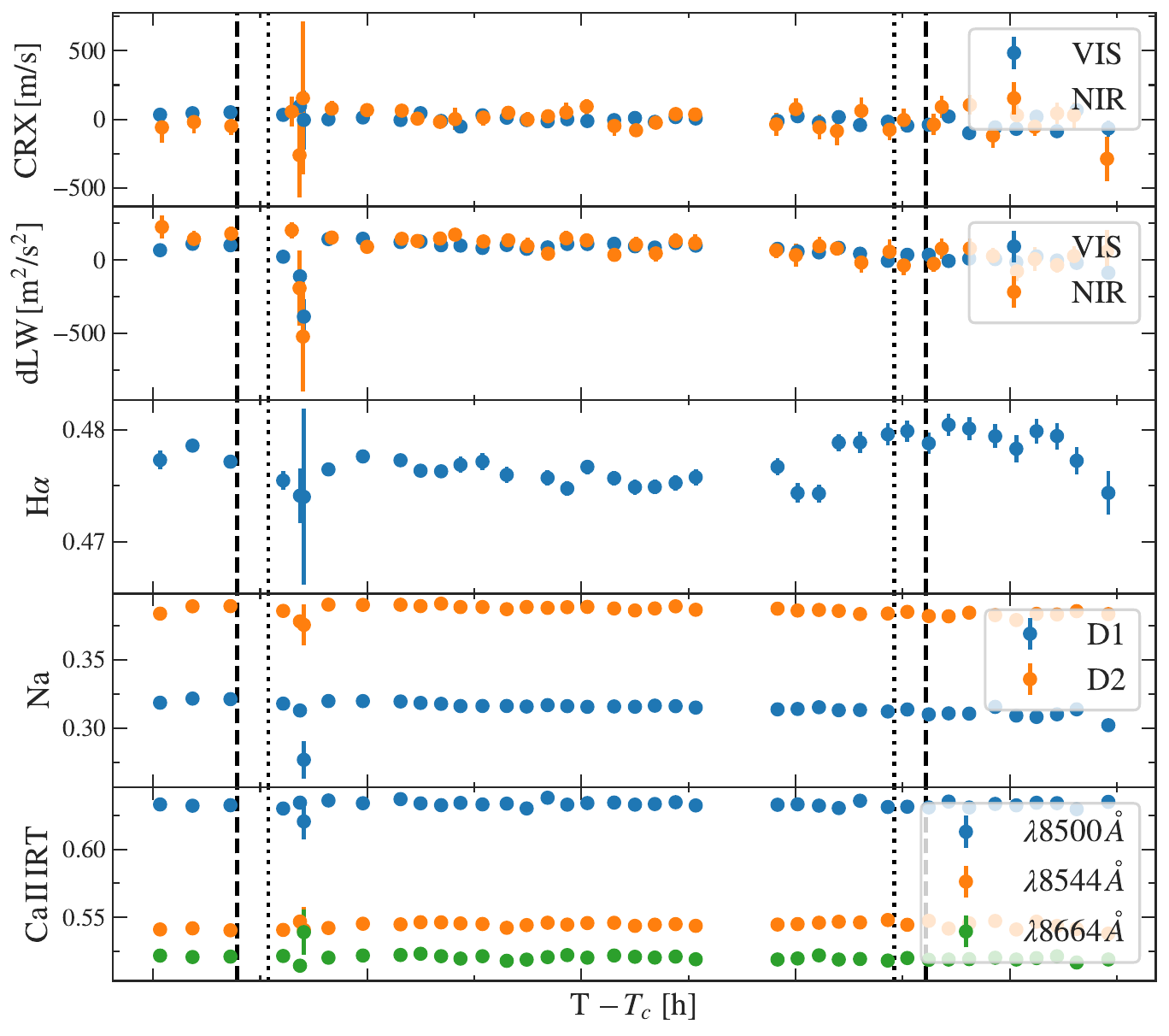}
    \end{subfigure}
    \hfill
    \begin{subfigure}{0.48\textwidth}
         \centering
         \includegraphics[width=\textwidth]{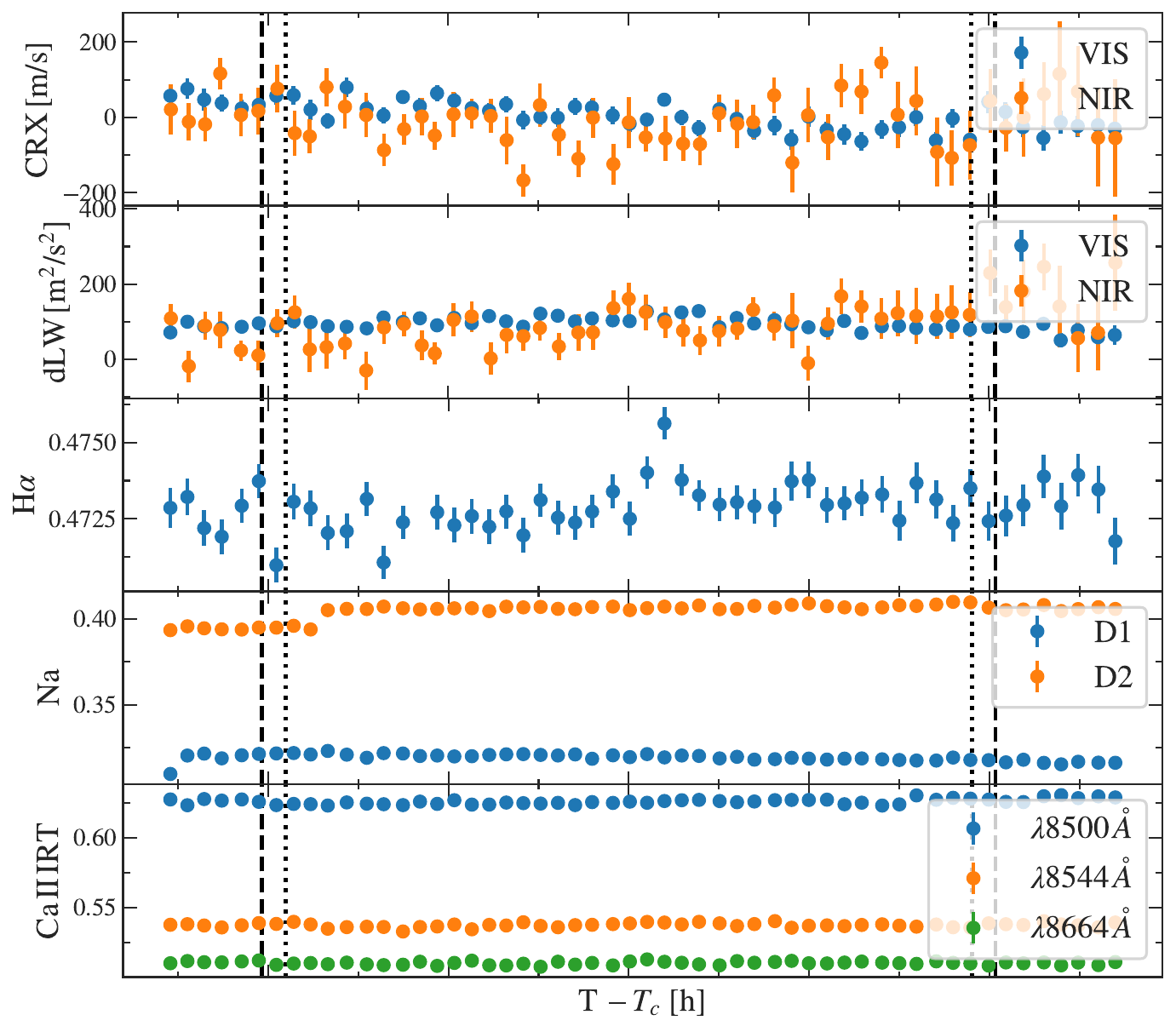}
    \end{subfigure}
    \caption{\label{Fig: serval 1}
    Time evolution of the activity indicators for K2-100\,b (top left), V1298\,Tau\,c (top right), TOI-2048\,b (middle left), HD\,63433\,b (1 November 2021, middle right; and 27 November 2022, bottom left), and HD\,63433\,c (bottom right). The vertical dashed and dotted lines represent the different contacts during the transit.
    }
\end{figure*}

\begin{figure*}[h!]
    \centering
    \begin{subfigure}{0.49\textwidth}
         \centering
         \includegraphics[width=\textwidth]{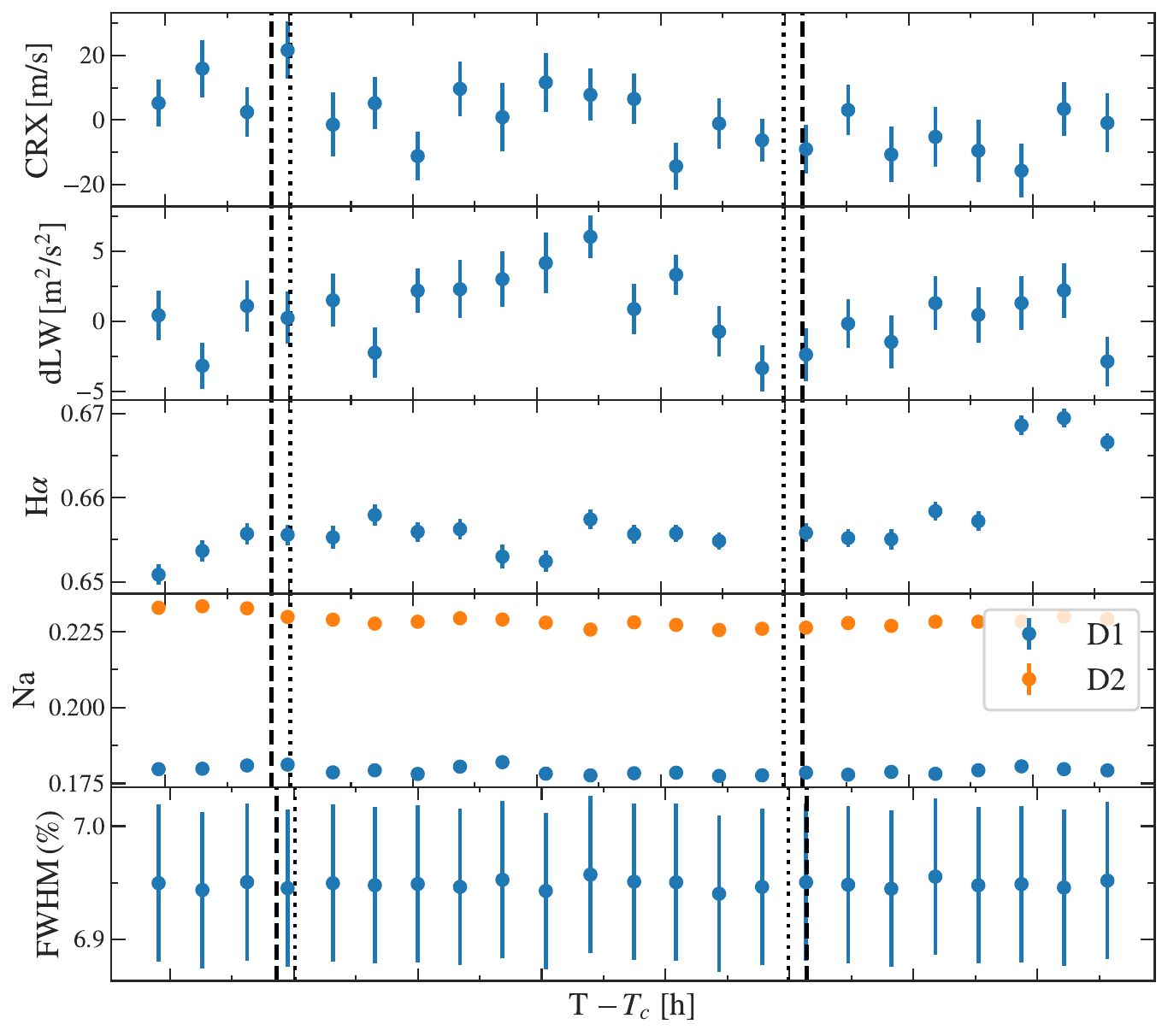}
    \end{subfigure}
    \hfill
    \begin{subfigure}{0.49\textwidth}
         \centering
         \includegraphics[width=\textwidth]{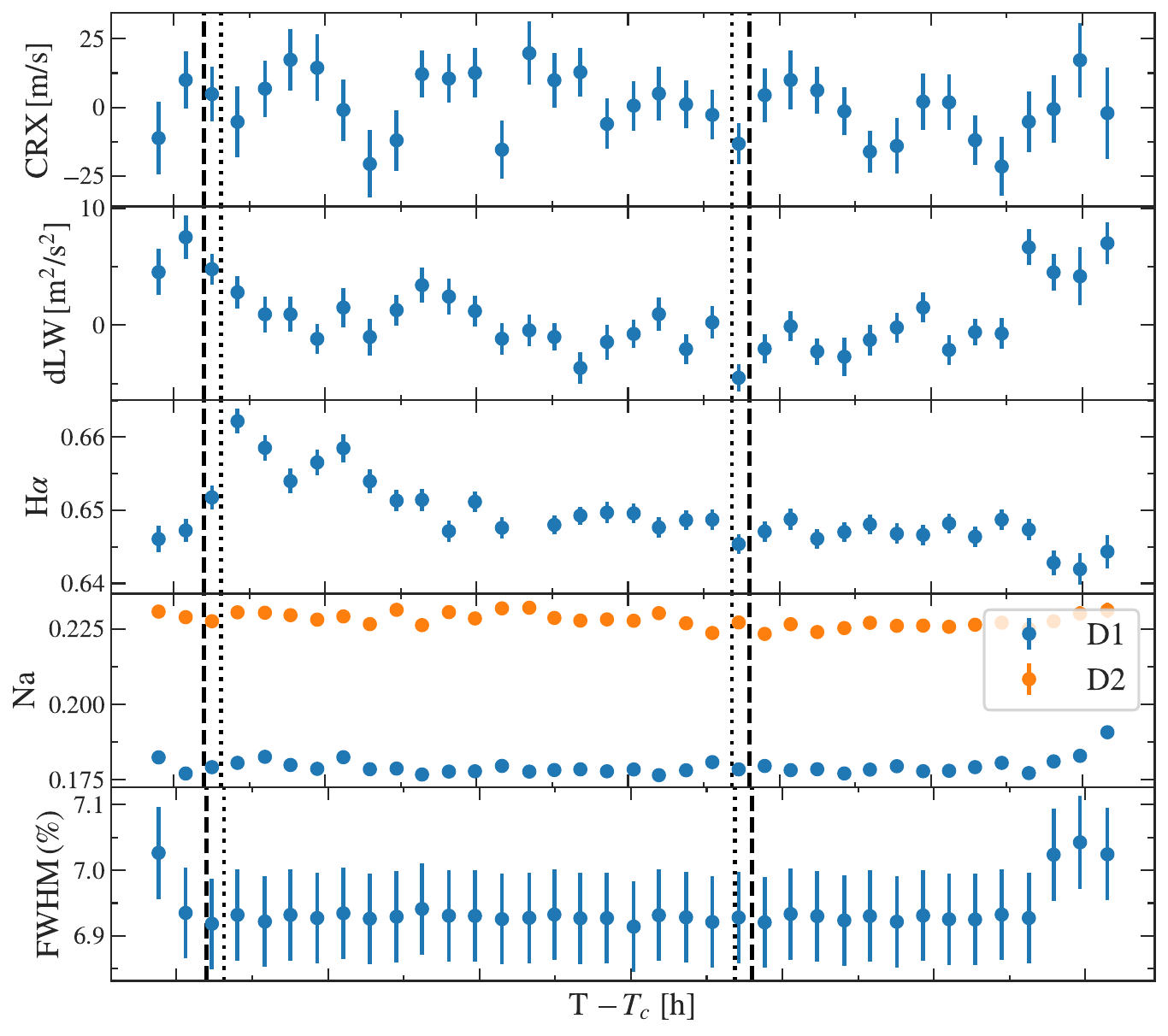}
    \end{subfigure}
    \hfill
    \begin{subfigure}{0.49\textwidth}
         \centering
         \includegraphics[width=\textwidth]{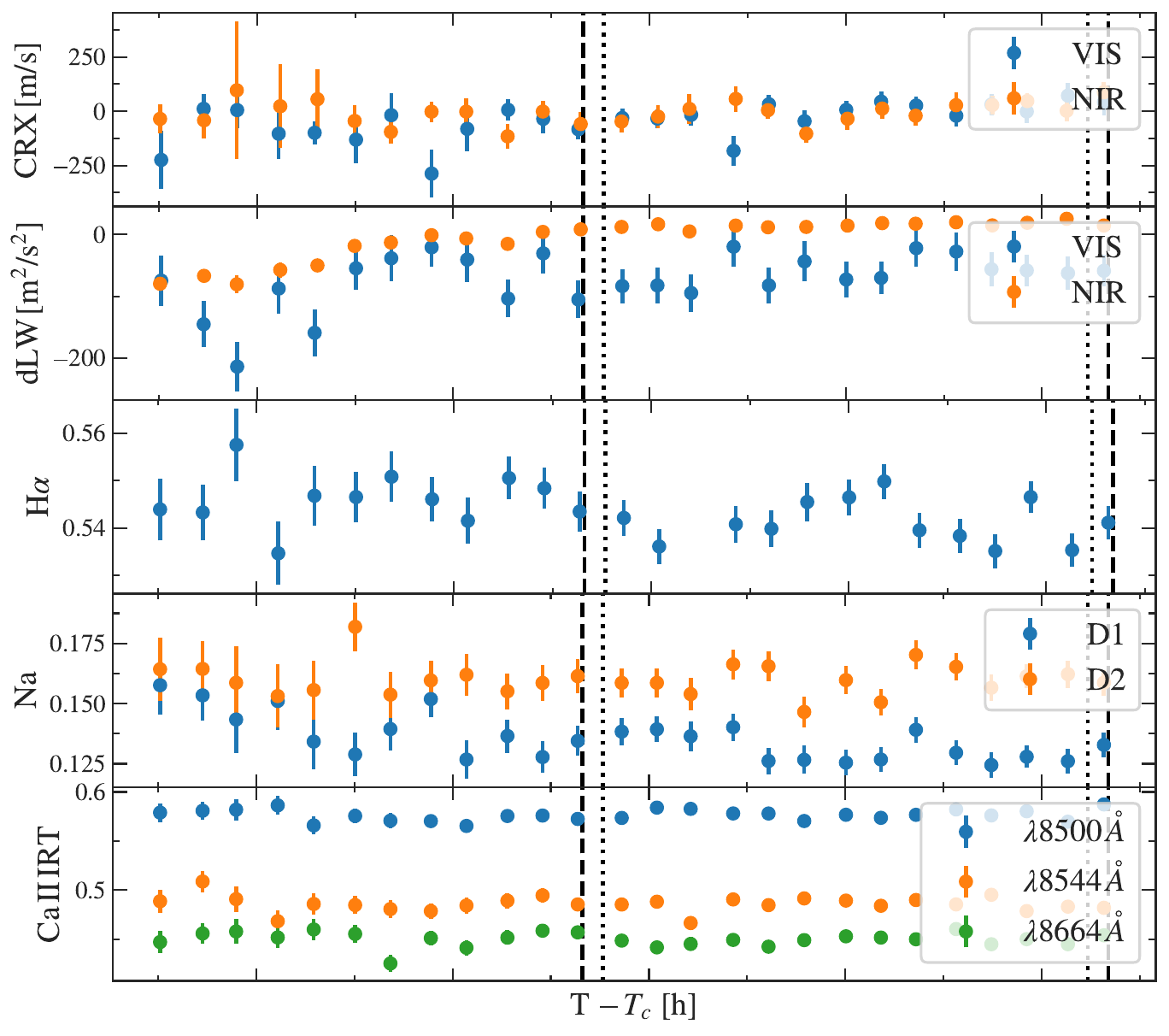}
    \end{subfigure}
    \hfill
    \begin{subfigure}{0.49\textwidth}
         \centering
         \includegraphics[width=\textwidth]{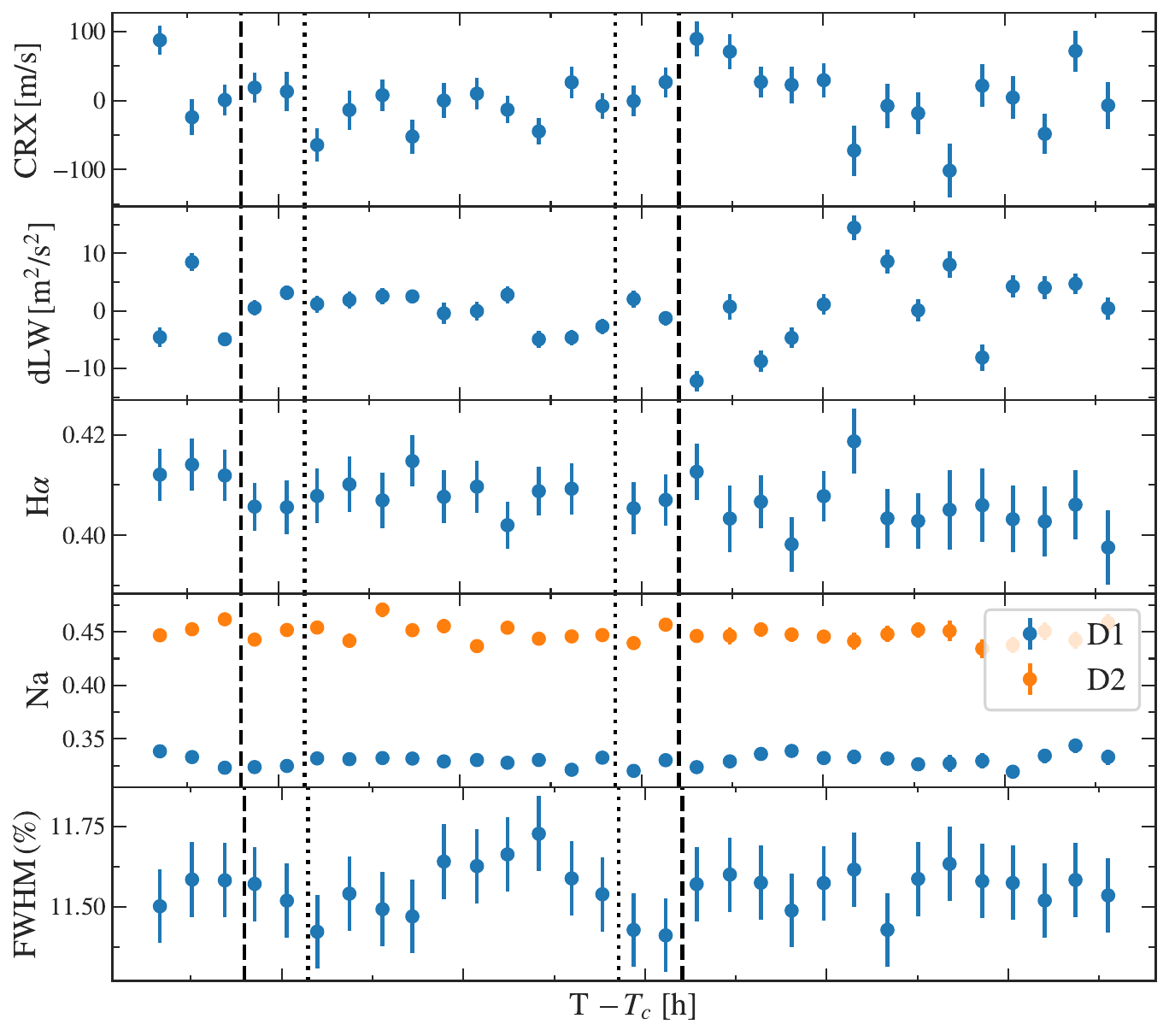}
    \end{subfigure}
    \hfill
    \begin{subfigure}{0.49\textwidth}
         \centering
         \includegraphics[width=\textwidth]{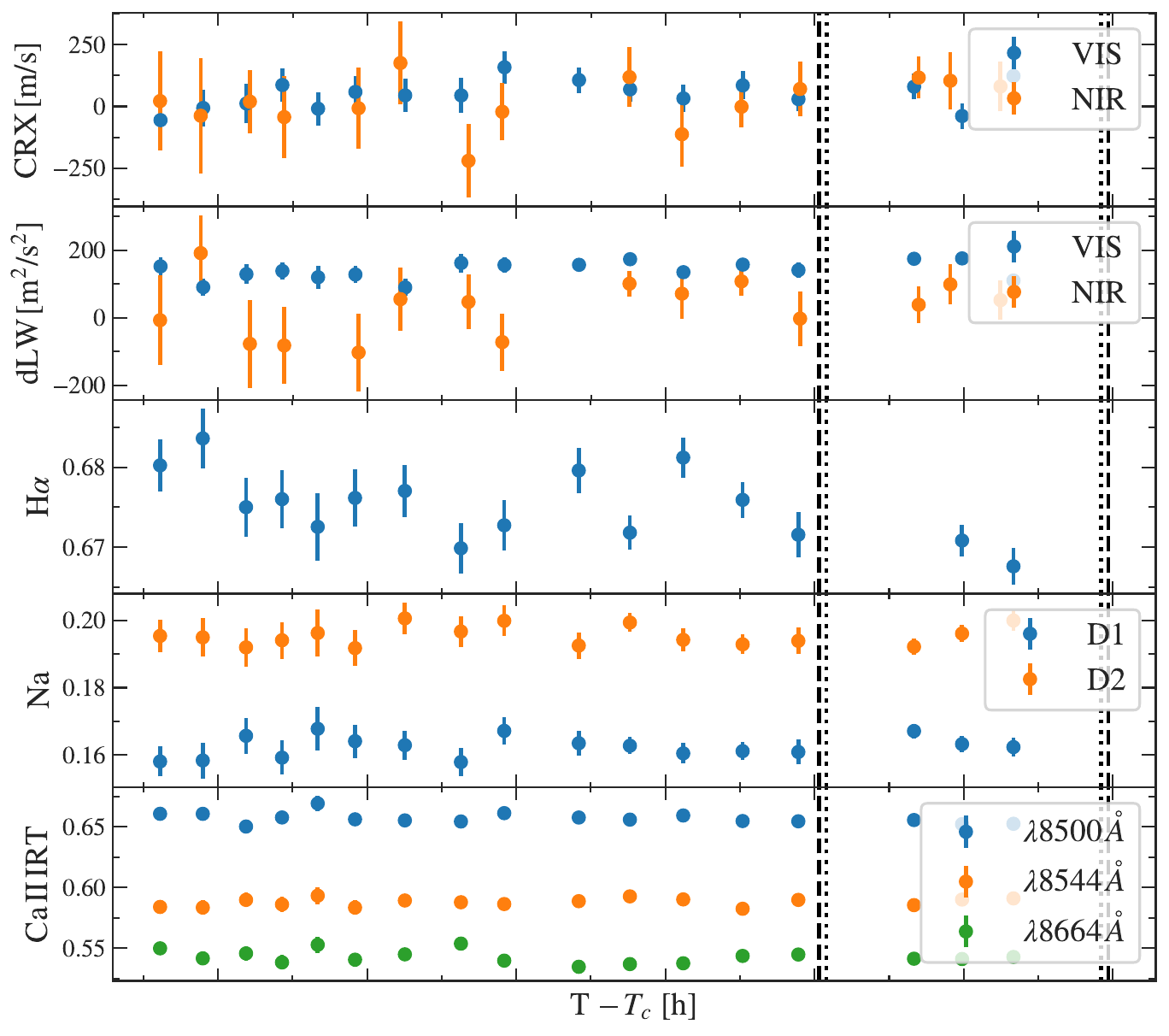}
    \end{subfigure}
    \hfill
    \begin{subfigure}{0.49\textwidth}
         \centering
         \includegraphics[width=\textwidth]{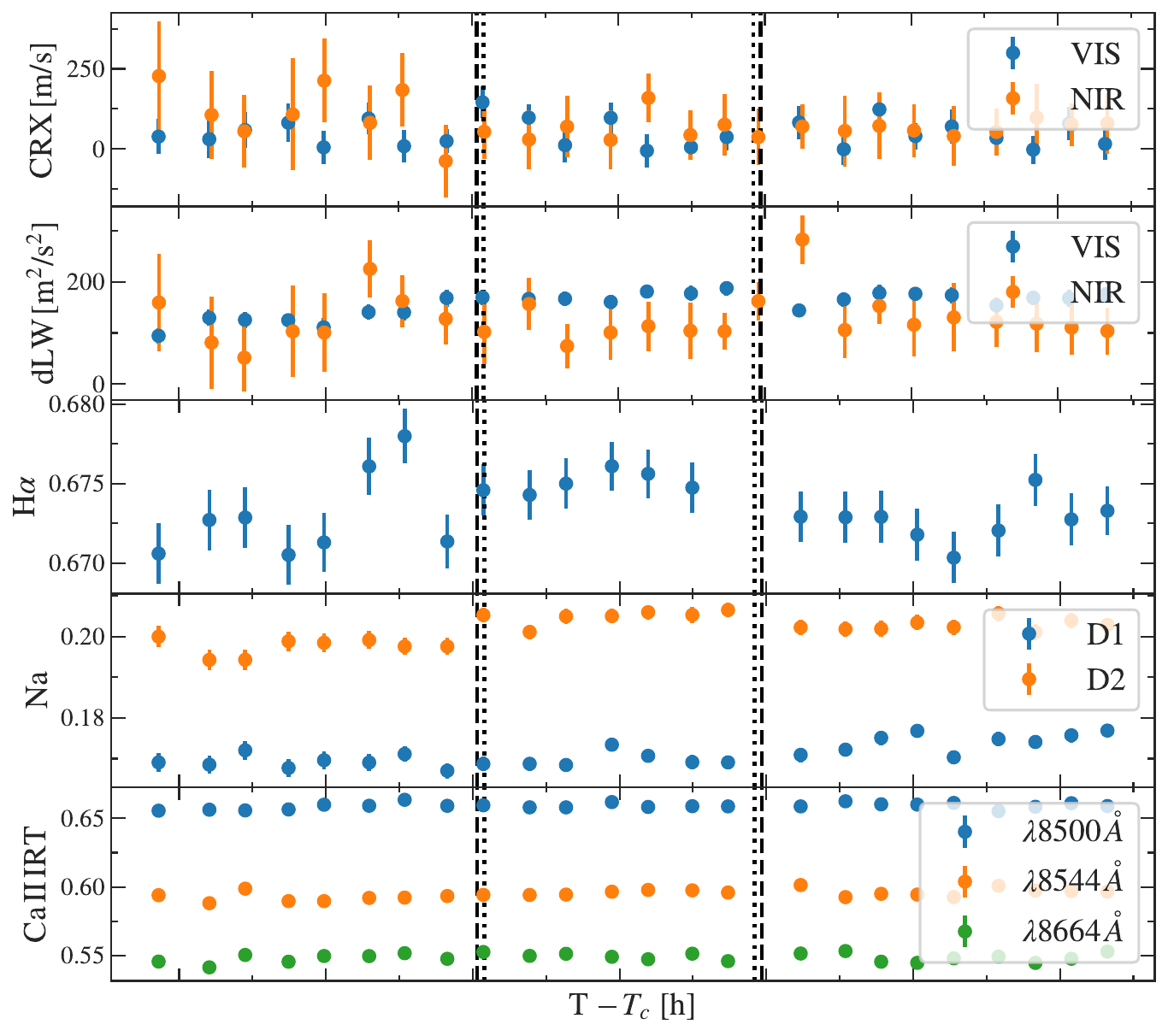}
    \end{subfigure}
    \caption{\label{Fig: serval 2}
    Same as Fig.\,\ref{Fig: serval 1}, but for HD\,73583\,b (top left), HD\,73583\,b (top right), K2-77\,b (middle left), TOI-2046\,b (middle right), and TOI-1807\,b (16 December 2021, bottom left; and 23 December 2022, bottom right).
    }
\end{figure*}

\begin{figure*}[h!]
    \centering
    \begin{subfigure}{0.49\textwidth}
         \centering
         \includegraphics[width=\textwidth]{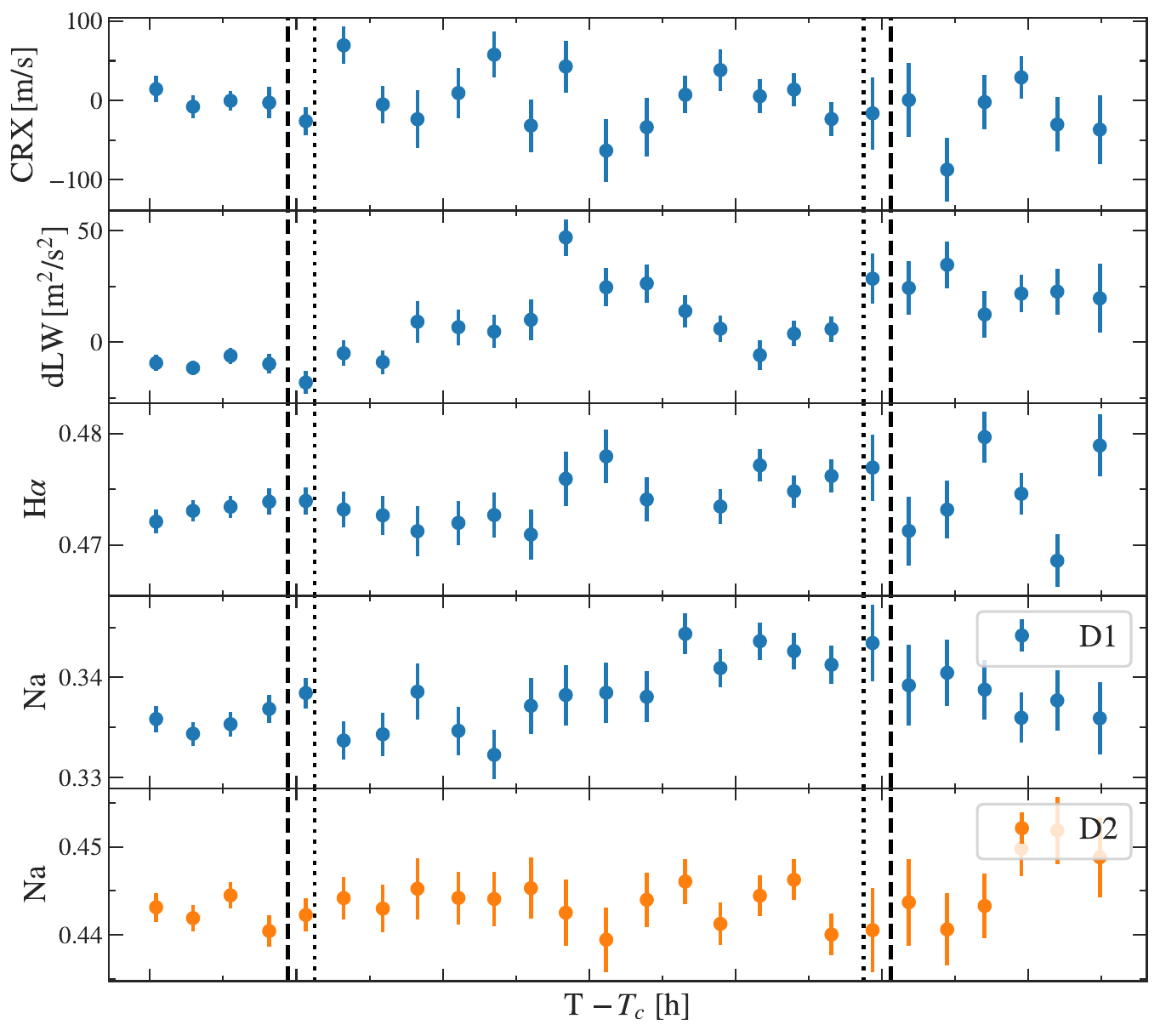}
    \end{subfigure}
    \hfill
    \begin{subfigure}{0.49\textwidth}
         \centering
         \includegraphics[width=\textwidth]{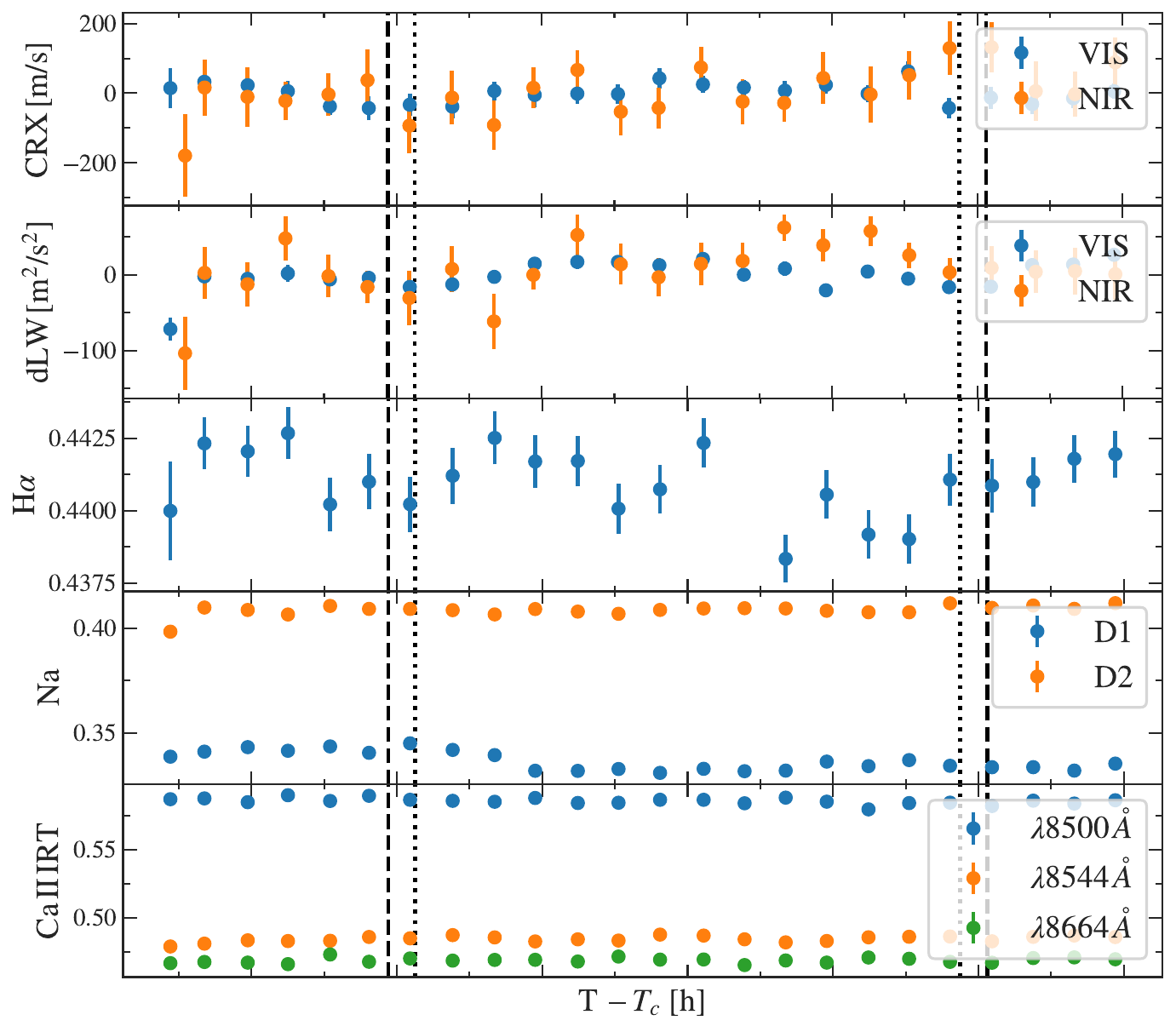}
    \end{subfigure}
    \hfill
    \begin{subfigure}{0.49\textwidth}
         \centering
         \includegraphics[width=\textwidth]{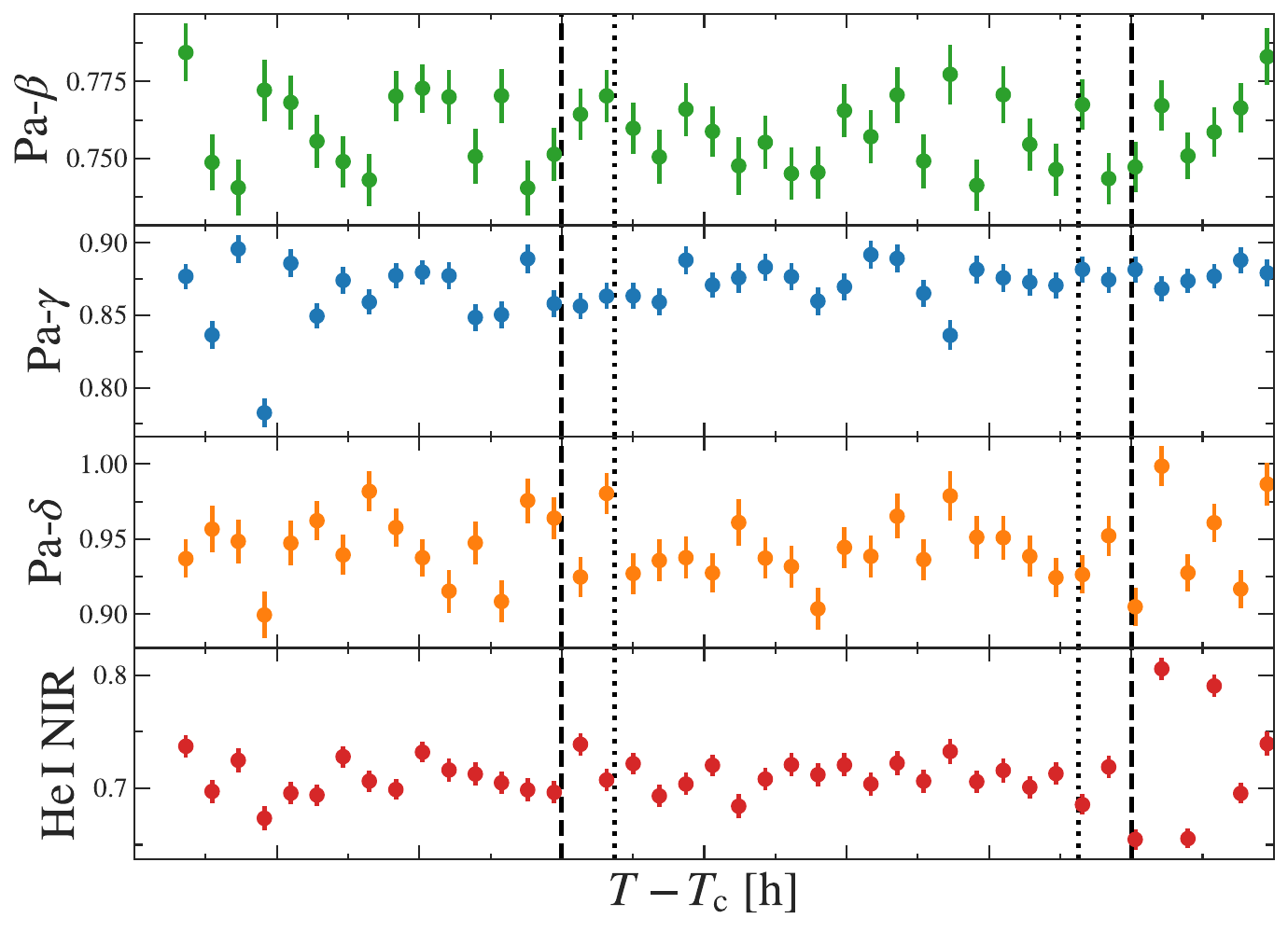}
    \end{subfigure}
    \hfill
    \begin{subfigure}{0.49\textwidth}
         \centering
         \includegraphics[width=\textwidth]{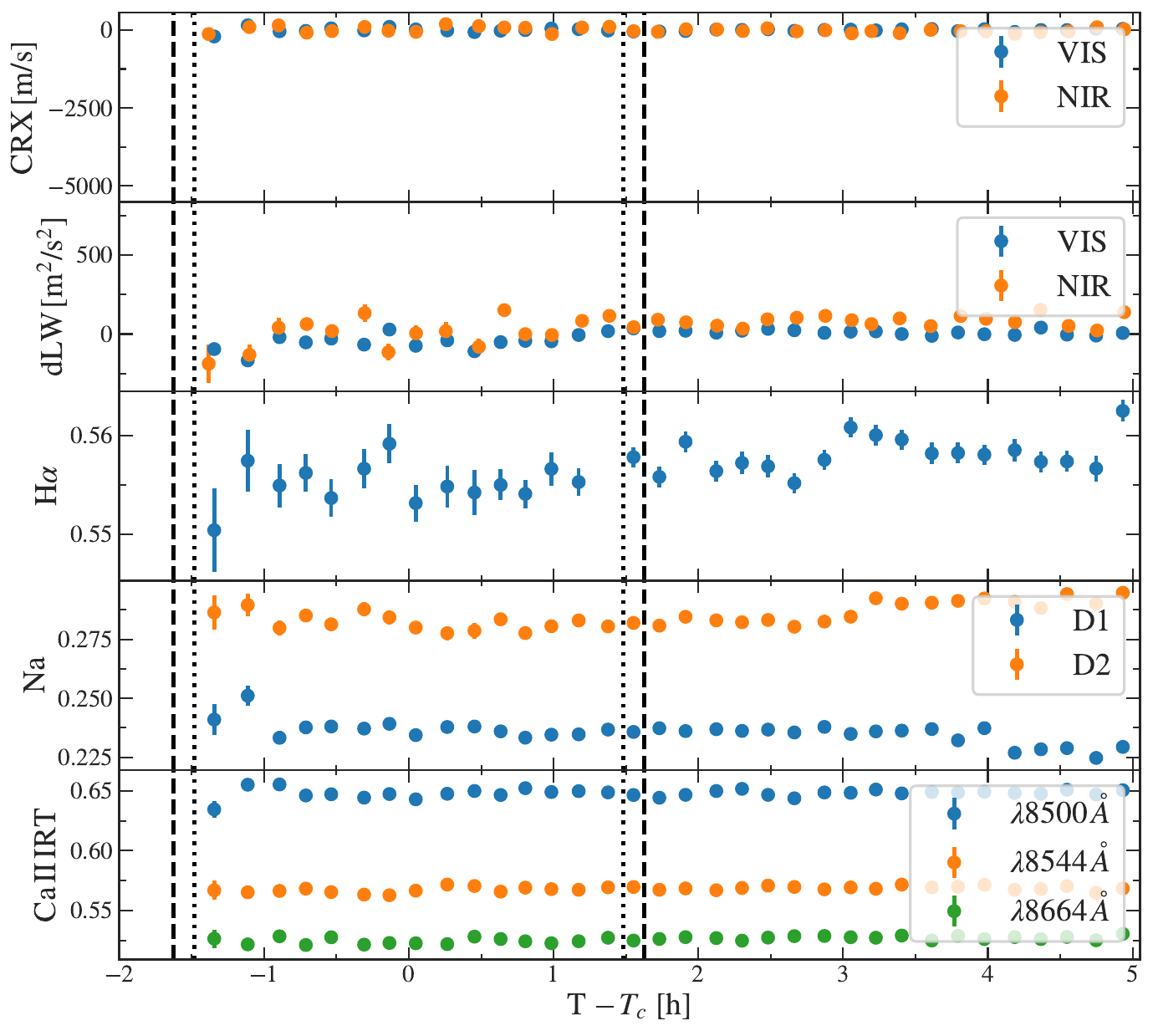}
    \end{subfigure}
    \hfill
    \begin{subfigure}{0.485\textwidth}
         \centering
         \includegraphics[width=\textwidth]{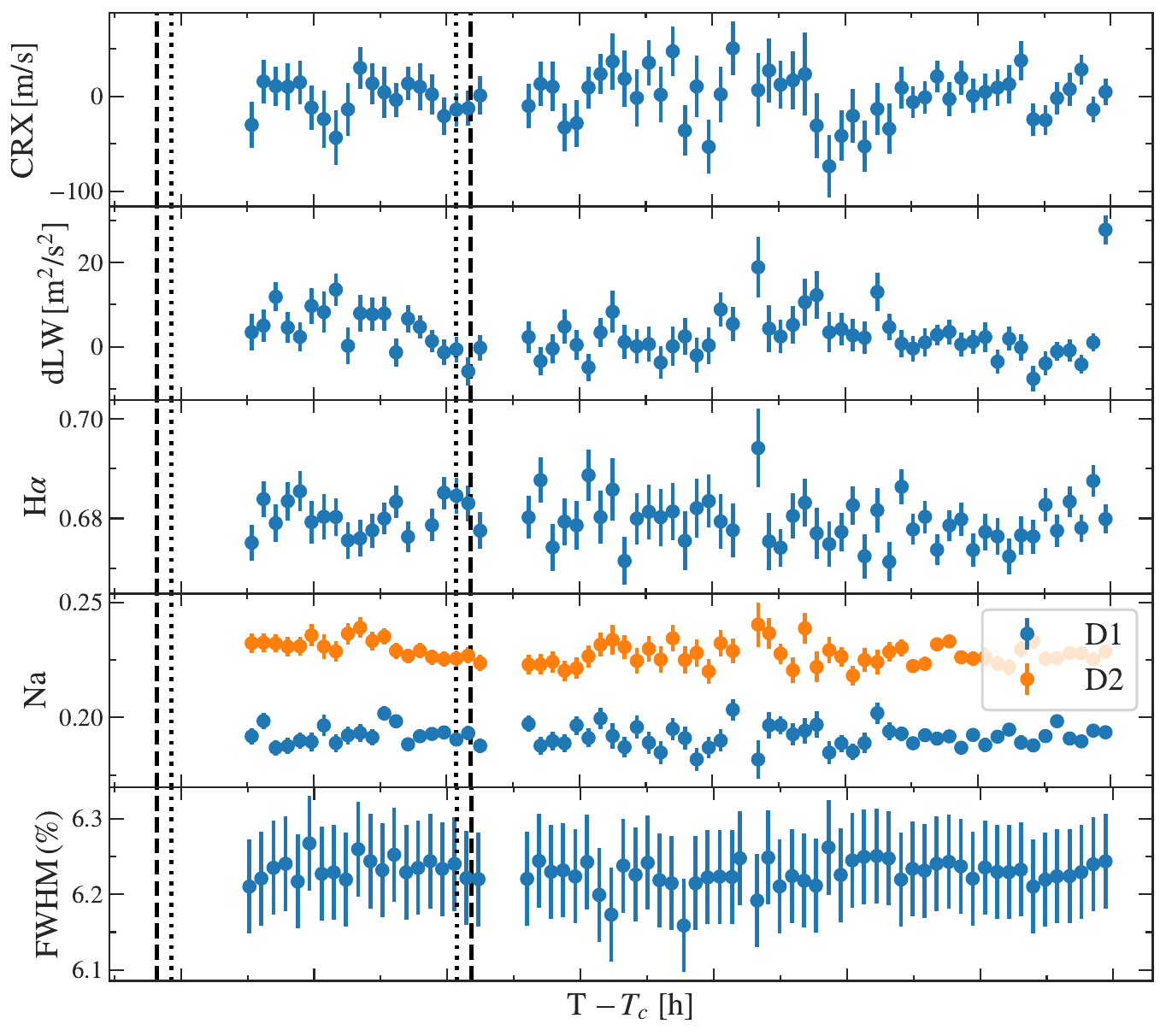}
    \end{subfigure}
    \hfill
    \begin{subfigure}{0.485\textwidth}
         \centering
         \includegraphics[width=\textwidth]{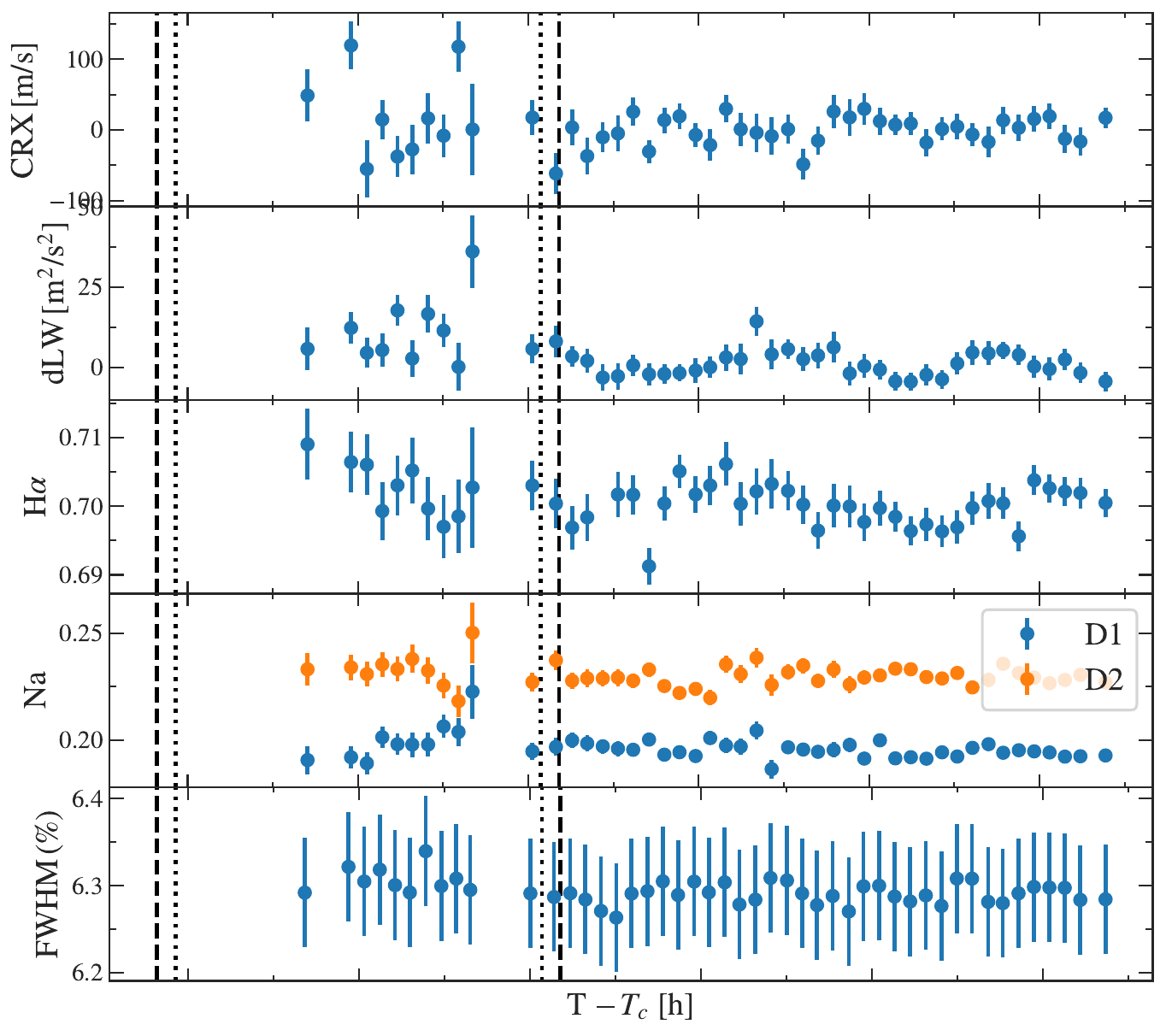}
    \end{subfigure}
    \caption{\label{Fig: serval 2}
    Same as Fig.\,\ref{Fig: serval 1}, but for TOI-1136\,d (14 May 2021, top left; and 30 January 2023, top right), TOI-1268\,b (middle left), and TOI-2076\,b (middle right), and TOI-2018\,b (9 April 2022, bottom left; and 15 June 2022, bottom right).
    }
\end{figure*}

\begin{figure}[]
    \centering
    \includegraphics[width=0.45\textwidth]{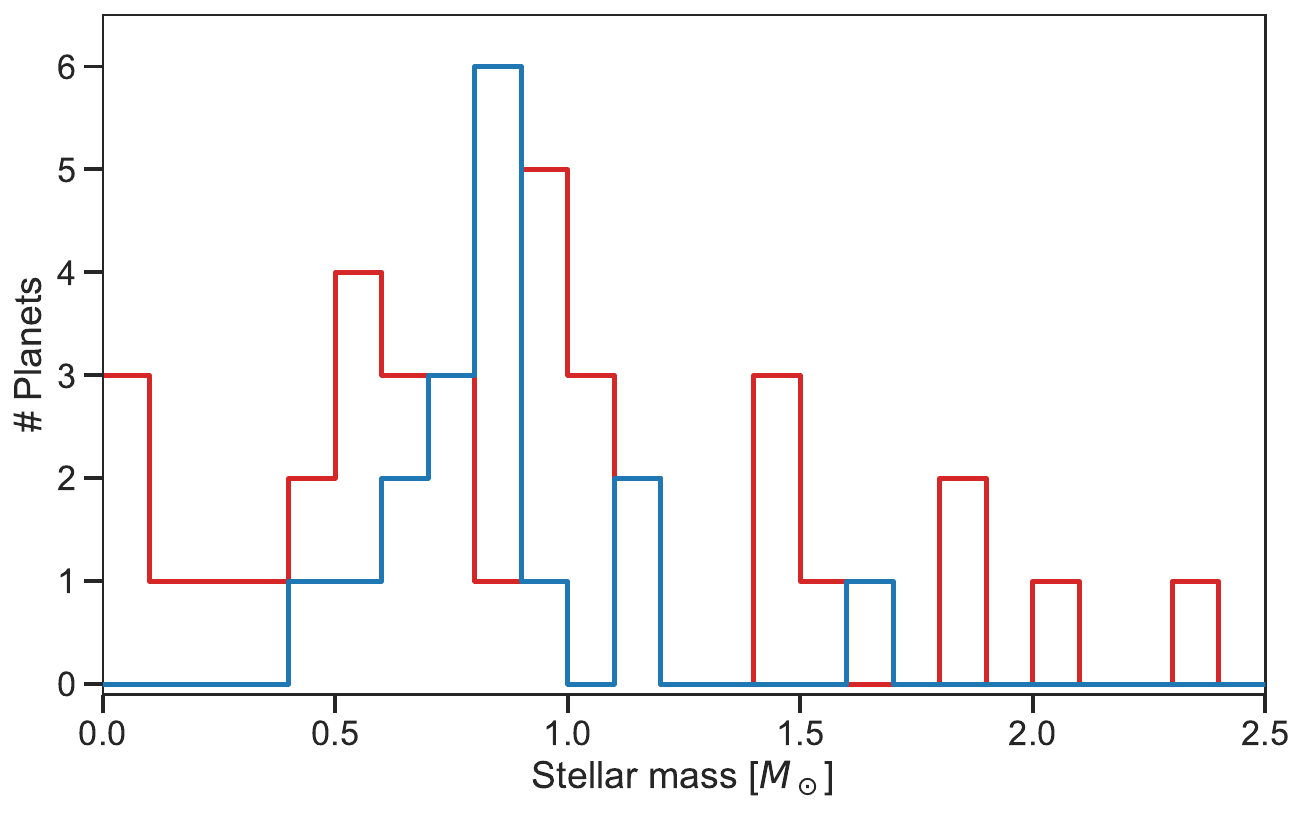}
    \caption{\label{Fig: histogram Stellar mass} Histogram of \ion{He}{i} triplet observations (detections: blue, non-detections: red) as a function of stellar mass.}
\end{figure}


\section{V1298\,Tau extra material}
\label{App: V1298 material}

\begin{figure}[h!]
    \centering
     \begin{subfigure}{0.49\textwidth}
         \centering
         \includegraphics[width=\textwidth]{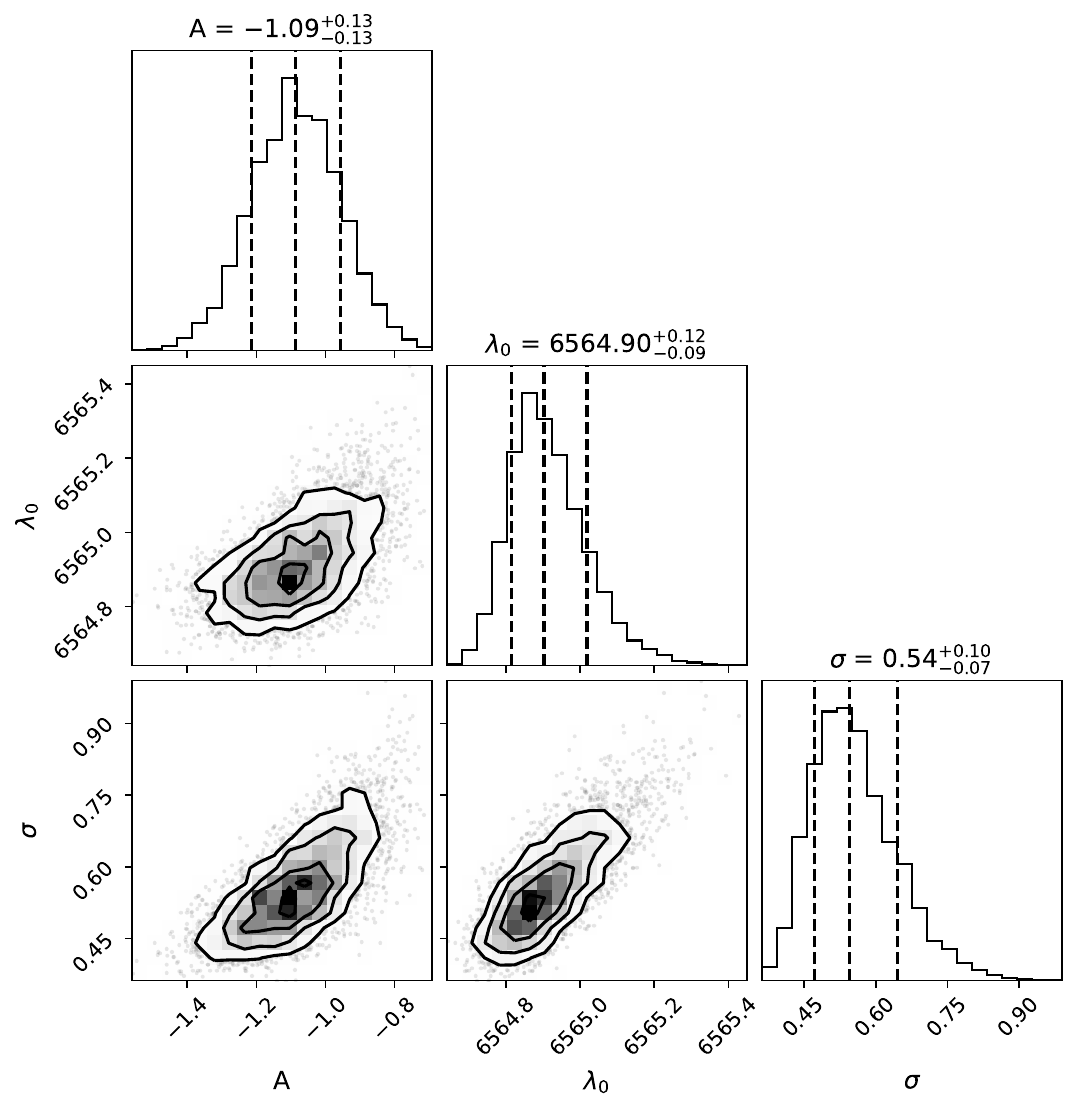}
     \end{subfigure}
     \hfill
     \begin{subfigure}{0.49\textwidth}
         \centering
         \includegraphics[width=\textwidth]{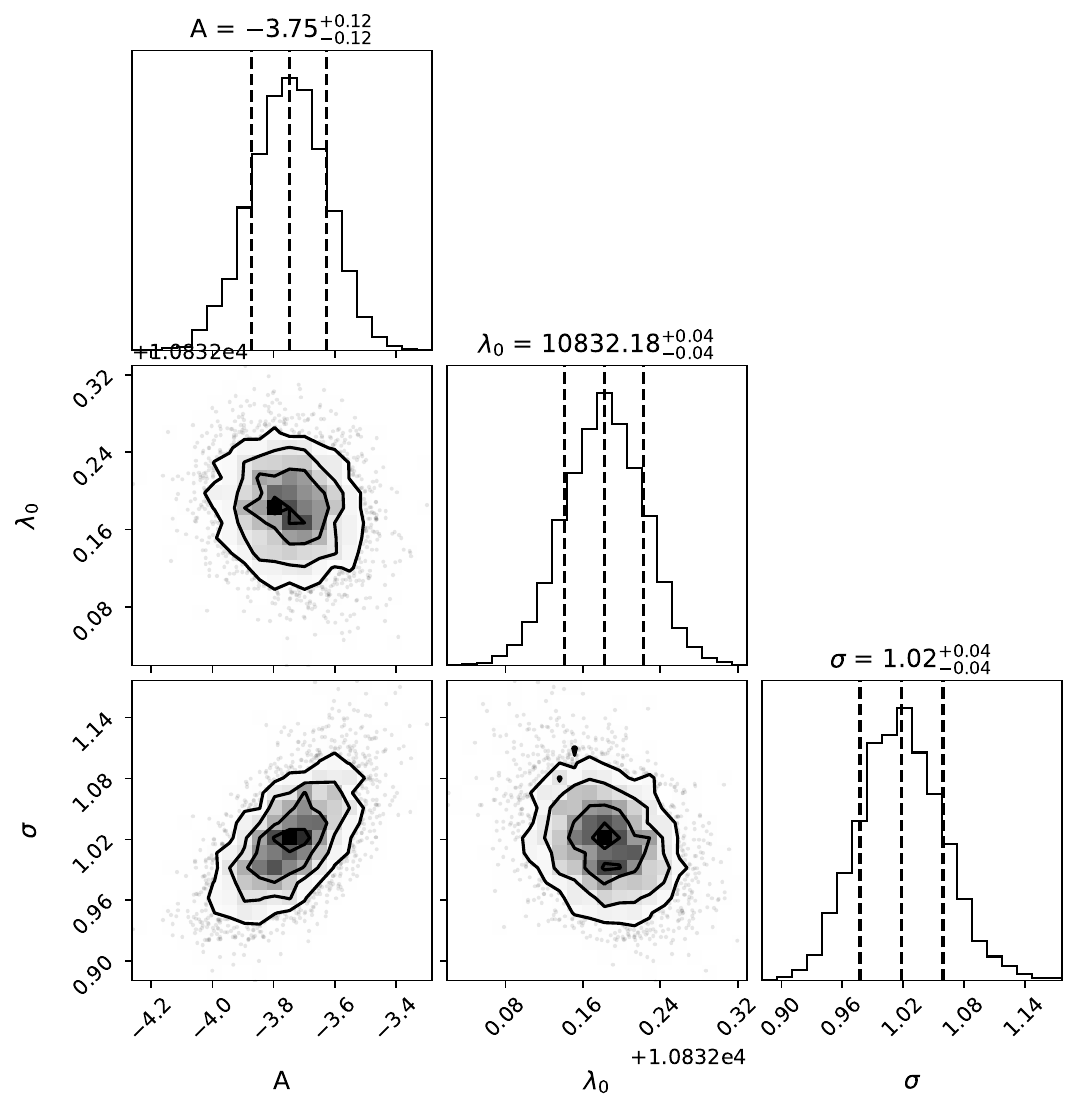}
     \end{subfigure}
    \caption{\label{Fig: V1298Tau_c nested Halpha Helium}
    Corner plot for the nested sampling posterior distribution of V1298 Tau\,c H$\alpha$ (left) and \ion{He}{i} (right) feature.
    }
\end{figure}

\begin{table}[h!]
\caption[width=\textwidth]{
\label{table - V1298Tau_c Ha priors and posteriors}
Prior and posterior distributions from the nested sampling fitting for V1298 Tau\,c H$\alpha$ feature (see Fig.\,\ref{Fig: TS V1298Tau c}). Prior label $\mathcal{U}$ represents uniform distribution.
}
\centering

\begin{tabular}{lcc}

\hline \hline 
\noalign{\smallskip} 

Parameter & Prior & Posterior \vspace{0.05cm}\\
\hline
\noalign{\smallskip}

Absorption [\%] & $\mathcal{U}(-2, 0)$ & $-$1.10$\pm$0.13  \vspace{0.05cm} \\ 
$\lambda_0$ [\AA] & $\mathcal{U}(6560,6570)$ & 6564.90$^{+0.11}_{-0.10}$ \vspace{0.05cm} \\ 
$\sigma$ [\AA] &  $\mathcal{U}(0,1)$ & 0.55$^{+0.10}_{-0.08}$  \vspace{0.05cm} \\ 

$\Delta$v [km\,s$^{-1}$] & -- & 12$^{+5}_{-4}$ \\
FWHM [\AA] & -- &  1.30$^{+0.24}_{-0.17}$ \\
EW [m\AA] &  -- & 14.8$^{+1.7}_{-1.6}$  \vspace{0.05cm} \\ 

\noalign{\smallskip}
\hline
\end{tabular}

\end{table}

\begin{table}[h!]
\caption[width=\textwidth]{
\label{table - V1298Tau_c He priors and posteriors}
Prior and posterior distributions from the nested sampling fitting for V1298 Tau\,c \ion{He}{I} feature (see Fig.\,\ref{Fig: TS V1298Tau c}). Prior label $\mathcal{U}$ represents uniform distribution.
}
\centering

\begin{tabular}{lcc}

\hline \hline 
\noalign{\smallskip} 

Parameter & Prior & Posterior \vspace{0.05cm}\\
\hline
\noalign{\smallskip}

Absorption [\%] & $\mathcal{U}(-4.5,4.5)$ &  $-$3.75$\pm$0.12  \vspace{0.05cm} \\ 
$\lambda_0$ [\AA] & $\mathcal{U}(10830,10835)$ & 10832.18$\pm$0.04 \vspace{0.05cm} \\ 
$\sigma$ [\AA] &  $\mathcal{U}(0.0,2)$ & 1.02$\pm$0.04  \vspace{0.05cm} \\ 

$\Delta$v [km\,s$^{-1}$] & -- &  $-$28.7$\pm$1.1 \\
FWHM [\AA] & -- &  2.40$\pm$0.10 \\
EW [m\AA] &  -- & 95.9$^{+3.2}_{3.0}$  \vspace{0.05cm} \\ 

\noalign{\smallskip}
\hline
\end{tabular}

\end{table}

\begin{figure*}[]
    \centering
    \includegraphics[width=0.45\textwidth]{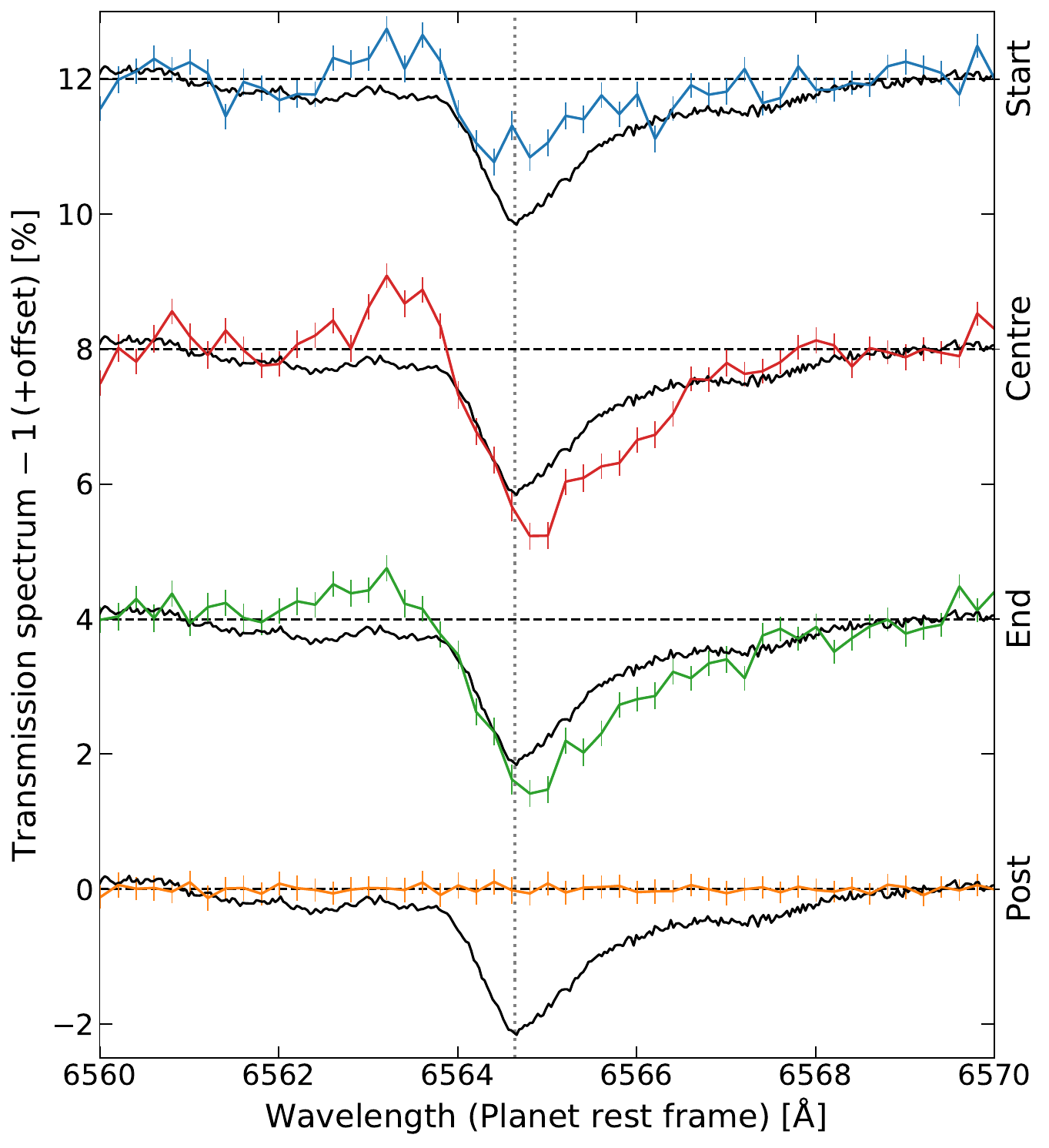}
    \includegraphics[width=0.45\textwidth]{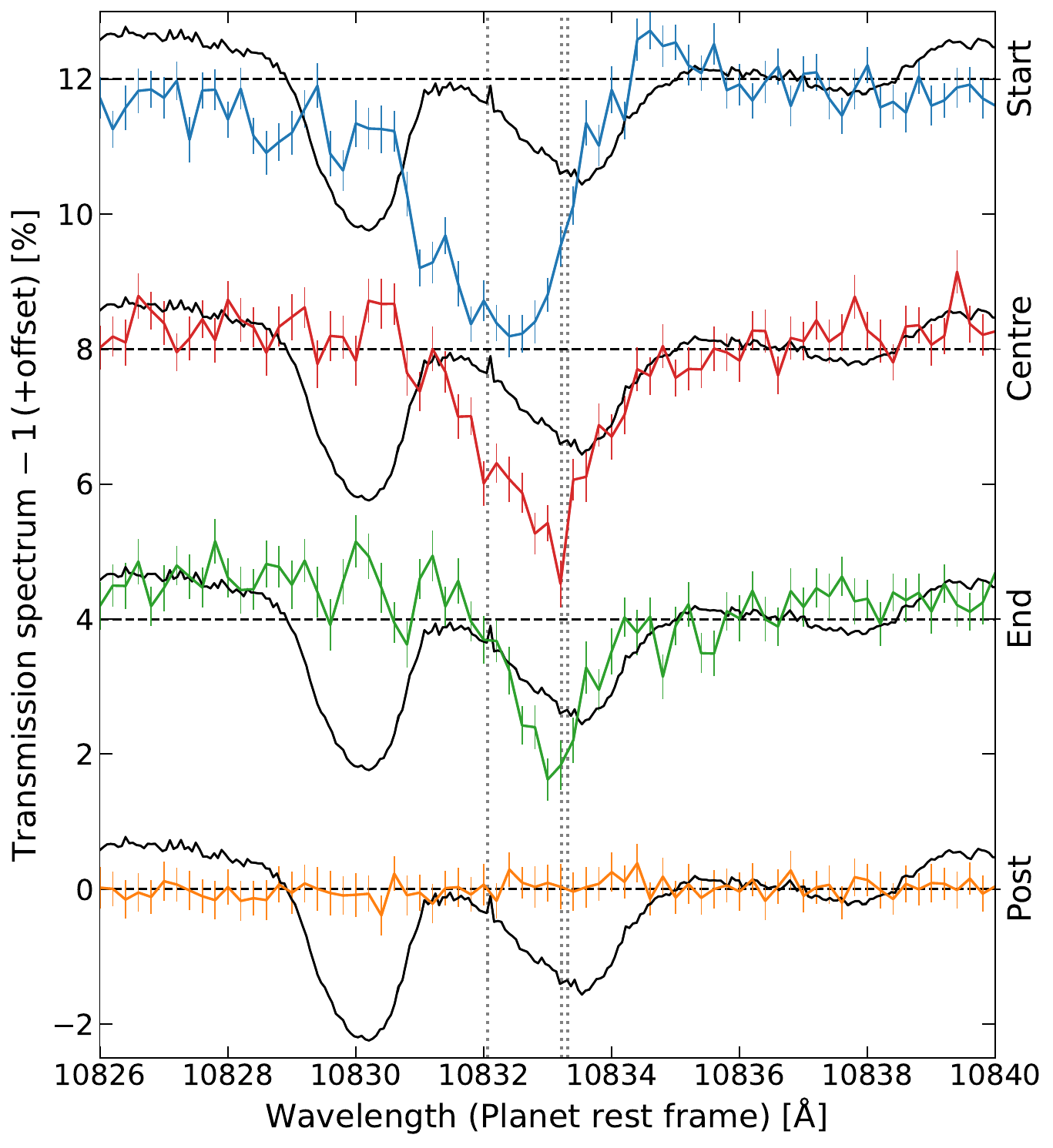}
    \caption{\label{Fig: V1298Tau_line_profile}
    TS around the H$\alpha$ (left) and \ion{He}{I} triplet (right) lines for the transit phases from top to bottom (consecutively offset): start (blue), centre (red), end (green), and post-transit (orange). Master-Out spectrum is overplotted in black along with the individual TS for comparison. The dotted vertical lines indicate the lines positions. The wavelengths in this figure are given in vacuum.
    }
\end{figure*}

\clearpage

\section{TOI-2048 extra material}
\label{App: TOI-2048 juliet}

We analysed the TESS data from Sectors\,16, 23, 24, 50, and 51 with the procedure described in Sect.\,\ref{Sect: Photommetric observations}. We used the \texttt{celerite} GP quasi-periodic kernel to account for the young star variability and we adopted the stellar parameters from \citet{TOI2048b_Newton} for a proper comparison.

\begin{figure*}[h!]
    \centering
    \includegraphics[width=1\linewidth]{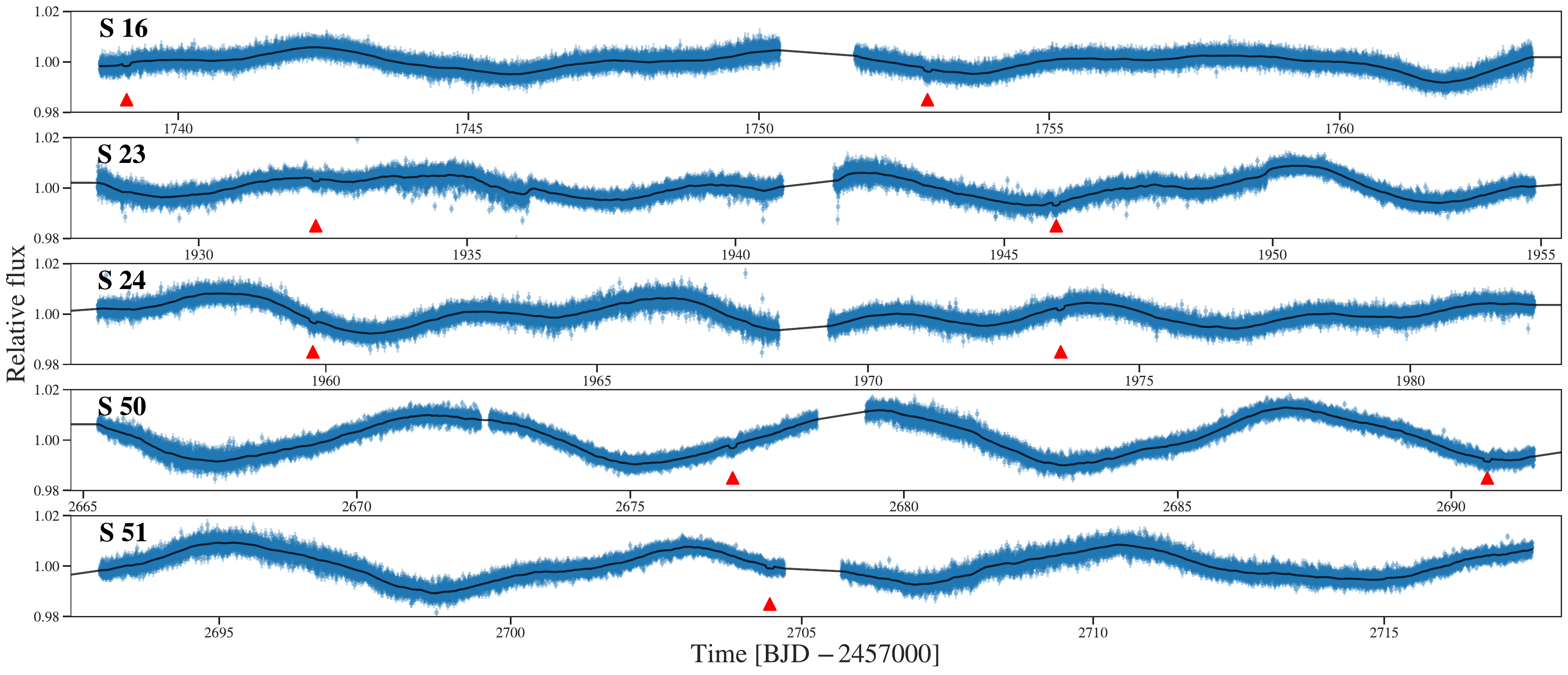}
    \caption{\label{Fig: TOI-2048 TESS fit}
    TOI-2048 two-minute cadence SAP TESS photometry from Sectors 16, 23, 24, 50, and 51 along with the transit plus GP model. The upward-pointing red triangles mark the transits for TOI-2048\,b.
    }
\end{figure*}

\begin{figure}
    \centering
    \includegraphics[width=\hsize]{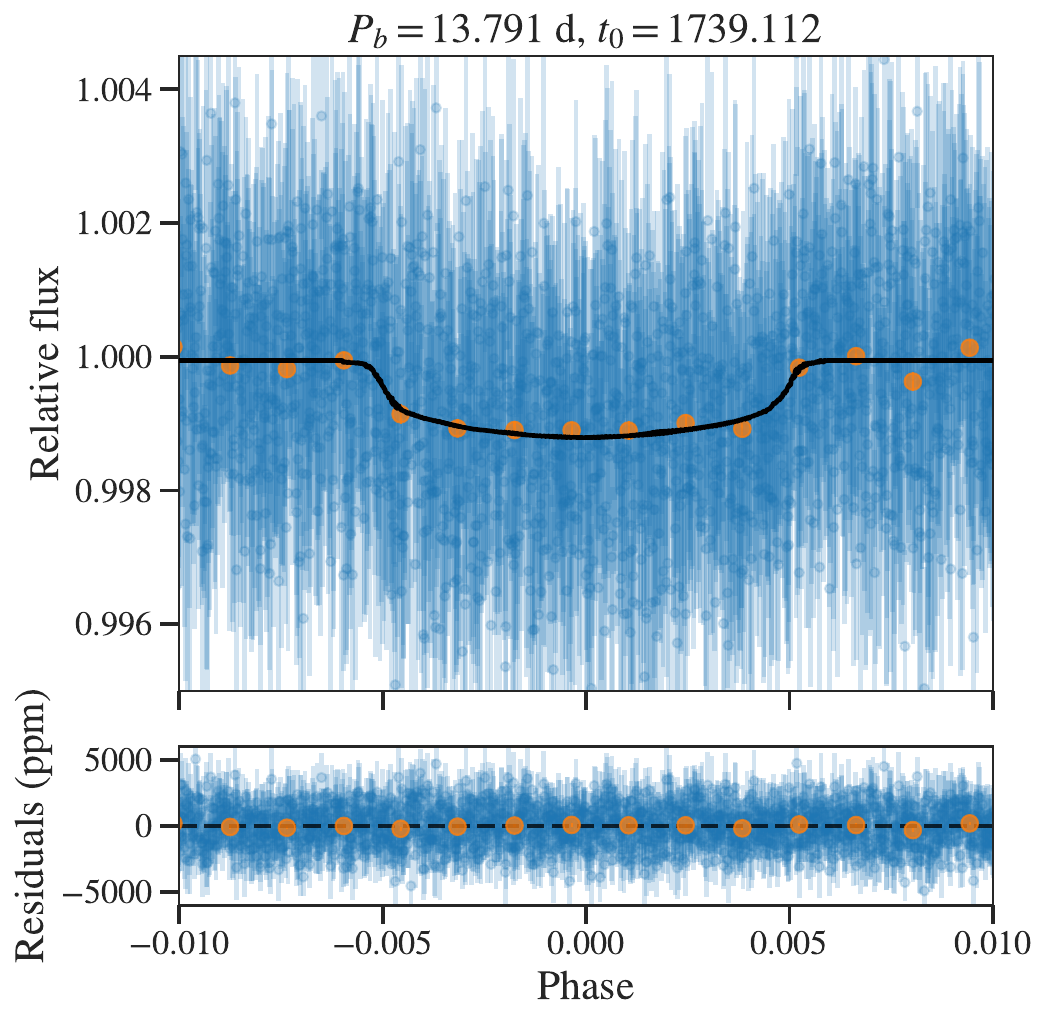}
    \caption{\label{Fig: TOI-2048 phase folded}
    TOI-2048 TESS photometry (blue dots with error bars) phase-folded to the period $P$ and central time of transit $t_0$ (shown above the panel, $t_0$ units are BJD\,$-$\,2\,457\,000) derived from the \texttt{juliet} fit. The black line is the best transit model for TOI-2048\,b. The orange points show binned photometry for visualisation.
    }
\end{figure}

The fitted parameters with their prior and posterior values, and the derived parameters for TOI-2048\,b are shown in Table\,\ref{table - TOI-2048 juliet priors and posteriors}. The TESS data along with the best transiting and GP models is shown in Fig.\,\ref{Fig: TOI-2048 TESS fit} and TOI-2048\,b phase folded transit is shown in Fig.\,\ref{Fig: TOI-2048 phase folded}. We obtained similar results using the \texttt{celerite} GP exponential kernel.

To derive an estimation of TOI-2048\,b's semi-amplitude $K_{\star}$, first we forecasted its mass using the probabilistic mass-radius relationship for sub-Neptune-sized planets (${R}_{\mathrm{p}}$\,<\,4\,${R}_\oplus$) of \citet{Wolfgang_2016}. We predicted a planetary mass of $\sim$9\,$\pm$\,2\,${M}_\oplus$, which is translated into $K_{\star}$\,$\simeq$\,2.8\,$\pm$\,0.6\,m\,s$^{-1}$ using the equation
\begin{equation}
K_{\star} = 28.4\,{\rm m\,s^{-1}} ~ ( P_{\mathrm{pl}}/ {\rm year} )^{-1/3} ~  ( M_{\mathrm{pl}} / {\rm M_{\mathrm{Jup}}} ) ~ ( M_{\star}/{\rm M_{\odot}} )^{-2/3}
\label{eq: Semi-amplitude}
\end{equation}

\begin{table}[h!]
\caption[width=\textwidth]{
\label{table - TOI-2048 juliet priors and posteriors}
Prior and posterior distributions from the \texttt{juliet} fitting for TOI-2048\,b. Prior labels $\mathcal{U}$, $\mathcal{N}$, $\mathcal{F}$, and $\mathcal{J}$ represents uniform, normal, fixed, and Jeffrey's distribution, respectively.
}
\centering

\begin{tabular}{lcc}

\hline \hline 
\noalign{\smallskip} 

Parameter & Prior & Posterior \vspace{0.05cm}\\
\hline
\noalign{\smallskip}

$P$ [d] & $\mathcal{N}(13.7905,0.0001)$ & 13.790546\,(55)  \vspace{0.05cm} \\ 
$t_0$\,$^{(a)}$ & $\mathcal{N}(1739.11,0.01)$ & 1739.1123\,(27)  \vspace{0.05cm} \\ 
\textit{ecc} & $\mathcal{F }( 0)$ & --  \vspace{0.05cm} \\ 
$\omega$ (deg) & $\mathcal{F }( 90)$ & --  \vspace{0.05cm} \\ 
$r_{1}$ & $\mathcal{U }(0,1 )$ & 0.54$\pm$0.12  \vspace{0.05cm} \\ 
$r_{2}$ & $\mathcal{ U}(0,1 )$ & 0.0306$\pm$0.0018  \vspace{0.05cm} \\ 
$\rho_{\star}$ [kg\,m$^{-3}$]  & $\mathcal{ N }(2367.0,500.0 )$ & 2552$^{+350}_{-420}$  \vspace{0.05cm} \\
$\mu_{\textit{TESS}}$ (ppm) & $\mathcal{N }(0.0,0.1)$ & $-$100$^{+200}_{-180}$  \vspace{0.05cm} \\ 
$\sigma_{\textit{TESS}}$ (ppm) & $\mathcal{J }( 10^{-6}, 10^{6})$ & 5600$^{+14000}_{-5600 }$  \vspace{0.05cm} \\ 
$q_{1,\textit{TESS}}$ & $\mathcal{U }(0,1 )$ & 0.51$\pm$0.28  \vspace{0.05cm} \\ 
$q_{2,\textit{TESS}}$ & $\mathcal{ U}(0,1 )$ & 0.46$\pm$0.30  \vspace{0.05cm} \\ 
GP$_\mathrm{B}$ (ppm) & $\mathcal{J }(10^{-6}, 10^{6})$ & 33$^{+16}_{-10}$  \vspace{0.05cm} \\ 
GP$_\mathrm{L}$ [d]  & $\mathcal{J }(10^{-3}, 10^{3})$ & 43.5$^{+20}_{-12}$  \vspace{0.05cm} \\ 
GP$_\mathrm{C}$ (ppm) & $\mathcal{J }(10^{-6}, 10^{6})$ & 200$^{+ 40000}_{-180}$  \vspace{0.05cm} \\ 
GP$_\mathrm{P_{\mathrm{rot}}}$ [d] & $\mathcal{N }(8,2)$ & 7.48$\pm$0.12  \vspace{0.05cm} \\ 

\noalign{\smallskip} 
\hline 
\noalign{\smallskip} 
\multicolumn{3}{c}{\textit{ Derived planetary parameters }} \\ 
\noalign{\smallskip} 
$p = {R}_{\mathrm{p}}/{R}_{\star}$  & & 0.0306$\pm$0.0018 \vspace{0.05cm} \\  
$b =(a_{\rm p}/{R}_{\star}) \cos{ i_{\mathrm{p}} }$  & &  0.30$^{+0.16}_{-0.18}$ \vspace{0.05cm} \\  
$i_{\mathrm{p}}$ (deg)  & & 89.41$\pm$0.35 \vspace{0.05cm} \\  
$T_{14}$ [h]  &  & 3.5$\pm$0.1 \vspace{0.05cm} \\  
$T_{12}$ [min]  &  & 6.9$^{+1.2}_{-0.7}$  \vspace{0.05cm} \\  
$R_{\mathrm{p}}$ [${R}_\oplus$]  & & 2.60$\pm$0.20 \vspace{0.05cm} \\  
$a_{\mathrm{p}}$ [AU]  &  & 0.1078$\pm$0.0080 \vspace{0.05cm} \\  
$T_{\mathrm{eq}}$ [K]\,$^{(b)}$  & &  675$^{+22}_{-16}$  \vspace{0.05cm} \\

\noalign{\smallskip}
\hline
\end{tabular}

\tablefoot{
$^{(a)}$ Central time of transit ($t_0$) units are BJD\,$-$\,2\,457\,000.
$^{(b)}$ Equilibrium temperatures were calculated assuming zero Bond albedo.}
\end{table}

\clearpage

\section{HD\,63433 system extra material}
\label{App: HD63433 plots}

\begin{figure}[h!]
    \centering
    \includegraphics[width=\hsize]{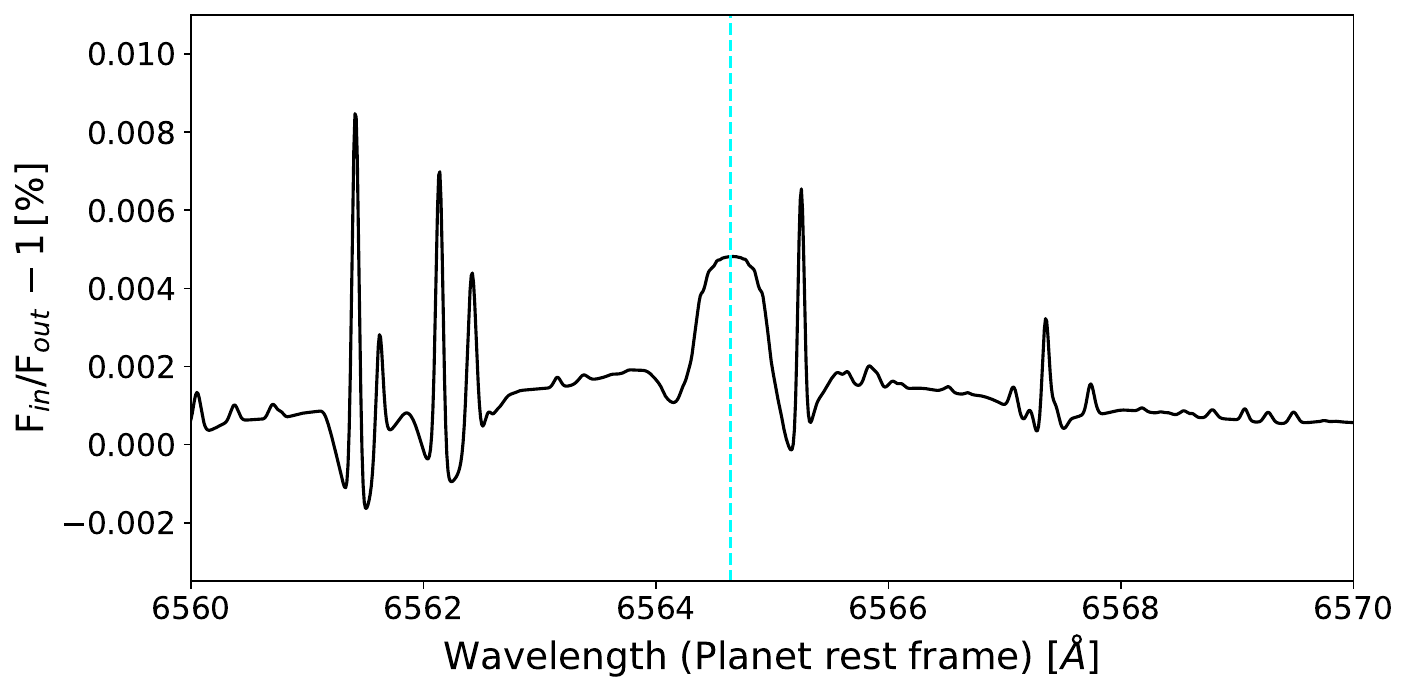}
    \includegraphics[width=\hsize]{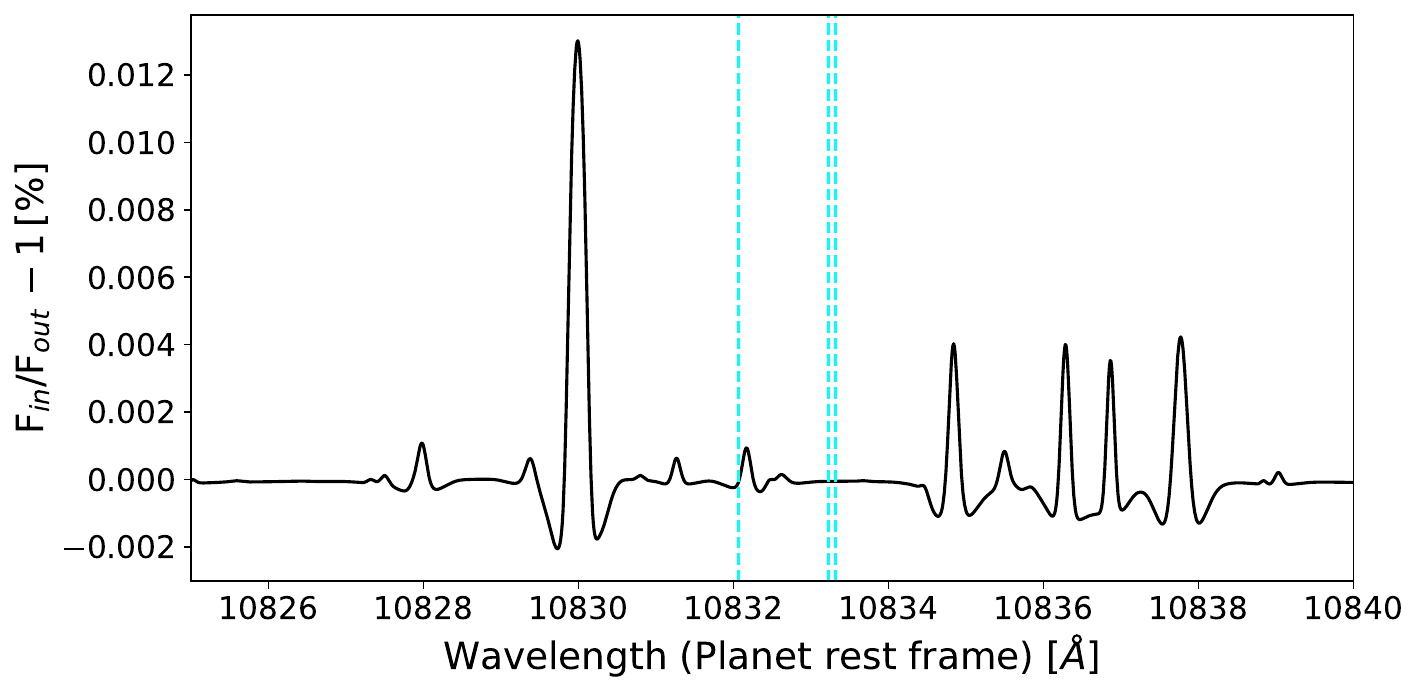}
    \caption{\label{Fig: HD63433b RM}
    Contribution of the RM and CLV effects to the HD 63433\,b transmission spectra around the H$\alpha$ (top) and \ion{He}{I} NIR triplet (bottom) lines. The vertical cyan lines mark the position of the lines of interest.
    }
\end{figure}

\begin{table}[h!]
\caption[width=\textwidth]{
\label{table - HD63433b priors and posteriors}
Prior and posterior distributions from the nested sampling fitting for HD 63433\,b H$\alpha$ feature (see Fig.\,\ref{Fig: TS HD63433 B}). Prior label $\mathcal{U}$ represents uniform distribution.
}
\centering

\begin{tabular}{lcc}

\hline \hline 
\noalign{\smallskip} 

Parameter & Prior & Posterior \vspace{0.05cm}\\
\hline
\noalign{\smallskip}

Absorption [\%] & $\mathcal{U}(-3, 3)$ & $-$1.21$^{+0.22}_{-0.24}$  \vspace{0.05cm} \\ 
$\lambda_0$ [\AA] & $\mathcal{U}(6560.0,6570.0)$ & 6564.60$\pm$0.03 \vspace{0.05cm} \\ 
$\sigma$ [\AA] &  $\mathcal{U}(0.01,0.5)$ & 0.15$^{+0.04}_{-0.03}$  \vspace{0.05cm} \\ 

$\Delta$v [km\,s$^{-1}$] & -- & $-$2.0$\pm$1.5 \\
FWHM [\AA] & -- &  0.35$^{+0.09}_{-0.07}$ \\
EW [m\AA] &  -- & 4.6$\pm$0.8  \vspace{0.05cm} \\ 

\noalign{\smallskip}
\hline
\end{tabular}

\end{table}

\begin{figure}[h!]
    \centering
    \includegraphics[width=\hsize]{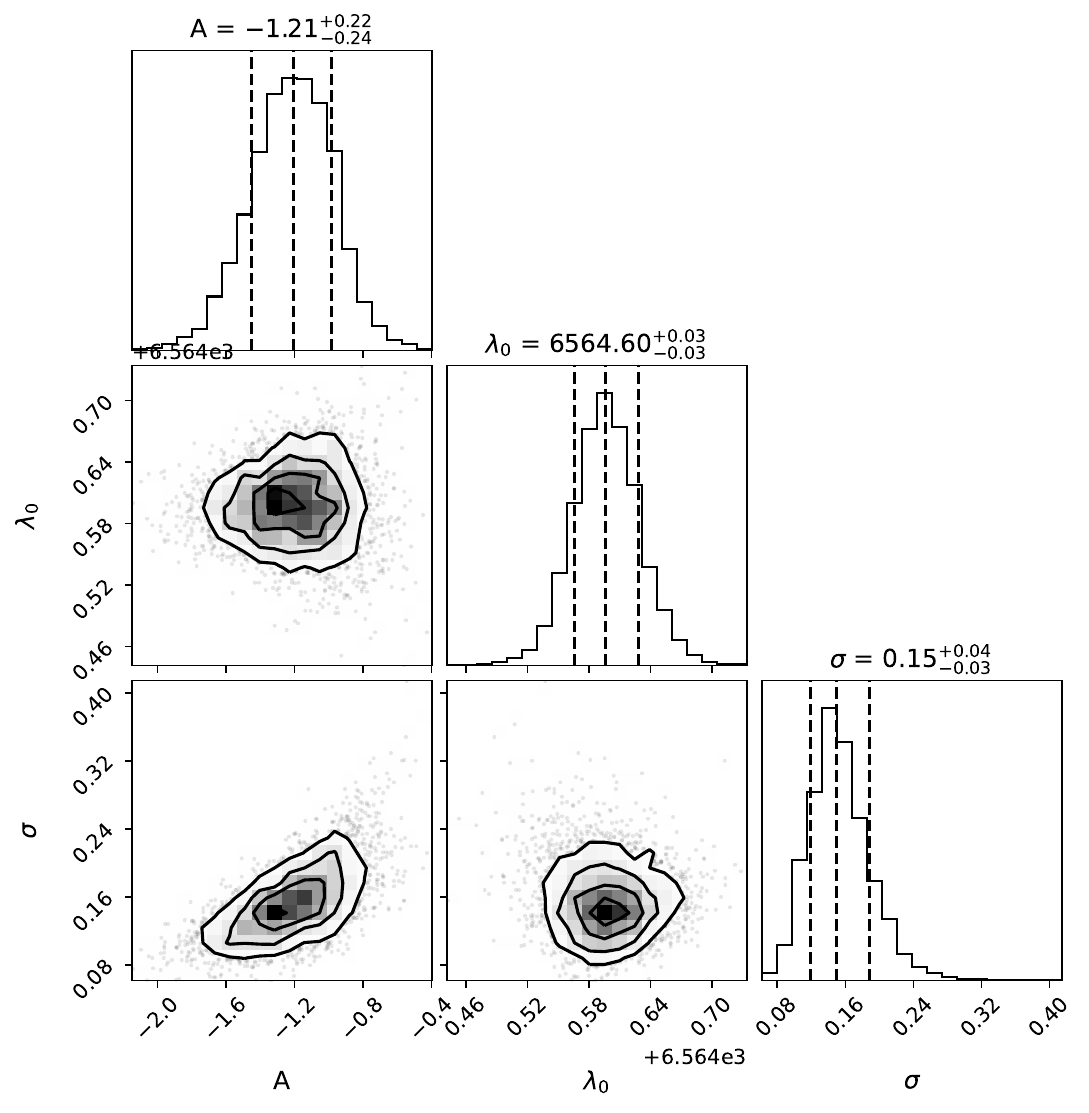}
    \caption{\label{Fig: HD63433b nested Halpha}
    Corner plot for the nested sampling posterior distribution of HD 63433\,b H$\alpha$ feature from partial transit.
    }
\end{figure}

\clearpage

\section{HD\,73583 system extra material}
\label{App: HD73583 juliet}

We analysed \hd73 TESS data from Sectors\,8, 34, and 61. Because our purpose for this system is not to study the young stellar activity but derive precise ephemeris, we used the \texttt{celerite} GP exponential kernel to account for the stellar variability and rotation. We adopted the stellar parameters used in \citet{Barragan_TOI560b}.

\begin{figure*}[h!]
    \centering
    \includegraphics[width=1\linewidth]{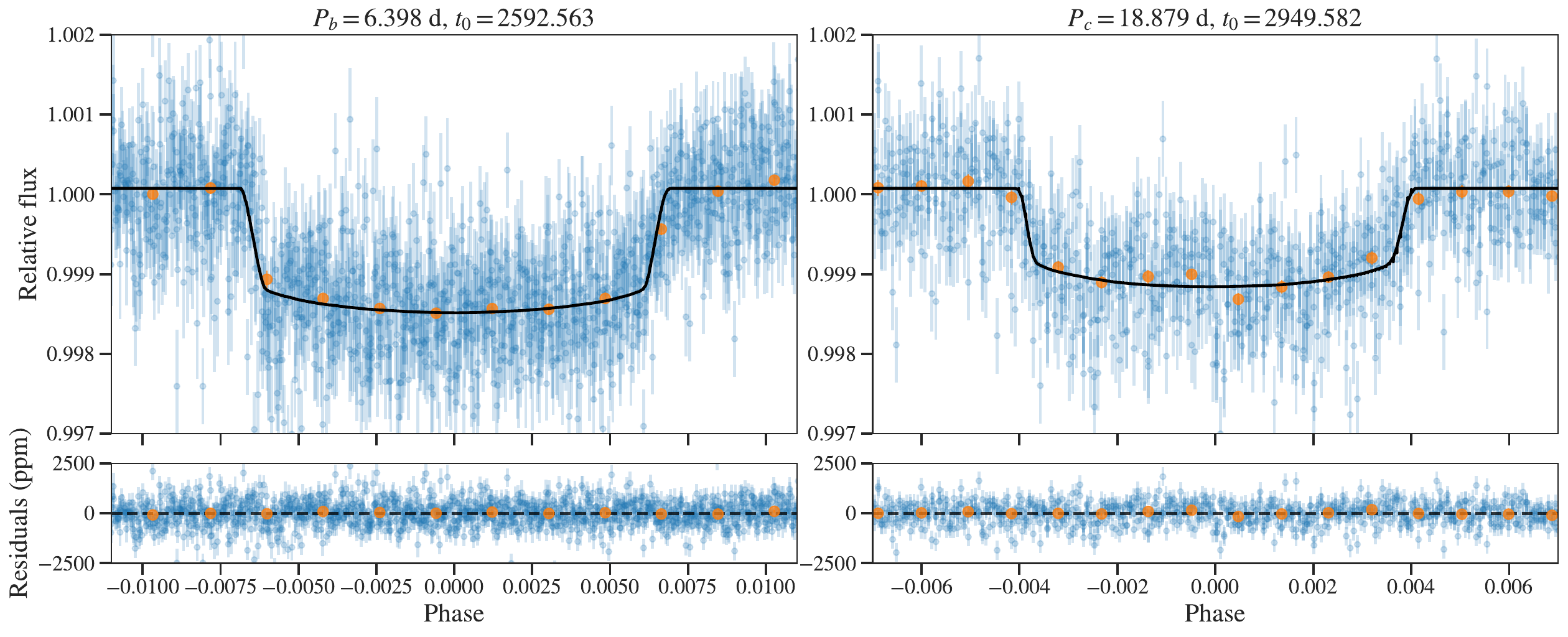}
    \caption{\label{Fig: HD73583 phase folded}
    \hd73 \textit{TESS} photometry (blue dots with error bars) phase-folded to the period $P$ and central time of transit $t_0$ (shown above each panel, $t_0$ units are BJD\,$-$\,2\,457\,000) derived for planet b (left), and c (right) from the \texttt{juliet} fit. The black line is the best transit model for each planet. The orange points show binned photometry for visualisation.
    }
\end{figure*}

\begin{figure*}[h!]
    \centering
    \includegraphics[width=1\linewidth]{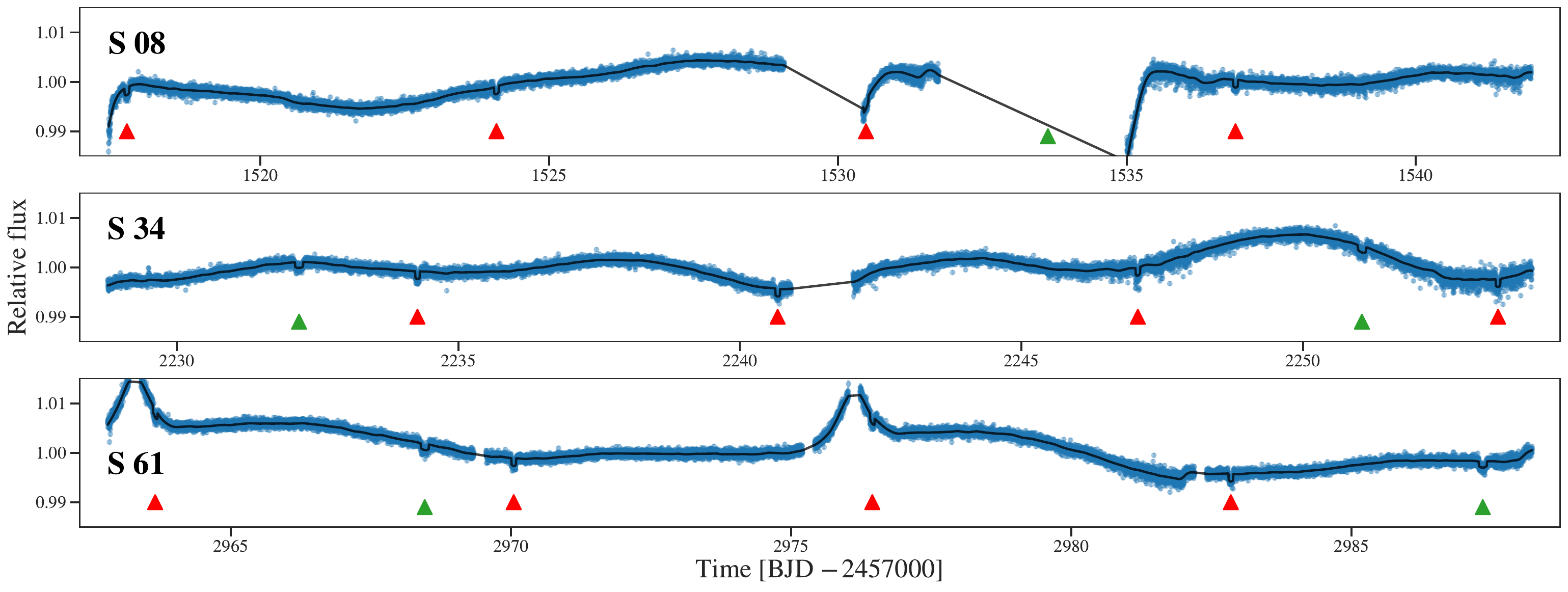}
    \caption{\label{Fig: HD73583 TESS fit}
    \hd73 2 min cadence TESS photometry from Sectors\,8, 34, and 61 along with the transit plus GP model. The upward-pointing red and green triangles mark the transit times for \hd73\,b and c, respectively.
    }
\end{figure*}

The fitted parameters with their prior and posterior values, and the derived parameters for \hd73\,b and c are shown in Table\,\ref{table - HD73583 juliet priors and posteriors}. The TESS data along with the best transiting and GP models is shown in Fig.\,\ref{Fig: HD73583 TESS fit} and \hd73\,b and c phase folded transits are shown in Fig.\,\ref{Fig: HD73583 phase folded}.

\begin{table}[h!]
\caption[width=\textwidth]{
\label{table - HD73583 juliet priors and posteriors}
Prior and posterior distributions from the \texttt{juliet} fitting for \hd73\,b and c. Prior labels $\mathcal{U}$, $\mathcal{N}$, $\mathcal{F}$, and $\mathcal{J}$ represents uniform, normal, fixed, and Jeffrey's distribution, respectively.
}
\centering

\begin{tabular}{lcc}

\hline \hline 
\noalign{\smallskip} 

Parameter & Prior & Posterior \vspace{0.05cm}\\
\hline
\noalign{\smallskip}

$P_b$ [d] & $\mathcal{N}(6.398,0.01)$ & 6.3980580\,(26)  \vspace{0.05cm} \\ 
$t_{0,b}$\,$^{(a)}$ & $\mathcal{N}(2592.55,0.1)$ & 2592.56287$^{+0.00025}_{-0.00022}$  \vspace{0.05cm} \\ 
\textit{ecc$_b$} & $\mathcal{F }(0)$ & --  \vspace{0.05cm} \\ 
$\omega_b$ (deg) & $\mathcal{F }( 90)$ & --  \vspace{0.05cm} \\ 
$r_{1,b}$ & $\mathcal{U }(0,1 )$ & 0.718$^{+0.009}_{-0.012}$  \vspace{0.05cm} \\ 
$r_{2,b}$ & $\mathcal{ U}(0,1 )$ & 0.0387$\pm$0.0004  \vspace{0.05cm} \\ 
$P_c$ [d] & $\mathcal{N}(18.8797,0.01)$ & 18.879300\,(48)  \vspace{0.05cm} \\ 
$t_{0,c}$\,$^{(a)}$ & $\mathcal{N}(2949.6,0.1)$ & 2949.58243$^{+0.00077}_{-0.00094}$  \vspace{0.05cm} \\ 
\textit{ecc$_c$} & $\mathcal{F }(0)$ & --  \vspace{0.05cm} \\ 
$\omega_c$ (deg) & $\mathcal{F }( 90)$ & --  \vspace{0.05cm} \\ 
$r_{1,c}$ & $\mathcal{U }(0,1 )$ & 0.353$^{+0.015}_{-0.012}$  \vspace{0.05cm} \\ 
$r_{2,c}$ & $\mathcal{ U}(0,1 )$ & 0.03301$\pm$0.00077  \vspace{0.05cm} \\ 
$\rho_{\star}$ [kg\,m$^{-3}$]  & $\mathcal{ N }(3500.0,500.0)$ & 3750$^{+160}_{-110}$  \vspace{0.05cm} \\
$\mu_{\textit{TESS}}$ (ppm) & $\mathcal{N }(0.0,0.1)$ & 2.0$^{+1.6}_{-1.4}$\,$\times$10$^{3}$  \vspace{0.05cm} \\ 
$\sigma_{\textit{TESS}}$ (ppm) & $\mathcal{J }( 10^{-6}, 10^{6})$ & 436.3$\pm$2.2 \vspace{0.05cm} \\ 
$q_{1,\textit{TESS}}$ & $\mathcal{U }(0,1 )$ & 0.12$^{+0.06}_{-0.04}$  \vspace{0.05cm} \\ 
$q_{2,\textit{TESS}}$ & $\mathcal{ U}(0,1 )$ & 0.54$^{+0.24}_{-0.27}$  \vspace{0.05cm} \\ 
GP$_\mathrm{\sigma}$ (ppm) & $\mathcal{J}(10^{-6}, 10^{6})$ & 8.5$^{+1.2}_{-0.9}$\,$\times$10$^{3}$ \vspace{0.05cm} \\ 
GP$_\mathrm{\rho}$ [d]  & $\mathcal{J}(10^{-3}, 10^{3})$ & 2.04$^{+0.20}_{-0.17}$  \vspace{0.05cm} \\

\noalign{\smallskip} 
\hline 
\noalign{\smallskip} 
\multicolumn{3}{c}{\textit{ Derived planetary parameters for} \hd73\,b } \\ 
\noalign{\smallskip} 
$p_b = {R}_{\mathrm{p}}/{R}_{\star}$  & & 0.0387$\pm$0.0004 \vspace{0.05cm} \\  
$b_b =(a_{\rm p}/{R}_{\star}) \cos{ i_{\mathrm{p}} }$  & &  0.577$^{+0.013}_{-0.018}$ \vspace{0.05cm} \\  
$i_{\mathrm{p}, b}$ (deg)  & & 88.35$^{+0.07}_{-0.05}$  \vspace{0.05cm} \\  
$T_{14, b}$ [h]  &  & 2.100$^{+0.015}_{-0.013}$ \vspace{0.05cm} \\  
$T_{12, b}$ [min]  &  & 6.93$^{+0.17}_{-0.22}$  \vspace{0.05cm} \\  
$R_{\mathrm{p}, b}$ [${R}_\oplus$]  & & 2.78$\pm$0.09 \vspace{0.05cm} \\  
$a_{\mathrm{p}, b}$ [AU]  &  & 0.0618$\pm$0.0020  \vspace{0.05cm} \\  
$T_{\mathrm{eq}, b}$ [K]\,$^{(b)}$  & &  710$\pm$18  \vspace{0.05cm} \\  

\noalign{\smallskip} 
\hline 
\noalign{\smallskip} 
\multicolumn{3}{c}{\textit{ Derived planetary parameters for} \hd73\,c } \\ 
\noalign{\smallskip} 
$p_c = {R}_{\mathrm{p}}/{R}_{\star}$  & & 0.0330$\pm$0.0008 \vspace{0.05cm} \\  
$b_c =(a_{\rm p}/{R}_{\star}) \cos{ i_{\mathrm{p}} }$  & &  0.030$^{+0.022}_{-0.017}$ \vspace{0.05cm} \\  
$i_{\mathrm{p}, c}$ (deg)  & & 89.96$\pm$0.03  \vspace{0.05cm} \\  
$T_{14, c}$ [h]  &  & 3.60$^{+0.03}_{-0.05}$ \vspace{0.05cm} \\  
$T_{12, c}$ [min]  &  & 6.90$\pm$0.20  \vspace{0.05cm} \\  
$R_{\mathrm{p}, c}$ [${R}_\oplus$]  & & 2.38$\pm$0.09 \vspace{0.05cm} \\  
$a_{\mathrm{p}, c}$ [AU]  &  & 0.1270$\pm$0.0040  \vspace{0.05cm} \\  
$T_{\mathrm{eq}, c}$ [K]\,$^{(b)}$  & &  495$\pm$12  \vspace{0.05cm} \\

\noalign{\smallskip}
\hline
\end{tabular}

\tablefoot{
$^{(a)}$ Central time of transit ($t_0$) units are BJD\,$-$\,2\,457\,000.
$^{(b)}$ Equilibrium temperatures were calculated assuming zero Bond albedo.}
\end{table}

\begin{figure*}[h!]
    \centering
     \begin{subfigure}{0.49\textwidth}
         \centering
         \includegraphics[width=\textwidth]{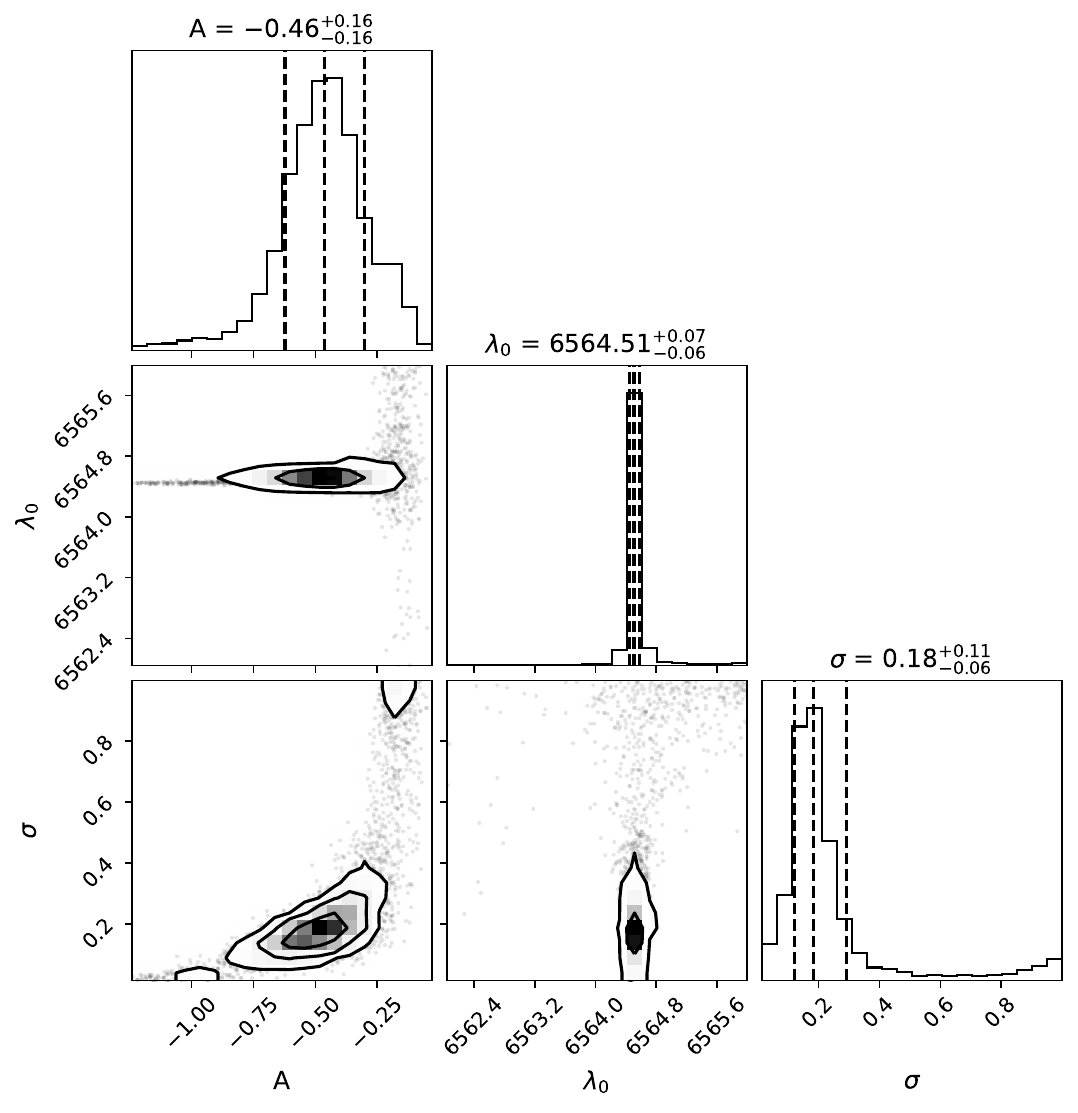}
     \end{subfigure}
     \hfill
     \begin{subfigure}{0.49\textwidth}
         \centering
         \includegraphics[width=\textwidth]{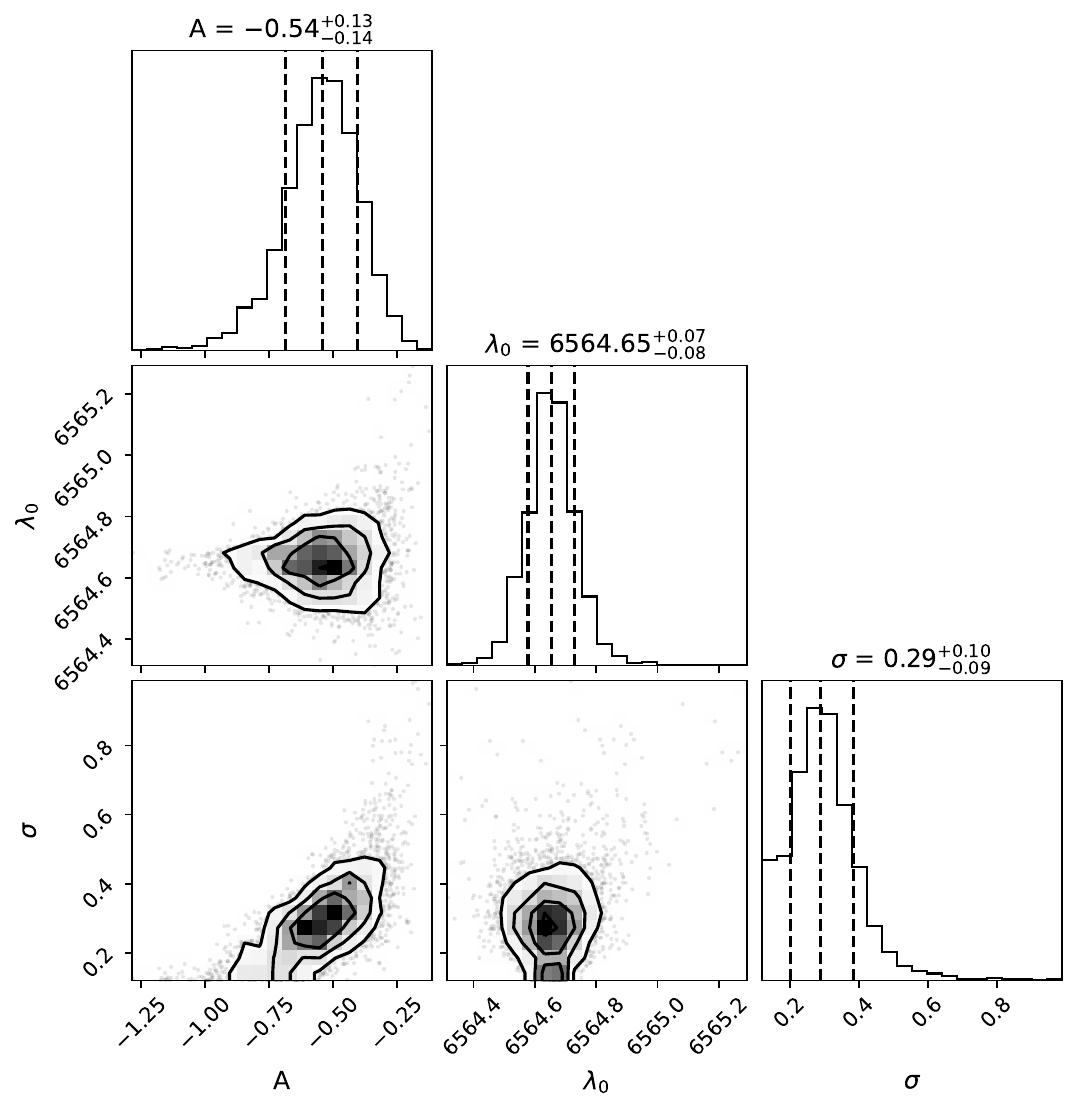}
     \end{subfigure}
    \caption{\label{Fig: HD73583 b and c nested Ha}
    Corner plot for the nested sampling posterior distribution of \hd73\,b (left) and \hd73\,c (right) H$\alpha$ features.
    }
\end{figure*}

\begin{table}[h!]
\caption[width=\textwidth]{
\label{table - HD73583b Ha priors and posteriors}
Prior and posterior distributions from the nested sampling fitting for \hd73\,b H$\alpha$ feature (see Fig.\,\ref{Fig: TS HD 73583 B}). Prior label $\mathcal{U}$ represents uniform distribution.
}
\centering

\begin{tabular}{lcc}

\hline \hline 
\noalign{\smallskip} 

Parameter & Prior & Posterior \vspace{0.05cm}\\
\hline
\noalign{\smallskip}

Absorption [\%] & $\mathcal{U}(-3, 3)$ & $-$0.46$\pm$0.16  \vspace{0.05cm} \\ 
$\lambda_0$ [\AA] & $\mathcal{U}(6562,6566)$ & 6564.51$^{+0.07}_{-0.06}$ \vspace{0.05cm} \\ 
$\sigma$ [\AA] &  $\mathcal{U}(0,1)$ & 0.18$^{+0.11}_{-0.06}$  \vspace{0.05cm} \\ 

$\Delta$v [km\,s$^{-1}$] & -- & $-$6.1$^{+3.2}_{-2.5}$ \\
FWHM [\AA] & -- &  0.43$^{+0.26}_{-0.14}$ \\
EW [m\AA] &  -- & 2.1$^{+0.7}_{-0.6}$  \vspace{0.05cm} \\ 

\noalign{\smallskip}
\hline
\end{tabular}

\end{table}

\begin{table}[h!]
\caption[width=\textwidth]{
\label{table - HD73583c Ha priors and posteriors}
Prior and posterior distributions from the nested sampling fitting for \hd73\,c H$\alpha$ feature (see Fig.\,\ref{Fig: TS HD 73583 C}). Prior label $\mathcal{U}$ represents uniform distribution.
}
\centering

\begin{tabular}{lcc}

\hline \hline 
\noalign{\smallskip} 

Parameter & Prior & Posterior \vspace{0.05cm}\\
\hline
\noalign{\smallskip}

Absorption [\%] & $\mathcal{U}(-3, 3)$ & $-$0.54$^{+0.13}_{-0.14}$  \vspace{0.05cm} \\ 
$\lambda_0$ [\AA] & $\mathcal{U}(6562,6566)$ & 6564.65$^{+0.07}_{-0.08}$ \vspace{0.05cm} \\ 
$\sigma$ [\AA] &  $\mathcal{U}(0.12,1)$ & 0.30$^{+0.10}_{-0.09}$  \vspace{0.05cm} \\ 

$\Delta$v [km\,s$^{-1}$] & -- & 0.6$\pm$3.4 \\
FWHM [\AA] & -- &  0.68$^{+0.22}_{-0.21}$ \\
EW [m\AA] &  -- & 3.8$^{+1.0}_{-0.9}$  \vspace{0.05cm} \\ 

\noalign{\smallskip}
\hline
\end{tabular}

\end{table}

\clearpage

\section{K2-77\,b planetary parameters}
\label{App: K2-77 juliet}

We analysed the K2-77\,b TESS data from Sectors\,5, 42, 43, and 44. Because we did not see a clear rotation modulation in the light curves, we used the \texttt{celerite} GP exponential kernel to account for the stellar variability. We adopted the stellar parameters used in \citet{K2-77_Gaidos}.

\begin{figure}[h!]
    \centering
    \includegraphics[width=\hsize]{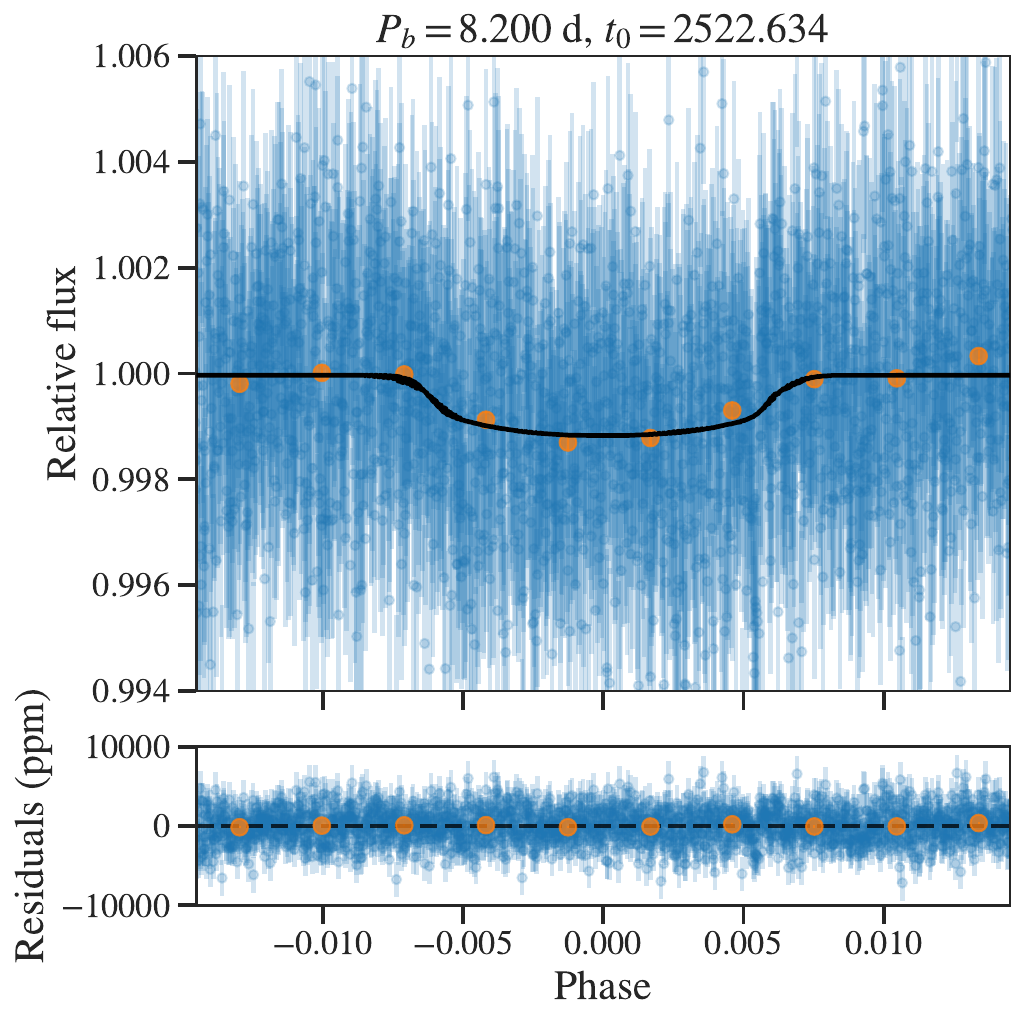}
    \caption{\label{Fig: K2-77 phase folded}
    K2-77 TESS photometry (blue dots with error bars) phase-folded to the period $P$ and central time of transit $t_0$ (shown above the panel, $t_0$ units are BJD\,$-$\,2\,457\,000) derived from the \texttt{juliet} fit. The black line is the best transit model for K2-77\,b. The orange points show binned photometry for visualisation.
    }
\end{figure}

\begin{figure*}[h!]
    \centering
    \includegraphics[width=1\linewidth]{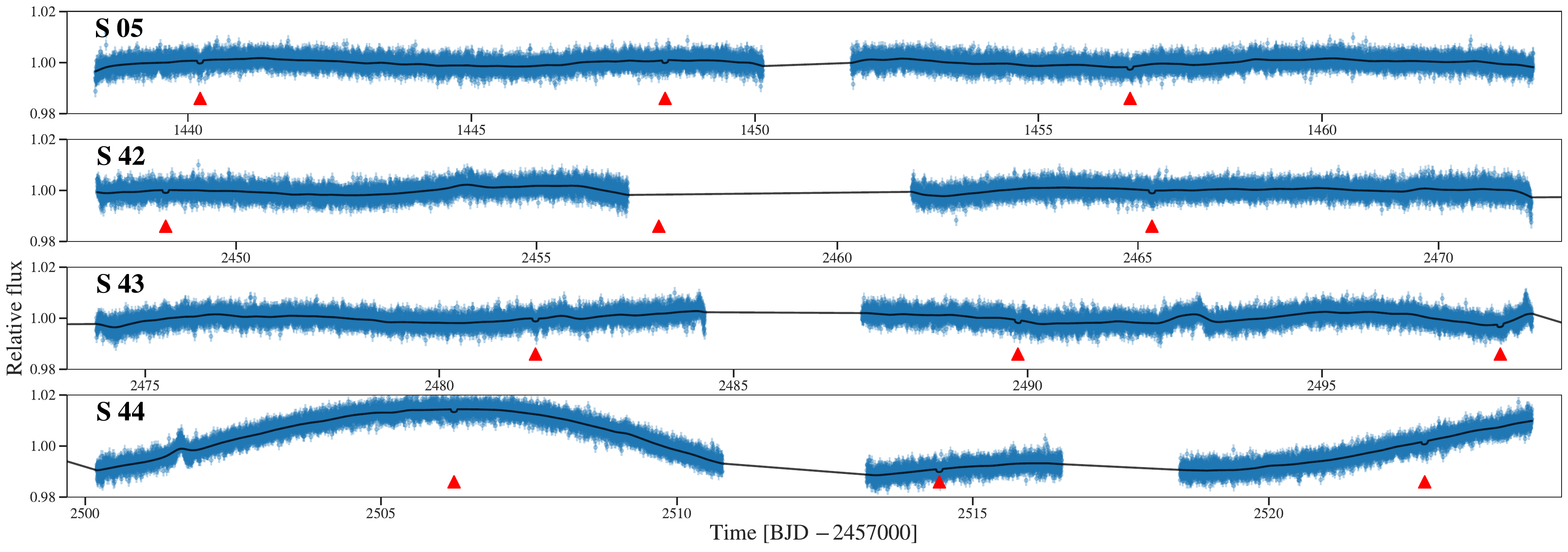}
    \caption{\label{Fig: K2-77 TESS fit}
    K2-77 2 min cadence TESS photometry from Sectors 5, 42, 43, and 44 along with the transit plus GP model. The upward-pointing red triangles mark the transits for K2-77\,b.
    }
\end{figure*}

The fitted parameters with their prior and posterior values, and the derived parameters for K2-77\,b are shown in Table\,\ref{table - K2-77 juliet priors and posteriors}. The TESS data along with the best transiting and GP models is shown in Fig.\,\ref{Fig: K2-77 TESS fit} and K2-77\,b phase folded transit is shown in Fig.\,\ref{Fig: K2-77 phase folded}.

As we did in App.\,\ref{App: TOI-2048 juliet}, we forecasted K2-77\,b mass and we computed the estimated $K_{\star}$ using Eq.\,\ref{eq: Semi-amplitude}. We predicted a planetary mass of $\sim$9\,$\pm$\,2\,${M}_\oplus$, and a semi-amplitude $K_{\star}$ of $\sim$3.4\,$\pm$\,0.8\,m\,s$^{-1}$. \cite{K2-77_Gaidos} only reported an upper limit to K2-77\,b's mass of 1.9\,${M}_{J}$.

\begin{table}[h!]
\caption[width=\textwidth]{
\label{table - K2-77 juliet priors and posteriors}
Prior and posterior distributions from the \texttt{juliet} fitting for K2-77\,b. Prior labels $\mathcal{U}$, $\mathcal{N}$, $\mathcal{F}$, and $\mathcal{J}$ represents uniform, normal, fixed, and Jeffrey's distribution, respectively.
}
\centering

\begin{tabular}{lcc}

\hline \hline 
\noalign{\smallskip} 

Parameter & Prior & Posterior \vspace{0.05cm}\\
\hline
\noalign{\smallskip}

$P$ [d] & $\mathcal{N}(8.2,0.01)$ & 8.200139\,(60)  \vspace{0.05cm} \\ 
$t_0$\,$^{(a)}$ & $\mathcal{N}(2522.6,0.1)$ & 2522.6338$^{+0.0037}_{-0.0033}$  \vspace{0.05cm} \\ 
\textit{ecc} & $\mathcal{F }(0)$ & --  \vspace{0.05cm} \\ 
$\omega$ (deg) & $\mathcal{F }( 90)$ & --  \vspace{0.05cm} \\ 
$r_{1}$ & $\mathcal{U }(0,1 )$ & 0.66$^{+0.09}_{-0.17}$  \vspace{0.05cm} \\ 
$r_{2}$ & $\mathcal{ U}(0,1 )$ & 0.0309$^{+0.0022}_{-0.0025}$  \vspace{0.05cm} \\ 
$\rho_{\star}$ [kg\,m$^{-3}$]  & $\mathcal{ N }(2577.0,500.0)$ & 2700$^{+450}_{-460}$  \vspace{0.05cm} \\
$\mu_{TESS}$ (ppm) & $\mathcal{N }(0.0,0.1)$ & 300$^{+810}_{-770}$  \vspace{0.05cm} \\ 
$\sigma_{TESS}$ (ppm) & $\mathcal{J }( 10^{-6}, 10^{6})$ & 4.5$^{+1700}_{-4.5}$\,$\times$10$^{3}$ \vspace{0.05cm} \\ 
$q_{1,TESS}$ & $\mathcal{U }(0,1 )$ & 0.55$^{+0.30}_{-0.34}$  \vspace{0.05cm} \\ 
$q_{2,TESS}$ & $\mathcal{ U}(0,1 )$ & 0.52$^{+0.31}_{-0.33}$  \vspace{0.05cm} \\ 
GP$_\mathrm{\sigma}$ (ppm) & $\mathcal{J }(10^{-6}, 10^{6})$ & 4.2$^{+0.5}_{-0.4}$\,$\times$10$^{3}$ \vspace{0.05cm} \\ 
GP$_\mathrm{\rho}$ [d]  & $\mathcal{J }(10^{-3}, 10^{3})$ & 1.65$^{+0.17}_{-0.15}$  \vspace{0.05cm} \\

\noalign{\smallskip} 
\hline 
\noalign{\smallskip} 
\multicolumn{3}{c}{\textit{ Derived planetary parameters }} \\ 
\noalign{\smallskip} 
$p = {R}_{\mathrm{p}}/{R}_{\star}$  & & 0.0309$^{+0.0022}_{-0.0025}$ \vspace{0.05cm} \\  
$b =(a_{\rm p}/{R}_{\star}) \cos{ i_{\mathrm{p}} }$  & &  0.50$^{+0.13}_{-0.25}$ \vspace{0.05cm} \\  
$i_{\mathrm{p}}$ (deg)  & & 88.7$^{+0.7}_{-0.4}$  \vspace{0.05cm} \\  
$T_{14}$ [h]  &  & 2.68$\pm$0.22 \vspace{0.05cm} \\  
$T_{12}$ [min]  &  & 6.3$^{+1.5}_{-1.1}$  \vspace{0.05cm} \\  
$R_{\mathrm{p}}$ [${R}_\oplus$]  & & 2.55$\pm$0.20 \vspace{0.05cm} \\  
$a_{\mathrm{p}}$ [AU]  &  & 0.0751$^{+0.0045}_{-0.0050}$  \vspace{0.05cm} \\  
$T_{\mathrm{eq}}$ [K]\,$^{(b)}$  & &  760$^{+25}_{-21}$  \vspace{0.05cm} \\

\noalign{\smallskip}
\hline
\end{tabular}

\tablefoot{
$^{(a)}$ Central time of transit ($t_0$) units are BJD\,$-$\,2\,457\,000.
$^{(b)}$ Equilibrium temperatures were calculated assuming zero Bond albedo.}
\end{table}

\clearpage

\section{TOI-1807\,b extra material}
\label{App: TOI-1807 juliet}

We analysed TOI-1807 TESS data from Sectors\,22, 23, and 49. Because our purpose is not to model the young stellar activity, but derive precise ephemerides, we used the unspecific \texttt{celerite} GP exponential kernel to account for the stellar variability and rotation. We adopted the stellar parameters used in \citet{TOI-1807_Nardiello} for a proper comparison.

\begin{figure}[h!]
    \centering
    \includegraphics[width=\hsize]{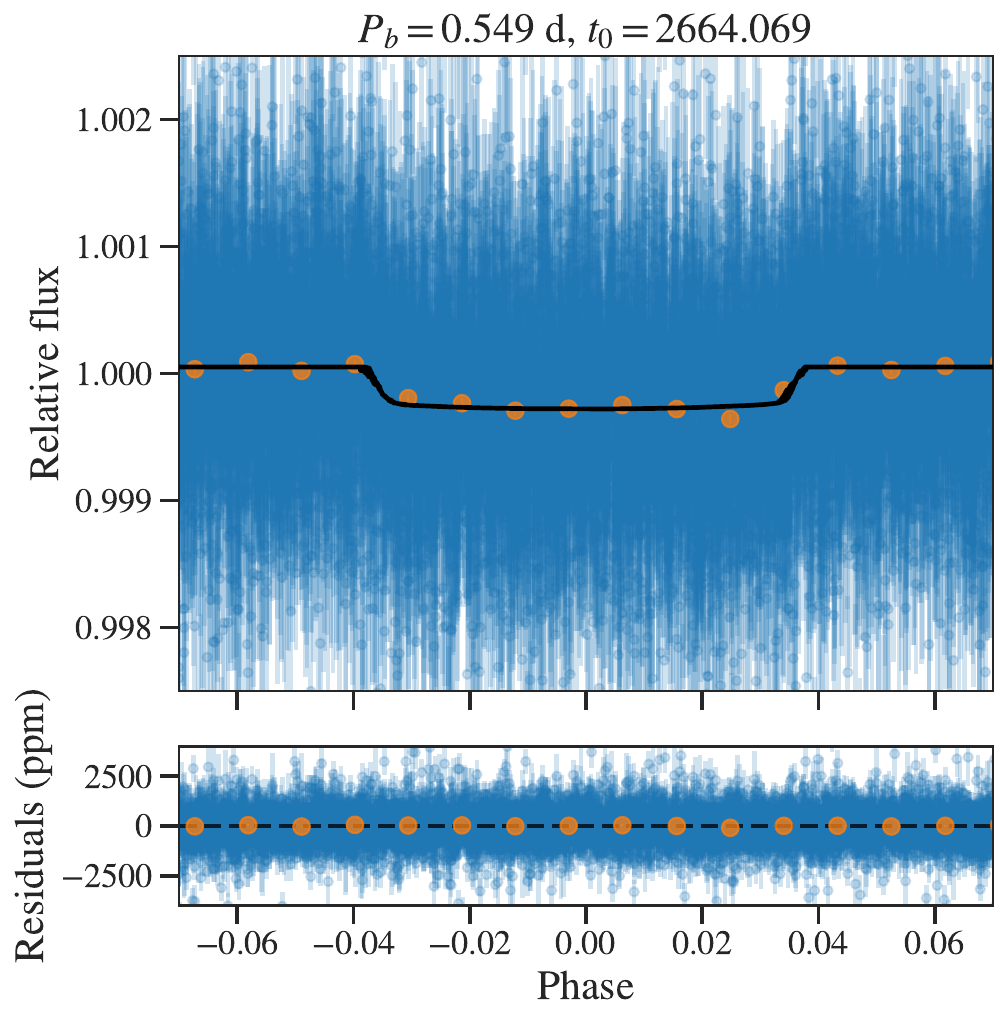}
    \caption{\label{Fig: TOI-1807 phase folded}
    TOI-1807 TESS photometry (blue dots with error bars) phase-folded to the period $P$ and central time of transit $t_0$ (shown above the panel, $t_0$ units are BJD\,$-$\,2\,457\,000) derived from the \texttt{juliet} fit. The black line is the best transit model for TOI-1807\,b. The orange points show binned photometry for visualisation.
    }
\end{figure}

\begin{figure*}[h!]
    \centering
    \includegraphics[width=1\linewidth]{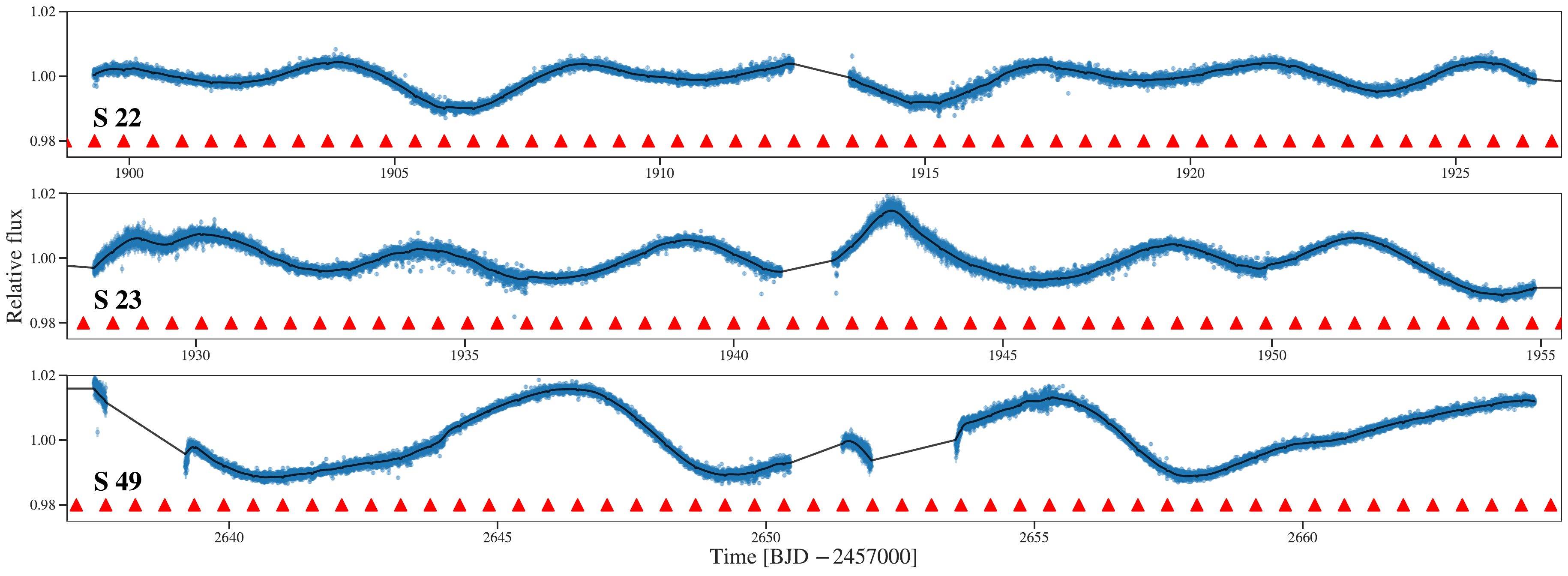}
    \caption{\label{Fig: TOI-1807 TESS fit}
    TOI-1807 2 min cadence TESS photometry from Sectors\,22, 23, and 49 along with the transit plus GP model. The upward-pointing red triangles mark the transit times for the ultra-short period TOI-1807\,b.
    }
\end{figure*}

The fitted parameters with their prior and posterior values, and the derived parameters for TOI-1807\,b are shown in Table\,\ref{table - TOI-1807 juliet priors and posteriors}. The TESS data along with the best transiting and GP models is shown in Fig.\,\ref{Fig: TOI-1807 TESS fit} and TOI-1807\,b phase folded transit is shown in Fig.\,\ref{Fig: TOI-1807 phase folded}.

\begin{table}[h!]
\caption[width=\textwidth]{
\label{table - TOI-1807 juliet priors and posteriors}
Prior and posterior distributions from the \texttt{juliet} fitting for TOI-1807\,b. Prior labels $\mathcal{U}$, $\mathcal{N}$, $\mathcal{F}$, and $\mathcal{J}$ represents uniform, normal, fixed, and Jeffrey's distribution, respectively.
}
\centering

\begin{tabular}{lcc}

\hline \hline 
\noalign{\smallskip} 

Parameter & Prior & Posterior \vspace{0.05cm}\\
\hline
\noalign{\smallskip}

$P$ [d] & $\mathcal{N}(0.54929,0.001)$ & 0.54937084\,(65)  \vspace{0.05cm} \\ 
$t_0$\,$^{(a)}$ & $\mathcal{N}(2664.07,0.1)$ & 2664.06930$^{+0.00073}_{-0.00075}$  \vspace{0.05cm} \\ 
\textit{ecc} & $\mathcal{F }(0)$ & --  \vspace{0.05cm} \\ 
$\omega$ (deg) & $\mathcal{F }( 90)$ & --  \vspace{0.05cm} \\ 
$r_{1}$ & $\mathcal{U }(0,1 )$ & 0.70$^{+0.05}_{-0.06}$  \vspace{0.05cm} \\ 
$r_{2}$ & $\mathcal{ U}(0,1 )$ & 0.01776$^{+0.00045}_{-0.00052}$  \vspace{0.05cm} \\ 
$\rho_{\star}$ [kg\,m$^{-3}$]  & $\mathcal{ N }(3300.0,600.0)$ & 3430$^{+530}_{-550}$  \vspace{0.05cm} \\
$\mu_{\textit{TESS}}$ (ppm) & $\mathcal{N }(0.0,0.1)$ & $-$0.6$^{+2.0}_{-2.0}$\,$\times$10$^{3}$  \vspace{0.05cm} \\ 
$\sigma_{\textit{TESS}}$ (ppm) & $\mathcal{J }( 10^{-6}, 10^{6})$ & 500$^{+4.3}_{-4.1}$ \vspace{0.05cm} \\ 
$q_{1,\textit{TESS}}$ & $\mathcal{U }(0,1 )$ & 0.15$^{+0.24}_{-0.11}$  \vspace{0.05cm} \\ 
$q_{2,\textit{TESS}}$ & $\mathcal{ U}(0,1 )$ & 0.26$^{+0.36}_{-0.20}$  \vspace{0.05cm} \\ 
GP$_\mathrm{\sigma}$ (ppm) & $\mathcal{J}(10^{-6}, 10^{6})$ & 8.7$^{+1.5}_{-1.1}$\,$\times$10$^{3}$ \vspace{0.05cm} \\ 
GP$_\mathrm{\rho}$ [d]  & $\mathcal{J}(10^{-3}, 10^{3})$ & 2.30$^{+0.28}_{-0.25}$  \vspace{0.05cm} \\

\noalign{\smallskip} 
\hline 
\noalign{\smallskip} 
\multicolumn{3}{c}{\textit{ Derived planetary parameters }} \\ 
\noalign{\smallskip} 
$p = {R}_{\mathrm{p}}/{R}_{\star}$  & & 0.01776$^{+0.00045}_{-0.00052}$ \vspace{0.05cm} \\  
$b =(a_{\rm p}/{R}_{\star}) \cos{ i_{\mathrm{p}} }$  & &  0.550$^{+0.077}_{-0.100}$ \vspace{0.05cm} \\  
$i_{\mathrm{p}}$ (deg)  & & 81.7$\pm$0.18  \vspace{0.05cm} \\  
$T_{14}$ [h]  &  & 0.970$\pm$0.022 \vspace{0.05cm} \\  
$T_{12}$ [min]  &  & 1.45$^{+0.25}_{-0.18}$  \vspace{0.05cm} \\  
$R_{\mathrm{p}}$ [${R}_\oplus$]  & & 1.33$\pm$0.08 \vspace{0.05cm} \\  
$a_{\mathrm{p}}$ [AU]  &  & 0.0121$\pm$0.0009  \vspace{0.05cm} \\  
$T_{\mathrm{eq}}$ [K]\,$^{(b)}$  & &  1720$^{+55}_{-50}$  \vspace{0.05cm} \\

\noalign{\smallskip}
\hline
\end{tabular}

\tablefoot{
$^{(a)}$ Central time of transit ($t_0$) units are BJD\,$-$\,2\,457\,000.
$^{(b)}$ Equilibrium temperatures were calculated assuming zero Bond albedo.}
\end{table}

\clearpage

\section{TOI-1136\,d extra material}
\label{App: TOI-1136 extra}

\begin{figure}[h!]
    \centering
    \includegraphics[width=\hsize]{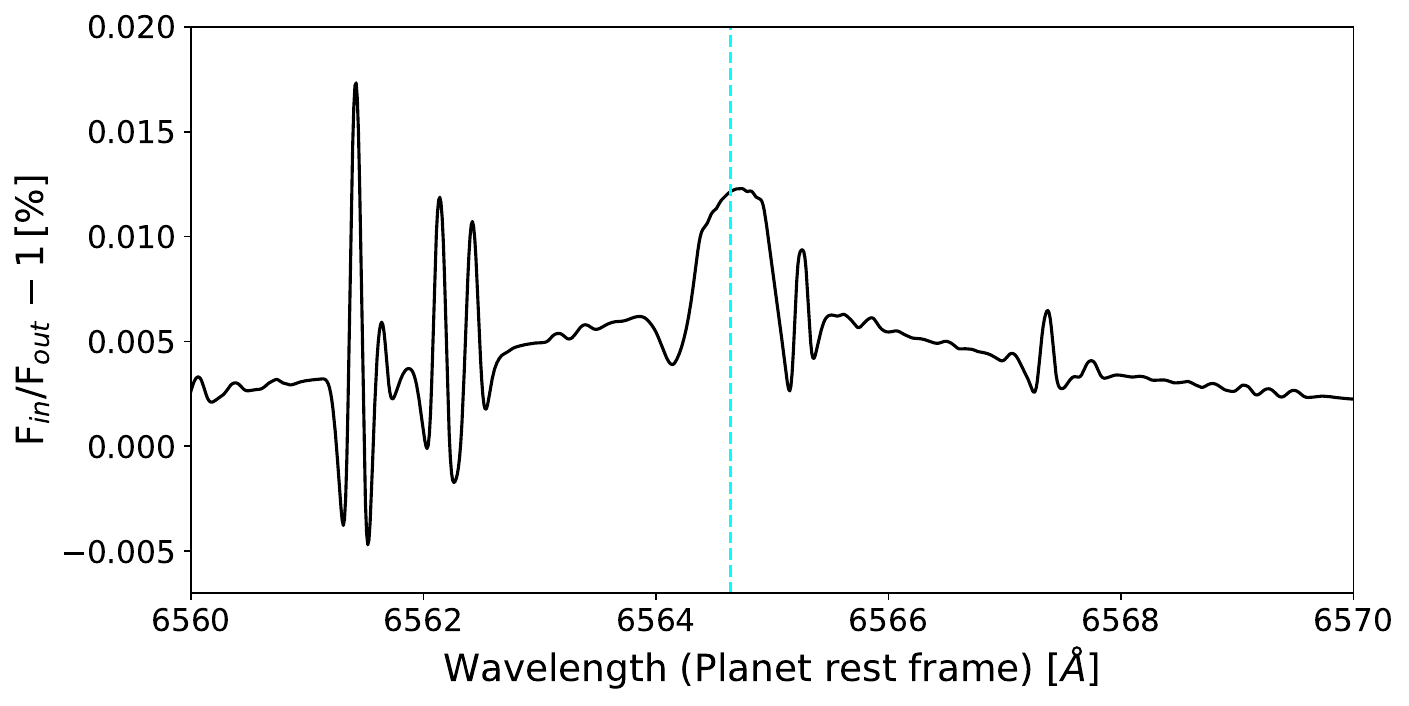}
    \includegraphics[width=\hsize]{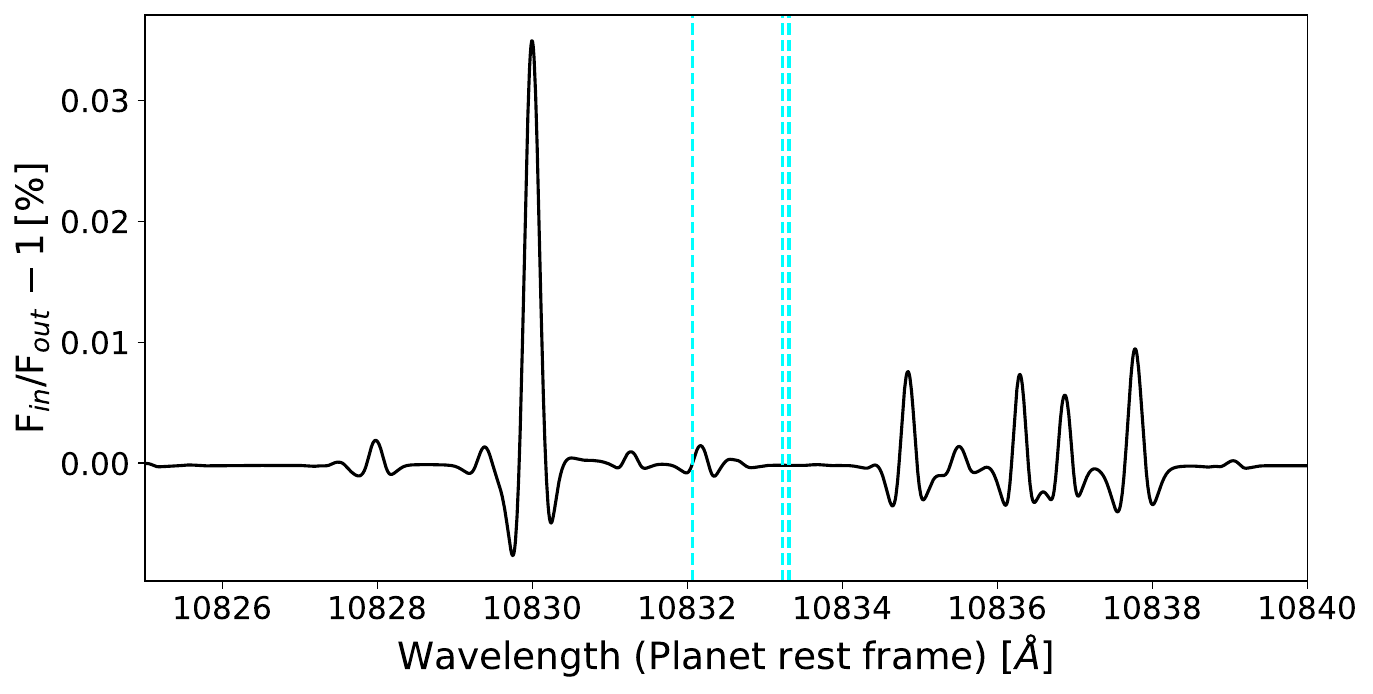}
    \caption{\label{Fig: TOI-1136d RM}
    Contribution of the RM and CLV effects to the  TOI-1136\,d transmission spectra around the H$\alpha$ (top) and \ion{He}{I} NIR triplet (bottom) lines. The vertical cyan lines mark the position of the lines of interest.
    }
\end{figure}

\begin{table}[h]
\caption[width=\textwidth]{
\label{table - TOI-1136d Ha priors and posteriors}
Prior and posterior distributions from the nested sampling fitting for TOI-1136\,d H$\alpha$ feature (see Fig.\,\ref{Fig: TS TOI-1136 detections}). Prior label $\mathcal{U}$ represents uniform distribution.
}
\centering

\begin{tabular}{lcc}

\hline \hline 
\noalign{\smallskip} 

Parameter & Prior & Posterior \vspace{0.05cm}\\
\hline
\noalign{\smallskip}

Absorption [\%] & $\mathcal{U}(-3, 3)$ & $-$1.12$^{+0.12}_{-0.13}$  \vspace{0.05cm} \\ 
$\lambda_0$ [\AA] & $\mathcal{U}(6562,6566)$ & 6564.464$^{+0.024}_{-0.022}$ \vspace{0.05cm} \\ 
$\sigma$ [\AA] &  $\mathcal{U}(0,1)$ & 0.186$^{+0.030}_{-0.025}$  \vspace{0.05cm} \\ 

$\Delta$v [km\,s$^{-1}$] & -- & $-$8.0$\pm$1.0 \\
FWHM [\AA] & -- &  0.45$^{+0.07}_{-0.06}$ \\
EW [m\AA] &  -- & 5.20$^{+0.57}_{-0.53}$  \vspace{0.05cm} \\ 

\noalign{\smallskip}
\hline
\end{tabular}

\end{table}

\begin{figure}[h]
    \centering
    \includegraphics[width=\hsize]{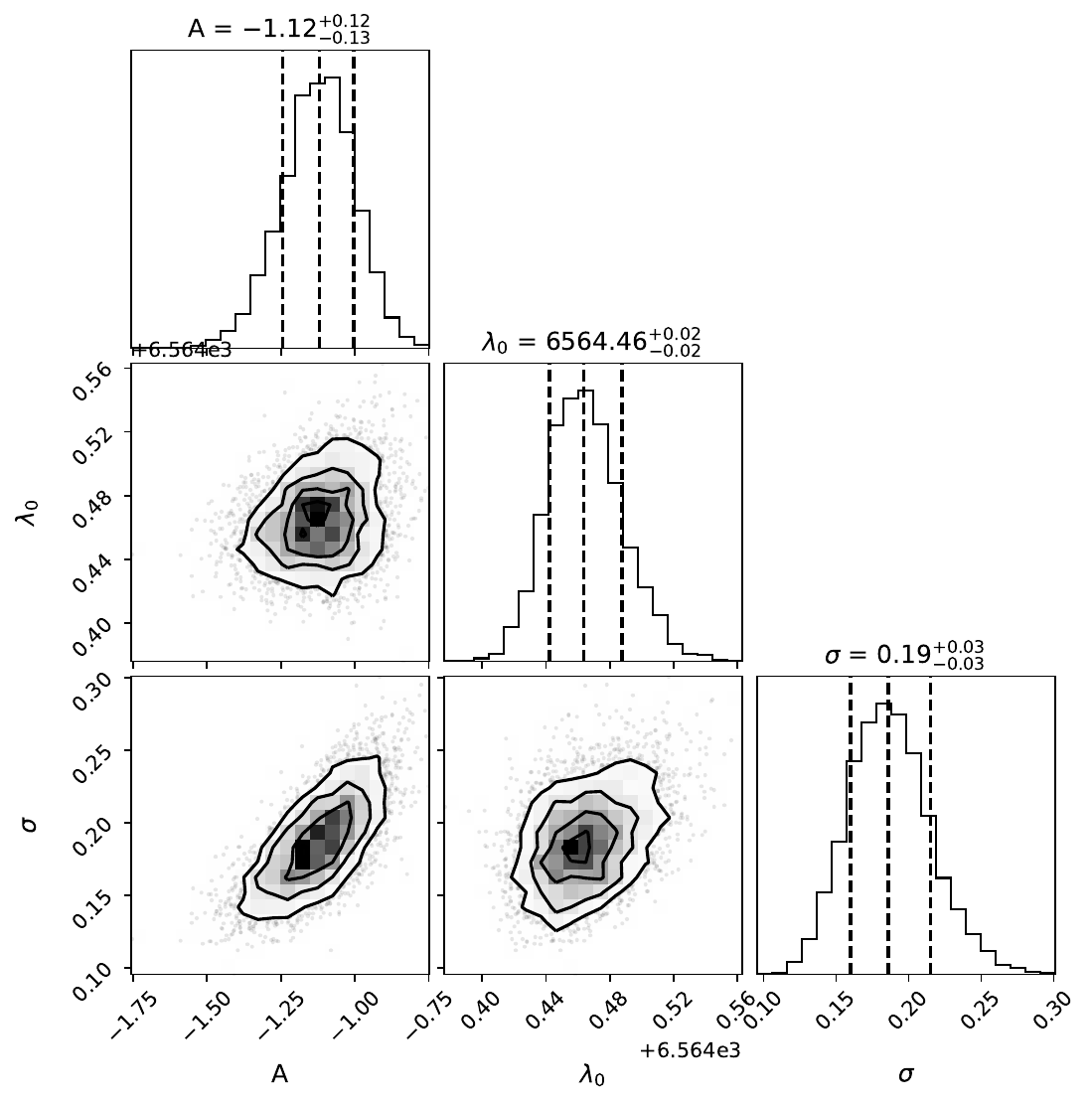}
    \caption{\label{Fig: TOI-1136d nested Ha}
    Corner plot for the nested sampling posterior distribution of TOI-1136\,d H$\alpha$ feature.
    }
\end{figure}

\clearpage

\section{TOI-1268\,b extra material}
\label{App: TOI-1268}

\begin{figure}[h!]
    \centering
    \includegraphics[width=\hsize]{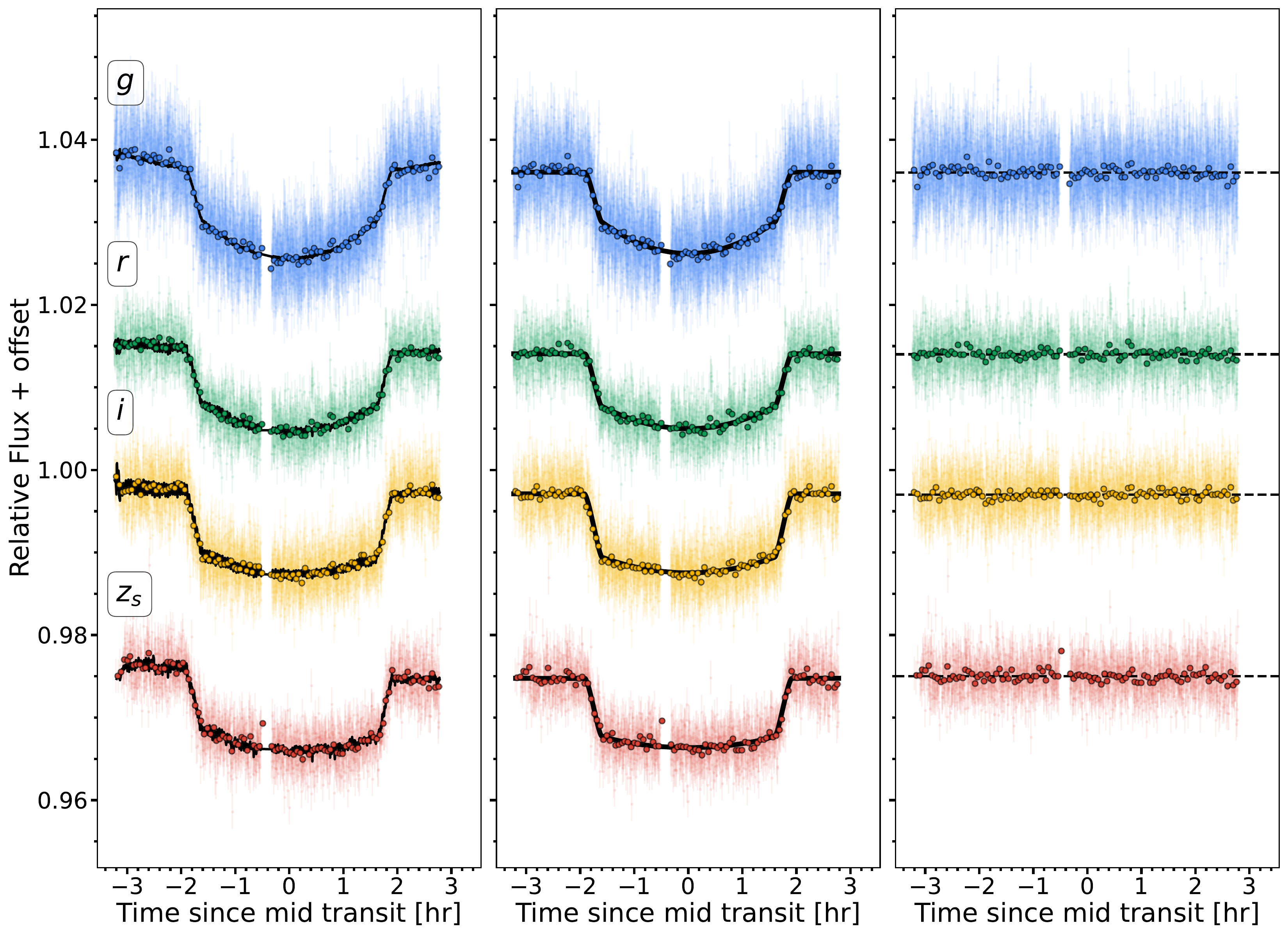}
    \caption{MuSCAT2 ground-based transit observation of TOI-1268\,b. Raw light curve (left panel), detrended light curve (centre panel), and residuals (right panel) for the Sloan $g$ (blue), $r$ (green), $i$ (yellow), $z_s$ (red) filters. The points show the individual observations, the circles represent binned data points, and the best-fit model is shown by the black line.}
    \label{Fig: MUSCAT2 TOI-1268b}
\end{figure}

\begin{table}[h!]
\caption[width=\textwidth]{
\label{table - TOI-1268 He priors and posteriors}
Prior and posterior distributions from the nested sampling fitting for TOI-1268\,b \ion{He}{I} signal (see Fig.\,\ref{Fig: TS TOI-1268}). Prior label $\mathcal{U}$ represents uniform distribution.
}
\centering

\begin{tabular}{lcc}

\hline \hline 
\noalign{\smallskip} 

Parameter & Prior & Posterior \vspace{0.05cm}\\
\hline
\noalign{\smallskip}

Absorption [\%] & $\mathcal{U}(-3, 3)$ & $-$2.00$^{+0.15}_{-0.16}$  \vspace{0.05cm} \\ 
$\lambda_0$ [\AA] & $\mathcal{U}(10831,10834)$ & 10832.760$\pm$0.050 \\ 
$\sigma$ [\AA] &  $\mathcal{U}(0,1.5)$ & 0.39$^{+0.036}_{-0.034}$  \vspace{0.05cm} \\ 

$\Delta$v [km\,s$^{-1}$] & -- & $-$12.8$\pm$1.3 \\
FWHM [\AA] & -- &  0.91$^{+0.9}_{-0.08}$ \vspace{0.05cm} \\
EW [m\AA] &  -- & 19.1$^{+1.8}_{-1.9}$  \vspace{0.05cm} \\ 

\noalign{\smallskip}
\hline
\end{tabular}

\end{table}

\begin{figure}[h!]
    \centering
    \includegraphics[width=\hsize]{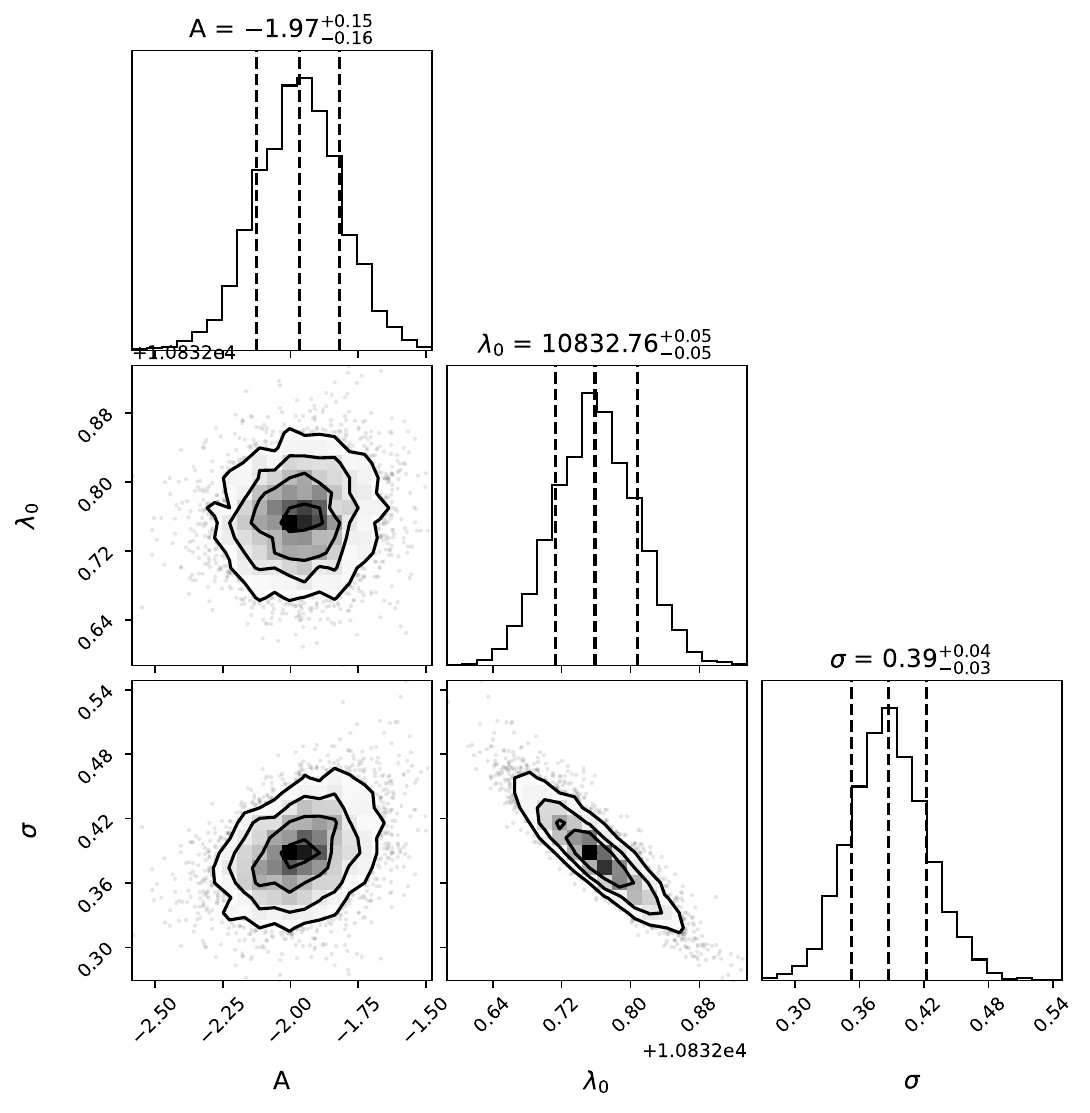}
    \caption{\label{Fig: TOI-1268 nested He}
    Corner plot for the nested sampling posterior distribution of TOI-1268\,b \ion{He}{i} signal.
    }
\end{figure}

\clearpage

\section{TOI-2076\,b extra material}
\label{App: TOI-2076}

\begin{table}[h!]
\caption[width=\textwidth]{
\label{table - TOI-2076 juliet priors and posteriors}
Prior and posterior distributions from the \texttt{juliet} fitting for TOI-2076\,b data from LCO (see Fig.\,\ref{Fig: TOI-2076 LCO data}). Prior labels $\mathcal{U}$, $\mathcal{N}$, $\mathcal{F}$, and $\mathcal{J}$ represents uniform, normal, fixed, and Jeffrey's distribution, respectively.
}
\centering
\resizebox{0.48\textwidth}{!}{%

\begin{tabular}{lcc}

\hline \hline 
\noalign{\smallskip} 

Parameter & Prior & Posterior \vspace{0.05cm}\\
\hline
\noalign{\smallskip}

$P$ [d] & $\mathcal{N}(10.355183,0.000065)$ & 10.355184\,(58)  \vspace{0.05cm} \\ 
$t_0$\,$^{(a)}$ & $\mathcal{U}(3079.47,3079.62)$ & 3079.5495$^{+0.0061}_{-0.0045}$  \vspace{0.05cm} \\ 
\textit{ecc} & $\mathcal{F}(0)$ & --  \vspace{0.05cm} \\ 
$\omega$ (deg) & $\mathcal{F }( 90)$ & --  \vspace{0.05cm} \\ 
$b =(a_{\rm p}/{R}_{\star}) \cos{ i_{\mathrm{p}} }$ & $\mathcal{N}(0.149,0.089)$ & 0.158$^{+0.071}_{-0.074}$  \vspace{0.05cm} \\ 
$p = {R}_{\mathrm{p}}/{R}_{\star}$ & $\mathcal{N}(0.02998,0.00035)$ & 0.03003\,(30)  \vspace{0.05cm} \\ 
$\rho_{\star}$ [kg\,m$^{-3}$]  & $\mathcal{ N }(2544,120)$ & 2535$\pm$100  \vspace{0.05cm} \\
$\mu_{\textit{LCO}}$ (ppm) & $\mathcal{N }(0.0,0.1)$ & $-$0.6$^{+11}_{-25}$\,$\times$10$^{3}$  \vspace{0.05cm} \\ 
$\sigma_{\textit{LCO}}$ (ppm) & $\mathcal{J }( 10^{-6}, 10^{6})$ & 1175$^{+76}_{-72}$ \vspace{0.05cm} \\ 
$q_{1,\textit{LCO}}$ & $\mathcal{U }(0,1 )$ & 0.57$^{+0.30}_{-0.35}$  \vspace{0.05cm} \\ 
GP$_\mathrm{\sigma}$ (ppm) & $\mathcal{J}(10^{-6}, 10^{6})$ & 13$^{+80}_{-12}$\,$\times$10$^{3}$ \vspace{0.05cm} \\ 
GP$_\mathrm{\rho}$ [d]  & $\mathcal{J}(10^{-3}, 10^{3})$ & 2.5$^{+25}_{-2.3}$  \vspace{0.05cm} \\

\noalign{\smallskip} 
\hline 
\noalign{\smallskip} 
\multicolumn{3}{c}{\textit{ Derived planetary parameters }} \\ 
\noalign{\smallskip} 
$i_{\mathrm{p}}$ (deg)  & & 89.63$\pm$0.18  \vspace{0.05cm} \\  
$T_{14}$ [h]  &  & 3.31$\pm$0.06 \vspace{0.05cm} \\  
$R_{\mathrm{p}}$ [${R}_\oplus$]  & & 2.523$\pm$0.032 \vspace{0.05cm} \\

\noalign{\smallskip}
\hline
\end{tabular}
}
\tablefoot{
$^{(a)}$ Central time of transit ($t_0$) units are BJD\,$-$\,2\,457\,000.
$^{(b)}$ Equilibrium temperatures were calculated assuming zero Bond albedo.}
\end{table}

\begin{figure}[h!]
    \centering
    \includegraphics[width=\hsize]{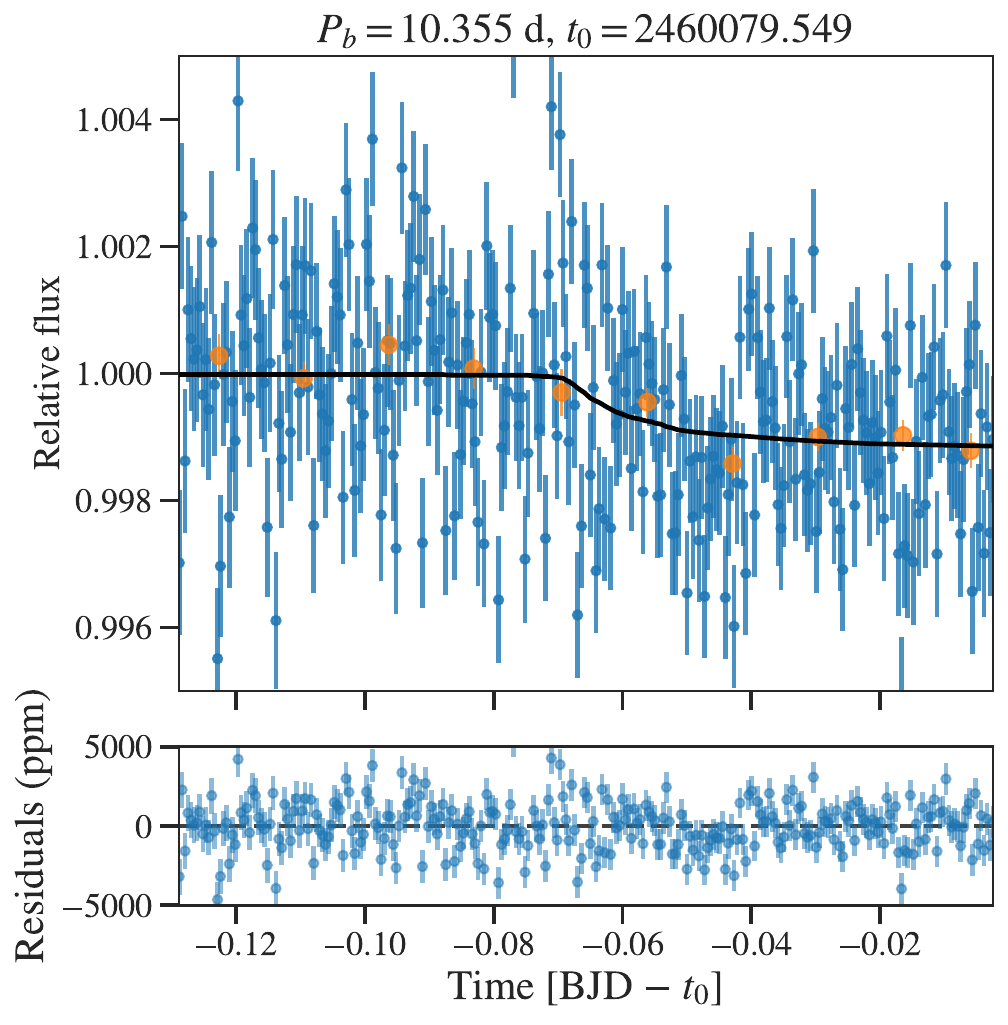}
    \caption{\label{Fig: TOI-2076 LCO data}
    TOI-2076 LCO photometric data (blue dots with error bars). The period $P$ and central time of transit $t_0$ derived from the \texttt{juliet} fit are shown above the panel. The black line is the best transit model for TOI-2076\,b. The orange points show binned photometry for visualisation. The GP contribution was removed from the data.
    }
\end{figure}

\clearpage

\section{TOI-1683\,b extra material}
\label{App: TOI-1683 juliet}

\begin{table}[h!]
\caption[width=\textwidth]{
\label{table - TOI-1683 juliet priors and posteriors}
Prior and posterior distributions from the \texttt{juliet} fitting for TOI-1683\,b. Prior labels $\mathcal{U}$, $\mathcal{N}$, $\mathcal{F}$, and $\mathcal{J}$ represents uniform, normal, fixed, and Jeffrey's distribution, respectively.
}
\centering

\begin{tabular}{lcc}

\hline \hline 
\noalign{\smallskip} 

Parameter & Prior & Posterior \vspace{0.05cm}\\
\hline
\noalign{\smallskip}

$P$ [d] & $\mathcal{N}(3.057,0.001)$ & 3.057541$^{+0.000014}_{-0.000010}$  \vspace{0.05cm} \\ 
$t_0$\,$^{(a)}$ & $\mathcal{N}(2522.7,0.1)$ & 2522.7001$^{+0.0012}_{-0.0010}$  \vspace{0.05cm} \\ 
\textit{ecc} & $\mathcal{F }( 0)$ & --  \vspace{0.05cm} \\ 
$\omega$ (deg) & $\mathcal{F }( 90)$ & --  \vspace{0.05cm} \\ 
$r_{1}$ & $\mathcal{U }(0,1 )$ & 0.799$^{+0.030}_{-0.040}$   \vspace{0.05cm} \\ 
$r_{2}$ & $\mathcal{ U}(0,1 )$ & 0.0319$\pm$0.0014  \vspace{0.05cm} \\ 
$\rho_{\star}$ [kg\,m$^{-3}$]  & $\mathcal{ N }(3800.0,600.0 )$ & 3900$^{+570}_{-580}$  \vspace{0.05cm} \\
$\mu_{\textit{TESS}}$ (ppm) & $\mathcal{N }(0.0,0.1)$ & 510$\pm$3900  \vspace{0.05cm} \\ 
$\sigma_{\textit{TESS}}$ (ppm) & $\mathcal{J }( 10^{-6}, 10^{6})$ & 334$^{+11}_{-12}$  \vspace{0.05cm} \\ 
$q_{1,\textit{TESS}}$ & $\mathcal{U }(0,1 )$ & 0.62$\pm$0.25  \vspace{0.05cm} \\ 
$q_{2,\textit{TESS}}$ & $\mathcal{ U}(0,1 )$ & 0.56$^{+28}_{-34}$ \vspace{0.05cm} \\ 
GP$_\mathrm{B}$ (ppm) & $\mathcal{J }(10^{-6}, 10^{6})$ & 62$^{+60}_{-25}$  \vspace{0.05cm} \\ 
GP$_\mathrm{L}$ [d]  & $\mathcal{J }(10^{-3}, 10^{3})$ & 100$^{+100}_{-42}$  \vspace{0.05cm} \\ 
GP$_\mathrm{C}$ (ppm) & $\mathcal{J }(10^{-6}, 10^{6})$ & 500$^{+ 33000}_{-500}$  \vspace{0.05cm} \\ 
GP$_\mathrm{P_{\mathrm{rot}}}$ [d] & $\mathcal{U }(1,50)$ & 20.0$\pm$1.0  \vspace{0.05cm} \\

\noalign{\smallskip} 
\hline 
\noalign{\smallskip} 
\multicolumn{3}{c}{\textit{ Derived planetary parameters }} \\ 
\noalign{\smallskip} 
$p = {R}_{\mathrm{p}}/{R}_{\star}$  & & 0.0319$\pm$0.0014 \vspace{0.05cm} \\  
$b =(a_{\rm p}/{R}_{\star}) \cos{ i_{\mathrm{p}} }$  & &  0.70$^{+0.05}_{-0.06}$ \vspace{0.05cm} \\  
$i_{\mathrm{p}}$ (deg)  & & 86.80$\pm$0.38 \vspace{0.05cm} \\  
$T_{14}$ [h]  &  & 1.43$^{+0.07}_{-0.05}$ \vspace{0.05cm} \\  
$T_{12}$ [min]  &  & 5.0$^{+0.8}_{-0.6}$  \vspace{0.05cm} \\  
$R_{\mathrm{p}}$ [${R}_\oplus$]  & & 2.21$\pm$0.13 \vspace{0.05cm} \\  
$a_{\mathrm{p}}$ [AU]  &  & 0.0368$\pm$0.0023 \vspace{0.05cm} \\ 
$T_{\mathrm{eq}}$ [K]\,$^{(b)}$  & &  910$\pm$30  \vspace{0.05cm} \\

\noalign{\smallskip}
\hline
\end{tabular}

\tablefoot{
$^{(a)}$ Central time of transit ($t_0$) units are BJD\,$-$\,2\,457\,000.
$^{(b)}$ Equilibrium temperatures were calculated assuming zero Bond albedo.}
\end{table}

\begin{figure}[h!]
    \centering
    \includegraphics[width=\hsize]{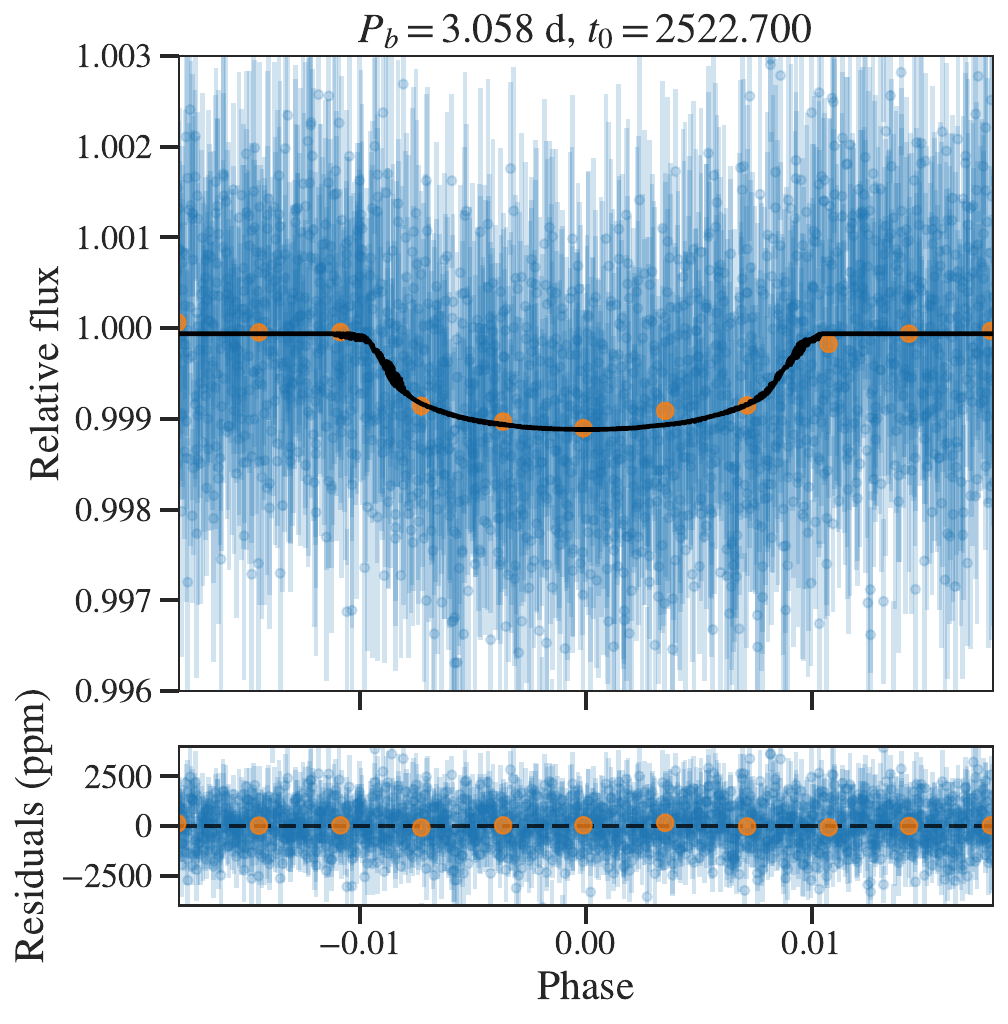}
    \caption{\label{Fig: TOI-1683 phase folded}
    TOI-1683 TESS photometry (blue dots with error bars) phase-folded to the period $P$ and central time of transit $t_0$ (shown above the panel, $t_0$ units are BJD\,$-$\,2\,457\,000) derived from the \texttt{juliet} fit. The black line is the best transit model for TOI-1683\,b. The orange points show binned photometry for visualisation.
    }
\end{figure}

\begin{figure*}[h!]
    \centering
    \includegraphics[width=1\linewidth]{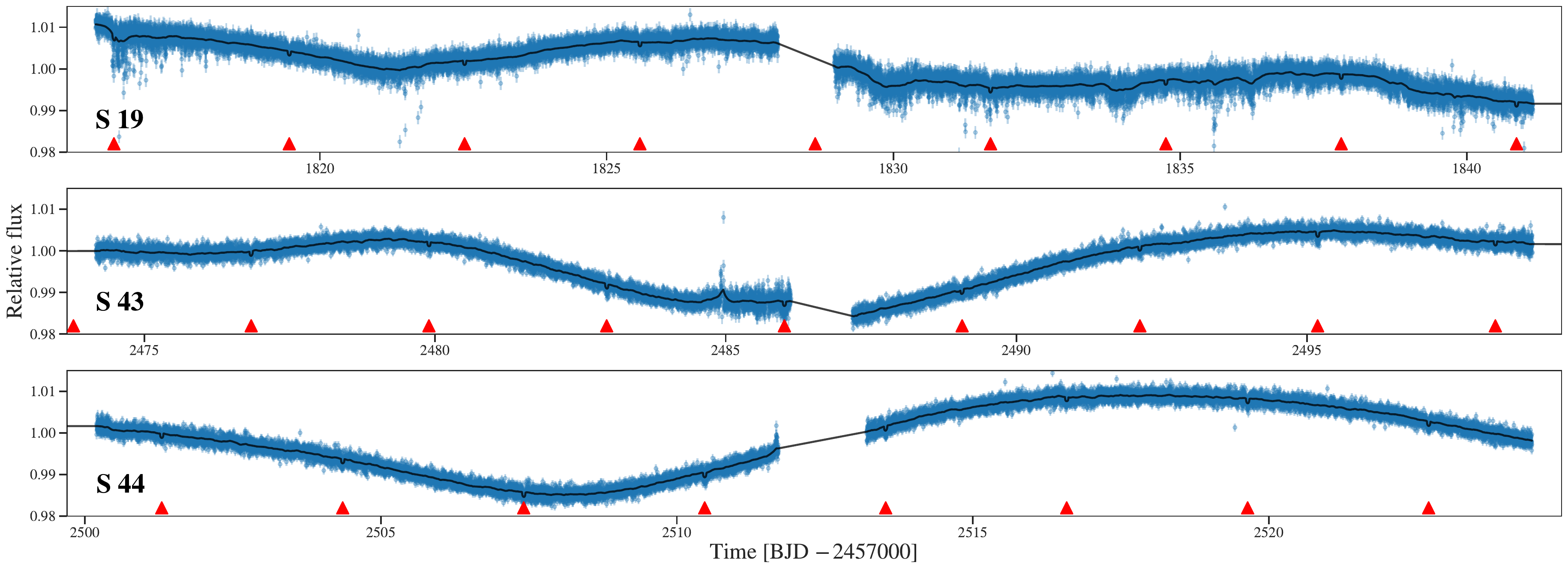}
    \caption{\label{Fig: TOI-1683 TESS fit}
    TOI-1683 2 min cadence TESS photometry from Sectors\,19, 43, and 44 along with the transit plus GP model. The upward-pointing red triangles mark the transit times for the TOI-1683\,b.
    }
\end{figure*}

\begin{figure}[h!]
\includegraphics[width=1\linewidth]{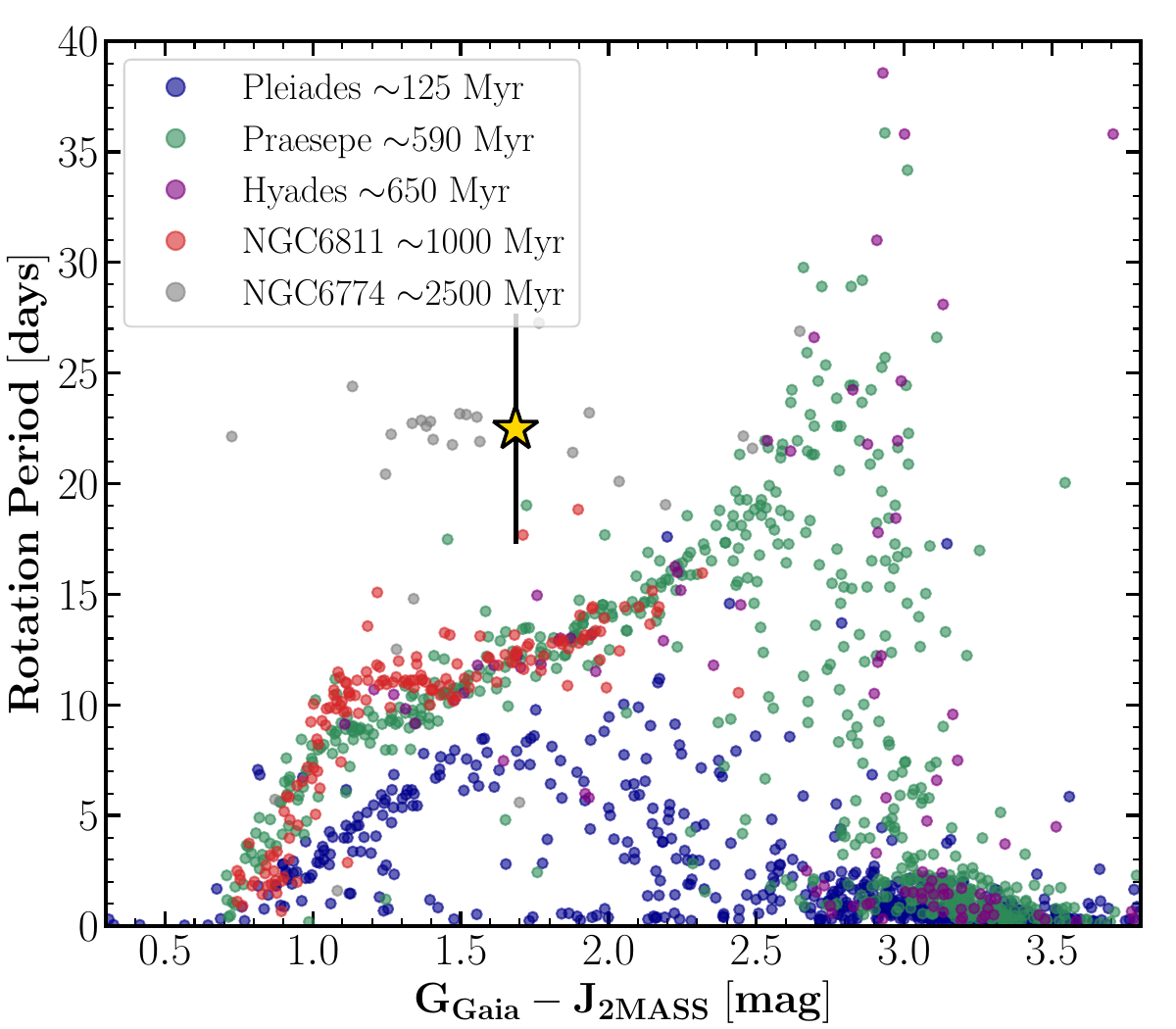}
\caption{Rotation period distribution as a function of colour $G$--$J$ for the Pleiades ($\sim$125\,Myr; \citealp{rebull2016}), Praesepe ($\sim$590\,Myr; \citealp{douglas2017}), Hyades ($\sim$650\,Myr; \citealp{douglas2019}), NGC 6811 ($\sim$1000\,Myr; \citealp{curtis2019}), and NGC 6774 ($\sim$2500\,Myr; \citealp{gruner2020}) clusters. The gold star represents TOI-1683. 
\label{fig:Prot}}
\end{figure}

\begin{figure}[h!]
\includegraphics[width=1\linewidth]{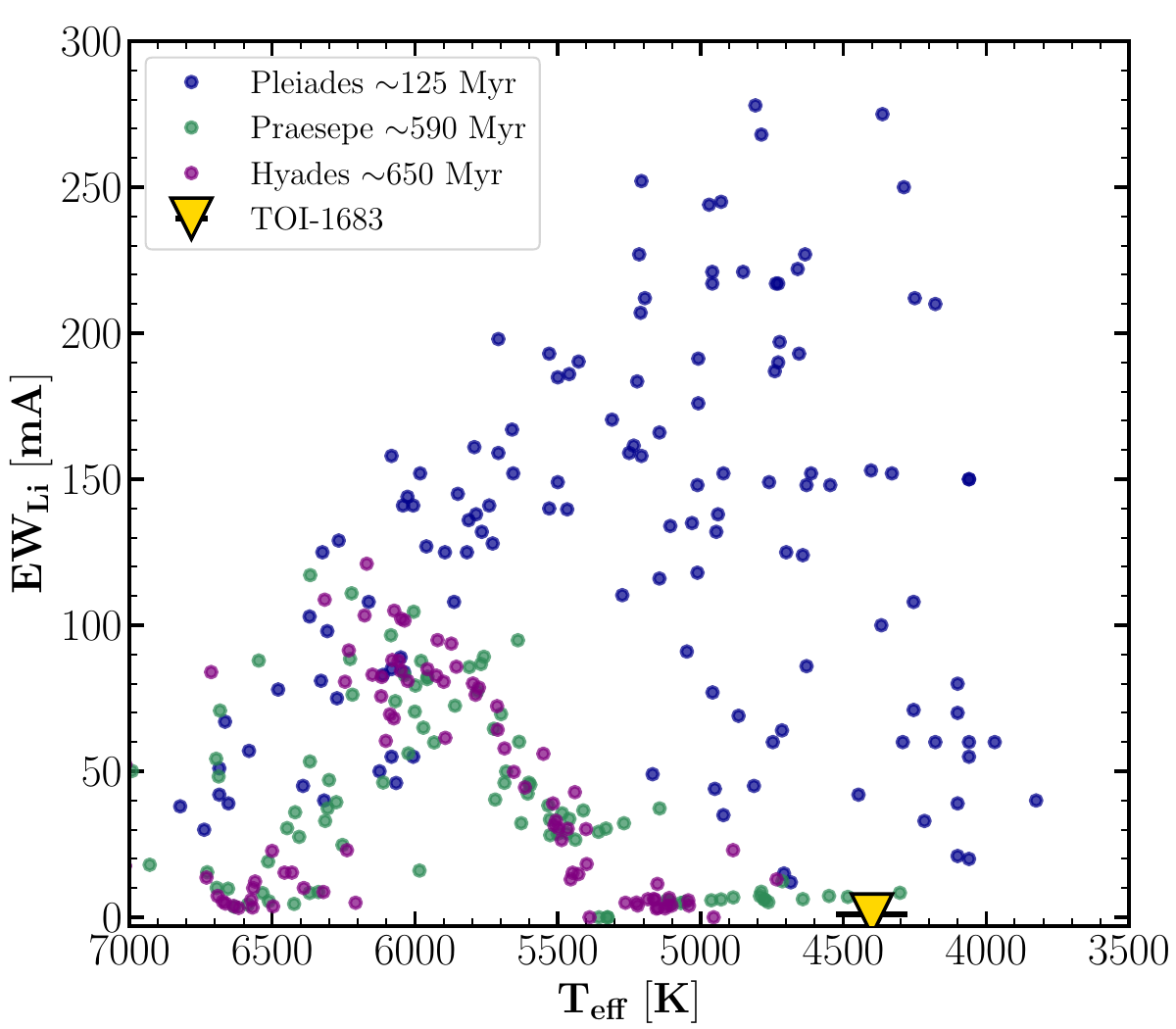}
\caption{Equivalent width distribution of \ion{Li}{i} as a function of the effective temperature for the Pleiades ($\sim$125\,Myr; \citealp{bouvier18}), Praesepe ($\sim$590\,Myr), and Hyades ($\sim$650\,Myr; \citealp{cummings17}). The gold triangle represents TOI-1683.
\label{fig:EW_Li}}
\end{figure}

\begin{figure}[h!]
\includegraphics[width=1\linewidth]{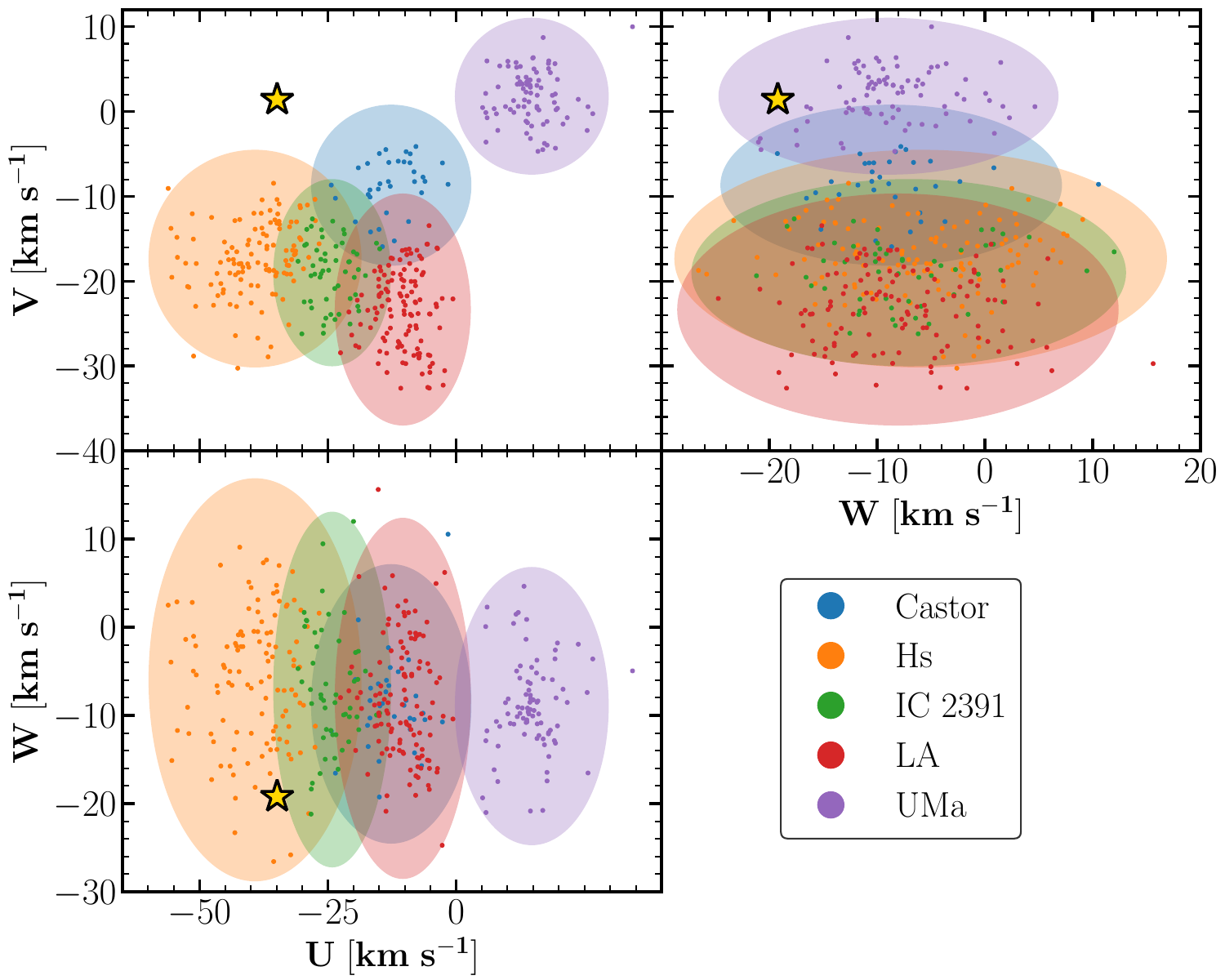}
\caption{$UVW$ velocity diagram for TOI-1683 (gold star). The members of the Castor moving group, Hyades supercluster (Hs), IC 2391 supercluster, Local Association (LA), and Ursa Major group (UMa) from \cite{montes2001b} are included. The ellipses represent the 3$\sigma$ values of the $UVW$ for each Young Moving Group.  
\label{fig:UVW}}
\end{figure}

We analysed TOI-1683 TESS data from Sectors\,19, 43, and 44. Because our purpose is to model the stellar rotation, we used the \texttt{celerite} GP quasi-periodic kernel. We adopted the stellar parameters used in \citet{Zhang_young_planets} for a proper comparison. The fitted parameters with their prior and posterior values, and the derived parameters for the planet candidate TOI-1683\,b are shown in Table\,\ref{table - TOI-1683 juliet priors and posteriors}. The TESS data along with the best transiting and GP models is shown in Fig.\,\ref{Fig: TOI-1683 TESS fit} and TOI-1683\,b phase folded transit is shown in Fig.\,\ref{Fig: TOI-1683 phase folded}.

We computed the GLS periodogram \citep[generalised Lomb-Scargle]{GLS} in the light curve of the three available TESS Sectors. We derived a stellar rotation period of 22.5\,$\pm$\,5.2\,days, which is consistent with the GP$_{\rm P_{rot}}$ value from the photometric fit (GP$_{\rm P_{rot}}$\,20\,$\pm$\,1\,d).
We estimated TOI-1683's age using the gyrochronology methods from \citet[Eq.\,12, 13, 14 and parameters from Table\,10]{Mamajek_2008}, \citet[Eq.\,1]{Schlaufman_2010}, and \citet[python package \texttt{gyro-interp}\footnote{\url{https://gyro-interp.readthedocs.io/en/latest/index.html}}]{Bouma23_gyrointerp}, and the derived ages are 1450$^{+700}_{-550}$\,Myr, 2300$^{+1300}_{-900}$\,Myr, and 2600\,$\pm$\,1200\,Myr, respectively. All the three methods are consistent with an age older than 1\,Gyr.
Figure\,\ref{fig:Prot} shows the distribution of rotation periods as a function of the $G-J$ colour for various young clusters. From this qualitative analysis, it is evident that TOI-1683 appears to be older than the NGC 6811 cluster (which has an age $\sim$1\,Gyr), and it is positioned above the sequence of NGC 6774 ($\sim$2.5 Gyr).

On the other hand, the presence of atmospheric absorption lines, such as \ion{Li}{I} at 6709.61\,\AA, can serve as a valuable age indicator for the stars. We searched for the presence of \ion{Li}{I} in the co-added spectrum generated by the \texttt{serval} from the CARMENES spectra. However, no clear \ion{Li}{I} feature was detected in the spectrum. Therefore, we set an upper limit for the \ion{Li}{I} equivalent width of 1\,m$\AA$ at 3$\sigma$. As Figure\,\ref{fig:EW_Li} shows, this upper limit suggests that the star is older than the Hyades or Praesepe clusters, which have ages between 590--650\,Myr.  Lastly, to investigate the kinematic properties of TOI-1683 and to determine if it shares any characteristics with known clusters, moving groups, or associations, we calculated the $UVW$ galactocentric space velocities using the astrometry and systemic velocity data from $Gaia$. Figure\,\ref{fig:UVW} places TOI-1683 outside of any shown young moving group or cluster. Additionally, we conducted a search in the literature of cluster catalogs, and TOI-1683 was not found to be associated with any of these known stellar groups. These results suggest that it is not part of any well-established stellar association or cluster. Based on our analysis, we adopt an age older than 1\,Gyr for TOI-1683 and, therefore, it is not a young object in terms of planet formation and evolution.

The age resulting of combining the three gyrochronology methods is 2000$^{+1300}_{-900}$\,Myr, which is the age used in this work.
However, \citet{Zhang_young_planets} claim that TOI-1683 is a young star ($<$1\,Gyr) with an age of 500\,$\pm$\,150\,Myr based on its gyrochronology analysis. The main difference with our analyses is that they derived a rotation period of 11.3\,$\pm$\,1.5 days, which is consistent with the half of the rotation period that we derived. Looking at a single TESS sector it is possible to confuse the stellar rotation period with one of its harmonics when it is of the order of the duration of the TESS sector ($\sim$28 days). This could be the case of TOI-1683 Sector\,19. However, the consecutive Sectors\,43 and 44 make it clear that the stellar rotation period of TOI-1683 is greater than 11.3 days.

\clearpage

\section{TOI-2018\,b extra material}
\label{App: TOI-2018 material}

\begin{figure}[h!]
    \centering
    \includegraphics[width=\hsize]{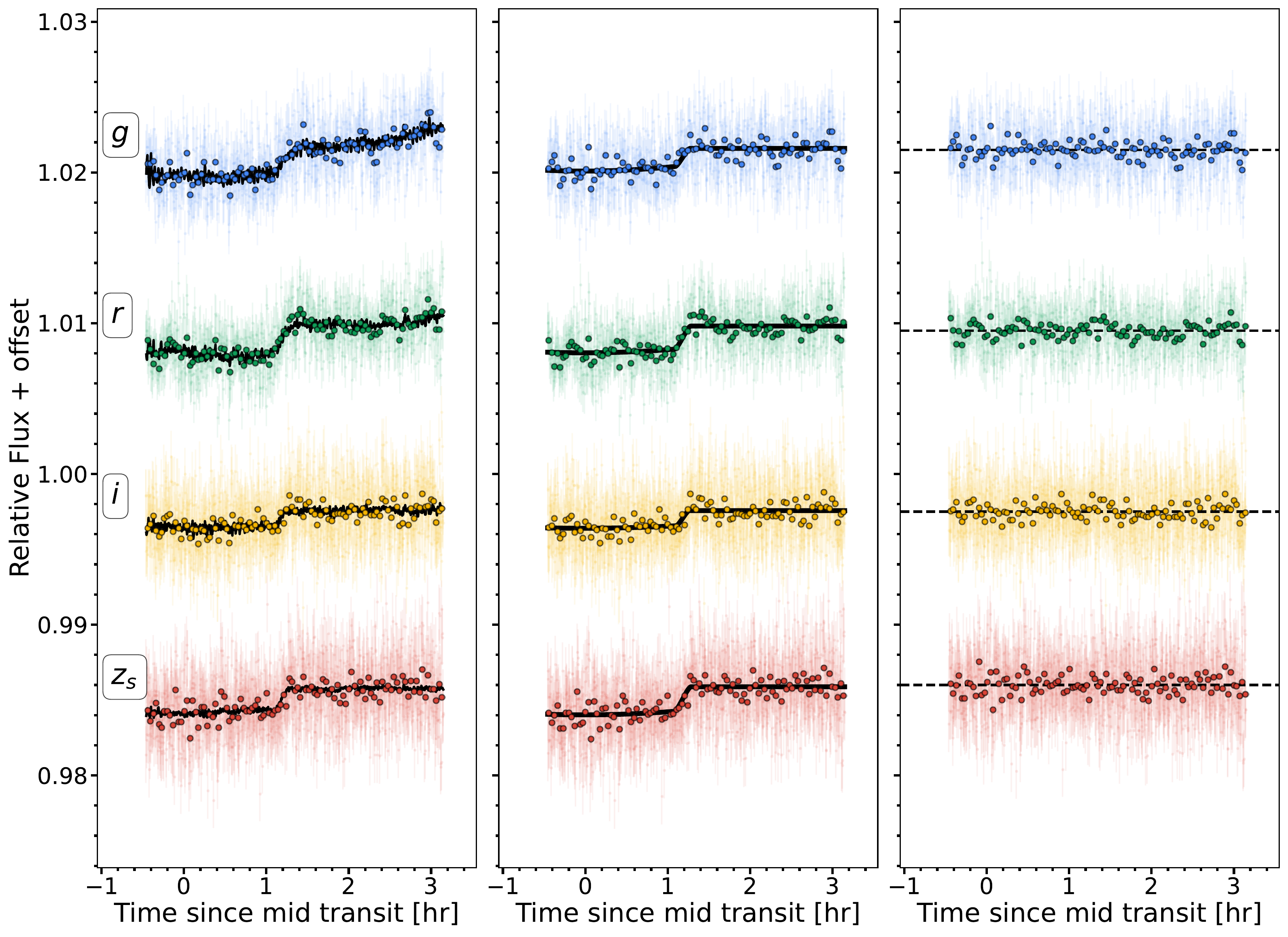}
    \caption{MuSCAT2 ground-based transit observation of TOI-2018\,b. Raw light curve (left panel), detrended light curve (centre panel), and residuals (right panel) for the Sloan $g$ (blue), $r$ (green), $i$ (yellow), $z_s$ (red) filters. The points show the individual observations, the circles represent binned data points, and the best-fit model is shown by the black line.}
    \label{Fig: MUSCAT2 TOI-2018b}
\end{figure}

\begin{table}[]
\caption[width=\textwidth]{
\label{table - TOI-2018 He priors and posteriors N1}
Prior and posterior distributions from the nested sampling fitting for TOI-2018\,b \ion{He}{I} signal from the first night. Prior label $\mathcal{U}$ represents uniform distribution.
}
\centering

\begin{tabular}{lcc}

\hline \hline 
\noalign{\smallskip} 

Parameter & Prior & Posterior \vspace{0.05cm}\\
\hline
\noalign{\smallskip}

Absorption [\%] & $\mathcal{U}(-3, 3)$ & $-$1.10$^{+0.26}_{-0.30}$  \vspace{0.05cm} \\ 
$\lambda_0$ [\AA] & $\mathcal{U}(10831,10834)$ & 10833.61$\pm$0.07 \\ 
$\sigma$ [\AA] &  $\mathcal{U}(0,1.5)$ & 0.24$^{+0.07}_{-0.06}$  \vspace{0.05cm} \\ 

$\Delta$v [km\,s$^{-1}$] & -- & 8.4$\pm$1.9 \\
FWHM [\AA] & -- &  0.57$^{+0.17}_{-0.15}$ \vspace{0.05cm} \\
EW [m\AA] &  -- & 6.6$^{+1.6}_{-1.5}$  \vspace{0.05cm} \\ 

\noalign{\smallskip}
\hline
\end{tabular}

\end{table}

\begin{table}[]
\caption[width=\textwidth]{
\label{table - TOI-2018 He priors and posteriors N2}
Prior and posterior distributions from the nested sampling fitting for TOI-2018\,b \ion{He}{I} signal from the second night. Prior label $\mathcal{U}$ represents uniform distribution.
}
\centering

\begin{tabular}{lcc}

\hline \hline 
\noalign{\smallskip} 

Parameter & Prior & Posterior \vspace{0.05cm}\\
\hline
\noalign{\smallskip}

Absorption [\%] & $\mathcal{U}(-3, 3)$ & $-$1.15$^{+0.33}_{-0.50}$  \vspace{0.05cm} \\ 
$\lambda_0$ [\AA] & $\mathcal{U}(10831,10834)$ & 10833.21$^{+0.14}_{-0.20}$ \\ 
$\sigma$ [\AA] &  $\mathcal{U}(0,1.5)$ & 0.50$^{+0.30}_{-0.20}$  \vspace{0.05cm} \\ 

$\Delta$v [km\,s$^{-1}$] & -- & $-$0.2$^{+4.0}_{-5.5}$ \\
FWHM [\AA] & -- &  1.20$^{+0.65}_{-0.45}$ \vspace{0.05cm} \\
EW [m\AA] &  -- & 14.7$^{+4.9}_{-4.5}$  \vspace{0.05cm} \\ 

\noalign{\smallskip}
\hline
\end{tabular}

\end{table}

\begin{figure}[]
    \centering
     \begin{subfigure}{0.49\textwidth}
         \centering
         \includegraphics[width=\textwidth]{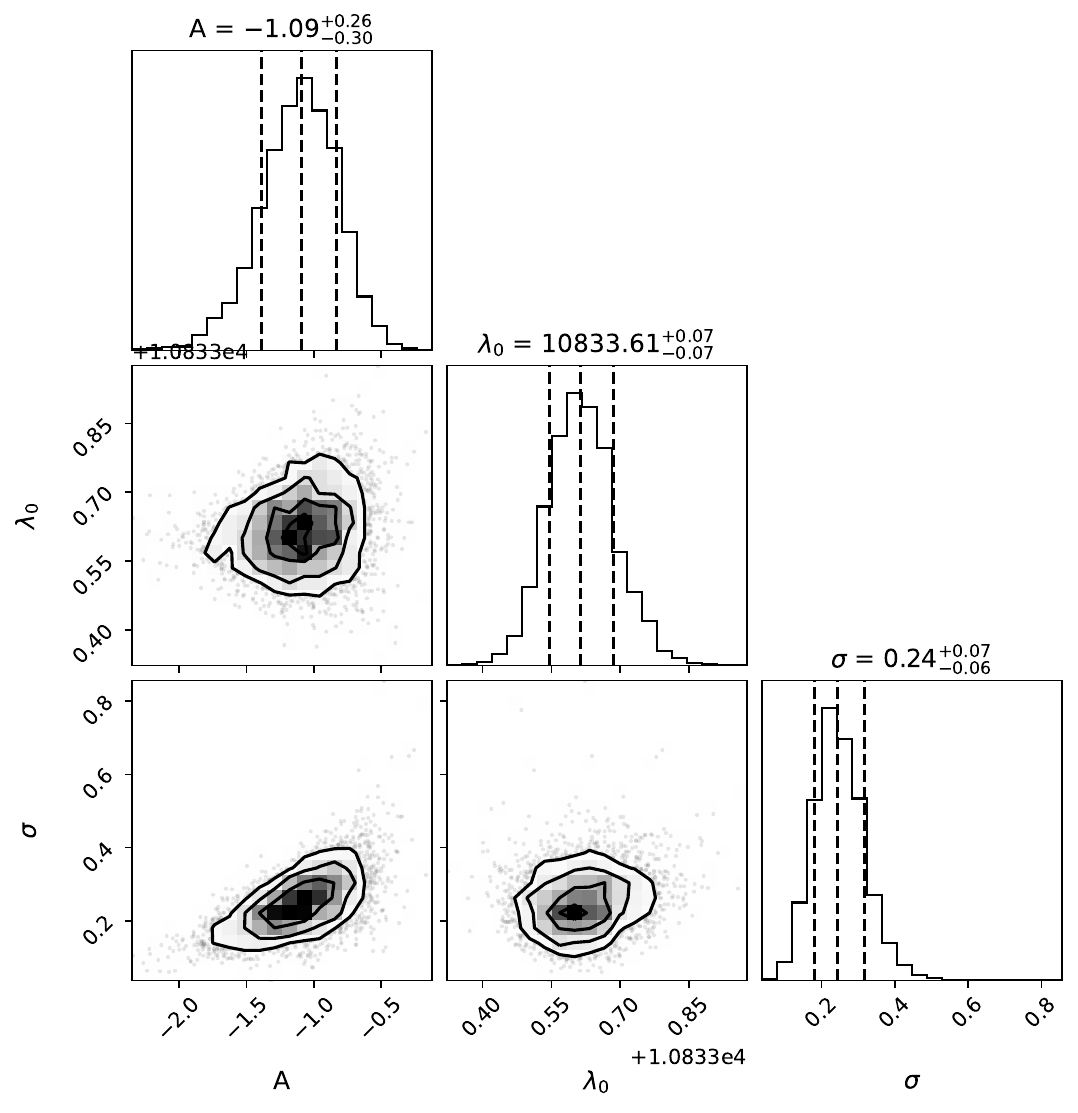}
     \end{subfigure}
     \hfill
     \begin{subfigure}{0.49\textwidth}
         \centering
         \includegraphics[width=\textwidth]{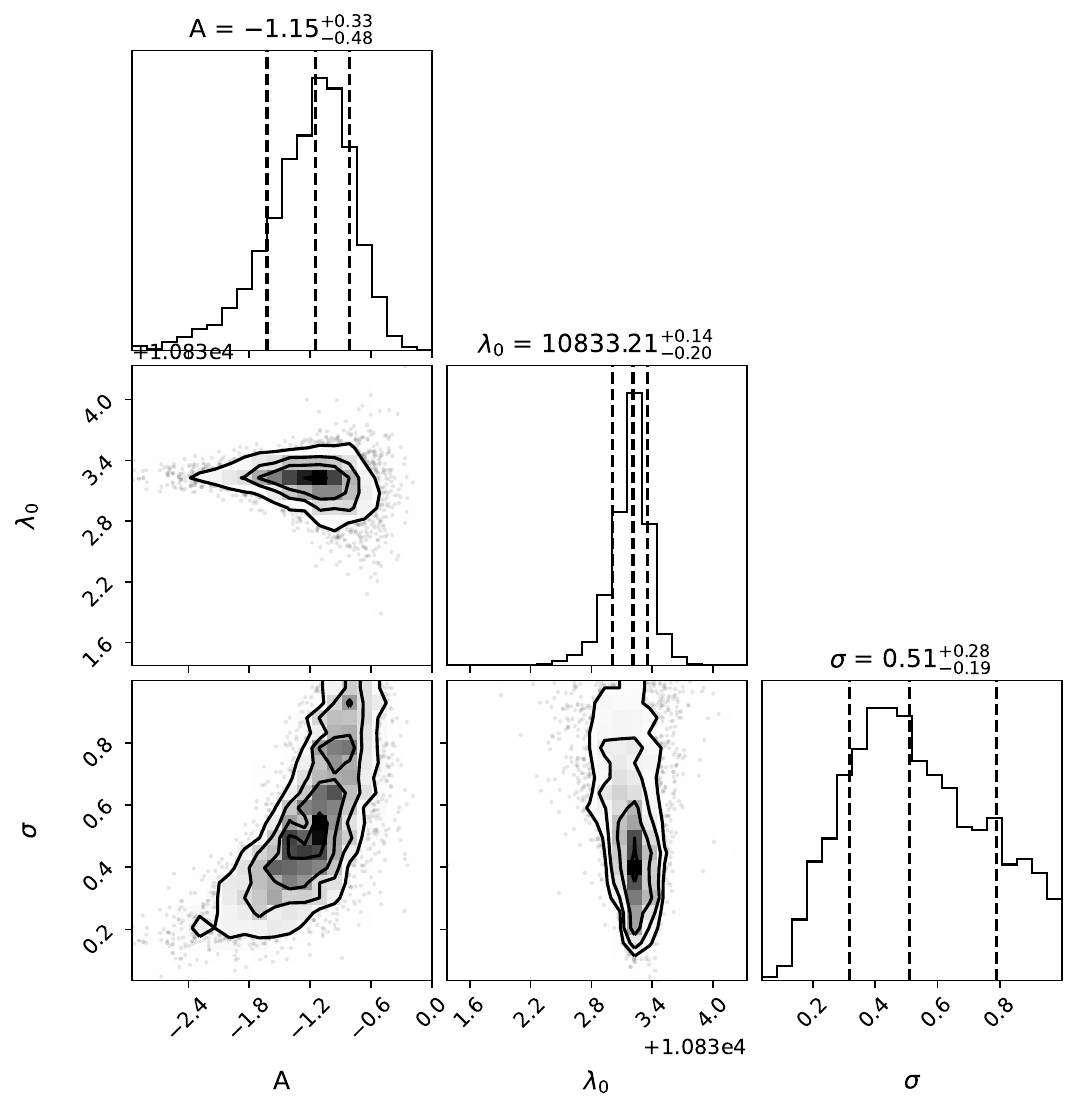}
     \end{subfigure}
    \caption{\label{Fig: TOI-2018 nested He N1 N2}
    Corner plot for the nested sampling posterior distribution of TOI-2018\,b \ion{He}{i} signals for first (top) and second (bottom) nights.
    }
\end{figure}

\begin{figure}[]
    \centering
    \includegraphics[width=\hsize]{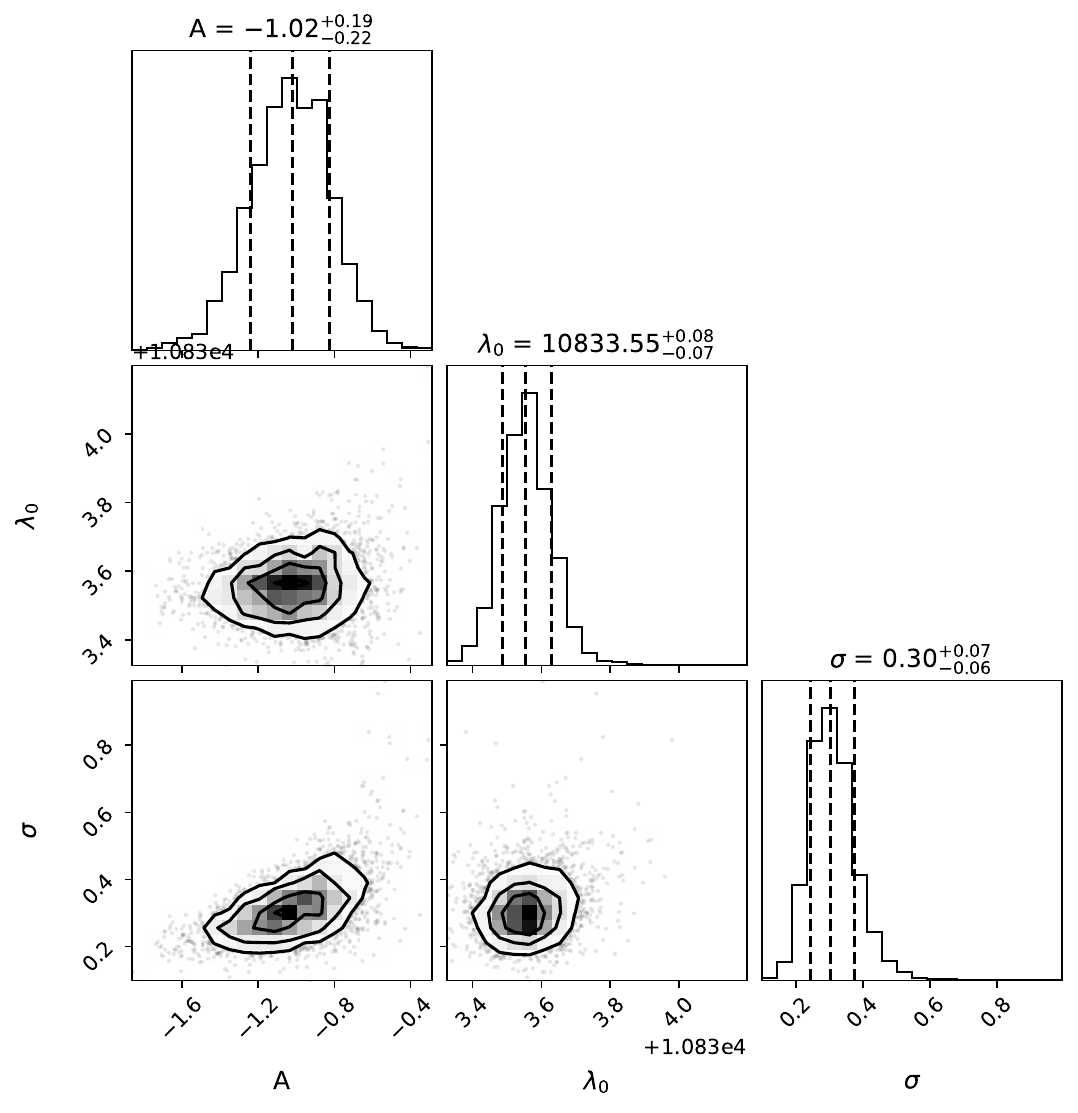}
    \caption{\label{Fig: TOI-2018 nested He}
    Corner plot for the nested sampling posterior distribution of TOI-2018\,b \ion{He}{i} signals from the combined nights.
    }
\end{figure}

\begin{table}[]
\caption[width=\textwidth]{
\label{table - TOI-2018 He priors and posteriors}
Prior and posterior distributions from the nested sampling fitting for TOI-2018\,b \ion{He}{I} signal from the combined nights. Prior label $\mathcal{U}$ represents uniform distribution.
}
\centering

\begin{tabular}{lcc}

\hline \hline 
\noalign{\smallskip} 

Parameter & Prior & Posterior \vspace{0.05cm}\\
\hline
\noalign{\smallskip}

Absorption [\%] & $\mathcal{U}(-3, 3)$ & $-$1.02$^{+0.19}_{-0.22}$  \vspace{0.05cm} \\ 
$\lambda_0$ [\AA] & $\mathcal{U}(10831,10834)$ & 10833.55$^{+0.08}_{-0.07}$ \\ 
$\sigma$ [\AA] &  $\mathcal{U}(0,1.5)$ & 0.30$^{+0.07}_{-0.06}$  \vspace{0.05cm} \\ 

$\Delta$v [km\,s$^{-1}$] & -- & 6.8$^{+2.1}_{-1.9}$ \\
FWHM [\AA] & -- &  0.71$^{+0.17}_{-0.14}$ \vspace{0.05cm} \\
EW [m\AA] &  -- & 7.8$\pm$1.5  \vspace{0.05cm} \\

\noalign{\smallskip}
\hline
\end{tabular}

\end{table}

\clearpage

\section{\ion{He}{i} database information}
\label{App: He data base}

As \citet{TOI-2134b_Zhang} remarked, the compilation of consistent parameters and r tudy the \ion{He}{i} triplet is a challenging work. We thank the effort done by \citet{TOI-2134b_Zhang} to construct Table\,3 therein. We based our \ion{He}{i} database (Table\,\ref{table - DATABASE}) on the compilations by \citet{WASP-52_He_NIRSPEC}, \citet{HAT-P-32b_Ha_He_Czesla2022}, \citet{WASP-80_He_Fossati, Fossati2023_He_table}, \citet{GIANO_He_survey}, \citet{ TOI-2134b_Zhang}, and J.\,Sanz-Forcada, priv. comm., which we further complemented.

Because the \ion{He}{i} results are given in terms of the equivalent width (EW) or absorption peak of the line, we firstly needed a relation between both magnitudes to convert one to the other.
The EW for a spectral line is the width of a rectangle, with a height of the continuum level, such that the area of the spectral line is equal to the area of the rectangle. Usually, the planetary spectral lines are fitted using a Gaussian profile as G($\lambda$)\,=\,$A \exp{(-(\lambda - \lambda_0)^2/(2\sigma^2))}$ where $A$ is the amplitude of the signal, $\lambda_0$ is the central position of the line, and $\sigma$ is the standard deviation. Thus, the area under the curve described by G($\lambda$) can be computed as
\begin{equation}
\int_{-\infty}^{+\infty} G(\lambda) ~ d\lambda = A \sqrt{2 \pi \sigma^2}
\label{eq: AREA GAUSSIAN}
\end{equation}
If the spectra are normalised to 1, we can calculate EW\,=\,$\mid$\,$A$\,$\mid$\,$\sqrt{2 \pi \sigma^2}$, where $A$ is in units of normalised flux. EW is in the units of $\sigma$.
Then, we calculated from the papers where EW and absorption values are reported (e.g. \citealp{GIANO_Helium, Nuria_2021_WASP-76b, Zhang_TOI560b, Zhang_young_planets, Orell2022, Orell2023}) a master-$\sigma$ as the mean value. We obtained master-$\sigma$\,=\,0.43\,m$\AA$.

\onecolumn{

\tiny{

\begin{landscape}
\centering
\begin{longtable}{ c c c c c c c c c c c c c c }

\caption{Exoplanet database of \ion{He}{i} observation by means of high-resolution spectroscopy from this work and the literature (possibly incomplete). We complemented the planets inspected for \ion{He}{i} with information about H$\alpha$ and/or Ly$\alpha$ studies (H flag column). \fxuv\ wavelength range is $\lambda$\,=\,5--504\,$\AA$.
\label{table - DATABASE}
}\\ %

\noalign{\smallskip} \noalign{\smallskip}
\hline\hline
\noalign{\smallskip} 

 Planet & P & $R_{\rm p}$ & $M_{\rm p}$ & $T_{\rm eq}$ & Age & $M_{\star}$ & $R_{\star}$ & $T_{\rm eff}$ & H & \ion{He}{i} & $D_{\rm He}$ & $EW_{\rm He}$ & $F_{\rm XUV}$\,$^{b}$ \\ 
 & [d] & [R$_{\oplus}$] & [M$_{\oplus}$] & [K] & [Gyr] & [M$_{\sun}$] & [R$_{\sun}$] & [K] & flag\,$^{a}$ & flag\,$^{a}$ & [\%] & [m$\AA$] & [W$\,$m$^{-2}$] \\

 \noalign{\smallskip}
 \hline
 \noalign{\smallskip}		
 \endfirsthead

\caption{continued.}\\ %
\hline\hline
\noalign{\smallskip}		

 Planet & P & $R_{\rm p}$ & $M_{\rm p}$ & $T_{\rm eq}$ & Age & $M_{\star}$ & $R_{\star}$ & $T_{\rm eff}$ & H & \ion{He}{i} & $D_{\rm He}$ & $EW_{\rm He}$ & $F_{\rm XUV}$\,$^{b}$ \\ 
 & [d] & [R$_{\oplus}$] & [M$_{\oplus}$] & [K] & [Gyr] & [M$_{\sun}$] & [R$_{\sun}$] & [K] & flag\,$^{a}$ & flag\,$^{a}$ & [\%] & [m$\AA$] & [W$\,$m$^{-2}$] \\ 

 \noalign{\smallskip}
 \hline
 \noalign{\smallskip}
 \endhead

 \noalign{\smallskip}
 \hline
 \endfoot

\hline \noalign{\smallskip} 
 55 Cnc e & 0.7365 & 1.875$\pm$0.029 & 7.99$^{+0.32}_{-0.33}$ & 1958$\pm$15 & 10.2$\pm$2.5 & 0.905$\pm$0.015 & 0.943$\pm$0.01 & 5172$\pm$18 & -- & ND &  $<$0.025 &  $<$0.27 & 4.57\,$^{ 1}$ \\ 
 \noalign{\smallskip} \hline \noalign{\smallskip} 
 AU Mic b & 8.463 & 4.16$\pm$0.18 & 11.7$\pm$5.0 & 593$\pm$21 & 0.022$\pm$0.003 & 0.5$\pm$0.03 & 0.75$\pm$0.03 & 3700$\pm$100 & -- & ND &  $<$0.34 &  $<$3.7 & 22.9\,$^{ 1}$ \\ 
 \noalign{\smallskip} \hline \noalign{\smallskip} 
 GJ436 b & 2.6441 & 4.19$\pm$0.1 & 23.14$\pm$0.76 & 686$\pm$10 & 6.0$^{+4.0}_{-3.0}$ & 0.445$\pm$0.044 & 0.449$\pm$0.019 & 3479$\pm$60 & Y\,$^{\rm L}$ & ND &  $<$0.41 &  $<$1.45 & 0.145\,$^{ 1}$ \\ 
 \noalign{\smallskip} \hline \noalign{\smallskip} 
 GJ806 b & 0.9263 & 1.331$\pm$0.023 & 1.9$\pm$0.17 & 940$\pm$10 & 4.0$^{+4.0}_{-3.0}$ & 0.413$\pm$0.011 & 0.4144$\pm$0.0038 & 3600$\pm$16 & ND\,$^{\rm H}$ & ND &  $<$0.7 &  $<$7.4 & 1.62\,$^{ 1}$ \\ 
 \noalign{\smallskip} \hline \noalign{\smallskip} 
 GJ1214 b & 1.5804 & 2.742$^{+0.05}_{-0.053}$ & 8.17$\pm$0.43 & 596$\pm$19 & 3.0$^{+7.0}_{-0.0}$ & 0.178$\pm$0.01 & 0.215$\pm$0.008 & 3250$\pm$100 & -- & ND &  $<$2.1$^{+0.45}_{-0.5}$ &  $<$28.9$^{+9.4}_{-8.5}$ & 0.2951\,$^{ 1}$ \\ 
 \noalign{\smallskip} \hline \noalign{\smallskip} 
 GJ3470 b & 3.3366 & 4.04$\pm$0.11 & 11.44$\pm$0.64 & 733$\pm$23 & 1.65$\pm$1.4 & 0.476$\pm$0.019 & 0.474$\pm$0.014 & 3725$\pm$54 & Y\,$^{\rm L}$ & Y & 1.5$\pm$0.3 & 20.72$\pm$1.3 & 1.44\,$^{ 2}$ \\ 
 \noalign{\smallskip} \hline \noalign{\smallskip} 
 GJ9827 b & 1.209 & 1.529$\pm$0.058 & 4.87$\pm$0.37 & 1114$^{+44}_{-26}$ & 10.0$^{+3.0}_{-5.0}$ & 0.593$\pm$0.018 & 0.579$\pm$0.018 & 4294$\pm$52 & ND\,$^{\rm B}$ & ND &  $<$0.21 &  $<$1.86 & 0.72\,$^{ 1}$ \\ 
 \noalign{\smallskip} \hline \noalign{\smallskip} 
 GJ9827 d & 6.2018 & 1.955$\pm$0.075 & 3.42$\pm$0.62 & 646$^{+26}_{-15}$ & 10.0$^{+3.0}_{-5.0}$ & 0.593$\pm$0.018 & 0.579$\pm$0.018 & 4294$\pm$52 & ND\,$^{\rm B}$ & ND &  $<$0.3 &  $<$1.24 & 0.081\,$^{ 1}$ \\ 
 \noalign{\smallskip} \hline \noalign{\smallskip} 
 HD63433 b & 7.1079 & 2.141$^{+0.09}_{-0.067}$ & 5.5$\pm$2.3 & 967$^{+19}_{-16}$ & 0.414$\pm$0.023 & 0.956$\pm$0.022 & 0.934$\pm$0.029 & 5553$\pm$56 & ND\,$^{\rm B}$ & ND &  $<$0.34 &  $<$2.0 & 10.3\,$^{ 3}$ \\ 
 \noalign{\smallskip} \hline \noalign{\smallskip} 
 HD63433 c & 20.5438 & 2.69$^{+0.11}_{-0.09}$ & 15.5$^{+3.9}_{-3.8}$ & 679$^{+13}_{-11}$ & 0.414$\pm$0.023 & 0.956$\pm$0.022 & 0.934$\pm$0.029 & 5553$\pm$56 & Y\,$^{\rm L}$ & ND &  $<$0.4 &  $<$4.2 & 2.5\,$^{ 3}$ \\ 
 \noalign{\smallskip} \hline \noalign{\smallskip} 
 HD73583 b & 6.398 & 2.79$\pm$0.1 & 10.2$^{+3.4}_{-3.1}$ & 721$\pm$21 & 0.49$\pm$0.19 & 0.71$\pm$0.02 & 0.66$\pm$0.02 & 4511$\pm$110 & ND\,$^{\rm H}$ & Y & 0.72$\pm$0.08 & 8.6$\pm$0.6 & 3.1\,$^{ 3}$ \\ 
 \noalign{\smallskip} \hline \noalign{\smallskip} 
 HD73583 c & 18.8797 & 2.39$\pm$0.1 & 9.7$^{+1.8}_{-1.7}$ & 503$\pm$15 & 0.49$\pm$0.19 & 0.71$\pm$0.02 & 0.66$\pm$0.02 & 4511$\pm$110 & ND\,$^{\rm H}$ & ND &  $<$0.5 &  $<$5.3 & 0.7\,$^{ 3}$ \\ 
 \noalign{\smallskip} \hline \noalign{\smallskip} 
 HD89345 b & 11.8144 & 6.86$\pm$0.14 & 35.6$\pm$3.2 & 1053$\pm$14 & 9.4$^{+0.4}_{-1.3}$ & 1.12$\pm$0.04 & 1.657$\pm$0.02 & 5499$\pm$73 & -- & NC &  $<$0.7 &  $<$7.4 & 0.244\,$^{ 4}$ \\ 
 \noalign{\smallskip} \hline \noalign{\smallskip} 
 HD97658 b & 9.4893 & 2.247$^{+0.098}_{-0.095}$ & 7.55$^{+0.83}_{-0.79}$ & 757$^{+12}_{-13}$ & 6.0$\pm$1.0 & 0.77$\pm$0.05 & 0.741$\pm$0.024 & 5170$\pm$50 & -- & ND &  $<$0.21 &  $<$0.9 & 0.128\,$^{ 1}$ \\ 
 \noalign{\smallskip} \hline \noalign{\smallskip} 
 HD189733 b & 2.2186 & 12.76$\pm$0.3 & 357.0$\pm$14.0 & 1200$\pm$20 & 6.8$\pm$5.2 & 0.806$\pm$0.048 & 0.756$\pm$0.018 & 5040$\pm$50 & Y\,$^{\rm B}$ & Y & 0.75$\pm$0.03 & 12.76$\pm$0.4 & 16.75\,$^{ 5}$ \\ 
 \noalign{\smallskip} \hline \noalign{\smallskip} 
 HD209458 b & 3.5247 & 15.23$^{+0.16}_{-0.21}$ & 217.7$^{+4.8}_{-4.4}$ & 1449$\pm$12 & 4.0$\pm$2.0 & 1.119$\pm$0.033 & 1.155$\pm$0.015 & 6065$\pm$50 & Y\,$^{\rm B}$ & Y & 0.91$\pm$0.1 & 5.252$\pm$0.5 & 1.004\,$^{ 6}$ \\ 
 \noalign{\smallskip} \hline \noalign{\smallskip} 
 HD235088 b & 7.4341 & 2.045$\pm$0.075 & 7.0$\pm$2.0 & 805$^{+13}_{-12}$ & 0.65$^{+0.15}_{-0.05}$ & 0.843$\pm$0.05 & 0.789$\pm$0.021 & 5037$\pm$14 & -- & Y & 0.91$^{+0.11}_{-0.1}$ & 9.5$^{+1.1}_{-1.0}$ & 1.854\,$^{ 7}$ \\ 
 \noalign{\smallskip} \hline \noalign{\smallskip} 
 HAT-P-3 b & 2.8997 & 10.2$\pm$0.4 & 189.1$\pm$7.6 & 1170$\pm$17 & 2.9$^{+4.9}_{-2.7}$ & 0.925$\pm$0.046 & 0.85$\pm$0.021 & 5190$\pm$80 & -- & ND &  $<$1.9 &  $<$20.2 & 7.968\,$^{ 4}$ \\ 
 \noalign{\smallskip} \hline \noalign{\smallskip} 
 HAT-P-11 b & 4.8878 & 4.36$\pm$0.06 & 27.7$\pm$3.1 & 832$\pm$10 & 6.5$^{+5.9}_{-4.1}$ & 0.802$\pm$0.028 & 0.683$\pm$0.009 & 4780$\pm$50 & Y\,$^{\rm L}$ & Y & 1.08$\pm$0.05 & 12.4$\pm$2.4 & 2.109\,$^{ 8}$ \\ 
 \noalign{\smallskip} \hline \noalign{\smallskip} 
 HAT-P-18 b & 5.508 & 11.15$\pm$0.58 & 62.6$\pm$4.1 & 852$\pm$28 & 12.4$^{+1.4}_{-6.4}$ & 0.77$\pm$0.031 & 0.749$\pm$0.037 & 4803$\pm$80 & -- & Y & 0.7$\pm$0.16 & 29.21$\pm$1.0 & 0.7\,$^{ 9}$ \\ 
 \noalign{\smallskip} \hline \noalign{\smallskip} 
 HAT-P-32 b & 2.15 & 20.05$\pm$0.28 & 185.9$\pm$9.9 & 1786$\pm$26 & 2.7$\pm$0.8 & 1.16$\pm$0.041 & 1.219$\pm$0.016 & 6269$\pm$64 & Y\,$^{\rm H}$ & Y & 5.3$\pm$0.1 & 114.0$\pm$4.0 & 163.0\,$^{ 10}$ \\ 
 \noalign{\smallskip} \hline \noalign{\smallskip} 
 HAT-P-33 b & 3.4745 & 18.9$\pm$0.5 & 229.0$^{+41.0}_{-38.0}$ & 1782$\pm$28 & 2.3$\pm$0.3 & 1.42$\pm$0.15 & 1.91$\pm$0.26 & 6460$^{+300}_{-290}$ & -- & ND &  $<$1.4 &  $<$14.9 & 6.195\,$^{ 4}$ \\ 
 \noalign{\smallskip} \hline \noalign{\smallskip} 
 HAT-P-49 b & 2.6916 & 15.84$^{+1.4}_{-0.86}$ & 550.0$\pm$65.0 & 2131$^{+69}_{-42}$ & 1.5$\pm$0.2 & 1.543$\pm$0.051 & 1.833$\pm$0.138 & 6820$\pm$52 & -- & ND &  $<$0.6 &  $<$6.4 & 14.51\,$^{ 4}$ \\ 
 \noalign{\smallskip} \hline \noalign{\smallskip} 
 HAT-P-57 b & 2.4653 & 15.84$\pm$0.61 & $<$590 & 2200$\pm$26 & 1.0$^{+0.67}_{-0.51}$ & 1.47$\pm$0.12 & 1.5$\pm$0.05 & 7500$\pm$250 & NC\,$^{\rm H}$ & NC &  $<$1.0 &  $<$10.6 &  --  \\ 
\noalign{\smallskip} \hline \noalign{\smallskip} 
HAT-P-67 b & 4.8101 & 23.37$^{+1.1}_{-0.8}$ & 108.0$^{+79.0}_{-60.0}$ & 1903$\pm$35 & 1.24$^{+0.24}_{-0.22}$ & 1.642$\pm$0.1 & 2.65$\pm$0.12 & 6406$^{+65}_{-61}$ & NC\,$^{\rm L}$ & Y & 10.0$\pm$0.1 & 140.0$\pm$10.0 &  --  \\
 \noalign{\smallskip} \hline \noalign{\smallskip} 
 HAT-P-70 b & 2.7443 & 21.0$^{+1.7}_{-1.1}$ & $<$2155 & 2562$\pm$50 & 0.60$^{+0.38}_{-0.2}$ & 1.890$\pm$0.013 & 1.86$\pm$0.12 & 8450$\pm$540 & Y\,$^{\rm H}$ & -- & -- & -- & -- \\
 \noalign{\smallskip} \hline \noalign{\smallskip}
 LTT9779 b & 0.7920 & 4.72$\pm$0.23 & 28.32$^{+0.78}_{-0.81}$ & 1978$\pm$19 & 2$^{+1.3}_{-0.9}$ & 1.02$\pm$0.03 & 0.949$\pm$0.006 & 5443$\pm$14 & -- & ND &  $<$0.2 &  $<$3.79 &  4.8\,$^{ 15}$  \\
 \noalign{\smallskip} \hline \noalign{\smallskip}
 MASCARA-2 b & 3.4741 & 20.51$\pm$0.78 & $<$1075 & 2262$\pm$73 & 0.2$^{+0.1}_{-0.05}$ & 1.89$\pm$0.06 & 1.6$\pm$0.06 & 8980$^{+90}_{-130}$ & Y\,$^{\rm H}$ & ND &  $<$0.5 &  $<$5.3 &  --  \\
 \noalign{\smallskip} \hline \noalign{\smallskip} 
 TOI-1136 d & 12.5194 & 4.626$^{+0.076}_{-0.072}$ & 8.0$^{+2.4}_{-1.9}$ & 840$\pm$12 & 0.7$\pm$0.15 & 1.022$\pm$0.027 & 0.968$\pm$0.036 & 5770$\pm$50 & Y\,$^{\rm H}$ & ND &  $<$0.5 &  $<$5.4 &  --  \\ 
 \noalign{\smallskip} \hline \noalign{\smallskip} 
 TOI-1235 b & 3.4447 & 1.694$^{+0.08}_{-0.077}$ & 5.9$^{+0.62}_{-0.61}$ & 775$^{+14}_{-13}$ & 5.0$^{+5.0}_{-4.4}$ & 0.63$\pm$0.024 & 0.619$\pm$0.019 & 3997$\pm$51 & -- & ND &  $<$0.09 &  $<$1.44 &  --  \\ 
 \noalign{\smallskip} \hline \noalign{\smallskip} 
 TOI-1268 b & 8.1577 & 9.1$\pm$0.6 & 96.0$\pm$13.0 & 919$\pm$30 & 0.245$\pm$0.14 & 0.96$\pm$0.04 & 0.92$\pm$0.06 & 5300$\pm$100 & -- & Y & 1.97$^{+0.16}_{-0.15}$ & 19.1$^{+1.9}_{-1.8}$ & 7.2\,$^{ 1}$ \\ 
 \noalign{\smallskip} \hline \noalign{\smallskip} 
 TOI-1431 b & 2.6502 & 16.7$\pm$0.56 & 992.0$\pm$57.0 & 2370$\pm$70 & 0.29$^{+0.32}_{-0.19}$ & 1.895$\pm$0.1 & 1.923$\pm$0.068 & 7690$^{+400}_{-250}$ & ND\,$^{\rm H}$ & ND &  $<$0.4 &  $<$4.2 &  --  \\ 
 \noalign{\smallskip} \hline \noalign{\smallskip} 
 TOI-1683 b & 3.0575 & 2.3$\pm$0.3 & 8.0$\pm$3.0 & 927$\pm$35 & 2.0$^{+1.3}_{-0.9}$ & 0.69$\pm$0.09 & 0.636$\pm$0.03 & 4539$\pm$100 & -- & Y & 0.84$\pm$0.17 & 8.5$\pm$1.6 & 7.4\,$^{ 3}$ \\ 
 \noalign{\smallskip} \hline \noalign{\smallskip} 
 TOI-1728 b & 3.4914 & 4.62$\pm$0.09 & 26.8$^{+5.4}_{-5.1}$ & 757$^{+16}_{-15}$ & 7.1$\pm$4.6 & 0.646$\pm$0.023 & 0.6243$\pm$0.01 & 3980$^{+31}_{-32}$ & -- & ND &  $<$1.1 &  $<$11.7 &  --  \\ 
 \noalign{\smallskip} \hline \noalign{\smallskip} 
 TOI-1807 b & 0.5494 & 1.37$\pm$0.09 & 2.57$\pm$0.5 & 2100$^{+39}_{-40}$ & 0.3$\pm$0.08 & 0.75$\pm$0.025 & 0.68$\pm$0.015 & 4757$^{+51}_{-50}$ & ND\,$^{\rm H}$ & ND &  $<$0.38 &  $<$4.0 & 3.05\,$^{ 1}$ \\ 
 \noalign{\smallskip} \hline \noalign{\smallskip} 
 TOI-2018 b & 7.4356 & 2.268$\pm$0.069 & 9.2$\pm$2.1 & 642$\pm$12 & 2.4$^{+0.2}_{-0.8}$ & 0.57$\pm$0.02 & 0.62$\pm$0.01 & 4174$^{+34}_{-42}$ & NC\,$^{\rm H}$ & Y & 1.02$^{+0.19}_{-0.22}$ & 7.8$\pm$1.5 & 1.56\,$^{ 1}$ \\ 
 \noalign{\smallskip} \hline \noalign{\smallskip} 
 TOI-2046 b & 1.4972 & 16.1$\pm$1.2 & 731.0$\pm$89.0 & 2000$\pm$55 & 0.4$^{+0.22}_{-0.3}$ & 1.13$\pm$0.19 & 1.21$\pm$0.07 & 6200$\pm$100 & NC\,$^{\rm H}$ & NC &  $<$2.9 &  $<$30.5 & 22.45\,$^{ 1}$ \\ 
 \noalign{\smallskip} \hline \noalign{\smallskip} 
 TOI-2048 b & 13.7905 & 2.6$\pm$0.2 & 9.0$\pm$3.0 & 675$^{+22}_{-16}$ & 0.3$\pm$0.05 & 0.83$\pm$0.03 & 0.79$\pm$0.04 & 5185$\pm$60 & NC\,$^{\rm H}$ & NC &  $<$1.0 &  $<$10.2 &  --  \\ 
 \noalign{\smallskip} \hline \noalign{\smallskip} 
 TOI-2076 b & 10.3557 & 2.52$\pm$0.056 & 9.0$\pm$3.0 & 800$\pm$13 & 0.34$\pm$0.08 & 0.824$\pm$0.036 & 0.77$\pm$0.006 & 5200$\pm$70 & ND\,$^{\rm H}$ & Y & 1.01$\pm$0.05 & 10.0$\pm$0.7 & 6.7\,$^{ 3}$ \\ 
 \noalign{\smallskip} \hline \noalign{\smallskip} 
 TOI-2134 b & 9.2292 & 2.69$\pm$0.16 & 9.13$^{+0.78}_{-0.76}$ & 666$\pm$8 & 3.8$^{+5.5}_{-2.7}$ & 0.744$\pm$0.027 & 0.709$\pm$0.017 & 4580$\pm$54 & -- & Y & 0.38$\pm$0.05 & 3.3$\pm$0.3 & 0.46\,$^{ 3}$ \\ 
 \noalign{\smallskip} \hline \noalign{\smallskip} 
 TOI-2136 b & 7.8519 & 2.2$\pm$0.07 & 4.7$^{+3.1}_{-2.6}$ & 378$\pm$13 & 4.6$\pm$1.0 & 0.3272$\pm$0.0082 & 0.344$\pm$0.0099 & 3373$\pm$108 & -- & ND &  $<$1.44 &  $<$7.8 &  --  \\ 
 \noalign{\smallskip} \hline \noalign{\smallskip} 
 TOI-3757 b & 3.4388 & 12.0$^{+0.4}_{-0.5}$ & 85.3$^{+8.8}_{-8.7}$ & 759$\pm$13 & 7.1$\pm$4.5 & 0.64$\pm$0.02 & 0.62$\pm$0.01 & 3913$\pm$56 & -- & NC &  $<$6.9 &  $<$73.0 &  --  \\ 
 \noalign{\smallskip} \hline \noalign{\smallskip} 
 TRAPPIST-1 b & 1.5109 & 1.086$\pm$0.035 & 0.85$\pm$0.72 & 400$\pm$7 & 7.6$\pm$2.2 & 0.0802$\pm$0.0073 & 0.117$\pm$0.0036 & 2550$\pm$50 & -- & ND &  $<$0.33 &  $<$3.467 & 0.7244\,$^{ 1}$ \\ 
 \noalign{\smallskip} \hline \noalign{\smallskip} 
 TRAPPIST-1 e & 6.0996 & 0.918$\pm$0.039 & 0.62$\pm$0.58 & 251$\pm$4 & 7.6$\pm$2.2 & 0.0802$\pm$0.0073 & 0.117$\pm$0.0036 & 2550$\pm$50 & -- & ND &  $<$1.07 &  $<$10.458 & 0.112\,$^{ 1}$ \\ 
 \noalign{\smallskip} \hline \noalign{\smallskip} 
 TRAPPIST-1 f & 9.2067 & 1.045$\pm$0.038 & 0.68$\pm$0.18 & 219$\pm$4 & 7.6$\pm$2.2 & 0.0802$\pm$0.0073 & 0.117$\pm$0.0036 & 2550$\pm$50 & -- & ND &  $<$0.38 &  $<$4.143 & 0.0645\,$^{ 1}$ \\ 
 \noalign{\smallskip} \hline \noalign{\smallskip} 
 NGTS-5 b & 3.357 & 12.73$\pm$0.26 & 73.0$\pm$12.0 & 952$\pm$24 & 5.0$^{+8.0}_{-3.5}$ & 0.661$\pm$0.065 & 0.739$\pm$0.014 & 4987$\pm$41 & -- & ND &  $<$1.02$^{+0.48}_{-0.46}$ &  $<$10.8$^{+5.1}_{-4.9}$ & 3.2\,$^{ 9}$ \\ 
 \noalign{\smallskip} \hline \noalign{\smallskip} 
 K2-25 b & 3.4846 & 3.43$\pm$0.12 & 28.5$^{+8.5}_{-8.3}$ & 345$^{+8}_{-7}$ & 0.725$\pm$0.075 & 0.294$\pm$0.021 & 0.295$\pm$0.02 & 3180$\pm$60 & ND\,$^{\rm L}$ & ND &  $<$1.7 &  $<$17.0 &  --  \\ 
 \noalign{\smallskip} \hline \noalign{\smallskip} 
 K2-77 b & 8.1998 & 2.3$\pm$0.16 & $<$600 & 770$\pm$30 & 0.12$^{+0.78}_{-0.02}$ & 0.8$\pm$0.12 & 0.76$\pm$0.03 & 4970$\pm$45 & NC\,$^{\rm H}$ & NC &  $<$2.7 &  $<$28.0 & 13.45\,$^{ 1}$ \\ 
 \noalign{\smallskip} \hline \noalign{\smallskip} 
 K2-100 b & 1.6739 & 3.88$\pm$0.16 & 21.8$\pm$6.2 & 1841$\pm$41 & 0.7$\pm$0.1 & 1.15$\pm$0.05 & 1.24$\pm$0.05 & 5945$\pm$110 & ND\,$^{\rm H}$ & ND &  $<$1.3 &  $<$5.7 & 141.253\,$^{ 1}$ \\ 
 \noalign{\smallskip} \hline \noalign{\smallskip} 
 K2-105 b & 8.267 & 3.59$^{+0.11}_{-0.07}$ & 30.0$\pm$19.0 & 814$\pm$12 & 5.0$^{+8.0}_{-4.4}$ & 1.05$\pm$0.02 & 0.97$\pm$0.01 & 5636$^{+49}_{-52}$ & -- & NC &  $<$2.33 &  $<$24.7 & 14.69\,$^{4 }$ \\ 
 \noalign{\smallskip} \hline \noalign{\smallskip} 
 K2-136 c & 17.307 & 3.0$\pm$0.13 & 18.1$^{+1.8}_{-1.9}$ & 425$^{+10}_{-33}$ & 0.65$\pm$0.07 & 0.742$\pm$0.02 & 0.677$\pm$0.027 & 4500$\pm$50 & -- & NC &  $<$2.3 &  $<$25.0 & 0.5888\,$^{ 1}$ \\ 
 \noalign{\smallskip} \hline \noalign{\smallskip} 
 KELT-9 b & 1.4811 & 21.701$\pm$0.053 & 920.0$\pm$110.0 & 3900$^{+182}_{-174}$ & 0.45$^{+0.14}_{-0.13}$ & 2.32$\pm$0.16 & 2.418$\pm$0.058 & 9600$\pm$400 & Y\,$^{\rm H}$ & ND &  $<$0.33 &  $<$1.17 & 0.15\,$^{ 5}$ \\ 
 \noalign{\smallskip} \hline \noalign{\smallskip} 
 Kepler-25 c & 12.7204 & 5.217$^{+0.07}_{-0.065}$ & 15.2$^{+1.3}_{-1.6}$ & 992$\pm$8 & 2.75$\pm$0.3 & 1.26$\pm$0.03 & 1.34$\pm$0.01 & 6354$\pm$27 & -- & NC &  $<$1.86 &  $<$19.8 & 1.019\,$^{ 4}$ \\ 
 \noalign{\smallskip} \hline \noalign{\smallskip} 
 Kepler-68 b & 5.3988 & 2.31$^{+0.06}_{-0.09}$ & 8.3$^{+2.2}_{-2.4}$ & 1280$\pm$90 & 6.3$\pm$1.7 & 1.079$\pm$0.051 & 1.243$\pm$0.019 & 5793$\pm$74 & -- & NC &  $<$0.72 &  $<$7.6 & 1.176\,$^{ 4}$ \\ 
 \noalign{\smallskip} \hline \noalign{\smallskip} 
 V1298Tau c & 8.2489 & 5.2$\pm$0.39 & $<$76 & 968$\pm$31 & 0.023$\pm$0.004 & 1.17$\pm$0.06 & 1.278$\pm$0.07 & 5050$\pm$100 & NC\,$^{\rm H}$ & NC &  $<$3.75 &  $<$95.8 & 151.356\,$^{ 1}$ \\ 
 \noalign{\smallskip} \hline \noalign{\smallskip} 
 V1298Tau b & 24.1399 & 9.77$\pm$0.65 & 203.0$\pm$60.0 & 677$\pm$22 & 0.023$\pm$0.004 & 1.17$\pm$0.06 & 1.278$\pm$0.07 & 5050$\pm$100 & -- & NC &  $<$1.7 &  $<$19.0 & 87.09\,$^{ 1}$ \\ 
 \noalign{\smallskip} \hline \noalign{\smallskip} 
 WASP-11 b & 3.7225 & 11.1$\pm$0.25 & 156.0$\pm$8.0 & 992$\pm$14 & 7.6$^{+6.0}_{-3.5}$ & 0.81$\pm$0.04 & 0.772$\pm$0.015 & 4900$\pm$65 & -- & NC &  $<$1.56 &  $<$16.6 & 1.9\,$^{ 11}$ \\ 
 \noalign{\smallskip} \hline \noalign{\smallskip} 
 WASP-12 b & 1.0914 & 21.71$\pm$0.63 & 466.0$\pm$25.0 & 2592$\pm$57 & 2.0$^{+0.8}_{-1.0}$ & 1.434$\pm$0.11 & 1.657$\pm$0.046 & 6300$^{+200}_{-100}$ & Y\,$^{\rm H}$ & ND &  $<$0.5 &  $<$5.3 & 3.18\,$^{ 12}$ \\
 \noalign{\smallskip} \hline \noalign{\smallskip} 
 WASP-39 b & 4.0553 & 14.34$\pm$0.45 & 89.0$\pm$10.0 & 1166$\pm$14 & 8.5$^{+4.0}_{-3.4}$ & 0.913$\pm$0.047 & 0.939$\pm$0.022 & 5485$\pm$50 & -- & NC &  $<$2.04 &  $<$21.7 & 1.2\,$^{ 11}$ \\ 
 \noalign{\smallskip} \hline \noalign{\smallskip} 
 WASP-47 d & 9.0305 & 3.567$\pm$0.045 & 14.2$\pm$1.3 & 919$\pm$13 & 6.5$^{+2.6}_{-1.2}$ & 1.04$\pm$0.031 & 1.137$\pm$0.013 & 5552$\pm$75 & -- & NC &  $<$3.29 &  $<$34.9 & 0.577\,$^{ 4}$ \\ 
 \noalign{\smallskip} \hline \noalign{\smallskip} 
 WASP-48 b & 2.1436 & 18.7$\pm$1.1 & 328.0$^{+15.0}_{-14.0}$ & 2035$\pm$52 & 7.9$^{+2.0}_{-1.6}$ & 1.19$\pm$0.05 & 1.75$\pm$0.09 & 6000$\pm$150 & -- & ND &  $<$0.25$\pm$0.21 &  $<$2.7$\pm$2.2 &  --  \\ 
 \noalign{\smallskip} \hline \noalign{\smallskip} 
 WASP-52 b & 1.7498 & 14.04$\pm$0.3 & 137.9$\pm$7.6 & 1315$\pm$26 & 0.4$^{+0.3}_{-0.2}$ & 0.804$\pm$0.05 & 0.786$\pm$0.016 & 5000$\pm$100 & Y\,$^{\rm H}$ & Y & 3.44$\pm$0.31 & 39.583$\pm$1.4 & 24.8\,$^{ 13}$ \\ 
 \noalign{\smallskip} \hline \noalign{\smallskip} 
 WASP-69 b & 3.8681 & 11.85$\pm$0.53 & 82.6$\pm$5.4 & 963$\pm$18 & 2.0$\pm$0.5 & 0.826$\pm$0.029 & 0.813$\pm$0.028 & 4700$\pm$50 & NC\,$^{\rm H}$ & Y & 3.59$\pm$0.19 & 28.31$\pm$0.9 & 4.17\,$^{ 5}$ \\ 
 \noalign{\smallskip} \hline \noalign{\smallskip} 
 WASP-76 b & 1.8099 & 20.78$^{+0.86}_{-0.85}$ & 284.1$^{+4.4}_{-4.1}$ & 2160$\pm$40 & 5.3$^{+6.1}_{-2.9}$ & 1.458$\pm$0.021 & 1.756$\pm$0.071 & 6329$\pm$25 & Y\,$^{\rm H}$ & ND &  $<$0.88 &  $<$1.7 & 112.2018\,$^{ 1}$ \\ 
 \noalign{\smallskip} \hline \noalign{\smallskip} 
 WASP-77 b & 1.36 & 13.79$^{+0.35}_{-0.33}$ & 530.0$^{+22.0}_{-20.0}$ & 1715$^{+26}_{-25}$ & 6.2$^{+4.0}_{-3.5}$ & 0.903$\pm$0.06 & 0.91$\pm$0.025 & 5617$\pm$72 & ND\,$^{\rm H}$ & ND &  $<$0.8 &  $<$8.4 & 13.182\,$^{ 1}$ \\ 
 \noalign{\smallskip} \hline \noalign{\smallskip} 
 WASP-80 b & 3.0679 & 11.2$^{+0.35}_{-0.34}$ & 171.0$\pm$11.0 & 825$\pm$19 & $\lesssim$0.2 & 0.577$\pm$0.05 & 0.586$\pm$0.018 & 4143$^{+92}_{-94}$ & -- & ND &  $<$0.85 &  $<$2.48 & 1.6595\,$^{ 1}$ \\ 
 \noalign{\smallskip} \hline \noalign{\smallskip} 
 WASP-107 b & 5.7215 & 10.54$\pm$0.22 & 35.0$\pm$3.2 & 770$\pm$60 & 8.3$\pm$4.3 & 0.69$\pm$0.05 & 0.66$\pm$0.02 & 4430$\pm$120 & -- & Y & 7.26$\pm$0.24 & 87.152$\pm$7.6 & 2.664\,$^{ 1}$ \\ 
 \noalign{\smallskip} \hline \noalign{\smallskip} 
 WASP-127 b & 4.1781 & 15.36$\pm$0.45 & 57.2$\pm$6.4 & 1400$\pm$24 & 11.41$\pm$1.8 & 1.08$\pm$0.03 & 1.39$\pm$0.03 & 5620$\pm$85 & ND\,$^{\rm H}$ & ND &  $<$0.48 &  $<$6.8 & 0.058\,$^{ 14}$ \\ 
 \noalign{\smallskip} \hline \noalign{\smallskip} 
 WASP-177 b & 3.0717 & 17.7$^{+7.4}_{-4.0}$ & 161.0$\pm$12.0 & 1142$\pm$32 & 9.7$\pm$3.9 & 0.876$\pm$0.038 & 0.885$\pm$0.046 & 5017$\pm$70 & -- & ND &  $<$1.28$\pm$0.30 &  $<$6.8$\pm$1.6 & 3.5\,$^{ 13
 }$ \\ 
 \noalign{\smallskip} \hline \noalign{\smallskip} 
 WASP-189 b & 2.724 & 18.15$\pm$0.24 & 632.0$^{+60.0}_{-44.0}$ & 2641$\pm$34 & 0.75$\pm$0.13 & 2.03$\pm$0.066 & 2.36$\pm$0.03 & 8000$\pm$80 & NC\,$^{\rm H}$ & ND &  $<$0.3 &  $<$3.2 &  --  \\ 
 
 \noalign{\smallskip} \hline

\end{longtable}

\tablefoot{ 
$^{a}$ Y: Detection, ND: Non Detection, NC: Non Conclusive. For H, we indicate if the detection/non-detection comes from H$\alpha$ (H), Ly$\alpha$ (L), or both (B).
$^{b}$ $F_{\rm XUV}$ from:
$^{1}$\,Sanz-Forcada in prep.,
$^{2}$\,\citet{He_GJ3470b_Enric2020}, 
$^{3}$\,\citet{TOI-2134b_Zhang}, 
$^{4}$\,\citet{GIANO_He_survey}, 
$^{5}$\,\citet{Nortmann_WASP-69_He}, 
$^{6}$\,\citet{HD209458_He_Alonso}, 
$^{7}$\,\citet{Orell2023}, 
$^{8}$\,\citet{HAT-P-11b_He_Allart}, 
$^{9}$\,\citet{Vissapragada2022_Neptune_Desert}, 
$^{10}$\,\citet{HAT-P-32b_Ha_He_Czesla2022}, 
$^{11}$\,\citet{Allart_He_survey}, 
$^{12}$\,\citet{WASP-12b_Czesla24}, 
$^{13}$\,\citet{WASP-52_He_NIRSPEC},
$^{14}$\,\citet{WASP-127b_He_DosSantos}, and 
$^{15}$\,\citet{LTT9779b_He_Vissapragada2024}
}

\tablebib{Data compilation from NASA Exoplanet Archive and ExoAtmospheres database. Some references in particular are as follows. 55 Cnc e: \cite{55Cnc_star, 55Cnce_He} ; AU Mic b: \cite{AUMIC_star, AUMic_masses, AU_Mic_Enric, AU_Mic_Hirano_He, AU_Mic_Lyalpha} ; GJ436 b: \cite{Nortmann_WASP-69_He, GJ436b_Ha, GJ436b_Lya, Ehrenreich2015_GJ436b_Lya}   ; GJ806 b: \cite{GJ806b_Enric}   ; GJ1214 b: \cite{Kasper_2020, Orell2022, Spake_He_GJ1214} ; GJ3470 b: \cite{Bourrier2018_Lya_GJ3470b, He_GJ3470b_Enric2020, GJ3470b_He_Ninan2020}   ; GJ9827 b: \cite{Krishnamurthy2023_TOI1235b_GJ9827bd, GJ9827b_d_Ilaria}   ; GJ9827 d: \cite{Krishnamurthy2023_TOI1235b_GJ9827bd, GJ9827b_d_Ilaria, Kasper_2020}  ; HD63433 b: \cite{HD63433_Mann_discovery, Manu_HD63433, HD63433_Zhang}   ; HD63433 c: \cite{HD63433_Mann_discovery, Manu_HD63433, HD63433_Zhang}   ; HD73583 b: \cite{Barragan_TOI560b, Zhang_TOI560b, Zhang_young_planets}   ; HD73583 c: \cite{Barragan_TOI560b} ; HD89345 b: \cite{GIANO_He_survey}   ; HD97658 b: \cite{Kasper_2020}   ; HD189733 b: \cite{HD189733b_Halpha, HD189733b_He, GIANO_Helium, HD189733b_He_variability_Zhang2022, Allart_He_survey, HD189733b_Lya}   ; HD209458 b: \cite{VidalMadjar2003_Lya, HD209458_He_Alonso}   ; HD235088 b: \cite{Orell2023}   ; HAT-P-3 b: \cite{GIANO_He_survey}   ; HAT-P-11 b: \cite{HAT-P-11b_He_Allart, HAT-P-11b_Lya, Allart_He_survey}   ; HAT-P-18 b: \cite{HAT-P-18b_He, HAT-P-18b_He_2}   ; HAT-P-32 b: \cite{HAT-P-32b_Ha_He_Czesla2022, HAT-P-32b_He_Tail}    ; HAT-P-33 b:  \cite{GIANO_He_survey}  ; HAT-P-49 b: \cite{GIANO_He_survey}   ; HAT-P-57 b: \cite{HAT-P-57_Hartmann, UHJ_Monika}   ; HAT-P-67 b: \cite{HAT-P-67b_Ha_He_carmenes, HAT-P-67b_He_toroide}   ; HAT-P-70 b: \cite{HAT-P-70_Halpha}  ; LTT9779 b: \cite{LTT9779b_discovery, LTT9779b_He_Vissapragada2024} ; MASCARA-2 b: \cite{MASCARA-2b_Lund, MASCARA-2b_Talens, MASCARA2_Nuria}   ; TOI-1136 d: \cite{TOI-1136_system}   ; TOI-1235 b: \cite{Krishnamurthy2023_TOI1235b_GJ9827bd}   ; TOI-1268 b: \cite{TOI-1268_Subjak2022}   ; TOI-1431 b: \cite{TOI-1431_2021, UHJ_Monika}   ; TOI-1683 b: \cite{Zhang_young_planets}   ; TOI-1728 b: \cite{TOI-1728b_planet_He}   ; TOI-1807 b: \cite{TOI-1807_Nardiello, TOI-1807_TOI-2076_Gaidos2022}   ; TOI-2018 b: \cite{TOI-2018_Dai23}   ; TOI-2046 b: \cite{TOI-2046b_Kabath}    ; TOI-2048 b: \cite{TOI2048b_Newton}   ; TOI-2076 b: \cite{TOI-2076_Osborn22, Zhang_young_planets, TOI-1807_TOI-2076_Gaidos2022}   ; TOI-2134 b: \cite{TOI-2134b_Zhang}   ; TOI-2136 b: \cite{TOI-2136b_planet_He}   ; TOI-3757 b: \cite{TOI-3757b_planet_He}   ; TRAPPIST-1 b: \cite{TRAPPIST-1_system, TRAPPISTb_e_f_He}   ; TRAPPIST-1 e: \cite{TRAPPIST-1_system, TRAPPISTb_e_f_He}   ; TRAPPIST-1 f: \cite{TRAPPIST-1_system, TRAPPISTb_e_f_He}   ; NGTS-5 b: \cite{Vissapragada2022_Neptune_Desert}   ; K2-25 b: \cite{K2-25_Discovery_Gaidos2016, K2-25_Gaidos, K2-25b_Lya}   ; K2-77 b: \cite{K2-77_Gaidos}   ; K2-100 b: \cite{K2-100_Stefansson_2018, K2-100_Barragan_2019, K2-100b_Gaidos_He}   ; K2-105 b: \cite{GIANO_He_survey}   ; K2-136 c: \cite{K2-136_system, K2-136c_Gaidos_He, K2-136_masses}   ; KELT-9 b: \cite{KELT-9_Ha, Nortmann_WASP-69_He}   ; Kepler-25 c: \cite{GIANO_He_survey}   ; Kepler-68 b: \cite{GIANO_He_survey}   ; V1298Tau c: \cite{V1298Tau_paper, V1298Tau_c_Alejandro, V1298Tau_c_Feinstein, V1298Tau_He_Vissapragada2021, V1298Tau_c_Halpha, V1298Tau_Gaidos}   ; V1298Tau b: \cite{V1298Tau_paper, V1298Tau_c_Alejandro, V1298Tau_He_Vissapragada2021, V1298Tau_Gaidos} ; WASP-11 b: \cite{Allart_He_survey}   ; WASP-12 b: \cite{WASP-12b_Ha, WASP-12b_He_HST, WASP-12b_Czesla24}  ; WASP-39 b: \cite{Allart_He_survey}   ; WASP-47 d: \cite{GIANO_He_survey}   ; WASP-48 b: \cite{WASP-49b_He}  ; WASP-52 b: \cite{WASP-52_Halpha_Chen, WASP-52_He_NIRSPEC, WASP-52_He_2020}   ; WASP-69 b: \cite{Nortmann_WASP-69_He}   ; WASP-76 b: \cite{WASP-76b_star, Nuria_2021_WASP-76b}  ; WASP-77 b: \cite{WASP-77b_star}, Khalafinejad et al. in prep. ; WASP-80 b: \cite{Salz2015_WASP80_age, WASP-80_He_Fossati}   ; WASP-107 b: \cite{Spake2018_WASP-107b_HST, WASP-107b_He_Allart2019, WASP-107b_He_Kirk2020, WASP-107b_He_Spake2021, WASP-107_star}   ; WASP-127 b: \cite{WASP-127_star, WASP-127b_He_DosSantos, WASP-127b_Ha_ESPRESSO, Allart_He_survey}   ; WASP-177 b: \cite{WASP-177_star, WASP-52_He_NIRSPEC, Vissapragada2022_Neptune_Desert}  ; WASP-189 b: \cite{WASP-189b_Anderson, WASP-189b_Lendl, UHJ_Monika}.}

\end{landscape}

}
}

\newpage


\end{appendix}

\end{document}